\documentclass[preprint]{aastex62}
\usepackage{mathrsfs}

\received{12-11-2018}
\revised{}
\accepted{03-29-2019}
\shorttitle{Do {\it Kepler} superflare stars really include slowly-rotating Sun-like stars ?}
\shortauthors{Notsu et al.}
\begin{document}

\title{Do {\it Kepler}  superflare stars really include slowly-rotating Sun-like stars ? \\
- Results using APO 3.5m telescope spectroscopic observations and {\it Gaia}-DR2 data - }

\correspondingauthor{Yuta Notsu}
\email{ynotsu@kwasan.kyoto-u.ac.jp}

\author[0000-0002-0412-0849]{Yuta Notsu}
\altaffiliation{JSPS Research Fellow DC1 (until March 31, 2019)}
\altaffiliation{JSPS Overseas Research Fellow (from April 1, 2019)}
\affil{Department of Astronomy, Kyoto University, Sakyo, Kyoto 606-8502, Japan}
\affil{Laboratory for Atmospheric and Space Physics, University of Colorado Boulder, 3665 Discovery Drive, Boulder, CO 80303, USA}

\author[0000-0003-0332-0811]{Hiroyuki Maehara}
\affil{Okayama Branch Office, Subaru Telescope, National Astronomical Observatory of Japan, NINS, Kamogata, Asakuchi, Okayama 719-0232, Japan}
\affil{Okayama Observatory, Kyoto University, Kamogata, Asakuchi, Okayama 719-0232, Japan}

\author{Satoshi Honda}
\affil{Nishi-Harima Astronomical Observatory, Center for Astronomy, University of Hyogo, Sayo, Hyogo 679-5313, Japan}

\author[0000-0002-6629-4182]{Suzanne L. Hawley}
\affil{Department of Astronomy, University of Washington, Seattle, WA 98195, USA}

\author[0000-0002-0637-835X]{James R. A. Davenport}
\affil{Department of Astronomy, University of Washington, Seattle, WA 98195, USA}

\author[0000-0002-1297-9485]{Kosuke Namekata}
\altaffiliation{JSPS Research Fellow DC1}
\affil{Department of Astronomy, Kyoto University, Sakyo, Kyoto 606-8502, Japan}

\author[0000-0003-2493-912X]{Shota Notsu}
\altaffiliation{JSPS Research Fellow DC1 (until March 31, 2019)}
\altaffiliation{JSPS Overseas Research Fellow (from April 1, 2019)}
\affil{Department of Astronomy, Kyoto University, Sakyo, Kyoto 606-8502, Japan}
\affil{Leiden Observatory, Leiden University, P.O. Box 9513, 2300 RA Leiden, The Netherlands}

\author{Kai Ikuta}
\affil{Department of Astronomy, Kyoto University, Sakyo, Kyoto 606-8502, Japan}

\author{Daisaku Nogami}
\affil{Department of Astronomy, Kyoto University, Sakyo, Kyoto 606-8502, Japan}

\author{Kazunari Shibata}
\affil{Astronomical Observatory, Kyoto University, Sakyo, Kyoto 606-8502, Japan}



\begin{abstract}

We report the latest view of {\it Kepler} solar-type (G-type main-sequence) superflare stars, 
including recent updates with Apache Point Observatory (APO) 3.5m telescope spectroscopic observations and {\it Gaia}-DR2 data.
First, we newly conducted APO3.5m spectroscopic observations 
of 18 superflare stars found from {\it Kepler} 1-min time cadence data.
More than half (43 stars) are confirmed to be ``single" stars, 
among 64 superflare stars in total that have been spectroscopically investigated so far 
in this APO3.5m and our previous Subaru/HDS observations.
The measurements of $v\sin i$ (projected rotational velocity) and chromospheric lines (Ca II H\&K and Ca II 8542\AA) 
support the brightness variation of superflare stars is caused by the rotation of a star with large starspots.
We then investigated the statistical properties of {\it Kepler} solar-type superflare stars 
by incorporating {\it Gaia}-DR2 stellar radius estimates.
As a result, the maximum superflare energy continuously decreases as the rotation period $P_{\mathrm{rot}}$ increases. 
Superflares with energies $\lesssim 5\times10^{34}$ erg
occur on old, slowly-rotating Sun-like stars ($P_{\mathrm{rot}}\sim$25 days) approximately once every 
2000--3000 years, 
while young rapidly-rotating stars with $P_{\mathrm{rot}}\sim$ a few days have superflares up to $10^{36}$ erg.
The maximum starspot area does not depend on the rotation period when the star is young, 
but as the rotation slows down, it starts to steeply decrease at $P_{\mathrm{rot}}\gtrsim$12 days for Sun-like stars.
These two decreasing trends are consistent since 
the magnetic energy stored around starspots explains the flare energy, 
but other factors like spot magnetic structure should also be considered.

\end{abstract}



\section{Introduction} \label{sec:intro}

Flares are energetic explosions in the stellar atmosphere, and are thought to occur 
by intense releases of magnetic energy stored around starspots (e.g., \citealt{Shibata2011}). 
Not only the Sun, but also many stars are known to show stellar magnetic activity including flares.
In particular, young rapidly-rotating stars, close binary stars, and dMe stars 
tend to show high magnetic activity levels, and magnetic fields of a few kG are considered to be distributed in large regions on the stellar surface
 (e.g., \citealt{Gershberg2005}; \citealt{Reid2005}; \citealt{Benz2010}; \citealt{Kowalski2010}; \citealt{Osten2016}).
They frequently have ``superflares", which have a total bolometric energy 10 -- 10$^{6}$ times more energetic ($\sim 10^{33-38}$ erg; \citealt{Schaefer2000}) 
than the largest solar flares ($\sim 10^{32}$ erg; \citealt{Emslie2012}).
In contrast, the Sun rotates slowly ($P_{\mathrm{rot}}\sim 25$ days) and the mean magnetic field is weak (a few Gauss).
Thus it has been thought that slowly-rotating Sun-like stars basically do not have high magnetic activity events like superflares.

Recently, however, many superflares on solar-type (G-type main-sequence) stars have been 
reported (\citealt{Maehara2012}\&\citeyear{Maehara2015}; \citealt{Shibayama2013}; 
\citealt{Candelaresi2014}; \citealt{Wu2015}; \citealt{Balona2015}; \citealt{Davenport2016}; \citealt{Doorsselaere2017}) 
by using the high-precision photometric data of the {\it Kepler} space telescope \citep{Koch2010} 
\footnote{
We note here for reference that superflares on solar-type stars have been recently reported not only with {\it Kepler} 
but also with X-ray space telescope observations (e.g., \citealt{Pye2015}) 
and ground-based photometric observations (e.g., \citealt{Jackman2018}), though currently the number of 
observed events are much smaller than the {\it Kepler} results.
} \footnote{
We also note here that now there are many flare studies using {\it Kepler} (and {\it K2}) data not only for solar-type stars discussed in this study,
but also for the other spectral-type stars such as K, M-dwarfs 
(e.g., \citealt{Walkowicz2011}; \citealt{Hawley2014}; \citealt{Candelaresi2014}; \citealt{Ramsay2015} \citealt{Davenport2016}; \citealt{Doorsselaere2017}; 
\citealt{Yang2017}; \citealt{Chang2018})
and brown dwarfs (e.g., \citealt{Gizis2013}; \citealt{Paudel2018}).
}. 
The analyses of {\it Kepler} data enabled us to discuss statistical properties of superflares since a large number of flare events were discovered. 
The frequency-energy distribution of superflares on solar-type stars shows a power-law distribution $dN/dE\approx E^{\alpha}$ 
with the index $\alpha=(-1.5)$ -- $(-1.9)$, and this distribution is consistent with that of solar flares 
(\citealt{Maehara2012}; \citealt{Shibayama2013}; \citealt{Maehara2015}).
Many superflare stars show quasi-periodic brightness variations with a typical period of from one day to a few tens of days
and a typical amplitude of 0.1\% -- 10\%.
They are assumed to be explained by the rotation of the star with fairly large starspots \citep{Notsu2013b}, 
and the starspot size $A_{\mathrm{spot}}$ and rotation period $P_{\mathrm{rot}}$ values can be estimated from these brightness variations. 

Using these $A_{\mathrm{spot}}$ values, 
we confirmed that the superflare energy is related to the total coverage of the starspots, 
and that the superflare energy can be explained by the magnetic energy stored around these large starspots (\citealt{Shibata2013}; \citealt{Notsu2013b}). 
We then found that energetic superflares with energy up to $10^{35}$ erg could occur 
on stars rotating as slowly as the Sun ($P_{\mathrm{rot}}\sim 25$ days), 
even though the frequency is low (once in a few thousand years), compared with rapidly-rotating stars \citep{Notsu2013b}.
We also investigated statistical properties of starspots in relation to superflare studies, 
and suggested that superflare events are well characterized by the existence of large starspots and 
the occurrence frequency of large starspots are also consistent with that of sunspots \citep{Maehara2017}.
In addition, we suggested 
that the Sun can generate a large magnetic flux that is sufficient for causing superflares 
on the basis of theoretical estimates \citep{Shibata2013}.
It is becoming important to find out
the maximum size of starspots and superflares that can be generated 
on Sun-like stars, 
not only in solar and stellar physics (e.g., \citealt{Aulanier2013}; \citealt{Shibata2013}; \citealt{Toriumi2017}; \citealt{Katsova2018}; \citealt{Schmieder2018}),
but also in solar-terrestrial physics and even exoplanet studies.
For example, extreme space weather events and their effects on our society
(\citealt{Royal2013}; \citealt{Tsurutani2014}; \citealt{Hudson2015}; \citealt{Schrijver2015}; \citealt{Takahashi2016}; 
\citealt{Eastwood2017}; \citealt{Riley2018}), 
a history of solar activity over $\sim$1000 years 
(e.g., \citealt{Miyake2012}\&\citeyear{Miyake2013}; \citealt{Hayakawa2017a}\&\citeyear{Hayakawa2017b}; \citealt{Usoskin2017}), 
and a potential habitability of various planets (e.g., \citealt{Segura2010}; \citealt{Airapetian2016}; \citealt{Atri2017}; 
\citealt{Lingam2017}). 

The results of {\it Kepler} described here are now also supported from spectroscopic studies. 
We have observed 50 solar-type superflare stars using Subaru/HDS 
(\citealt{Notsu2013a}, \citeyear{Notsu2015a}, \& \citeyear{Notsu2015b}; \citealt{Nogami2014}; \citealt{Honda2015}). 
We found that more than half (34 stars) of the 50 targets have no obvious evidence of being in a binary system, 
and the stellar atmospheric parameters (temperature, surface gravity, and metallicity) of these stars 
are in the range of those for ordinary solar-type stars \citep{Notsu2015a}. 
More importantly, \citet{Notsu2015b} supported the above interpretation that the quasi-periodic brightness variations of superflare stars 
are explained by the rotation of a star with large starspots, 
by measuring $v\sin i$ (projected rotational velocity) and the intensity of the Ca II 8542\AA~line. 
The existence of large starspots on superflare stars was further supported by \citet{Karoff2016}
using Ca II H\&K data from LAMOST telescope low-dispersion spectra. 
We also conducted lithium (Li) abundance analysis of these superflare stars \citep{Honda2015}. 
Li abundance is known to be a clue for investigating the stellar age (e.g., \citealt{Skumanich1972}; \citealt{Takeda2010}). 
Many of the superflare stars tend to show high Li abundance, which are also consistent with \citet{Wichmann2014}, 
but there are some objects that have low Li abundance and rotate slowly. 

These results from our previous spectroscopic observations support 
that even old slowly-rotating stars similar to the Sun can have large starspots and superflares. 
However, more spectroscopic observations are still needed for the following reasons: 
Among $\sim$300 solar-type superflare stars found from  {\it Kepler} data, only 34 stars ($\sim$10\%) have been confirmed to be single solar-type stars. 
This means that as for the remaining $\sim$90\% of the 300 solar-type stars, 
the statistical studies on superflares using {\it Kepler} data are not established in a strict sense.
In particular, the number of superflare stars rotating as slowly as the Sun ($P_{\mathrm{rot}}\sim$25 days) 
that were spectroscopically investigated is only a few in the above previous studies, 
and are not enough for considering whether the Sun can have superflares on the basis of spectroscopic data.

In this study, we have conducted new spectroscopic observations of solar-type superflare
stars that were found from  {\it Kepler} 1-min cadence data \citep{Maehara2015}. 
In this new study, we used the Echelle Spectrograph of the Apache Point Observatory (APO) 3.5m telescope.
This spectrograph has a wavelength coverage of 3200-10000\AA, 
which enabled us to take both Ca II H\&K (3968 \&~3934\AA) and Ca II 8542\AA~lines simultanenously.
In the first half of this paper (Sections \ref{sec:target-obs}, \ref{sec:overview-ana-results}, and \ref{sec:discussion-spec}), 
we report the results of these new spectroscopic observations.
We describe the selection of the target stars and the details of our observations in Section \ref{sec:target-obs}. 
In Section \ref{sec:overview-ana-results} (and Appendix \ref{sec:ana-results}), 
we check the binarity of the targets, and then estimate various stellar parameters on the basis of the spectroscopic data. 
In Section \ref{sec:discussion-spec}, we comment on the estimated stellar parameters including Li abundances, 
and discuss rotational velocity and chromospheric activities comparing with the quasi-periodic brightness variations of {\it Kepler} data. 

The recent {\it Gaia}-DR2 stellar radius data (e.g., \citealt{Berger2018}) have suggested 
the possibility of severe contaminations of subgiant stars in the classification of {\it Kepler} solar-type (G-type main-sequence) stars
used for the statistical studies of superflares described in the above.
The classification of solar-type stars in our previous studies (\citealt{Shibayama2013} and \citealt{Maehara2015})
were based on $T_{\mathrm{eff}}$ and $\log g$ values from the {\it Kepler} Input Catalog (KIC) (\citealt{Brown2011}; \citealt{Huber2014}).
However, most of those $T_{\mathrm{eff}}$ and $\log g$ values are based on photometric methods 
(see \citealt{Brown2011} and \citealt{Huber2014} for the details), 
and there can be large differences between the real and catalog values.
For example, \citet{Brown2011} reported uncertainties of $T_{\mathrm{eff}}$ and $\log g$ in the initial KIC are
$\pm$200 K and 0.4 dex, respectively.
There can be a severe contamination of subgiant stars in the sample of solar-type superflare stars in our previous studies.
In a strict sense, we cannot even deny an extreme possibility that all of the slowly-rotating Sun-like superflare stars in our previous studies
were the results of contaminations of subgiant stars.
In addition, the {\it Kepler} solar-type superflare stars discussed in our previous studies 
can include some number of binary stars. 
This is a problem especially for investigating whether 
even truly Sun-like single stars can have large superflares or not. 
On the basis of these current situations,
in Section \ref{sec:dis-Kepler}, 
we investigate the statistical properties of {\it Kepler} solar-type superflare stars originally described in our previous studies 
(\citealt{Maehara2012}, \citeyear{Maehara2015}, \& \citeyear{Maehara2017}; 
\citealt{Shibayama2013}; \citealt{Notsu2013b}), now incorporating {\it Gaia}-DR2 stellar radius estimates (reported in \citealt{Berger2018}) 
and the results of our new spectroscopic observations (Subaru and APO3.5m observations).

Finally in Sections \ref{subsec:summary-apo3.5m} -- \ref{subsec:summary-GaiaDR2},
we summarize the latest view of superflares on solar-type stars found from our series of studies using {\it Kepler} data,
by including the recent updates using Apache Point Observatory (APO) 3.5m telescope spectroscopic observations 
and {\it Gaia}-DR2 data.
We also mention implications for future studies in Section \ref{subsec:future}.

\section{APO 3.5m spectroscopic observations and data reduction}\label{sec:target-obs}
As target stars of our new spectroscopic observations, we selected all the 23 solar-type superflare stars 
reported in \citet{Maehara2015} on the basis of {\it Kepler} 1-min time cadence data.
The names of these 23 stars and their stellar parameters are listed in Table \ref{tab:DR25-gaia}. 
18 stars among these 23 stars are identified in \citet{Berger2018},
which is the catalog combining parallaxes from {\it Gaia} Data Release 2 ({\it Gaia}-DR2)
with DR25 {\it Kepler} Stellar Properties Catalog (KSPC) \citep{Mathur2017}.
We note that there are no slowly-rotating stars with their rotation period $P_{\rm{rot}}$ longer than 20 days in these 23 target stars
since no superflare stars with $P_{\rm{rot}}>20$ days were reported in \citet{Maehara2015}.
That might be related with the fact that the fraction of slowly-rotating stars is smaller in the {\it Kepler} 1-min cadence data 
compared with the long cadence data because of the initial target selection and observation time allocation of {\it Kepler} 
(See the second paragraph of page 5 of \citealt{Maehara2015} for the details). 

\begin{longrotatetable}
\begin{deluxetable*}{lccccccccccccc}
\tablecaption{Basic data of our target superflare stars. \label{tab:DR25-gaia}} 
\tabletypesize{\footnotesize}
\tablehead{
\colhead{Star name} & \colhead{Binarity\tablenotemark{a}} & \colhead{$P_{\mathrm{rot}}$\tablenotemark{b}} & 
\colhead{$N_{\mathrm{flare}}$\tablenotemark{b}} & \colhead{$K_{p}$\tablenotemark{c}} & 
\colhead{$T_{\mathrm{eff,DR25}}$ \tablenotemark{c}} & \colhead{$(\log g)_{\mathrm{DR25}}$\tablenotemark{c}} & 
\colhead{[Fe/H]$_{\mathrm{DR25}}$\tablenotemark{c}} & \colhead{$R_{\mathrm{DR25}}$\tablenotemark{c}} & 
\colhead{$d_{\mathrm{Gaia}}$\tablenotemark{d}} & \colhead{$R_{\mathrm{Gaia}}$\tablenotemark{d}} & 
\colhead{EvoFlg\tablenotemark{e}} & \colhead{BinFlg\tablenotemark{e}} \\
\colhead{} & \colhead{} & \colhead{(d)} & \colhead{} & \colhead{(mag)} & \colhead{(K)} & \colhead{(cm s$^{-2}$)} & 
\colhead{} & \colhead{($R_{\sun}$)} & \colhead{(pc)} & \colhead{($R_{\sun}$)} & \colhead{} & \colhead{} 
}
\startdata
KIC10532461 & no & 7.14 & 1 & 8.8 & $5689^{+189}_{-172}$ & $4.24^{+0.26}_{-0.19}$ & $-0.28^{+0.30}_{-0.25}$ & $1.15^{+0.37}_{-0.34}$ & $45.8^{+0.1}_{-0.1}$ & $0.78^{+0.06}_{-0.05}$ & 0 & 0 \\ 
KIC11652870 & no & 8.54 & 1 & 9.1 & $5516^{+149}_{-149}$ & $4.37^{+0.19}_{-0.20}$ & $-0.16^{+0.30}_{-0.30}$ & $0.98^{+0.27}_{-0.18}$ & $51.6^{+0.1}_{-0.1}$ & $0.78^{+0.06}_{-0.05}$ & 0 & 0 \\ 
KIC9139151 & no & 6.15 & 2 & 9.2 & $6299^{+62}_{-81}$ & $4.38^{+0.01}_{-0.01}$ & $0.08^{+0.10}_{-0.15}$ & $1.17^{+0.03}_{-0.04}$ & $102.7^{+0.3}_{-0.3}$ & $1.14^{+0.05}_{-0.05}$ & 0 & 0 \\ 
KIC4554830 & no & 7.73 & 1 & 10.3 & $5666^{+76}_{-76}$ & $4.16^{+0.01}_{-0.01}$ & $0.36^{+0.10}_{-0.15}$ & $1.43^{+0.07}_{-0.05}$ & $173.4^{+0.7}_{-0.7}$ & $1.44^{+0.06}_{-0.06}$ & 1 & 0 \\ 
KIC4742436\tablenotemark{f,g} & no & 2.34 & 1 & 10.6 & $5914^{+163}_{-133}$ & $4.11^{+0.33}_{-0.14}$ & $-0.70^{+0.35}_{-0.20}$ & $1.30^{+0.28}_{-0.42}$ & -- & -- & -- & -- \\ 
KIC4831454\tablenotemark{f} & no & 5.19 & 8 & 10.7 & $5479^{+146}_{-146}$ & $4.60^{+0.05}_{-0.11}$ & $-0.46^{+0.30}_{-0.30}$ & $0.74^{+0.13}_{-0.06}$ & $123.8^{+0.4}_{-0.4}$ & $0.91^{+0.07}_{-0.06}$ & 0 & 0 \\ 
KIC8656342 & no & 17.50 & 2 & 10.9 & $5959^{+80}_{-80}$ & $4.07^{+0.03}_{-0.03}$ & $-0.02^{+0.20}_{-0.15}$ & $1.59^{+0.10}_{-0.09}$ & $299.8^{+2.4}_{-2.4}$ & $1.70^{+0.07}_{-0.07}$ & 1 & 0 \\ 
KIC9652680\tablenotemark{f} & no & 1.47 & 7 & 11.2 & $5819^{+131}_{-160}$ & $4.57^{+0.03}_{-0.17}$ & $-0.32^{+0.30}_{-0.30}$ & $0.82^{+0.20}_{-0.07}$ & $220.2^{+1.1}_{-1.1}$ & $1.09^{+0.08}_{-0.07}$ & 0 & 0 \\ 
KIC6777146 & no & 7.21 & 1 & 11.3 & $6103^{+169}_{-169}$ & $4.34^{+0.15}_{-0.13}$ & $-0.56^{+0.30}_{-0.30}$ & $1.04^{+0.19}_{-0.19}$ & $331.6^{+2.9}_{-2.9}$ & $1.60^{+0.12}_{-0.11}$ & 1 & 0 \\ 
KIC8508009 & no & 2.95 & 1 & 11.5 & $6256^{+156}_{-188}$ & $4.37^{+0.09}_{-0.14}$ & $-0.26^{+0.25}_{-0.30}$ & $1.09^{+0.24}_{-0.15}$ & $285.9^{+1.8}_{-1.8}$ & $1.15^{+0.09}_{-0.08}$ & 0 & 0 \\ 
KIC11610797\tablenotemark{f} & no & 1.62 & 8 & 11.5 & $5868^{+78}_{-78}$ & $4.03^{+0.20}_{-0.09}$ & $-0.02^{+0.15}_{-0.15}$ & $1.67^{+0.27}_{-0.40}$ & $277.1^{+1.6}_{-1.6}$ & $1.17^{+0.05}_{-0.05}$ & 0 & 0 \\ 
KIC11253827 & no & 13.40 & 2 & 11.9 & $5593^{+111}_{-111}$ & $4.45^{+0.07}_{-0.10}$ & $0.02^{+0.15}_{-0.15}$ & $0.95^{+0.12}_{-0.08}$ & $223.7^{+1.8}_{-1.8}$ & $0.90^{+0.04}_{-0.04}$ & 0 & 0 \\ 
KIC6291837 & no & 14.30 & 1 & 12.4 & $6212^{+111}_{-136}$ & $4.31^{+0.10}_{-0.12}$ & $-0.04^{+0.15}_{-0.15}$ & $1.21^{+0.21}_{-0.14}$ & $523.2^{+15.8}_{-14.9}$ & $1.42^{+0.07}_{-0.07}$ & 0 & 0 \\ 
KIC11551430\tablenotemark{g} & RV(+VB) & 4.26 & 89 & 10.7 & $5648^{+113}_{-90}$ & $4.02^{+0.22}_{-0.11}$ & $-0.08^{+0.15}_{-0.10}$ & $1.61^{+0.30}_{-0.36}$ & -- & -- & -- & -- \\ 
KIC4543412 & RV & 2.16 & 13 & 11.2 & $5472^{+163}_{-147}$ & $4.44^{+0.14}_{-0.25}$ & $-0.30^{+0.30}_{-0.30}$ & $0.88^{+0.24}_{-0.14}$ & $148.2^{+0.6}_{-0.6}$ & $0.96^{+0.07}_{-0.06}$ & 0 & 0 \\ 
KIC11128041 & SB2+RV & 7.36 & 2 & 10.8 & $5913^{+133}_{-162}$ & $4.58^{+0.03}_{-0.14}$ & $-0.50^{+0.25}_{-0.30}$ & $0.80^{+0.16}_{-0.06}$ & $196.5^{+0.8}_{-0.8}$ & $1.21^{+0.09}_{-0.08}$ & 0 & 0 \\ 
KIC10338279\tablenotemark{g} & SB2 & 6.53 & 4 & 12.0 & $5615^{+152}_{-152}$ & $4.48^{+0.10}_{-0.14}$ & $-0.44^{+0.30}_{-0.30}$ & $0.84^{+0.19}_{-0.10}$ & -- & -- & -- & -- \\ 
KIC7093428\tablenotemark{g} & VB & 0.51 & 19 & 11.0 & $6181^{+167}_{-223}$ & $2.77^{+0.54}_{-0.18}$ & $0.07^{+0.25}_{-0.55}$ & $12.27^{+1.29}_{-7.32}$ & -- & -- & -- & -- \\ 
KIC6032920\tablenotemark{g} & -- & 3.16 & 6 & 13.5 & $5862^{+141}_{-159}$ & $4.58^{+0.03}_{-0.18}$ & $-0.52^{+0.30}_{-0.30}$ & $0.79^{+0.21}_{-0.07}$ & -- & -- & -- & -- \\ 
KIC10528093 & -- & 12.20 & 2 & 13.6 & $5334^{+159}_{-143}$ & $4.54^{+0.10}_{-0.07}$ & $-0.54^{+0.35}_{-0.30}$ & $0.75^{+0.08}_{-0.08}$ & $1497.0^{+37.3}_{-35.6}$ & $3.34^{+0.26}_{-0.24}$ & 1 & 0 \\ 
KIC10646889 & -- & 4.70 & 8 & 13.6 & $5674^{+152}_{-152}$ & $4.44^{+0.13}_{-0.17}$ & $-0.36^{+0.30}_{-0.30}$ & $0.90^{+0.23}_{-0.12}$ & $716.1^{+7.1}_{-7.0}$ & $1.35^{+0.10}_{-0.09}$ & 0 & 0 \\ 
KIC9655134 & -- & 15.50 & 1 & 13.6 & $5397^{+160}_{-160}$ & $4.55^{+0.09}_{-0.07}$ & $-0.66^{+0.35}_{-0.30}$ & $0.73^{+0.09}_{-0.08}$ & $457.6^{+4.3}_{-4.2}$ & $0.93^{+0.07}_{-0.06}$ & 0 & 0 \\ 
KIC10745663 & -- & 3.13 & 7 & 14.3 & $6014^{+162}_{-198}$ & $4.49^{+0.05}_{-0.21}$ & $-0.08^{+0.25}_{-0.35}$ & $0.96^{+0.30}_{-0.10}$ & $1480.6^{+38.0}_{-36.3}$ & $1.82^{+0.14}_{-0.13}$ & 1 & 0 \\ 
\enddata
\tablenotetext{a}{
For the details of this column, see Appendix \ref{subsec:ana-binarity}. 
``no" means that the star show no evidence of being in a binary system. 
``SB2" corresponds to stars that have double-lined profile. 
``RV" means that the star shows radial velocity changes.
``VB" means that the star has visual companion stars.
The bottom five stars (KIC6032920, KIC10528093, KIC10646889, KIC9655134, and KIC10745663) 
are much fainter compared with the other target stars. 
The data quality of our spectroscopic data for four of them (KIC6032920, KIC10528093, KIC10646889, and KIC9655134) 
are not enough to judge whether they show any evidence of binary system. 
As for the faintest KIC10745663, we have no spectroscopic data.}
\tablenotetext{b}{
Stellar rotation period ($P_{\mathrm{rot}}$) and number of superflares ($N_{\mathrm{flare}}$) reported in \citet{Maehara2015}.}
\tablenotetext{c}{
{\it Kepler} band magnitude ($K_{p}$), effective temperature ($T_{\mathrm{eff,DR25}}$), surface gravity ($(\log g)_{\mathrm{DR25}}$),
metallicity ([Fe/H]$_{\mathrm{DR25}}$), and stellar radius ($R_{\mathrm{DR25}}$) 
in DR25 {\it Kepler} Stellar Properties Catalog (KSPC) (\citealt{Mathur2017}).}
\tablenotetext{d}{
Stellar distance ($d_{\mathrm{Gaia}}$) and stellar radius ($R_{\mathrm{Gaia}}$) in \citet{Berger2018}.
They derived these values by combining parallaxes from {\it Gaia} Data Release 2 with the DR25 KSPC.}
\tablenotetext{e}{
Evolutionary Flag (EvoFlg) and Binary Flag (BinFlg) reported in \citet{Berger2018}, on the basis of $R_{\mathrm{Gaia}}$ values. 
The evolutionary flags are as follows: 0 = main-sequence dwarf and 1 = subgiant. 
The binary flags are as follows: 0 = no indication of binary, 1 = binary candidate based on $R_{\mathrm{Gaia}}$ only, 
2 = AO-detected binary only \citep{Ziegler2018}, and 3 = binary candidate based on $R_{\mathrm{Gaia}}$ and AO-detected binary.}
\tablenotetext{f}{
As described in Appendix \ref{subsec:ana-binarity}, 
we have also conducted spectroscopic observations of these four stars (KIC4742436, KIC4831454, KIC9652680, and KIC11610797) 
using Subaru telescope (\citealt{Notsu2015a} \& \citeyear{Notsu2015b}; \citealt{Honda2015}).
}
\tablenotetext{g}{
As for these five stars (KIC4742436, KIC11551430, KIC10338279, KIC7093428, and KIC6032920),
there are no $d_{\mathrm{Gaia}}$ and $R_{\mathrm{Gaia}}$ values reported in \citet{Berger2018}.}
\end{deluxetable*}
\end{longrotatetable}

Our new spectroscopic observations were carried out 
by using the ARC Echelle Spectrograph (ARCES: \citealt{Wang2003}) 
attached to the ARC 3.5 m Telescope at Apache Point Observatory (APO).
The wavelength resolution ($R=\lambda/\Delta\lambda$) is $\sim32,000$ and 
the spectral coverage is 3200-10000\AA.
Data reduction (bias subtraction, flat fielding, aperture determination, scattered light subtraction, spectral extraction, 
wavelength calibration, normalization by the continuum, and heliocentric radial velocity correction) 
\footnote{An ARCES data reduction manual by J. Thorburn is available at 
\url{http://astronomy.nmsu.edu:8000/apo-wiki/wiki/ARCES} .}
was conducted using the ECHELLE package of the IRAF\footnote{
IRAF is distributed by the National Optical Astronomy Observatories, 
which is operated by the Association of Universities for Research in Astronomy, Inc., 
under cooperate agreement with the National Science Foundation.
} and PyRAF\footnote{
PyRAF is part of the stsci$\_$python package of astronomical data analysis tools, 
and is a product of the Science Software Branch at the Space Telescope Science Institute.}
software. 

For the process of wavelength calibration, 
we took exposures of a thorium-argon (Th-Ar) lamp at the start and end 
of each half night of observation. 
However, ARCES spectral data has some systematic drift on CCD over a few hours.
This drift is most likely due to the thermal changes of the prisms with temperature, as described in Section 3.1 of \citet{Wang2003}.
This drift is a bit large in the early half night ($\sim$2 -- 3 km s$^{-1}$) compared with that in the latter half night ($\lesssim$1-2 km s$^{-1}$).
We remove this systematic drift by using the wavelength of $\sim$ 10 telluric absorption lines around 6890-6910\AA.

When we normalize the spectral orders around Ca II H\&K lines (3968\AA~and 3934\AA), as also done in \citet{Morris2017},
we fit the spectrum of an early-type standard star with a high-order polynomial to measure the blaze function,
and we then divide the spectra of target stars by the polynomial fit to normalize each spectral order.
We use this method since there are many absorption lines in the blue part of the spectra, 
and it is difficult to directly fit the continuum component of spectra of the target stars.
When normalizing red part of the spectra ($\gtrsim$5000\AA), which we use for the estimations of stellar parameters (except for 
Ca II H\&K line analyses) in this study,
we directly fit the target star spectra with a high-order polynomial to measure the blaze function.
After that, we shifted the normalized spectra in wavelength into the rest-frame by removing their radial velocities. 
We remove the radial velocity by maximizing the cross-correlation of our ARCES spectra with solar spectra in the wavelength range of 6212-6220\AA,
where we conduct $v\sin i$ measurements in Appendix \ref{subsec:ana-vsini}.

The observation date of each target superflare star and the obtained signal-to-noise ratio (S/N) are shown in Table \ref{tab:obs-RV}.
We selected 23 target superflare stars, and for the 18 brightest superflare stars, 
we obtained high enough signal-to-noise (S/N) data for scientific discussions in this paper. 
However, the other five stars (KIC6032920, KIC10528093, KIC10646889, KIC9655134, and KIC10745663) 
in the bottom of Table \ref{tab:DR25-gaia} are much fainter ($K_{p}\geq$13.5 mag) compared with the other 18 
relatively bright target stars ($K_{p}\leq$12.4 mag). 
The data quality of our spectroscopic data for four of them (KIC6032920, KIC10528093, KIC10646889, and KIC9655134) 
are not enough for detailed scientific discussions in this paper. 
As for KIC10745663, we only took a slit viewer image, and we have no spectroscopic data since it is too faint ($K_{p}$=14.3 mag).

\startlongtable
%\begin{longrotatetable}
\begin{deluxetable*}{ccccccccc}
\tablecaption{APO3.5m telescope observations of target superflare stars \label{tab:obs-RV}}
\tabletypesize{\footnotesize}
\tablehead{
\colhead{Star name} & \colhead{Remarks\tablenotemark{a}} & \colhead{Binarity\tablenotemark{b}} &
\colhead{date(UT)\tablenotemark{c}} & \colhead{time(UT)\tablenotemark{c}} & \colhead{RV\tablenotemark{d}} &
\colhead{S/N(8542)\tablenotemark{e}} & \colhead{S/N(H$\alpha$)\tablenotemark{e}} & \colhead{S/N(HK)\tablenotemark{e}} \\
\colhead{} & \colhead{} & \colhead{} & 
\colhead{} & \colhead{} & \colhead{(km s$^{-1}$)} &
 \colhead{}  & \colhead{}  & \colhead{} 
}
\startdata
KIC10532461 & & no & 2016-08-23 & 02:16:35.23 & 8.0 & 68 & 95 & 33 \\ 
-- & & -- & 2017-04-10 & 08:11:00.28 & 8.3 & 68 & 97 & 22 \\ 
-- & & -- & 2017-09-23 & 01:44:20.11 & 8.6 & 110 & 148 & 41 \\ 
-- & & -- & 2017-10-31 & 02:01:19.03 & 8.5 & 27 & 36 & 9 \\ 
-- & & -- & comb & -- & -- & 148 & 202 & 64 \\ 
KIC11652870 & & no & 2016-08-21 & 04:53:17.80 & -3.9 & 49 & 68 & 25 \\ 
-- & & -- & 2017-04-10 & 08:33:38.85 & -3.5 & 70 & 99 & 26 \\ 
-- & & -- & 2017-10-15 & 01:34:58.94 & -3.4 & 64 & 86 & 24 \\ 
-- & & -- & 2017-10-29 & 03:47:57.35 & -3.5 & 65 & 89 & 19 \\ 
-- & & -- & comb & -- & -- & 124 & 172 & 54 \\ 
KIC9139151 & & no & 2017-04-10 & 08:57:09.20 & -28.4 & 68 & 100 & 33 \\ 
-- & & -- & 2017-06-05 & 10:53:45.74 & -28.2 & 67 & 99 & 43 \\ 
-- & & -- & 2017-10-29 & 03:32:52.19 & -28.5 & 71 & 104 & 28 \\ 
-- & & -- & comb & -- & -- & 118 & 175 & 71 \\ 
KIC4554830 & & no & 2017-04-10 & 09:24:50.91 & -21.9 & 54 & 77 & 22 \\ 
-- & & -- & 2017-09-23 & 02:11:54.20 & -21.2 & 64 & 88 & 25 \\ 
-- & & -- & 2017-10-03 & 05:32:26.76 & -22.5 & 43 & 58 & 12 \\ 
-- & & -- & comb & -- & -- & 93 & 130 & 38 \\ 
KIC4742436 & & no & 2017-04-10 & 11:54:08.11 & -54.6 & 34 & 49 & 20 \\ 
-- & (1)  & -- & 2012-08-07 & 06:46:04.8 & -54.5 & 140 & 210 & -- \\ 
-- & (1)  & -- & 2012-08-08 & 06:04:19.2 & -54.6 & 140 & 190 & -- \\ 
-- & (1)  & -- & 2012-08-09 & 06:50:24.0 & -54 & 100 & 160 & -- \\ 
KIC4831454 & & no & 2017-04-10 & 10:30:13.61 & -27.4 & 45 & 63 & 21 \\ 
-- & (1)  & -- & 2012-08-07 & 07:03:21.6 & -26.4 & 150 & 220 & -- \\ 
-- & (1)  & -- & 2012-08-08 & 06:21:36.0 & -26.6 & 140 & 190 & -- \\ 
-- & (1)  & -- & 2012-08-09 & 07:12:00.0 & -27.2 & 110 & 170 & -- \\ 
KIC8656342 & & no & 2017-05-03 & 08:09:21.76 & -4.1 & 49 & 70 & 21 \\ 
-- & & -- & 2017-09-23 & 04:03:53.94 & -4.6 & 25 & 34 & 11 \\ 
-- & & -- & 2017-10-03 & 05:07:22.31 & -4.3 & 45 & 62 & 19 \\ 
-- & & -- & comb & -- & -- & 71 & 99 & 35 \\ 
KIC9652680 & & no & 2017-05-03 & 10:01:13.48 & -20.5 & 43 & 61 & 23 \\ 
-- & (1)  & -- & 2012-08-07 & 07:19:12.0 & -21.6 & 110 & 170 & -- \\ 
-- & (1)  & -- & 2012-08-08 & 06:41:45.6 & -20.5 & 90 & 150 & -- \\ 
-- & (1)  & -- & 2012-08-09 & 06:25:55.2 & -22.3 & 90 & 140 & -- \\ 
KIC6777146 & & no & 2017-05-03 & 10:34:00.37 & -7.3 & 41 & 58 & 25 \\ 
-- & & -- & 2017-10-03 & 01:51:57.37 & -6.9 & 41 & 56 & 19 \\ 
-- & & -- & 2017-10-29 & 03:07:44.74 & -7.7 & 41 & 57 & 17 \\ 
-- & & -- & comb & -- & -- & 71 & 99 & 42 \\ 
KIC8508009 & & no & 2017-05-03 & 11:06:38.23 & -28.8 & 36 & 52 & 22 \\ 
-- & & -- & 2017-10-03 & 02:24:30.58 & -28.8 & 37 & 51 & 18 \\ 
-- & & -- & 2017-10-07 & 06:30:24.29 & -28.8 & 27 & 38 & 8 \\ 
-- & & -- & comb & -- & -- & 58 & 81 & 34 \\ 
KIC11610797 & & no & 2017-06-05 & 07:42:39.45 & -6.3 & 36 & 51 & 20 \\ 
-- & (1)  & -- & 2012-08-07 & 08:32:38.4 & -10.3 & 110 & 170 & -- \\ 
-- & (1)  & -- & 2012-08-08 & 07:17:45.6 & -11.6 & 100 & 150 & -- \\ 
-- & (1)  & -- & 2012-08-09 & 07:58:04.8 & -10.4 & 80 & 130 & -- \\ 
KIC11253827 & & no & 2017-06-05 & 08:15:13.95 & -13.6 & 32 & 44 & 16 \\ 
-- & & -- & 2017-10-03 & 02:57:04.29 & -13.5 & 32 & 44 & 13 \\ 
-- & & -- & 2017-10-07 & 05:57:49.26 & -14.1 & 26 & 35 & 7 \\ 
-- & & -- & comb & -- & -- & 52 & 71 & 24 \\ 
KIC6291837 & & no & 2017-06-05 & 09:22:11.96 & -34.5 & 34 & 50 & 23 \\ 
-- & & -- & 2017-10-07 & 04:52:41.67 & -36.7 & 22 & 31 & 8 \\ 
-- & & -- & comb & -- & -- & 41 & 58 & 29 \\ 
KIC11551430A & (2) & RV(+VB) & 2017-04-10 & 11:03:00.04 & -14.9 & 43 & 60 & 20 \\ 
-- & & -- & 2017-09-23 & 02:46:47.94 & 5.4 & 53 & 71 & 21 \\ 
-- & & -- & 2017-10-03 & 04:34:48.04 & -54.0 & 53 & 72 & 20 \\ 
-- & & -- & comb & -- & -- & 86 & 116 & 41 \\ 
KIC11551430B & (2) & VB & 2017-04-10 & 11:28:42.47 & -25.5 & 11 & 16 & 4 \\ 
KIC4543412 & & RV & 2017-05-03 & 09:28:39.20 & 27.3 & 45 & 61 & 20 \\ 
-- & & -- & 2017-09-23 & 06:28:58.19 & 7.6 & 24 & 31 & 4 \\ 
-- & & -- & 2017-10-03 & 04:02:13.73 & 6.4 & 45 & 60 & 15 \\ 
-- & & -- & 2017-10-07 & 05:25:15.88 & -24.4 & 38 & 51 & 8 \\ 
-- & & -- & comb & -- & -- & 78 & 103 & 31 \\ 
KIC11128041 & & SB2+RV & 2017-05-03 & 07:40:28.97 & -- & 46 & 65 & 21 \\ 
-- & & -- & 2017-09-23 & 03:19:21.25 & -- & 42 & 57 & 18 \\ 
-- & & -- & 2017-10-29 & 04:03:06.11 & -- & 32 & 46 & 10 \\ 
KIC10338279 & & SB2 & 2017-06-05 & 08:49:33.49 & -- & 42 & 58 & 20 \\ 
KIC7093428 & (3) & VB & 2017-05-03 & 08:56:03.72 & -44.6 & 42 & 62 & 28 \\ 
KIC6032920 & & -- & 2017-12-10 & 02:30:50.58 & -58.3 & 14 & 21 & low S/N \\ 
KIC10528093 & & -- & 2017-12-10 & 03:46:14.03 & 34.0 & 10 & 16 & low S/N \\ 
KIC10646889 & & -- & 2017-12-10 & 03:10:31.49 & -45.1 & 10 & 16 & low S/N \\ 
KIC9655134 & & -- & 2017-12-10 & 04:20:18.10 & -12.6 & 9 & 15 & low S/N \\ 
KIC10745663 & & -- & No-spectra & No-spectra & No-spectra & No-spectra & No-spectra & No-spectra \\ 
\enddata
\tablenotetext{a}{
(1)
As described in Appendix \ref{subsec:ana-binarity}, 
we also conducted spectroscopic observations of these four stars (KIC4742436, KIC4831454, KIC9652680, and KIC11610797) 
using Subaru telescope (\citealt{Notsu2015a}). 
RV values of these observations are included here for reference.
\\
(2)
KIC11551430 is found to be visual binary (KIC11551430A and KIC1551430B), as described in Appendix \ref{subsec:ana-binarity}.
We took spectra of each component of visual binary separately, and measure parameters of each of them.
\\
(3)
As described in Appendix \ref{subsec:ana-binarity}, KIC7093428 has two fainter companion stars, and pixel data analyses suggest 
that flares occur on these companion stars. 
Since companion stars are faint, here we only measure parameters of the main (brightest) star for reference, 
though we do not use these values in the main discussion of this paper.
}
\tablenotetext{b}{
Same as in footnote $a$ of Table \ref{tab:DR25-gaia}. 
See also Appendix \ref{subsec:ana-binarity}.
}\tablenotetext{c}{
Observation date (Universal Time) (format: YYYY-MM-DD) and middle time (Universal Time) (format: hh:mm:ss) of each exposure.
``comb" corresponds to the co-added spectra that all APO spectra of each target star are combined into. 
}\tablenotetext{d}{
Radial velocity (See Appendix \ref{subsec:ana-binarity}).
}\tablenotetext{e}{
Signal-to-noise ratio of the spectral data around Ca II 8542 \AA (S/N(8542)), H$\alpha$ 6563\AA (S/N(H$\alpha$)), 
and Ca II H\&K 3934/3938\AA (S/N(HK)).}
\end{deluxetable*}
%\end{longrotatetable}

In addition to the above 23 target superflare stars, 
we repeatedly observed 
28 bright solar-type comparison stars as references of solar-type stars.
The name and basic parameters of these 28 comparison stars, observation date, and S/N of each observation are listed in Table \ref{tab:Smwo-Sapo} 
(The original sources of these stars are in footnotes of Table \ref{tab:Smwo-Sapo}).
All of these 28 stars were also observed in California Planet Search (CPS) program \citep{Isaacson2010}, 
and we use these solar-type stars to calibrate Ca II H\&K $S$-index in Appendix \ref{sec:measure-Sindex}.

%\startlongtable
\begin{longrotatetable}
\begin{deluxetable*}{lccccccccccccc}
\tablecaption{Observations of comparison stars to calibrate Ca II H\&K $S$-index. \label{tab:Smwo-Sapo}} 
\tabletypesize{\scriptsize}
\tablehead{
\colhead{Star name} & \colhead{Remarks\tablenotemark{a}} &  \colhead{$T_{\mathrm{eff}}$\tablenotemark{b}} & \colhead{$B-V$\tablenotemark{b}} & 
\colhead{$S_{\mathrm{MWO}}$\tablenotemark{c}} & \colhead{date(UT)\tablenotemark{d}} &\colhead{time(UT)\tablenotemark{d}} & 
\colhead{S/N(8542)\tablenotemark{e}} & \colhead{S/N(HK)\tablenotemark{e}} & \colhead{$r_{0}$(8542)} & 
\colhead{$S_{\mathrm{APO}}$} & \colhead{$S_{\mathrm{HK}}$} &\colhead{$\log \mathcal{F}^{+}_{\mathrm{HK}}$} & 
\colhead{$\log R^{+}_{\mathrm{HK}}$}  \\
\colhead{} & \colhead{} & \colhead{[K]} & \colhead{[mag]} & 
\colhead{} & \colhead{} & \colhead{} & 
\colhead{} & \colhead{} & \colhead{} & 
\colhead{} & \colhead{} & \colhead{(erg/(cm$^{2}$ s))} & 
\colhead{} 
}
\startdata
HIP9519\tablenotemark{e} & (1) & 5899 & 0.594 & 0.417 & 2016-11-07 & 08:01:15.75 & 134 & 64 & 0.50 & 0.0185 & 0.416 & 6.77 & -4.07 \\ 
-- & -- & -- & -- & -- & 2018-02-24 & 01:43:15.26 & 86 & 31 & 0.46 & 0.0167  & 0.376 & 6.71 & -4.12 \\ 
HIP19793\tablenotemark{e}  & (1) & 5833 & 0.620 & 0.347 & 2016-11-07 & 10:00:22.79 & 108 & 64 & 0.42 & 0.0165 & 0.372 & 6.66 & -4.16 \\ 
HIP21091\tablenotemark{e}  & (1) & 5857 & 0.601 & 0.352 & 2016-11-07 & 10:13:11.72 & 76 & 42 & 0.44 & 0.0163 & 0.367 & 6.69 & -4.14 \\ 
-- & -- & -- & -- & -- & 2018-01-05 & 08:29:18.26 & 102 & 21 & 0.44 & 0.0167  & 0.376 & 6.70 & -4.12 \\ 
HIP22175\tablenotemark{e}  & (1) & 5667 & 0.637 & 0.213 & 2016-11-07 & 10:25:47.00 & 116 & 58 & 0.26 & 0.0104 & 0.238 & 6.36 & -4.41 \\ 
-- & -- & -- & -- & -- & 2018-01-02 & 08:20:23.05 & 119 & 27 & 0.26 & 0.0104  & 0.238 & 6.36 & -4.41 \\ 
-- & -- & -- & -- & -- & 2018-02-24 & 02:18:39.39 & 278 & 37 & 0.26 & 0.0101  & 0.230 & 6.34 & -4.43 \\ 
HIP27980\tablenotemark{e}  & (1) & 5854 & 0.615 & 0.205 & 2016-11-07 & 10:53:12.14 & 105 & 58 & 0.23 & 0.0098 & 0.226 & 6.35 & -4.48 \\ 
-- & -- & -- & -- & -- & 2018-01-02 & 09:10:44.10 & 116 & 31 & 0.23 & 0.0091  & 0.208 & 6.29 & -4.54 \\ 
HIP35185\tablenotemark{e} & (1) & 5830 & 0.601 & 0.341 & 2016-11-07 & 11:16:44.46 & 74 & 43 & 0.41 & 0.0150 & 0.338 & 6.64 & -4.17 \\ 
-- & -- & -- & -- & -- & 2018-01-02 & 10:09:19.57 & 97 & 35 & 0.41 & 0.0153  & 0.346 & 6.65 & -4.16 \\ 
-- & -- & -- & -- & -- & 2018-02-01 & 07:48:08.24 & 95 & 38 & 0.43 & 0.0157  & 0.354 & 6.67 & -4.15 \\ 
-- & -- & -- & -- & -- & 2018-04-30 & 02:31:36.03 & 70 & 31 & 0.43 & 0.0158  & 0.357 & 6.67 & -4.14 \\ 
HIP37971\tablenotemark{e} & (1) & 5774 & 0.642 & 0.322 & 2016-11-07 & 11:28:03.37 & 69 & 38 & 0.38 & 0.0157 & 0.354 & 6.59 & -4.20 \\ 
-- & -- & -- & -- & -- & 2018-01-05 & 09:15:25.35 & 92 & 43 & 0.36 & 0.0148  & 0.334 & 6.56 & -4.24 \\ 
-- & -- & -- & -- & -- & 2018-04-30 & 02:44:36.92 & 77 & 36 & 0.39 & 0.0161  & 0.364 & 6.61 & -4.19 \\ 
HIP38228\tablenotemark{e} & (1) & 5631 & 0.657 & 0.406 & 2016-11-07 & 11:37:22.49 & 117 & 65 & 0.45 & 0.0181 & 0.407 & 6.65 & -4.10 \\ 
-- & -- & -- & -- & -- & 2018-01-04 & 08:32:28.21 & 130 & 59 & 0.42 & 0.0164  & 0.370 & 6.60 & -4.16 \\ 
-- & -- & -- & -- & -- & 2018-01-05 & 09:05:22.10 & 137 & 63 & 0.44 & 0.0173  & 0.389 & 6.63 & -4.13 \\ 
-- & -- & -- & -- & -- & 2018-02-01 & 07:59:52.56 & 160 & 73 & 0.48 & 0.0186  & 0.418 & 6.67 & -4.09 \\ 
-- & -- & -- & -- & -- & 2018-04-30 & 02:55:31.81 & 109 & 49 & 0.42 & 0.0167  & 0.377 & 6.61 & -4.14 \\ 
HIP51652\tablenotemark{e} & (1) & 5870 & 0.620 & 0.214 & 2016-11-07 & 12:12:29.36 & 72 & 33 & 0.25 & 0.0101 & 0.230 & 6.35 & -4.48 \\ 
-- & -- & -- & -- & -- & 2016-12-11 & 09:39:04.98 & 63 & 23 & 0.24 & 0.0099  & 0.226 & 6.33 & -4.49 \\ 
-- & -- & -- & -- & -- & 2018-01-05 & 11:09:21.87 & 102 & 50 & 0.24 & 0.0099  & 0.228 & 6.34 & -4.49 \\ 
HIP79068\tablenotemark{e} & (1) & 5809 & 0.615 & 0.235 & 2016-12-11 & 12:53:01.25 & 52 & 25 & 0.28 & 0.0120 & 0.272 & 6.48 & -4.33 \\ 
-- & -- & -- & -- & -- & 2018-01-05 & 13:02:29.03 & 71 & 32 & 0.27 & 0.0109  & 0.248 & 6.42 & -4.39 \\ 
HIP117184\tablenotemark{e} & (1) & 5890 & 0.609 & 0.354 & 2016-11-07 & 07:31:06.71 & 72 & 33 & 0.39 & 0.0153 & 0.346 & 6.64 & -4.20 \\ 
18Sco\tablenotemark{f} & (2) & 5794 & 0.618 & 0.170 & 2016-08-21 & 02:18:54.94 & 222 & 98 & 0.2 & 0.0074 & 0.172 & 6.13 & -4.68 \\ 
-- & -- & -- & -- & -- & 2016-08-23 & 02:00:15.72 & 198 & 102 & 0.23 & 0.0082  & 0.188 & 6.21 & -4.60 \\ 
-- & -- & -- & -- & -- & 2017-05-03 & 07:19:54.42 & 238 & 97 & 0.21 & 0.0082  & 0.190 & 6.21 & -4.59 \\ 
59Vir\tablenotemark{f} & (2) & 6017 & 0.573 & 0.324 & 2017-04-10 & 07:23:42.69 & 90 & 52 & 0.4 & 0.0154 & 0.347 & 6.70 & -4.17 \\ 
-- & -- & -- & -- & -- & 2017-12-31 & 12:53:01.38 & 212 & 106 & 0.38 & 0.0144  & 0.326 & 6.66 & -4.21 \\ 
-- & -- & -- & -- & -- & 2018-01-02 & 12:55:33.66 & 134 & 64 & 0.39 & 0.0144  & 0.327 & 6.66 & -4.21 \\ 
61Vir\tablenotemark{f} & (2) & 5565 & 0.678 & 0.169 & 2017-04-10 & 07:31:23.31 & 107 & 41 & 0.21 & 0.0074 & 0.171 & 6.03 & -4.70 \\ 
HIP100963\tablenotemark{f} & (2) & 5751 & 0.620 & 0.227 & 2016-10-12 & 01:01:03.01 & 160 & 98 & 0.23 & 0.0085 & 0.195 & 6.24 & -4.55 \\ 
-- & -- & -- & -- & -- & 2016-10-14 & 00:52:57.66 & 170 & 106 & 0.26 & 0.0088  & 0.203 & 6.27 & -4.52 \\ 
-- & -- & -- & -- & -- & 2017-06-05 & 07:19:50.33 & 160 & 69 & 0.23 & 0.0087  & 0.200 & 6.26 & -4.54 \\ 
-- & -- & -- & -- & -- & 2017-10-03 & 01:17:15.01 & 154 & 71 & 0.22 & 0.0085  & 0.196 & 6.24 & -4.55 \\ 
-- & -- & -- & -- & -- & 2017-10-07 & 01:11:38.75 & 133 & 56 & 0.24 & 0.0089  & 0.206 & 6.28 & -4.51 \\ 
-- & -- & -- & -- & -- & 2017-10-29 & 01:40:46.82 & 103 & 51 & 0.24 & 0.0088  & 0.202 & 6.27 & -4.53 \\ 
-- & -- & -- & -- & -- & 2017-10-31 & 00:37:45.39 & 116 & 55 & 0.27 & 0.0089  & 0.204 & 6.28 & -4.52 \\ 
-- & -- & -- & -- & -- & 2017-12-10 & 02:08:12.60 & 137 & 41 & 0.23 & 0.0086  & 0.199 & 6.26 & -4.54 \\ 
HD11131\tablenotemark{g} & (3) & 5735 & 0.654 & 0.299 & 2017-12-10 & 05:10:03.30 & 171 & 58 & 0.37 & 0.0143 & 0.324 & 6.52 & -4.27 \\ 
HD114710\tablenotemark{g} & (3) & 6006 & 0.572 & 0.222 & 2017-12-31 & 12:45:12.67 & 235 & 120 & 0.28 & 0.0104 & 0.238 & 6.46 & -4.40 \\ 
-- & -- & -- & -- & -- & 2018-01-02 & 12:40:47.75 & 154 & 76 & 0.27 & 0.0103  & 0.236 & 6.46 & -4.41 \\ 
-- & -- & -- & -- & -- & 2018-02-01 & 10:31:27.18 & 169 & 89 & 0.26 & 0.0099  & 0.227 & 6.43 & -4.44 \\ 
-- & -- & -- & -- & -- & 2018-04-23 & 05:59:54.81 & 135 & 67 & 0.26 & 0.0098  & 0.224 & 6.42 & -4.45 \\ 
-- & -- & -- & -- & -- & 2018-04-30 & 04:19:41.44 & 116 & 62 & 0.27 & 0.0103  & 0.237 & 6.46 & -4.41 \\ 
-- & -- & -- & -- & -- & 2018-05-05 & 09:21:05.68 & 180 & 67 & 0.26 & 0.0095  & 0.219 & 6.40 & -4.47 \\ 
HD129333\tablenotemark{g} & (3) & 5824 & 0.626 & 0.695 & 2017-10-03 & 01:29:23.45 & 130 & 36 & 0.75 & 0.0286 & 0.639 & 6.94 & -3.87 \\ 
-- & -- & -- & -- & -- & 2017-12-31 & 13:00:47.58 & 93 & 37 & 0.76 & 0.0309  & 0.688 & 6.98 & -3.84 \\ 
-- & -- & -- & -- & -- & 2018-01-02 & 12:47:38.75 & 92 & 35 & 0.74 & 0.0292  & 0.652 & 6.95 & -3.86 \\ 
-- & -- & -- & -- & -- & 2018-02-01 & 10:54:29.71 & 108 & 45 & 0.78 & 0.0321  & 0.715 & 7.00 & -3.82 \\ 
-- & -- & -- & -- & -- & 2018-02-24 & 06:54:57.88 & 89 & 29 & 0.76 & 0.0310  & 0.692 & 6.98 & -3.83 \\ 
-- & -- & -- & -- & -- & 2018-04-23 & 06:31:24.21 & 92 & 39 & 0.77 & 0.0318  & 0.709 & 6.99 & -3.82 \\ 
-- & -- & -- & -- & -- & 2018-04-30 & 04:30:19.90 & 120 & 51 & 0.79 & 0.0319  & 0.711 & 7.00 & -3.82 \\ 
-- & -- & -- & -- & -- & 2018-05-05 & 09:51:25.55 & 137 & 57 & 0.74 & 0.0293  & 0.653 & 6.95 & -3.86 \\ 
HD17925\tablenotemark{g} & (3) & 5183 & 0.862 & 0.659 & 2017-12-10 & 05:32:13.48 & 238 & 59 & 0.54 & 0.0281 & 0.626 & 6.55 & -4.06 \\ 
-- & -- & -- & -- & -- & 2018-02-24 & 01:53:43.39 & 170 & 33 & 0.57 & 0.0299  & 0.667 & 6.58 & -4.03 \\ 
HD1835\tablenotemark{g} & (3) & 5720 & 0.659 & 0.319 & 2017-10-29 & 04:18:11.31 & 212 & 66 & 0.4 & 0.0157 & 0.354 & 6.57 & -4.22 \\ 
-- & -- & -- & -- & -- & 2017-10-31 & 03:25:37.19 & 175 & 53 & 0.41 & 0.0156  & 0.352 & 6.56 & -4.22 \\ 
-- & -- & -- & -- & -- & 2017-12-10 & 04:58:32.18 & 148 & 32 & 0.39 & 0.0155  & 0.351 & 6.56 & -4.22 \\ 
HD190406\tablenotemark{g} & (3) & 5910 & 0.600 & 0.183 & 2017-09-23 & 01:28:27.37 & 224 & 98 & 0.24 & 0.0091 & 0.210 & 6.32 & -4.52 \\ 
-- & -- & -- & -- & -- & 2017-10-31 & 02:20:07.29 & 59 & 26 & 0.25 & 0.0092  & 0.211 & 6.32 & -4.52 \\ 
-- & -- & -- & -- & -- & 2017-10-31 & 03:09:10.26 & 281 & 107 & 0.24 & 0.0092  & 0.211 & 6.32 & -4.52 \\ 
-- & -- & -- & -- & -- & 2018-05-05 & 10:19:25.86 & 156 & 76 & 0.23 & 0.0085  & 0.195 & 6.26 & -4.58 \\ 
HD20630\tablenotemark{g} & (3) & 5654 & 0.681 & 0.379 & 2017-10-31 & 03:48:26.44 & 104 & 21 & 0.45 & 0.0182 & 0.410 & 6.61 & -4.15 \\ 
-- & -- & -- & -- & -- & 2017-10-31 & 05:51:54.28 & 201 & 31 & 0.46 & 0.0178  & 0.401 & 6.60 & -4.16 \\ 
-- & -- & -- & -- & -- & 2017-12-10 & 05:42:11.24 & 296 & 121 & 0.42 & 0.0170  & 0.382 & 6.57 & -4.19 \\ 
-- & -- & -- & -- & -- & 2018-01-05 & 07:18:21.90 & 245 & 34 & 0.41 & 0.0158  & 0.357 & 6.53 & -4.23 \\ 
-- & -- & -- & -- & -- & 2018-02-24 & 02:02:44.05 & 117 & 43 & 0.43 & 0.0171  & 0.384 & 6.58 & -4.19 \\ 
HD206860\tablenotemark{g} & (3) & 5954 & 0.587 & 0.332 & 2017-10-07 & 06:51:16.37 & 157 & 53 & 0.43 & 0.0153 & 0.345 & 6.67 & -4.18 \\ 
-- & -- & -- & -- & -- & 2017-10-31 & 02:35:20.55 & 136 & 63 & 0.44 & 0.0151  & 0.341 & 6.67 & -4.19 \\ 
-- & -- & -- & -- & -- & 2018-05-05 & 10:42:35.83 & 125 & 50 & 0.45 & 0.0160  & 0.362 & 6.70 & -4.15 \\ 
HD22049\tablenotemark{g} & (3) & 5140 & 0.881 & 0.501 & 2017-10-31 & 04:56:33.84 & 315 & 48 & 0.39 & 0.0204 & 0.457 & 6.36 & -4.24 \\ 
-- & -- & -- & -- & -- & 2018-02-24 & 01:59:09.42 & 191 & 49 & 0.4 & 0.0214  & 0.480 & 6.38 & -4.22 \\ 
HD30495\tablenotemark{g} & (3) & 5804 & 0.632 & 0.287 & 2018-01-02 & 07:34:44.74 & 180 & 33 & 0.34 & 0.0131 & 0.298 & 6.51 & -4.30 \\ 
-- & -- & -- & -- & -- & 2018-02-24 & 02:08:32.46 & 199 & 64 & 0.33 & 0.0126  & 0.286 & 6.48 & -4.33 \\ 
HD39587\tablenotemark{g} & (3) & 5930 & 0.594 & 0.379 & 2017-10-31 & 06:09:28.99 & 134 & 39 & 0.41 & 0.0147 & 0.333 & 6.64 & -4.21 \\ 
-- & -- & -- & -- & -- & 2017-12-31 & 08:19:44.48 & 243 & 114 & 0.4 & 0.0141  & 0.319 & 6.61 & -4.23 \\ 
-- & -- & -- & -- & -- & 2018-01-02 & 07:24:13.10 & 281 & 134 & 0.41 & 0.0144  & 0.325 & 6.63 & -4.22 \\ 
-- & -- & -- & -- & -- & 2018-01-10 & 07:32:03.90 & 64 & 34 & 0.41 & 0.0141  & 0.320 & 6.61 & -4.23 \\ 
-- & -- & -- & -- & -- & 2018-04-30 & 02:02:28.04 & 231 & 83 & 0.41 & 0.0133  & 0.302 & 6.58 & -4.26 \\ 
HD41593\tablenotemark{g} & (3) & 5296 & 0.814 & 0.467 & 2017-12-10 & 05:51:26.50 & 176 & 58 & 0.49 & 0.0230 & 0.515 & 6.52 & -4.12 \\ 
-- & -- & -- & -- & -- & 2017-12-31 & 08:37:11.85 & 157 & 55 & 0.49 & 0.0233  & 0.522 & 6.53 & -4.12 \\ 
-- & -- & -- & -- & -- & 2018-01-02 & 09:56:19.71 & 195 & 41 & 0.47 & 0.0228  & 0.510 & 6.52 & -4.13 \\ 
-- & -- & -- & -- & -- & 2018-01-05 & 08:46:02.36 & 193 & 65 & 0.49 & 0.0233  & 0.523 & 6.53 & -4.12 \\ 
-- & -- & -- & -- & -- & 2018-02-01 & 07:35:33.81 & 188 & 54 & 0.5 & 0.0240  & 0.538 & 6.55 & -4.10 \\ 
-- & -- & -- & -- & -- & 2018-04-30 & 02:20:09.28 & 138 & 37 & 0.47 & 0.0217  & 0.487 & 6.50 & -4.15 \\ 
HD72905\tablenotemark{g} & (3) & 5850 & 0.618 & 0.407 & 2017-12-31 & 11:51:52.03 & 233 & 94 & 0.44 & 0.0158 & 0.358 & 6.64 & -4.18 \\ 
-- & -- & -- & -- & -- & 2018-01-02 & 12:34:48.19 & 127 & 46 & 0.46 & 0.0166  & 0.373 & 6.67 & -4.16 \\ 
-- & -- & -- & -- & -- & 2018-01-04 & 07:18:42.04 & 78 & 36 & 0.46 & 0.0165  & 0.372 & 6.66 & -4.16 \\ 
-- & -- & -- & -- & -- & 2018-01-05 & 10:37:53.08 & 163 & 75 & 0.46 & 0.0162  & 0.365 & 6.65 & -4.17 \\ 
-- & -- & -- & -- & -- & 2018-01-10 & 07:56:29.39 & 67 & 31 & 0.47 & 0.0159  & 0.358 & 6.64 & -4.18 \\ 
-- & -- & -- & -- & -- & 2018-02-01 & 09:53:35.71 & 161 & 67 & 0.48 & 0.0171  & 0.385 & 6.68 & -4.14 \\ 
-- & -- & -- & -- & -- & 2018-02-24 & 06:46:08.98 & 129 & 60 & 0.48 & 0.0171  & 0.386 & 6.68 & -4.14 \\ 
-- & -- & -- & -- & -- & 2018-04-23 & 06:22:34.92 & 115 & 41 & 0.48 & 0.0172  & 0.387 & 6.69 & -4.14 \\ 
-- & -- & -- & -- & -- & 2018-04-30 & 04:41:56.76 & 168 & 68 & 0.49 & 0.0177  & 0.399 & 6.70 & -4.12 \\ 
\enddata
\tablenotetext{a}{
(1)
Active solar-type stars having strong X-ray luminosity, which we have also spectroscopically investigated in \citet{Notsu2017}.
\\
(2)
Bright solar-type (inactive or mildly active) comparison stars also observed in \citet{Notsu2017}.
\\
(3)
Other active solar-type stars that we have newly started monitoring spectroscopic observations in this study. 
These stars are also photometrically observed with TESS \citep{Ricker2015}, and are included in the 
target list of TESS Guest Investigator programs G011264 (PI: James Davenport) and G011299 (PI: Vladimir Airapetian) 
(cf. \url{https://heasarc.gsfc.nasa.gov/docs/tess/approved-programs.html}).
}
\tablenotetext{b}{
Stellar effective temperature ($T_{\mathrm{eff}}$) and stellar color ($B-V$). 
As for the first 15 stars with footnote $e$ or $f$, $T_{\mathrm{eff}}$ values reported in \citet{Notsu2017} are used here, 
and we newly derive $B-V$ values from the $T_{\mathrm{eff}}$ and metallicity values in \citet{Notsu2017}, by using Equation (2) of \citet{Alonso1996}.
As for the latter 13 stars with footnote $g$, $B-V$ values reported in Table 1 of \citet{Isaacson2010} are used here,
and $T_{\mathrm{eff}}$ values are derived from these $B-V$ values by using Equation (2) of \citet{Valenti2005}. 
}\tablenotetext{c}{
Mount Wilson S-index value (Median value of each target) reported in Table 1 of \citet{Isaacson2010}.
}\tablenotetext{d}{
Observation date (Universal Time) (format: YYYY-MM-DD) and middle time (Universal Time) (format: hh:mm:ss) of each exposure.
}\tablenotetext{e}{
Signal-to-noise ratio of the spectral data around Ca II 8542 \AA~(S/N(8542)) and Ca II H\&K 3934/3938\AA~(S/N(HK)).
}
\end{deluxetable*}
\end{longrotatetable}

\section{Analyses and results of the APO3.5m spectroscopic observations}\label{sec:overview-ana-results}

For the first step of our analysies we checked the binarity, as we have done in our previous studies (e.g., \citealt{Notsu2015a}; \citealt{Notsu2017}).
The details of the analyses of binarity are described in Appendix \ref{subsec:ana-binarity}. 
As a result, we regard 5 target stars as binary stars among the 18 superflare stars that we newly observed using the APO3.5m telescope in this study.
These five stars are shown in the second column of Table \ref{tab:DR25-gaia}. 
We treat the remaining 13 stars (among these 18 stars observed using APO3.5m) 
as “single stars” in this paper, since they do not show any evidence of binarity within the limits of our analyses. 
In the following, we conduct a detailed analyses for these 13 single stars.

We then estimated various stellar parameters of the target stars, using our spectroscopic data. 
The details of the analyses and results are described in Appendices \ref{subsec:ana-atmos} -- \ref{subsec:ana-Li} of this paper. 
We estimated stellar atmospheric parameters ($T_{\mathrm{eff}}$, $\log g$, and [Fe/H]) in Appendix  \ref{subsec:ana-atmos}, 
stellar radius ($R_{\mathrm{Gaia}}$ and $R_{\mathrm{spec}}$) in Appendix \ref{subsec:ana-rad}, 
and the projected rotational velocity ($v\sin i$) in Appendix \ref{subsec:ana-vsini}. 
We show measurement results of the intensity of Ca II 8542\AA~and H$\alpha$ lines in Appendix \ref{sec:measure-IRTHa}, 
Ca II H\&K $S$-index in Appendix \ref{sec:measure-Sindex}, and Ca II H\&K flux values in Appendix \ref{sec:measure-HKflux}.
We also describe the analysis of Lithium (Li) abundance of the target stars in Appendix \ref{subsec:ana-Li}. 
The resultant parameters are listed in Tables \ref{tab:spec-para} \& \ref{tab:act}.

\begin{longrotatetable}
%\startlongtable
\begin{deluxetable*}{lcccccccccccc}
\tablecaption{Parameters of the target stars estimated from our spectroscopic data. \label{tab:spec-para}} 
\tabletypesize{\footnotesize}
\tablehead{
\colhead{Star name} & \colhead{Remarks \tablenotemark{a}} & \colhead{Binarity \tablenotemark{b}} &
\colhead{$T_{\mathrm{eff}}$} & \colhead{$\log g$} & \colhead{$v_{\mathrm{t}}$} &
\colhead{[Fe/H]} &  \colhead{$v\sin i$} & \colhead{$R_{\mathrm{spec}}$ \tablenotemark{c}} & 
\colhead{$R_{\mathrm{Gaia}}$} & \colhead{$P_{\mathrm{rot}}$} & \colhead{$v_{\mathrm{lc}}$} &
\colhead{$A$(Li)} \\
\colhead{} & \colhead{} & \colhead{} & 
\colhead{(K)} & \colhead{(cm s$^{-2}$)} & \colhead{(km s$^{-1}$)} &
\colhead{} & \colhead{(km s$^{-1}$)} & \colhead{($R_{\sun}$)} & 
\colhead{($R_{\sun}$)} & \colhead{(d)} &  \colhead{(km s$^{-1}$)} &
\colhead{} 
}
\startdata 
KIC10532461 & -- & no & 5455$\pm$20 & 4.71$\pm$0.05 & 1.14$\pm$0.17 & -0.14$\pm$0.03 & 4.3$\pm$0.3 & 0.77\tablenotemark{c1} & $0.78^{+0.06}_{-0.05}$ & 7.14 & $5.5^{+0.9}_{-0.7}$ & 1.98 \\
KIC11652870 & -- & no & 5489$\pm$15 & 4.67$\pm$0.04 & 1.02$\pm$0.12 & -0.27$\pm$0.03 & $<$4 & 0.76 \tablenotemark{c1} & $0.78^{+0.06}_{-0.05}$ & 8.54 & $4.6^{+0.8}_{-0.6}$ & 2.29 \\ 
KIC9139151 & -- & no & 6063$\pm$18 & 4.31$\pm$0.04 & 1.13$\pm$0.11 & 0.09$\pm$0.03 & 4.5$\pm$0.2 & 1.24 & $1.14^{+0.05}_{-0.05}$ & 6.15 & $9.4^{+1.3}_{-1.0}$ & 2.76 \\ 
KIC4554830 & -- & no & 5642$\pm$30 & 4.22$\pm$0.07 & 1.05$\pm$0.17 & 0.35$\pm$0.05 & $<$4 & 1.34 & $1.44^{+0.06}_{-0.06}$ & 7.73 & $9.4^{+1.3}_{-1.0}$ & $<$1.0 \\ 
KIC4742436 & (1) & no & 5905$\pm$38 & 3.90$\pm$0.09 & 1.05$\pm$0.16 & -0.23$\pm$0.05 & $<$4 & 1.83 \tablenotemark{c1} & -- & 2.34 & $39.5^{+9.1}_{-8.7}$ & 2.44 \\ 
KIC4831454 & -- & no & 5637$\pm$25 & 4.65$\pm$0.06 & 1.21$\pm$0.17 & 0.04$\pm$0.04 & $<$4 & 0.88 & $0.91^{+0.07}_{-0.06}$ & 5.19 & $8.9^{+1.4}_{-1.1}$ & 2.79 \\ 
KIC8656342 & -- & no & 6028$\pm$33 & 4.11$\pm$0.07 & 1.22$\pm$0.14 & 0.02$\pm$0.05 & 4.6$\pm$0.2 & 1.57 & $1.70^{+0.07}_{-0.07}$ & 17.50 & $4.9^{+0.7}_{-0.5}$ & 2.57 \\ 
KIC9652680 & (2) & no & -- & -- & -- & -- & 38.2$\pm$0.1 & 0.88 & $1.09^{+0.08}_{-0.07}$ & 1.47 & $37.5^{+5.9}_{-4.3}$ & 3.43 \\ 
KIC6777146 & -- & no & 6158$\pm$38 & 4.39$\pm$0.08 & 1.37$\pm$0.19 & -0.02$\pm$0.06 & 7.6$\pm$0.1 & 1.11 & $1.60^{+0.12}_{-0.11}$ & 7.21 & $11.2^{+1.8}_{-1.3}$ & 2.57 \\ 
KIC8508009 & -- & no & 6301$\pm$35 & 4.51$\pm$0.07 & 1.13$\pm$0.17 & 0.18$\pm$0.05 & 4.5$\pm$0.2 & 1.19 \tablenotemark{c1} & $1.15^{+0.09}_{-0.08}$ & 2.95 & $19.7^{+3.1}_{-2.3}$ & 3.35 \\ 
KIC11610797 & -- & no & 6209$\pm$43 & 4.41$\pm$0.10 & 1.70$\pm$0.20 & 0.26$\pm$0.05 & 23.0$\pm$0.1 & 1.24 & $1.17^{+0.05}_{-0.05}$ & 1.62 & $36.5^{+4.9}_{-3.7}$ & 3.62 \\ 
KIC11253827 & -- & no & 5686$\pm$23 & 4.76$\pm$0.06 & 1.07$\pm$0.22 & 0.19$\pm$0.05 & $<$4 & 0.95 \tablenotemark{c4} & $0.90^{+0.04}_{-0.04}$ & 13.40 & $3.4^{+0.5}_{-0.4}$ & 1.34 \\ 
KIC6291837 & -- & no & 6270$\pm$48 & 4.29$\pm$0.09 & 1.48$\pm$0.23 & 0.00$\pm$0.07 & 7.9$\pm$0.1 & 1.30 & $1.42^{+0.07}_{-0.07}$ & 14.30 & $5.0^{+0.8}_{-0.6}$ & 2.71 \\ 
KIC11551430A & (1),(3) & RV(+VB) & 5589$\pm$40 & 3.95$\pm$0.10 & 1.88$\pm$0.25 & -0.20$\pm$0.07 & 18.8$\pm$0.1 & 1.81 & -- & 4.26 & $21.4^{+5.0}_{-4.8}$ & 2.33 \\ 
KIC11551430B & (3) & (VB) & 5943$\pm$98 & 5.44$\pm$0.27 & 0.81$\pm$0.50 & 0.22$\pm$0.10 & $<$4 & 1.08 \tablenotemark{c3} & -- & --  & -- & $<$2.3 \\ 
KIC4543412 & (4) & RV & 5655$\pm$73 & 4.92$\pm$0.18 & 1.95$\pm$0.38 & 0.07$\pm$0.10 & 18.3$\pm$0.1 & 0.97 \tablenotemark{c2} & $0.96^{+0.07}_{-0.06}$ & 2.16 & $22.5^{+3.5}_{-2.6}$ & $<$2.0 \\ 
KIC11128041 & (5) & SB2+RV & -- & -- & -- & -- & -- & -- & $1.21^{+0.09}_{-0.08}$ & 7.36 & $8.3^{+1.3}_{-1.0}$ & -- \\ 
KIC10338279 & (1),(5) & SB2 & -- & -- & -- & -- & -- & -- & -- & 6.53 & -- & -- \\ 
KIC7093428 & (1),(6) & VB & 6364$\pm$60 & 4.02$\pm$0.01 & 1.45$\pm$0.25 & -0.02$\pm$0.07 & 5.1$\pm$0.2 & 1.90 & -- & 0.51 & $187.8^{+43.0}_{-41.3}$ & $<$2.3 \\ 
\enddata
\tablenotetext{a}{
(1)
As for these four stars (KIC4742436, KIC11551430A, KIC10338279, and KIC7093428),
there are no stellar radius values $R_{\mathrm{Gaia}}$ reported in \citep{Berger2018} ({\it Gaia}-DR2).
For these four stars, $R_{\mathrm{spec}}$ (stellar radius estimated from $T_{\mathrm{eff}}$, $\log g$, and [Fe/H] in Appendix \ref{subsec:ana-rad}) 
values are used to calculate $v_{\mathrm{lc}}$ values, while $R_{\mathrm{Gaia}}$ values are used for the other stars.
\\
(2) We cannot estimate atmospheric parameters ($T_{\mathrm{eff}}$, $\log g$, $v_{\mathrm{t}}$, and [Fe/H]) 
of KIC9652680 since the rotational velocity of this star is large ($\gtrsim$30 km s$^{-1}$) 
and the spectral lines of these stars are too wide to estimate the atmospheric parameters in our way 
using equivalent widths of Fe I/II lines (see Appendix \ref{subsec:ana-atmos} for the details). 
When we estimate $v\sin i$, $R_{\mathrm{spec}}$, and $A$(Li) values listed in this table,
$T_{\mathrm{eff,DR25}}$, $(\log g)_{\mathrm{DR25}}$, and [Fe/H]$_{\mathrm{DR25}}$ (DR25 KSPC values) in Table \ref{tab:DR25-gaia} 
are used, and microturbulence velocity ($v_{\mathrm{t}}$) is assumed to be 1 km s$^{-1}$, as also done in \citet{Notsu2015a}.
\\
(3) KIC11551430 is found to be visual binary (KIC11551430A and KIC1551430B), as described in Appendix \ref{subsec:ana-binarity}.
We took spectra of each component of visual binary separately, and measured stellar parameters of each of them.
\\
(4)KIC4543412 shows radial velocity changes but has single-lined profiles. 
After shifting the wavelength value of each spectrum to the laboratory frame on the basis of the RV value of each observation,
we added up these shifted spectra to one co-added spectrum, as described in Appendix \ref{subsec:ana-binarity}. 
We measured stellar parameters by using this combined spectrum.
\\
(5)
We do not measure any parameters from the spectra of KIC11128041 and KIC10338279 since these stars show double-lined profiles.
\\
(6)As described in Appendix \ref{subsec:ana-binarity}, KIC7093428 has two fainter companion stars, and pixel data analyses suggest 
that flares occur on these companion stars. 
Since companion stars are faint, here we only measured stellar parameters of the main (brightest) star for reference, 
though we do not use these values in the main discussion of this paper.
}
\tablenotetext{b}{
Same as in footnote $a$ of Table \ref{tab:DR25-gaia}. 
See also Appendix \ref{subsec:ana-binarity}.
}
\tablenotetext{c}{
As for the six stars with the marks $c1$, $c2$, $c3$, or $c4$,
there are no suitable PARSEC isochrones within their original error range of 
$T_{\mathrm{eff}}$ and $\log g$ values ($\Delta T_{\mathrm{eff}}$ and $\Delta\log g$), as mentioned in Appendix \ref{subsec:ana-rad}.
We then enlarge error ranges to find the appropriate isochrone values.
As for the three stars with $c1$, we take into account $2\Delta T_{\mathrm{eff}}$ \& $2\Delta\log g$.
As for the stars with $c2$, $c3$, and $c4$, we take into account the error ranges ``$2\Delta T_{\mathrm{eff}}$ \& $3\Delta\log g$", 
``$2\Delta T_{\mathrm{eff}}$ \& $4\Delta\log g$", and ``$2\Delta T_{\mathrm{eff}}$ \& $5\Delta\log g$", respectively.
}
\end{deluxetable*}
\end{longrotatetable}

\begin{longrotatetable}
\begin{deluxetable*}{lcccccccccc}
\tablecaption{Activity indicators of the target stars. \label{tab:act}} 
\tabletypesize{\footnotesize}
\tablehead{
\colhead{Star name} & \colhead{Remarks\tablenotemark{a}} &\colhead{Binarity\tablenotemark{b}} &
\colhead{$\langle$BVAmp$\rangle$\tablenotemark{c}} & \colhead{$r_{0}$(8542)} & \colhead{$r_{0}$(H$\alpha$)} & 
\colhead{$S_{\mathrm{HK}}$} & \colhead{$\log \mathcal{F}^{'}_{\mathrm{HK}}$} & \colhead{$\log R^{'}_{\mathrm{HK}}$} &
\colhead{$\log \mathcal{F}^{+}_{\mathrm{HK}}$} & \colhead{$\log R^{+}_{\mathrm{HK}}$} 
\\
\colhead{} & \colhead{} & \colhead{} &
\colhead{(\%)} & \colhead{} & \colhead{} &
\colhead{} & \colhead{(erg/(cm$^{2}$ s))} & \colhead{} & 
\colhead{(erg/(cm$^{2}$ s))} & \colhead{}
}
\startdata
KIC10532461 & -- & no & $1.55^{+0.73}_{-0.74}$ & 0.51 & 0.33 & 0.465 & 6.70 & -4.00 & 6.66 & -4.04  \\ 
KIC11652870 & -- & no & $0.93^{+0.23}_{-0.25}$ & 0.46 & 0.32 & 0.413 & 6.68 & -4.03 & 6.64 & -4.07  \\ 
KIC9139151 & -- & no & $0.06^{+0.07}_{-0.02}$ & 0.22 & 0.19 & 0.177 & 6.40 & -4.49 & 6.27 & -4.61  \\ 
KIC4554830 & -- & no & $0.03^{+0.01}_{-0.01}$ & 0.19 & 0.18 & 0.164 & 6.11 & -4.65 & 5.93 & -4.83  \\ 
KIC4742436 & -- & no & $0.46^{+0.43}_{-0.24}$ & 0.28 & 0.30 & 0.216 & 6.52 & -4.32 & 6.43 & -4.41  \\ 
KIC4831454 & -- & no & $0.91^{+1.27}_{-0.50}$ & 0.53 & 0.32 & 0.480 & 6.77 & -3.99 & 6.73 & -4.02  \\ 
KIC8656342 & -- & no & $0.03^{+0.01}_{-0.01}$ & 0.22 & 0.20 & 0.175 & 6.39 & -4.48 & 6.26 & -4.61  \\ 
KIC9652680 & -- & no & $4.72^{+1.32}_{-2.06}$ & 0.65 & 0.51 & 0.509 & 6.95 & -3.87 & 6.92 & -3.90  \\ 
KIC6777146 & -- & no & $0.09^{+0.05}_{-0.03}$ & 0.26 & 0.23 & 0.179 & 6.47 & -4.44 & 6.35 & -4.56  \\ 
KIC8508009 & -- & no & $0.28^{+0.16}_{-0.09}$ & 0.38 & 0.28 & 0.278 & 6.74 & -4.21 & 6.69 & -4.27  \\ 
KIC11610797 & -- & no & $2.44^{+0.66}_{-0.95}$ & 0.63 & 0.47 & 0.498 & 6.99 & -3.93 & 6.96 & -3.96  \\ 
KIC11253827 & -- & no & $1.59^{+0.84}_{-0.83}$ & 0.39 & 0.26 & 0.351 & 6.61 & -4.17 & 6.55 & -4.22  \\ 
KIC6291837 & -- & no & $0.08^{+0.02}_{-0.02}$ & 0.26 & 0.22 & 0.195 & 6.56 & -4.38 & 6.47 & -4.47  \\ 
KIC11551430A & (1) & RV(+VB) & $2.65^{+1.42}_{-1.13}$ & 0.83 & 0.69 & 0.813 & 7.04 & -3.70 & 7.02 & -3.72  \\ 
KIC11551430B & (1) & VB & -- & 0.24 & 0.20 & 0.462 & 6.85 & -3.99 & 6.82 & -4.03  \\ 
KIC4543412 & (2) & RV & $4.77^{+2.02}_{-2.25}$ & 0.80 & 0.62 & 0.844 & 7.04 & -3.72 & 7.02 & -3.74  \\ 
KIC11128041 & (3) & SB2+RV & $0.48^{+0.28}_{-0.21}$ & -- & -- & -- & -- & -- & -- & --  \\ 
KIC10338279 & (3) & SB2 & $1.29^{+0.35}_{-0.40}$ & -- & -- & -- & -- & -- & -- & --  \\ 
KIC7093428 & (4) & VB & $0.13^{+0.04}_{-0.06}$ & 0.22 & 0.19 & 0.150 & 6.44 & -4.53 & 6.30 & -4.67  \\ 
\enddata
\tablenotetext{a}{
(1) KIC11551430 is found to be visual binary (KIC11551430A and KIC11551430B), as described in Appendix \ref{subsec:ana-binarity}.
We took spectra of each component of visual binary separately, and measure stellar parameters of each of them.
\\
(2)
KIC4543412 shows radial velocity changes but has single-lined profiles. 
After shifting the wavelength value of each spectrum to the laboratory frame on the basis of the RV value of each observation,
we added up these shifted spectra to one co-added spectrum, as described in Appendix \ref{subsec:ana-binarity}. 
We measured stellar parameters by using this combined spectrum.
\\
(3)
We do not measure any parameters from the spectra of KIC11128041 and KIC10338279 since these stars show double-lined profile.
\\
(4)
As described in Appendix \ref{subsec:ana-binarity}, KIC7093428 has two fainter companion stars, and pixel data analyses suggest 
that flares occur on these companion stars. 
Since companion stars are faint, here we only measured stellar parameters of the main (brightest) star for reference, 
though we do not use these values in the main discussion of this paper.
}
\tablenotetext{b}{
Same as in footnote $a$ of Table \ref{tab:DR25-gaia}. 
See also Appendix \ref{subsec:ana-binarity}.
}
\tablenotetext{c}{
The amplitude of the brightness variation. This value is calculated by taking the average of the amplitude value of each Quarter (Q2 -- Q16) data,
as also done in \citet{Notsu2015b}. 
We did not use Q0, Q1, and Q17 data since the duration of these three quarters is short (30 d) compared to 
those of the other 15 quarters ($\sim$90 d) \citep{Thompson2015}. 
The errors of $\langle$BVAmp$\rangle$ correspond to the maximum and minimum of the amplitude values of all Quarter (Q2 -- Q16) data.
}
\end{deluxetable*}
\end{longrotatetable}

\section{Discussions on the results from our APO3.5m spectroscopic observations}\label{sec:discussion-spec}
\subsection{Binarity}\label{sec:dis-binarity}

\begin{table}[ht!]
\begin{center} 
  \caption{Number of the ``single" and ``binary" stars}
  \label{tab:Nstar-Nsg-Nbin}
    \begin{tabular}{lccc}
      \hline
& Total  & Single & Binary \tablenotemark{a} \\
      \hline 
APO3.5m (This study) & 18 & 13 & 5(2) \\
Subaru/HDS \citep{Notsu2015a} & 50 & 34 & 16(4) \\
Total & 64\tablenotemark{b} & 43\tablenotemark{b} & 21(6)\\
    \hline     
    \end{tabular}
      \end{center}
\tablenotetext{a}{
Numbers in parentheses correspond to visual binary stars.
}\tablenotetext{b}{
Four single stars (KIC4742436, KIC4831454, KIC9652680, and KIC11610797) were observed
in the both studies (APO3.5m and Subaru/HDS).
} 
\end{table} 

In Section \ref{sec:overview-ana-results} and Appendix \ref{subsec:ana-binarity}, 
we described more than half (13 stars) of the 18 superflare stars that we newly conducted APO3.5m spectroscopic observations
have no obvious evidence of being in a binary system.
Combined with the results of the 50 stars that we observed with Subaru telescope in \citet{Notsu2015a}, 
we have conducted spectroscopic observations of 64 superflare stars in total (Table \ref{tab:Nstar-Nsg-Nbin}). 
Four stars (KIC4742436, KIC4831454, KIC9652680, and KIC11610797) 
among these 64 stars were observed in the both studies.
As a result, 43 stars among the total 64 solar-type superflare stars are classified as ``single" stars.
However, we need to remember here that we cannot completely exclude the possibility that 
some of these 43 ``single" superflare stars have companions since observations and analyses in this study are limited, 
as we have also described in detail in Section 4.1 of \citet{Notsu2015a}.

For example, as for the target stars with ``multiple" observations in this study, 
only those showing large radial velocity changes would likely be detected in the randomly spaced observations. 
Targets with longer period orbits would require more observations spaced accordingly.
Thus, only the short-period systems likely would be captured, and even then some would be missed by accident of poorly spaced observations 
(e.g., the case that the time differences between the observations correspond to the orbital period of the binary system).
We must note these points, but we consider that more detailed analyses of binarity are not really necessary 
for the overall discussion of stellar properties of superflare stars in this paper.

Two target stars KIC11551430 and KIC7093428 are classified as visual binary stars on the basis of the slit viewer images, Figures \ref{fig:SV-KIC11551430} \& \ref{fig:SV-KIC7093428} in Appendix \ref{subsec:ana-binarity}.
Pixel count data of these two stars suggested that 
superflares occur on the primary star KIC11551430A as for the visual binary system KIC11551430,
while flares occur on the fainter companion stars KIC7093428B or KIC7093428C as for the system KIC7093428.
Measurement results of rotation velocity $v\sin i$ and chromospheric line intensities in this study 
(listed in Tables \ref{tab:spec-para} \& \ref{tab:act})
support this suggestion.
As for KIC11551430, the primary star KIC11551430A rotates much more rapidly ($v\sin i\sim$18.8 km s$^{-1}$) 
than the companion star KIC11551430B ($v\sin i<$4 km s$^{-1}$).
Moreover, the primary KIC11551430A show strong chromospheric emissions 
while the companion KIC11551430B does not show any strong emissions (Figures \ref{fig:specCa8542}, \ref{fig:specHa}, \&~\ref{fig:specHK}).
The primary KIC11551430A can have much higher probability to generate superflares 
since superflare stars are generally well characterized with rapid rotation velocity and high chromospheric activity levels.
In contrast, the primary KIC7093428A has no strong chromospheric emissions (Figures \ref{fig:specCa8542}, \ref{fig:specHa}, \&~\ref{fig:specHK}),
and this primary star does not show any properties as superflare stars. 
It is therefore highly possible that flares occur 
on the fainter companion stars KIC7093428B or KIC7093428C mentioned above.

As seen in Table \ref{tab:DR25-gaia}, stars identified as binary stars tend to show larger number of flares.
For example, KIC11551430 and KIC4543412, which show radial velocity shifts, have $N_{\mathrm{flare}}$=89 and 13, respectively.
All the stars identified as ``single" stars have $N_{\mathrm{flare}}<10$ in Table \ref{tab:DR25-gaia}.
As a result, among the 187 superflares (on 23 stars) from {\it Kepler} 1-min cadence data in \citet{Maehara2015}, 
at least 127 events are found to occur on binary stars.
This means the large part of the superflare data of the Kepler 1-min cadence sample \citep{Maehara2015} 
are contaminated by binary stars (e.g., close binary stars).
The data from \citet{Maehara2015} are therefore not enough for investigating the possibility of superflares on Sun-like stars,
and we need to more investigations by increasing the number of single superflare stars (cf. See also the first paragraph of Section \ref{subsec:future}).

\subsection{Estimated atmospheric parameters}\label{sec:dis-atmos}

\begin{figure}[ht!]
\gridline{\fig{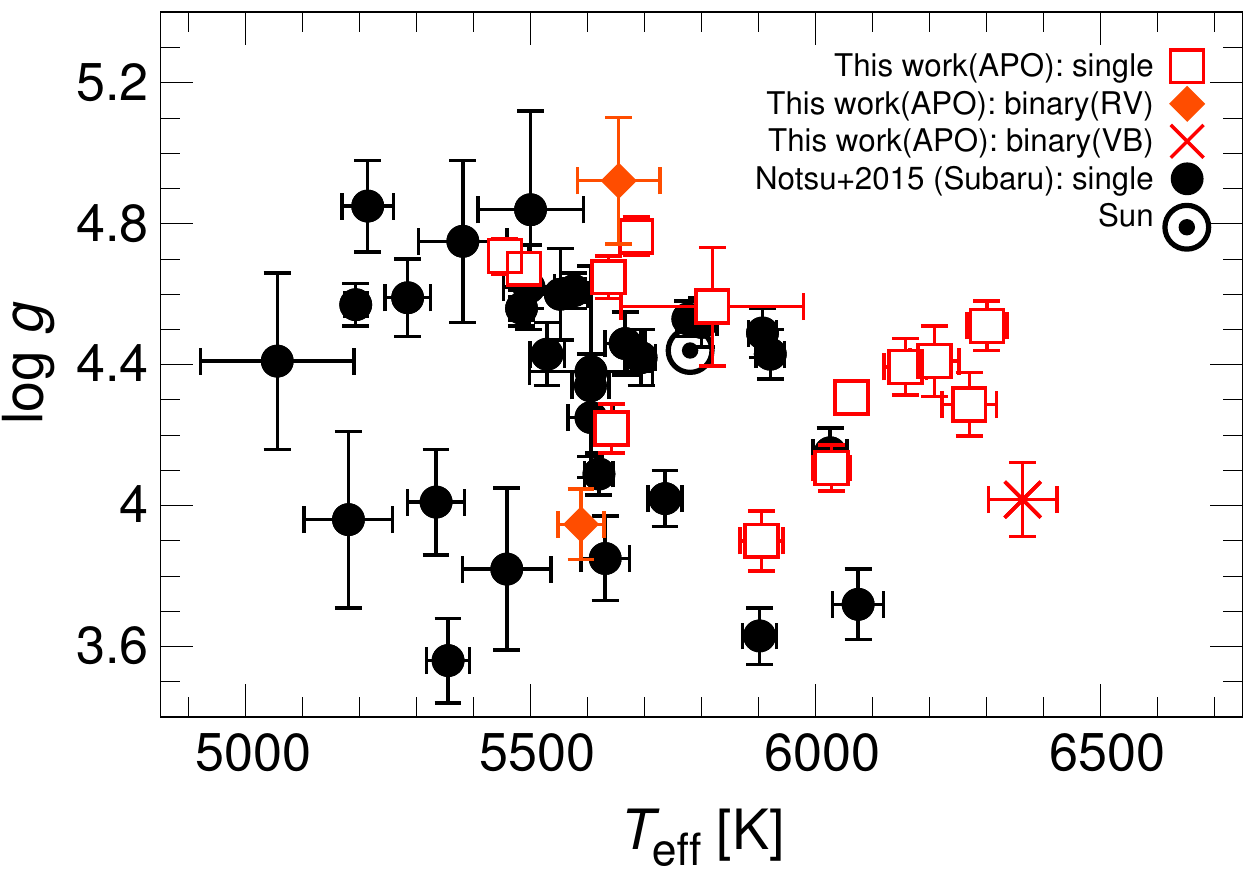}{0.50\textwidth}{(a)}
         \fig{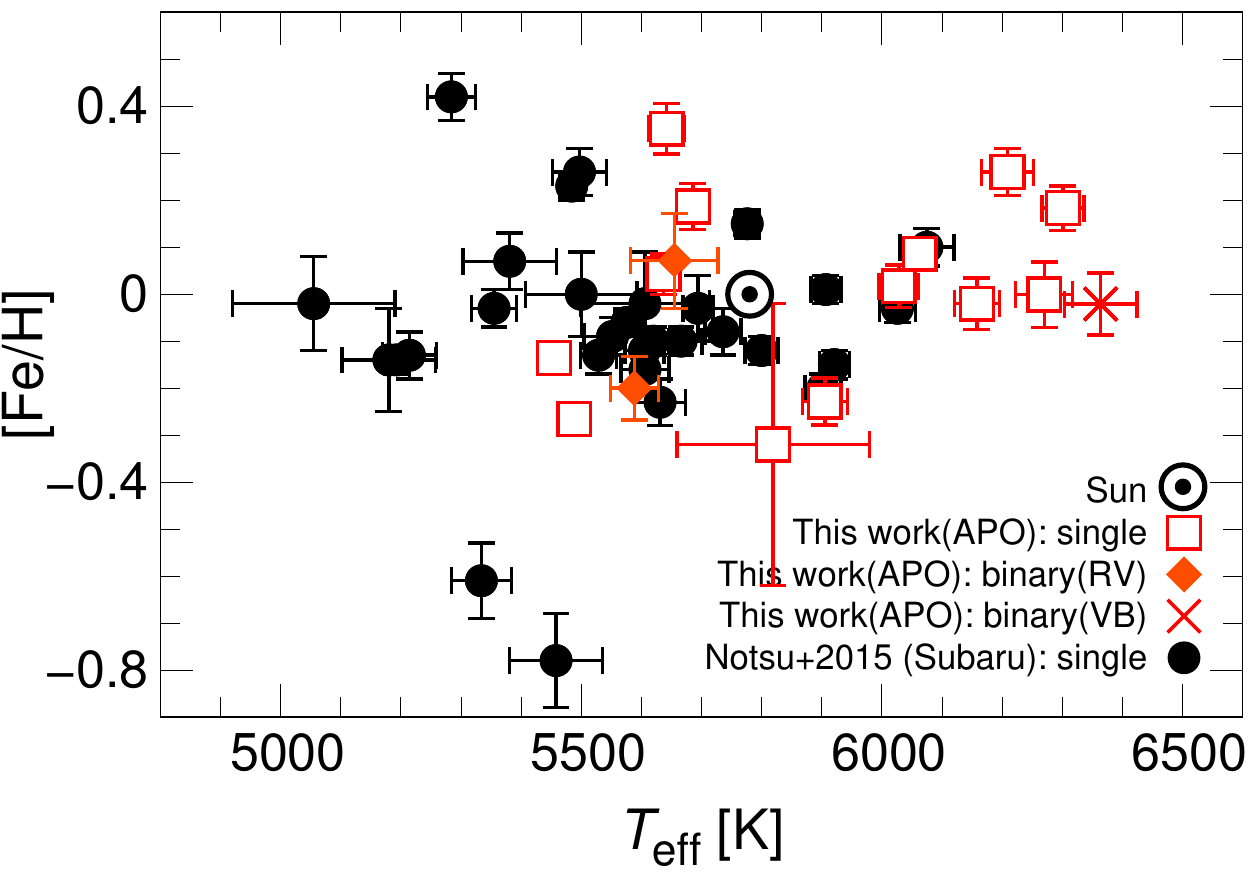}{0.50\textwidth}{(b)}}
\caption{
Temperature ($T_{\mathrm{eff}}$), surface gravity ($\log g$), and metallicity ([Fe/H]), estimated from spectroscopic observations.
The red open square points are the target superflare stars classified as single stars in Appendix \ref{subsec:ana-binarity}, 
and the orange diamonds correspond to the spectra of binary superflare stars 
that do not show any double-lined profiles (KIC11551430A and KIC4543412). 
A visual binary superflare star KIC7093428 (see Section \ref{sec:dis-binarity} for the details) is plotted for reference with the red cross mark. 
The single superflare stars that we investigated using Subaru telescope \citep{Notsu2015a},
excluding the four stars also investigated in this study (See footnote {\it f} of Table \ref{tab:DR25-gaia}), 
are plotted with the black circles. The open circle dot corresponds to the Sun.
\label{fig:Teff-logg-FeH}}
\end{figure}

We estimated the atmospheric parameters ($T_{\mathrm{eff}}$, $\log g$, and [Fe/H]) of the target superflare stars in Appendix \ref{subsec:ana-atmos}.
The measured values of the 13 single target stars of this observation are 
$T_{\mathrm{eff}}$= 5400 -- 6300K, $\log g$=3.9 -- 4.8, and [Fe/H]= $(-0.3)$ -- $(+0.3)$, respectively (Figure \ref{fig:Teff-logg-FeH}).
This indicates that the stellar parameters of these 13 single target superflare stars are roughly in the range of solar-type (G-type main-sequence) stars,
though the stars with $\log g\lesssim$4.0 are possibly sub-giant G-type stars 
(For the discussions on evolutionary state of the target stars, see also Section \ref{sec:dis-atmosHR}).
Compared with the 34 single target superflare stars that we observed in \citet{Notsu2015a}, 
the target stars tend to have a bit hotter $T_{\mathrm{eff}}$ values, and most of them are ``solar-analog" stars (early-type G-type main-sequence stars).
No clear ``metal-rich" or ``metal-poor" stars are included in the target stars of this observation.

\subsection{Rotational velocity} \label{sec:velocity}

\begin{figure}[ht!]
   \gridline{\fig{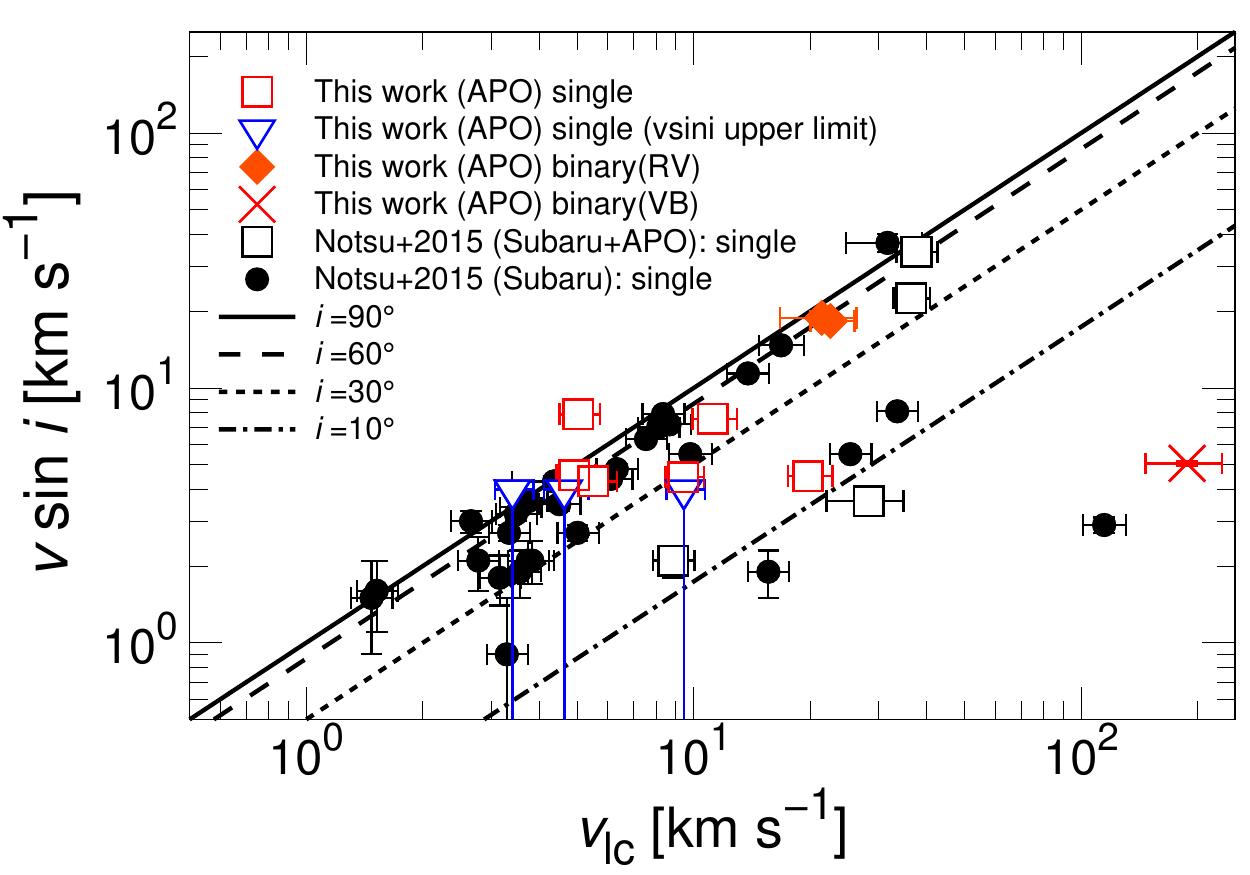}{0.55\textwidth}{}}
  \caption{
Projected rotational velocity ($v\sin i$) vs. the stellar rotational velocity ($v_{\mathrm{lc}}$)
estimated from the period of the brightness variation ($P_{\mathrm{rot}}$) and stellar radius ($R$). 
The solid line represents the case that our line of sight is perpendicular to the stellar rotation axis ($i$ = 90$^{\circ}$; $v\sin i= v_{\mathrm{lc}}$). 
We also plot three different lines, which correspond to smaller inclination angles ($i$ = 60$^{\circ}$, 30$^{\circ}$, 10$^{\circ}$). 
The open red square and blue downward triangle points are the target superflare stars classified as single stars in Appendix \ref{subsec:ana-binarity}, 
and the latter points (blue triangles) correspond to the slowly-rotating stars only with the upper limit of values $v\sin i$ ($v\sin i <4$ km s$^{-1}$) 
(cf. Appendix \ref{subsec:ana-vsini} for the details).
The orange diamonds correspond to the spectra of binary superflare stars 
that do not show any double-lined profiles (KIC11551430A and KIC4543412). 
A visual binary superflare star KIC7093428 (see Section \ref{sec:dis-binarity} for the details) is plotted for reference with the red cross mark.
The single superflare stars that we investigated using Subaru telescope \citep{Notsu2015b} are plotted with the black circles and black open squares. 
As for the four stars (KIC4742436, KIC4831454, KIC9652680, and KIC11610797) investigated both in \citet{Notsu2015b} and in this study (See footnote {\it f} of Table \ref{tab:DR25-gaia}),
the values in \citet{Notsu2015b} (Subaru) are only plotted in this table (black open squares), and the values from APO data are not used.  
This is because the wavelength resolution (i.e. accuracy of $v\sin i$ estimation) 
of Subaru/HDS data is higher than that of APO data (cf. Appendix \ref{subsec:ana-vsini}).
\label{fig:vlc-vsini}}
\end{figure}

As described above, we report the values of projected rotational velocity $v\sin i$, 
stellar radius $R$ ($R_{\mathrm{Gaia}}$ and $R_{\mathrm{spec}}$), 
and the rotation period $P_{\mathrm{rot}}$ from the brightness variation   
of the single target superflare stars (listed in Table \ref{tab:spec-para}).
Using the $R$ and $P_{\mathrm{rot}}$ values, we can estimate the rotational velocity ($v_{\mathrm{lc}}$):
\begin{eqnarray}\label{eq:vlc}
v_{\mathrm{lc}}=\frac{2\pi R}{P_{\mathrm{rot}}} \ ,
\end{eqnarray}
as also described in Section 4 of \citet{Notsu2015b}.
As for the $R$ values, we used $R=R_{\mathrm{Gaia}}$ as a first priority, 
and $R=R_{\mathrm{spec}}$ only for the stars without $R_{\mathrm{Gaia}}$ values.
As for the errors of $v_{\mathrm{lc}}$, we took the root sum squares of the two types of errors 
from $R$ (see the descriptions in Appendix \ref{subsec:ana-rad}) and $P_{\mathrm{rot}}$.
We here assume the possible errors of $P_{\mathrm{rot}}$ are about 10\%, as done in \citet{Notsu2015b}.
In this assumption, we very roughly consider typical possible differences between equatorial rotation period 
and the measured period caused by the solar-like differential rotations. For example, 
the solar latitudinal differential rotation has a magnitude of 11\% of the average rate from equator to midlatitudes ($\approx$45$^{\circ}$ latitude) 
(cf. \citealt{Benomar2018}). 
We must note here that the differences can be much larger if we observe starspots near the pole region,
and there can be solar-type stars with much larger differential rotation magnitudes \citep{Benomar2018}.
Moreover, it is sometimes hard to distinguish the correct $P_{\mathrm{rot}}$ values with $1/n \times P_{\mathrm{rot}}$ (n=2, 3, ...)  values in the periodogram,
especially when there are several starspots on the surface of the stars (cf. Figure 2 of \citealt{Notsu2013b}).
As a result, we must note here with caution that $P_{\mathrm{rot}}$ can have much larger error values, though 
the detailed investigations of each error value is beyond the scope of the overall discussions of $v\sin i$ vs. $v_{\mathrm{lc}}$ in Figure \ref{fig:vlc-vsini}.
The resultant values of $v_{\mathrm{lc}}$ are listed in Table \ref{tab:spec-para}.

We then plot $v\sin i$ as a function of $v_{\mathrm{lc}}$ in Figure \ref{fig:vlc-vsini}.
Not only the target superflare stars of this study 
(13 ``single" stars and two binary stars categorized as ``RV" and ``VB" in Appendix \ref{subsec:ana-binarity}), 
but also the 34 single superflare stars that we have observed in \citet{Notsu2015b} are plotted for reference. 
The $v_{lc}$ values of the latter 34 superflare stars are updated from the original ones in \citet{Notsu2015b}.
We newly recalculated $v_{lc}$ (cf. Equation \ref{eq:vlc}) of each target star by using $R_{\mathrm{Gaia}}$ value 
if the star has $R_{\mathrm{Gaia}}$ value reported in \citet{Berger2018}.

In Figure \ref{fig:vlc-vsini}, $v\sin i$ tends to be smaller than $v_{\mathrm{lc}}$, 
and such differences should be explained by the inclination angle effect, as mentioned in the previous studies 
(e.g., \citealt{Hirano2012}; \citealt{Notsu2015b}).
On the basis of $v\sin i$ and $v_{\mathrm{lc}}$, the stellar inclination angle ($i$) can be estimated by using the following relation:
\begin{eqnarray}\label{eq:inc}
i = \arcsin\left(\frac{v\sin i}{v_{\mathrm{lc}}}\right)\ .
\end{eqnarray}
In Figure \ref{fig:vlc-vsini}, we also show four lines indicating $i=90^{\circ}$($v\sin i=v_{\mathrm{lc}}$), $i=60^{\circ}$,
$i=30^{\circ}$, and $i=10^{\circ}$. 
First of all, almost all the stars (except for KIC6291837 with $v\sin i\sim$7.9 km s$^{-1}$ and $v_{\mathrm{lc}}\sim$5.0 km s$^{-1}$) 
in Figure \ref{fig:vlc-vsini},
the relation ``$v\sin i\lesssim v_{\mathrm{lc}}$" is satisfied.
This is consistent with the assumption that the brightness variation is caused by the rotation since the inclination 
angle effect can cause this relation ``$v\sin i\lesssim v_{\mathrm{lc}}$"
 if $v_{\mathrm{lc}}$ really corresponds to the rotation valosity (i.e. $v=v_{\mathrm{lc}}$).
This is also supported by the fact that the distribution of the data points in Figure \ref{fig:vlc-vsini} is not random,
but is concentrated between the lines of $i=90^{\circ}$ and $i=60^{\circ}$.
The distribution is expected to be much more random if the brightness variations have no relations with the stellar rotation. 
In addition, stars that are distributed in the lower right side of Figure \ref{fig:vlc-vsini} 
are expected to have small inclination angles and to be nearly pole-on stars.
Later in Section \ref{sec:spot-fene}, we see these inclination angle effects from another point of view
with the scatter plot of flare energy and starspot size.

Summarizing the results in this section, we can remark that rotation velocity values from spectroscopic results ($v\sin i$)
and those from {\it Kepler} brightness variation ($v_{\mathrm{lc}}$) are consistent, 
and this supports that the brightness variation of superflare stars is caused by the rotation.
This remark was already suggested in \citet{Notsu2015b}, 
but the conclusions are more strongly confirmed by the new spectroscopic observations and {\it Gaia}-DR2 stellar radius ($R_{\mathrm{Gaia}}$) values.

\subsection{Stellar chromospheric activity and starspots of superflare stars}  \label{sec:Ca-amp}

We measured the core intensity and flux values of chromospheric lines (Ca II 8542\AA, H$\alpha$ 6563\AA, and Ca II H\&K lines)
in Appendices \ref{sec:measure-IRTHa} -- \ref{sec:measure-HKflux}. 
These lines have been widely used for investigating stellar chromospheric activities 
(e.g., \citealt{Wilson1978}; \citealt{Linsky1979b}; \citealt{Noyes1984}; \citealt{Duncan1991}; \citealt{Baliunas1995}; \citealt{Hall2008}),
and are good indicators of stellar average magnetic fields (e.g., \citealt{Schrijver1989}; \citealt{Notsu2015b}).
Only the Ca II 8542\AA~line is mainly used in our previous studies using the Subaru telescope (\citealt{Notsu2013a} \& \citeyear{Notsu2015b}; \citealt{Nogami2014}).
In this study, Ca II H\&K lines are used with high dispersion spectra for the first time in our series of studies
of the {\it Kepler} solar-type superflare stars 
\footnote{We note here that \citet{Karoff2016} already investigated Ca II H\&K line intensity and flux values 
of superflare stars with LAMOST ``low-resolution" spectra.}.

\begin{figure}[ht!]
\gridline{\fig{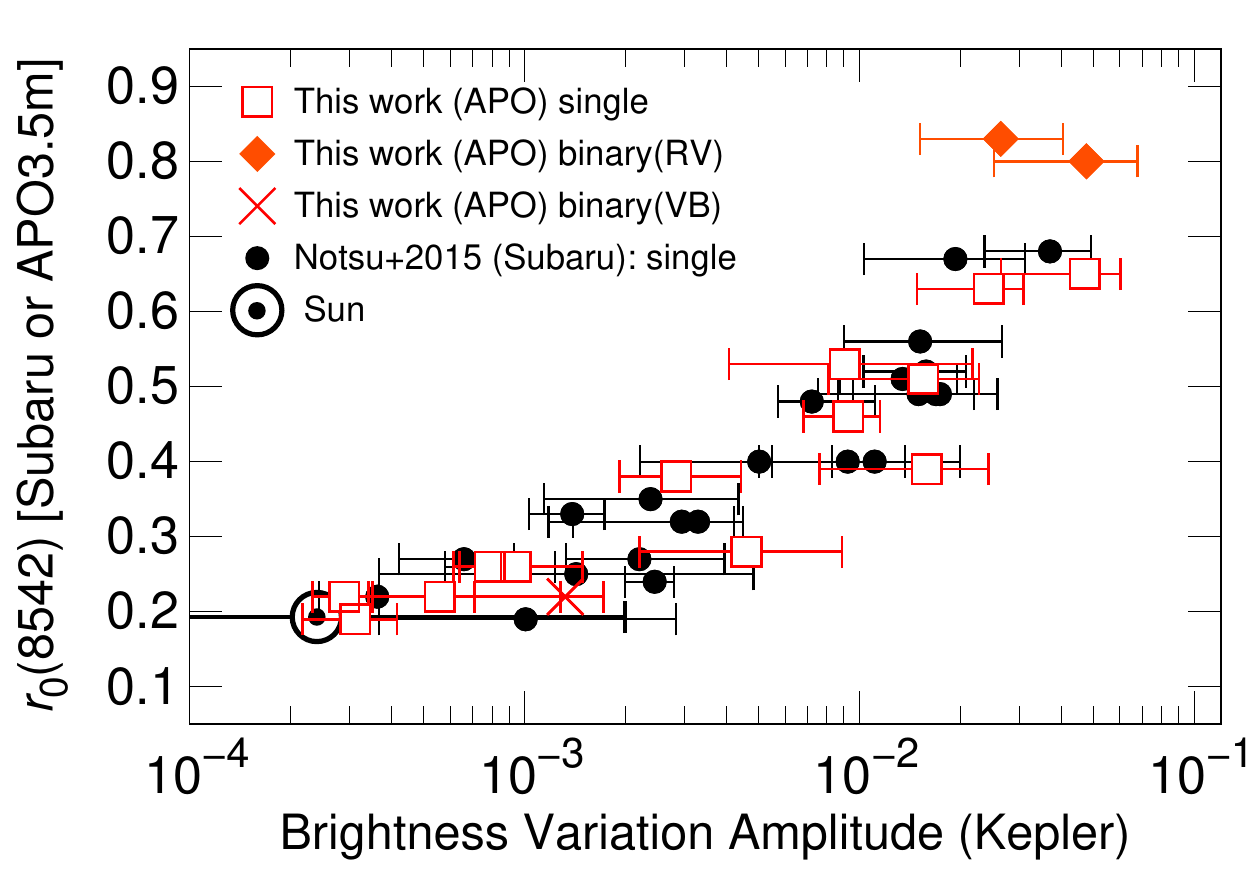}{0.5\textwidth}{(a)}
         \fig{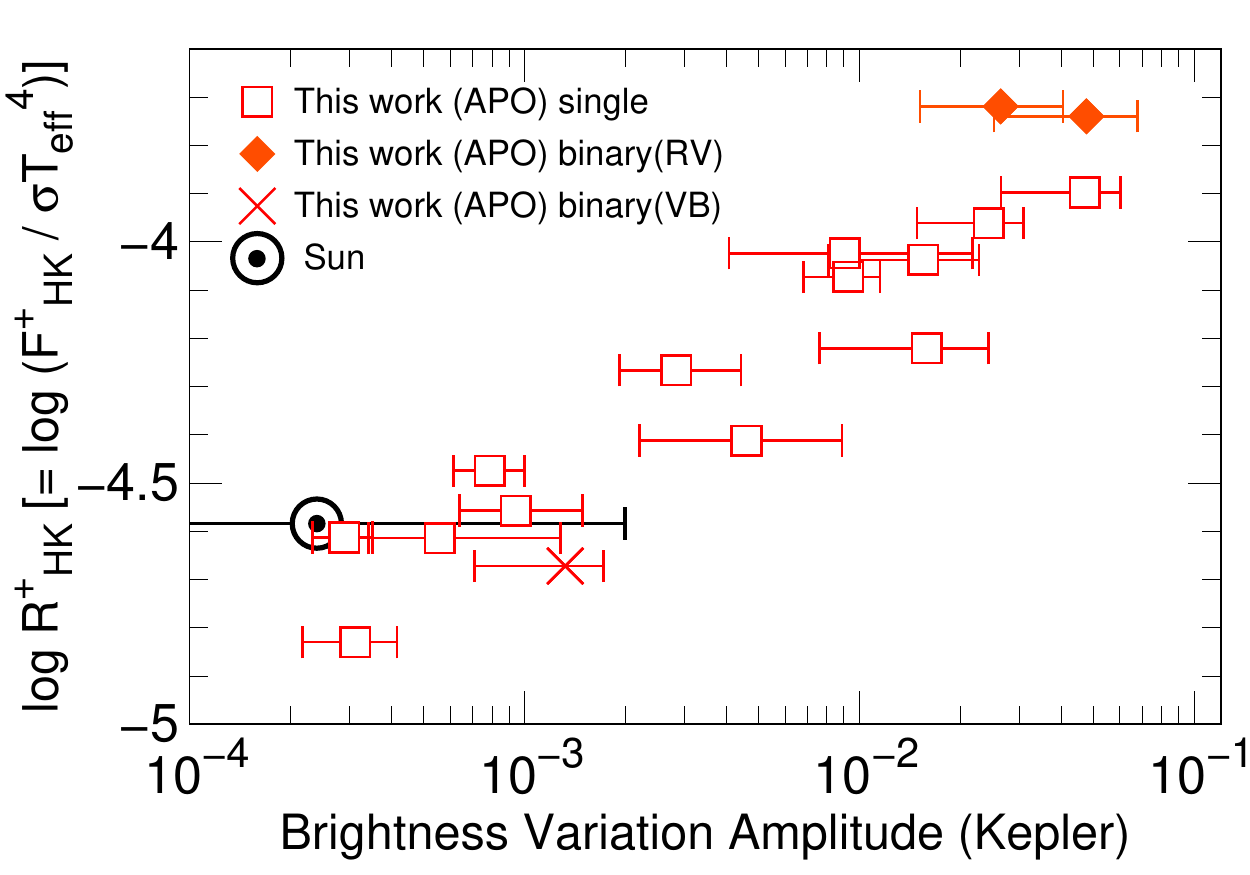}{0.5\textwidth}{(b)}}
  \caption{
\bf (a) \rm $r_{0}$(8542) (the normalized intensity at the center of Ca II 8542\AA~line) vs. 
$\langle$BVAmp$\rangle$ (the amplitude of stellar brightness variation of {\it Kepelr} data). 
\\
\bf (b) \rm Ca II H\&K activity index $\log \mathcal{R}^{+}_{\mathrm{HK}}$ ($= \mathcal{F}^{+}_{\mathrm{HK}} / \sigma T_{\mathrm{eff}}^{4}$) vs. $\langle$BVAmp$\rangle$.
\\
The red open square points are the target superflare stars classified as single stars in Appendix \ref{subsec:ana-binarity}, 
and the orange diamonds correspond to the spectra of binary superflare stars 
that do not show any double-lined profiles (KIC11551430A and KIC4543412). 
A visual binary superflare star KIC7093428 (see Section \ref{sec:dis-binarity} for the details) is plotted for reference with the red cross mark.
Only in (a), the single superflare stars that we investigated using Subaru telescope \citep{Notsu2015b}, 
excluding the four stars also investigated in this study (See footnote {\it f} of Table \ref{tab:DR25-gaia}), 
are also plotted with black circles. 
The solar value is plotted by using a circled dot point. The difference of Ca II index values between solar maximum and solar minimum is 
no larger than the size of this point (\citealt{Lockwood2007}).   
 \label{fig:CaAmp}}
  \end{figure}

In Figure \ref{fig:CaAmp}, we compare the Ca II 8542\AA~and Ca II H\&K index values ($r_{0}$(8542) and $\log \mathcal{R}^{+}_{\mathrm{HK}}$)
with the amplitude of the brightness variation of {\it Kepelr} data ($\langle$BVAmp$\rangle$).
The $r_{0}$(8542) index is the normalized intensity at the center of the Ca II 8542\AA~line (See Section \ref{sec:measure-IRTHa} for the details).
The $\log R^{+}_{\mathrm{HK}}$ index is a universal and ``pure" Ca II H\&K activity indicator introduced by \citet{Mittag2013},
and is defined as $\mathcal{R}^{+}_{\mathrm{HK}} = \mathcal{F}^{+}_{\mathrm{HK}} / \sigma T_{\mathrm{eff}}^{4}$ (Equation \ref{eq:RpHK}),
where $\mathcal{F}^{+}_{\mathrm{HK}}$ is the Ca II H\&K surface flux (unit : [erg cm$^{-2}$ s$^{-1}$]) with photospheric and ``basal" flux contributions removed
(See Appendix \ref{sec:measure-HKflux} for the details).
As described in Figure \ref{fig:8542oaoHKapo} in Appendix \ref{sec:measure-HKflux},
this $\log R^{+}_{\mathrm{HK}}$ index has a good correlation with the $r_{0}$(8542) index,
and $\log R^{+}_{\mathrm{HK}}$ can be more sensitive to the difference in the lower activity level region compared with $r_{0}$(8542).
$\langle$BVAmp$\rangle$ values are calculated by taking the average of the amplitude value of each Quarter (Q2 -- Q16) data,
as we have done in \citet{Notsu2015b}. 
The resultant values of $\langle$BVAmp$\rangle$ of the target stars of our APO3.5m observations are listed in Table \ref{tab:act}.
In Figure \ref{fig:CaAmp}, we also plot solar values as a reference, as done in \citet{Notsu2015b}, and we can see that 
as for most of the target superflare stars, Ca II 8542\AA~and Ca II H\&K index values are higher than the solar values.
This suggests that these superflare stars have higher chromospheric activities compared with the Sun.

More importantly, there is a rough positive correlation between Ca II index values and $\langle$BVAmp$\rangle$ in Figure \ref{fig:CaAmp}.
Assuming that the brightness variation of superflare stars is caused by the rotation of a star with starspots, 
the brightness variation amplitude ($\langle$BVAmp$\rangle$) corresponds to the starspot coverage of these stars. 
As mentioned above, Ca II index values are good indicators of stellar average magnetic field (or total magnetic flux).
Then, we can remark that there is a rough positive correlation between the starspot coverage from {\it Kepler} photometric data
and the stellar average magnetic field from spectroscopic data.
All the target stars expected to have large starspots on the basis of their large amplitude of the brightness variation 
show strong average magnetic field compared with the Sun. 
In other words, our assumption that the amplitude of the brightness variation corresponds to the spot coverage is supported, 
since the average magnetic field is considered to be caused by the existence of large starspots.
These results have been already confirmed in \citet{Notsu2015b} with Ca II 8542\AA~line, 
but in this study we confirmed the same conclusion with the following two updates: 
(i) Larger number of stars with Ca II 8542\AA~line data (Figure \ref{fig:CaAmp}a), 
(ii) Analyses using Ca II H\&K lines (Figure \ref{fig:CaAmp}b).

\subsection{Li abundances}

We also estimated Li abundances $A$(Li) of the target superflare stars in Appendix \ref{subsec:ana-Li}.
Li abundance is known to be a clue for investigating the stellar age of solar-type stars.
(e.g., \citealt{Skumanich1972}; \citealt{Takeda2010} \& \citeyear{Takeda2013}; \citealt{Honda2015}; \citealt{Notsu2017}).
The Li depletion is seen in the stars with $T_{\mathrm{eff}}\lesssim$5500K, 
while the stars with $T_{\mathrm{eff}}\gtrsim$6000K show no Li depletion.
This is because as the star becomes cooler, the convection zone in the stellar atmosphere evolves 
and the Li is transported to a deeper hotter zone, where Li is easily destroyed 
(p, $\alpha$ reactions: $^{7}$Li, $T\geq 2.5\times10^{6}$K; $^{6}$Li, $T\geq2.0\times10^{6}$K). 
The depletion of Li in the stellar surface caused by convective mixing increases as time passes, 
and we can remark that young stars tend to have high $A$(Li) values.

\begin{figure}[ht!]  
\gridline{
\fig{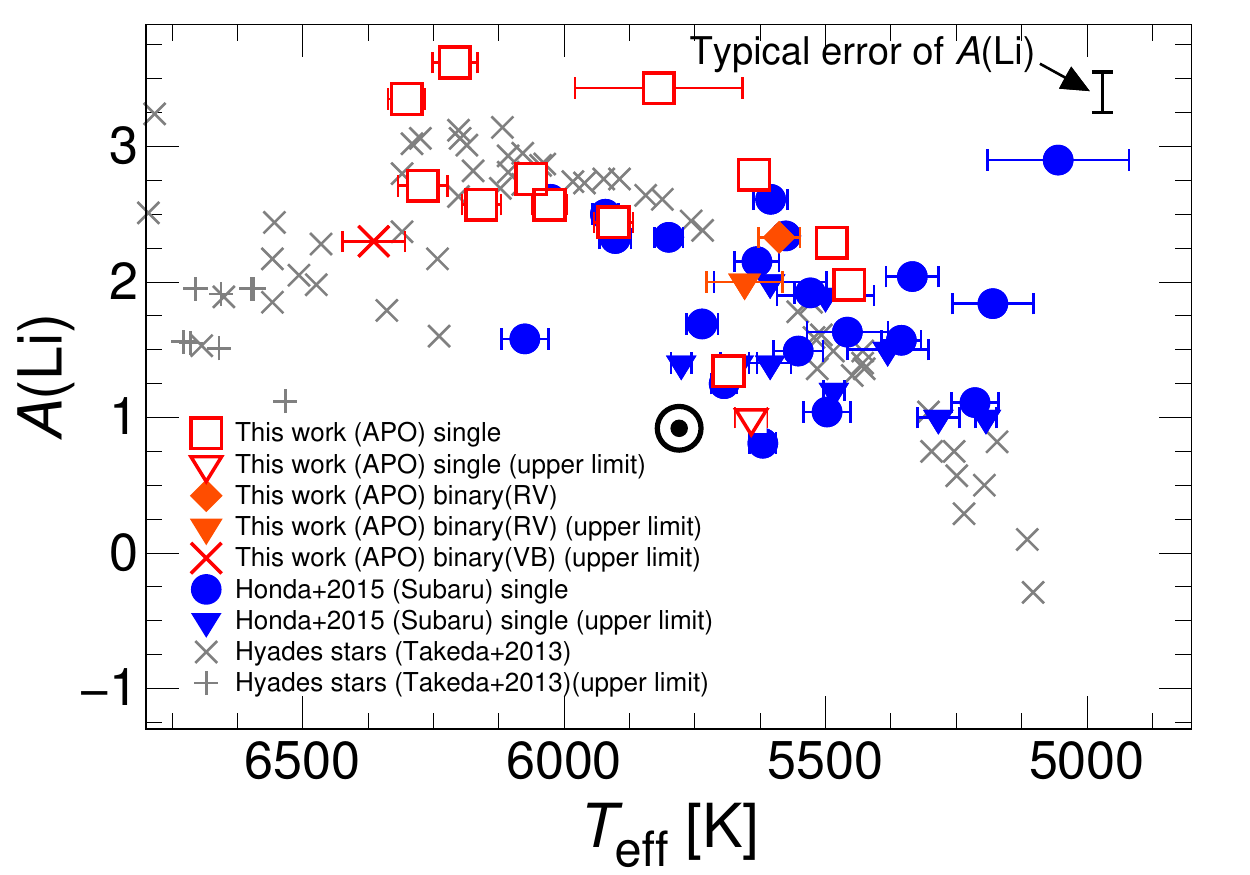}{0.55\textwidth}{}}
\caption{
 $A$(Li) vs. $T_{\mathrm{eff}}$ of the target superflare stars with the stars in the Hyades cluster. 
The age of Hyades is $6.25\times10^{8}$yr (e.g., \citealt{Perryman1998}). 
The red open square/triangle points are the target superflare stars classified as single stars in Appendix \ref{subsec:ana-binarity}, 
and the orange diamond/triangle points correspond to the spectra of binary superflare stars that do not show any double-lined profiles
(KIC11551430A and KIC4543412). 
A visual binary superflare star KIC7093428 (see Section \ref{sec:dis-binarity} for the details) is plotted for reference with the red cross mark.
The single superflare stars that we investigated using Subaru telescope \citep{Notsu2015b},
excluding the four stars also investigated in this study (See footnote {\it f} of Table \ref{tab:DR25-gaia}), 
are also plotted with blue circles/triangles. 
Among the above symbols, the triangles and the cross mark correspond to the upper limit values of $A$(Li) for the unmeasurable cases. 
We plot the data of the stars in Hyades cluster reported by \citet{Takeda2013} 
with gray cross marks (x marks and plus marks), and the plus marks correspond to the upper limit values of $A$(Li) for the unmeasurable cases.
The solar value is also plotted with a circled dot point for reference. 
Typical error value of $A$(Li) ($\sim$0.15 dex) mentioned in Appendix \ref{subsec:ana-Li} is shown with the error bar in the upper right of this figure. 
}
\label{fig:Teff-ALi}
\end{figure}

In Figure \ref{fig:Teff-ALi}, we plot $A$(Li) as a function of $T_{\mathrm{eff}}$ of the superflare stars with 
the ordinary stars in the Hyades cluster (taken from \citealt{Takeda2013})
Some of the superflare stars show high Li compared with the stars in the Hyades cluster (Figure \ref{fig:Teff-ALi}), 
and such stars are suggested to be younger than the Hyades cluster. 
The age of the Hyades cluster is estimated to be $6.25\times10^{8}$ yr (e.g., \citealt{Perryman1998}). 
It is reasonable that such young stars tend to have high activity levels and produce superflares. 
However, more than 10 target stars do not show higher A(Li) values compared with the Hyades,
and a few of them are as low as the solar value.
These results suggest that superflare stars include many young stars but also include old stars like our Sun, 
as also suggested in \citet{Honda2015}. 

\clearpage
\section{Statistical properties of {\it Kepler} solar-type superflare stars incorporating {\it Gaia}-DR2 data}\label{sec:dis-Kepler}

\subsection{Evolutionary state classifications of superflare stars}\label{sec:dis-atmosHR}

In this section, we investigate again the statistical discussions of {\it Kepler} solar-type superflare stars in our previous studies 
(\citealt{Maehara2012}, \citeyear{Maehara2015}, \& \citeyear{Maehara2017}; \citealt{Shibayama2013}; \citealt{Notsu2013b}), 
by incorporating {\it Gaia}-DR2 stellar radius estimates (reported in \citealt{Berger2018}) 
and the results of our spectroscopic observations (Subaru and APO3.5m observations).
In these previous studies, we reported 1547 superflare events on 279 solar-type (G-type main-sequence) stars 
from {\it Kepler} 30-min (long) time cadence data \citep{Shibayama2013}, 
and 189 superflares on 23 solar-type stars from  {\it Kepler} 1-min (short) time cadence data \citep{Maehara2015}
(Line 1 of Tables \ref{tab:Nstar-Nflare-30min} \& \ref{tab:Nstar-Nflare-1min}).
As also used in Appendix \ref{subsec:ana-rad}, 
\citet{Berger2018} reported the catalog of the 177,911 {\it Kepler} stars 
with the stellar radius estimates ($R_{\mathrm{Gaia}}$) estimated from {\it Gaia}-DR2 parallax values.
245 stars among the 279 solar-type superflare stars in \citet{Shibayama2013} 
and 18 stars among the 23 stars in \citet{Maehara2015} have $R_{\mathrm{Gaia}}$ values in this {\it Gaia}-DR2 catalog, respectively
(Line 2 of Tables \ref{tab:Nstar-Nflare-30min} \& \ref{tab:Nstar-Nflare-1min}).

\begin{figure}[ht!]
       \gridline{\fig{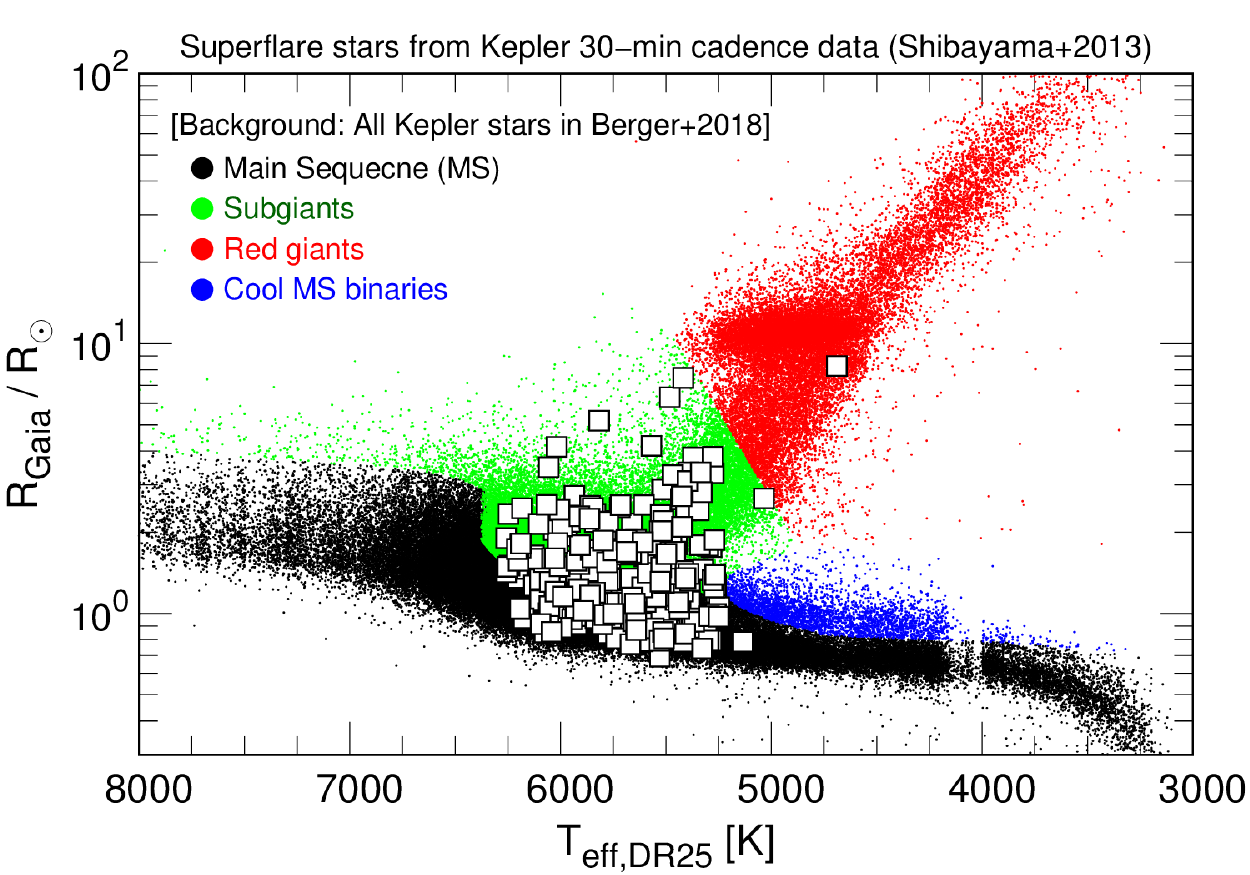}{0.48\textwidth}{(a)}
        \fig{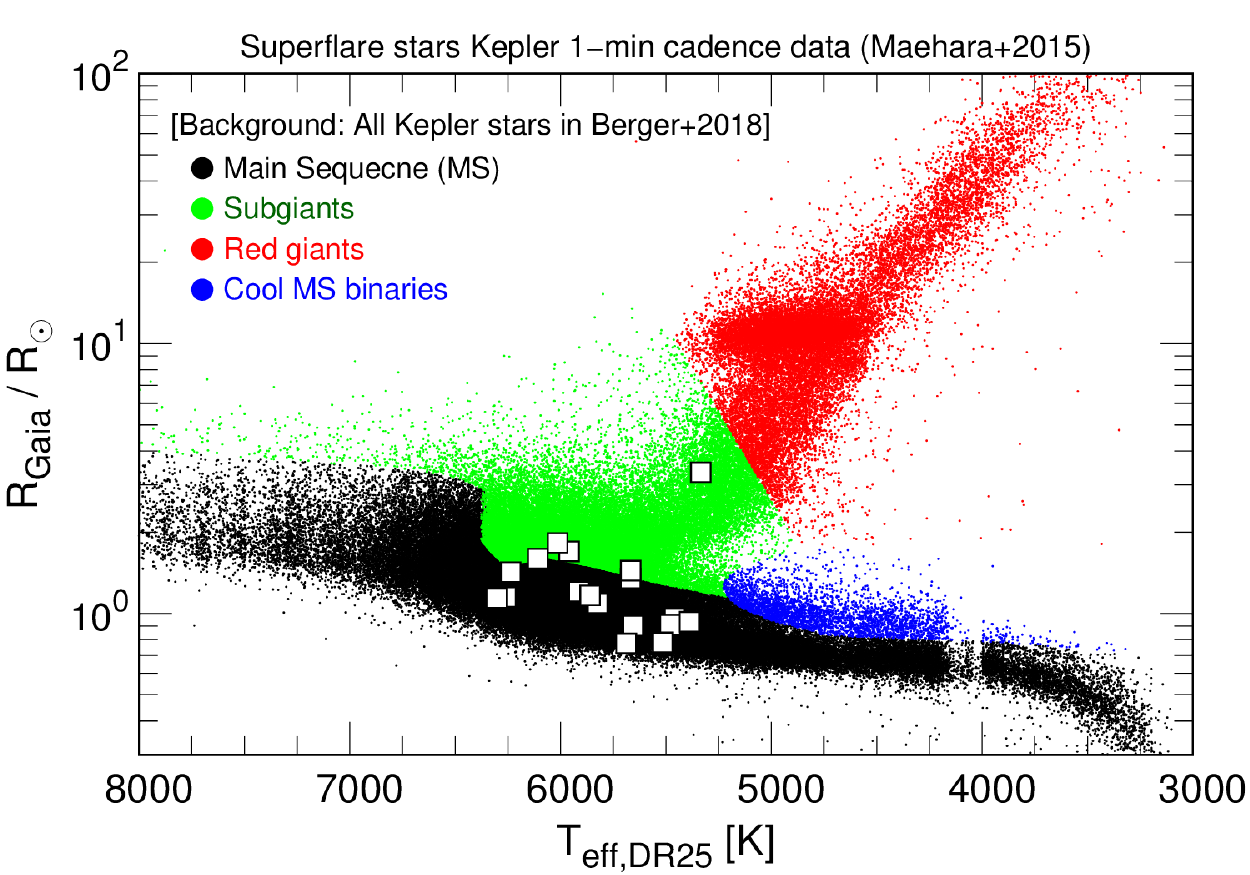}{0.48\textwidth}{(b)}}
       \gridline{\fig{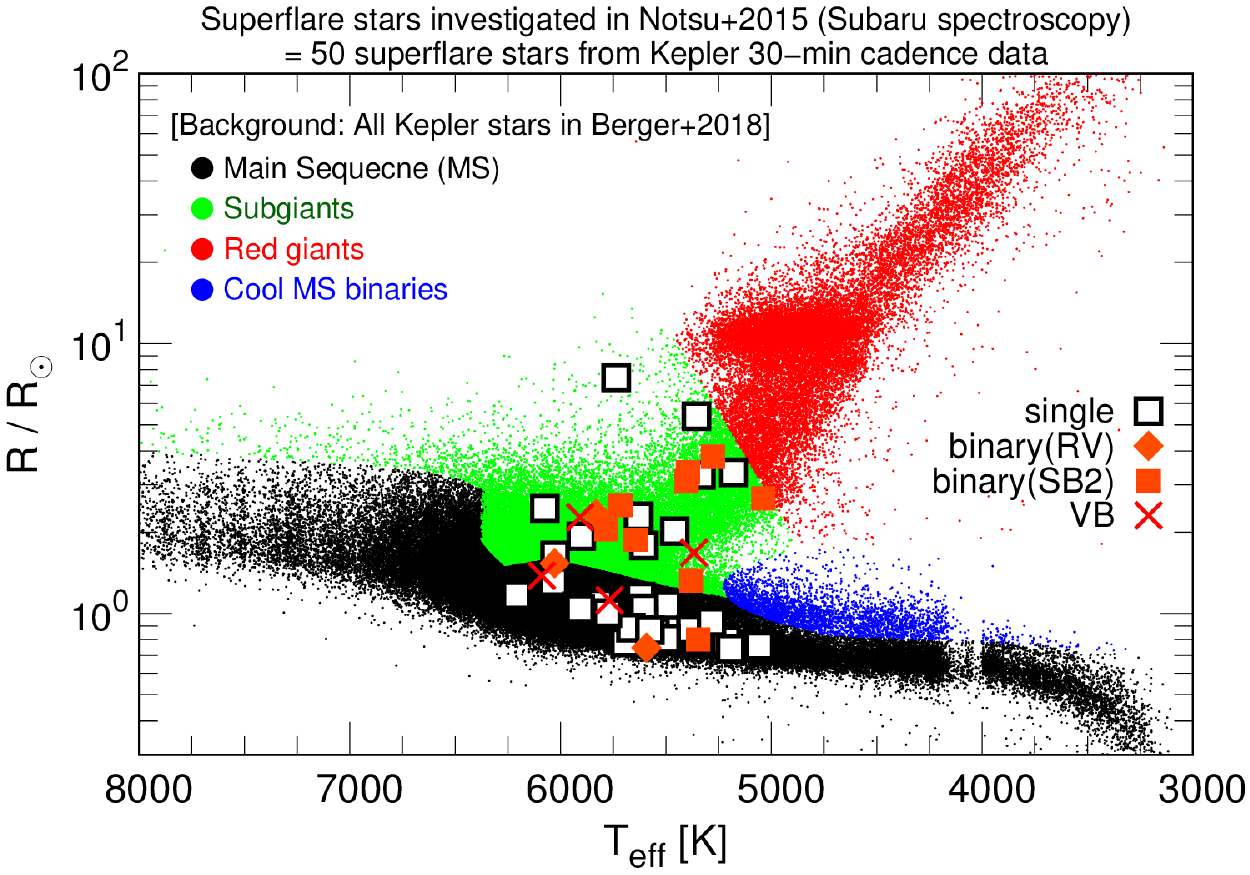}{0.48\textwidth}{(c)}
       \fig{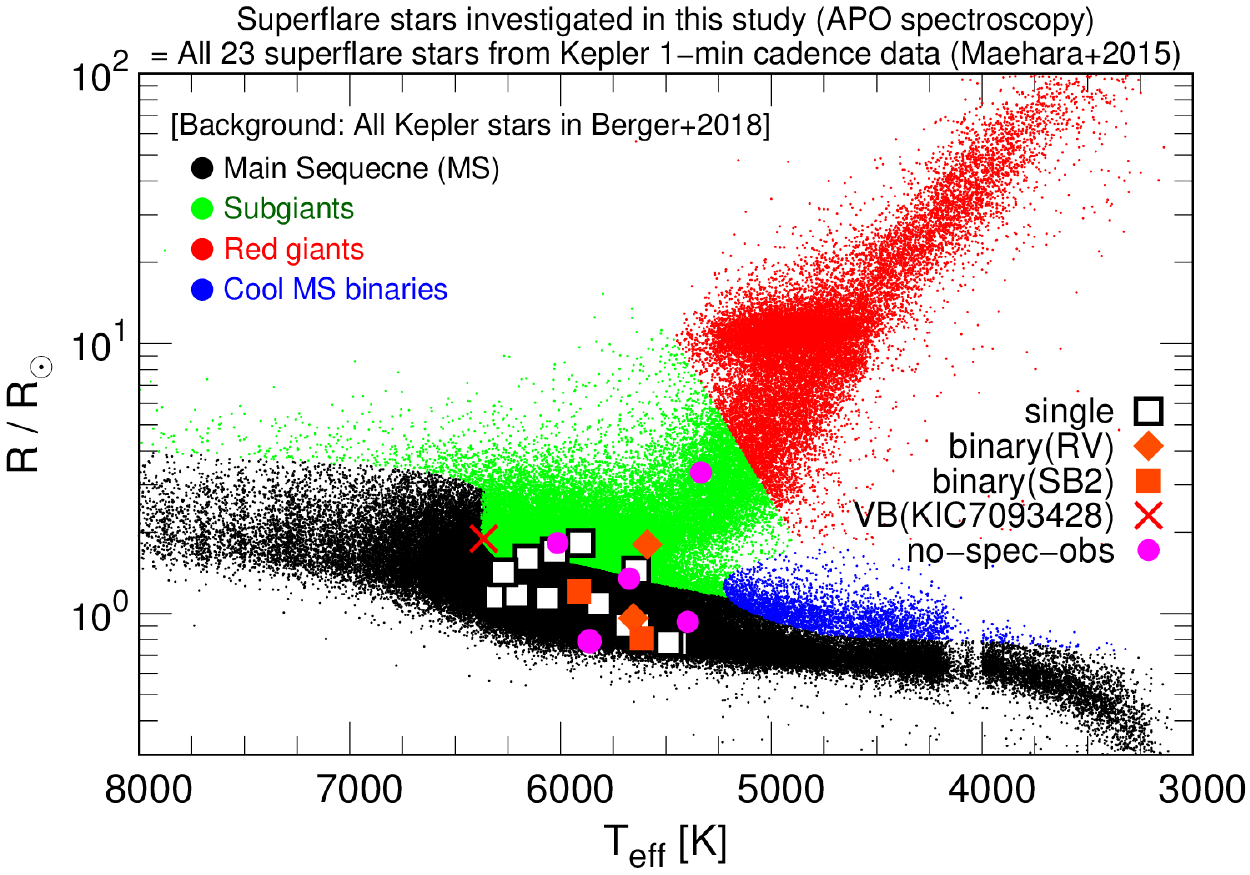}{0.48\textwidth}{(d)}} 
\caption{
Stellar radius ($R$) vs. effective temperature $T_{\mathrm{eff}}$ of superflare stars.
The background black, green, red, and blue circles are 
all the 177,911 {\it Kepler} stars listed in \citet{Berger2018} 
with $R_{\mathrm{Gaia}}$ (stellar radius based on {\it Gaia}-DR2 parallaxes) 
and $T_{\mathrm{eff, DR25}}$ (temperature from DR25 {\it Kepler} Stellar Properties Catalog: \citealt{Mathur2017}).
Each color corresponds to evolutionary state classification presented in \citet{Berger2018}:
main-sequence stars (black), subgiants (green), red giants (red), and cool main-sequence stars affected by binarity (blue).
\\
\bf (a) \rm
Superflare stars found from {\it Kepler} 30-min cadence data \citep{Shibayama2013} that have $R_{\mathrm{Gaia}}$ values in \citet{Berger2018} 
are plotted with the open square points. 
The horizontal and vertical axes are $R_{\mathrm{Gaia}}$ and $T_{\mathrm{eff, DR25}}$, resepctively.
\\
\bf (b) \rm Same as (a), but superflare stars from {\it Kepler} 1-min cadence data \citep{Maehara2015} are plotted.
\\
\bf (c) \rm Superflare stars that we have investigated with spectroscopic observations using Subaru telescope 
(\citealt{Notsu2015a}\&\citeyear{Notsu2015b}) are plotted.
The open squares are the target superflare stars classified as single stars 
(Classifications are basically the same as those used in Appendix \ref{subsec:ana-binarity} of this study).
The orange diamonds (RV) and squares (SB2) correspond to the stars classified as binary stars 
because of radial velocity shifts and double-lined profiles, respectively.
Visual binary stars are shown with the red cross marks.
As for the radius value in the vertical axis and the temperature value in the horizontal axis, we use different types of the values 
for each star. The details are described in the main text of Section \ref{sec:dis-atmosHR}.
\\
\bf (d) \rm The target superflare stars of this study, 
which are equal to all the 23 superflare stars from {\it Kepler} 1-min cadence data reported in \citet{Maehara2015}, are plotted.
The symbols and axes are used with the same way as (c).
Pink circles correspond to the five faint superflare stars that we did not get spectra with enough S/N in this study (cf. Section \ref{sec:target-obs}).
\label{fig:HR}}
\end{figure}

We plot these superflare stars on the stellar radius ($R$) vs. $T_{\mathrm{eff}}$ diagram in Figures \ref{fig:HR} (a)\&(b).
In these figures, all the {\it Kepler} stars reported in \citet{Berger2018} are also plotted for reference with the evolutionary state classifications 
(Main sequence (MS) / subgiants / red giants / cool main-sequence binaries).
In Figure \ref{fig:HR} (a), among the 245 stars found as solar-type superflare stars from {\it Kepler} 30-min time cadence data in \citet{Shibayama2013},
136 stars (55.5\%) are classified as main-sequence stars, 
while 108 stars (44.1\%) as subgiants and only one star (KIC4633721) as a red giant 
(Lines (3) -- (5) of Table \ref{tab:Nstar-Nflare-30min}).
Originally in \citet{Shibayama2013}, we used the $T_{\mathrm{eff}}$ and $\log g$ values
in the initial {\it Kepler} Input Catalog (KIC; \citealt{Brown2011})
and selected solar-type (G-type main-sequence) stars with the definition of 5100K$\leq T_{\mathrm{eff}}\leq$ 6000K and $\log g\geq 4.0$. 
$T_{\mathrm{eff}}$ and $\log g$ values in the initial KIC have large error values 
of $\pm$200K and 0.4 dex, respectively \citep{Brown2011}, and the reliability of each value can be low. 
In Figure \ref{fig:HR} (a), not only using the updated stellar radius $R_{\mathrm{Gaia}}$ from {\it Gaia}-DR2 data, 
but $T_{\mathrm{eff}}$ values are also updated with the latest DR25 {\it Kepler} Stellar Properties Catalog (DR25-KSPC: \citealt{Mathur2017}), 
which incorporates the revised method of $T_{\mathrm{eff}}$ estimation (cf. \citealt{Pinsonneault2012}).
As a result, large fraction (more than 40\%) of the stars that were originally identified as solar-type superflare stars in \citet{Shibayama2013} 
are now classified as subgiant or red giant stars.

In Figure \ref{fig:HR} (b), among the 18 stars found as solar-type superflare stars from {\it Kepler} 1-min time cadence data in \citet{Maehara2015},
13 stars are classified as main-sequence stars, while 5 stars are as subgiant stars (Lines (3) -- (5) of Table \ref{tab:Nstar-Nflare-1min}). 
In the case of these 18 stars, the revised KIC catalog \citep{Huber2014} were used 
when we selected solar-type stars in \citet{Maehara2015},
and it is possible that 
the errors of $T_{\mathrm{eff}}$ was somewhat smaller, compared with the case of the stars from \citet{Shibayama2013} in Figure \ref{fig:HR} (a).

In Figure \ref{fig:HR}(c), we plot 50 superflare stars found from {\it Kepler} 30-min cadence data 
that we have spectroscopically observed using Subaru telescope (\citealt{Notsu2015a}\&\citeyear{Notsu2015b}). 
As for the radius value in the vertical axis of these two figures, $R_{\mathrm{Gaia}}$ value is used if it exists.
If the star has no $R_{\mathrm{Gaia}}$ value in \citet{Berger2018}, 
$R_{\mathrm{spec}}$ values (in Table 4 of \citealt{Notsu2015a}) are used for single stars
and $R_{\mathrm{DR25}}$ values (from DR25-KSPC in the above) are used for binary stars.
As for the temperature value in the horizontal axis,
$T_{\mathrm{eff}}$ values estimated from spectroscopic data (in Table 4 of \citealt{Notsu2015a}) 
are used for single stars, while $T_{\mathrm{eff, DR25}}$ values are used for binary stars.
As a result, 24 stars are classified as main-sequence, 
while the remaining 10 stars \footnote{
KIC3626094, KIC6503434, KIC7420545, KIC8547383, KIC9412514,
KIC9459362, KIC10252382, KIC10528093, KIC11455711, and KIC11764567.
} are classified as subgiants, among all the 34 ``single" target stars.
Among the 16 ``binary" target stars, only 5 stars are classified as main-sequence stars,
while the remaining 11 stars \footnote{ 
KIC4045215, KIC4138557, KIC7264976, KIC7902097, KIC8479655,
KIC9653110, KIC9764192, KIC9764489, KIC10120296, KIC10453475, and KIC11560431.
} are as subgiants.

  \begin{table}[t!]
\begin{center} 
  \caption{Number of solar-type superflare stars ($N_{\mathrm{star}}$) 
  and superflares on these stars ($N_{\mathrm{flare}}$) found from {\it Kepler} 30-min time cadence data 
  (cf. Figures \ref{fig:HR}(a) and \ref{fig:spotEene-sunstar1})}\label{tab:Nstar-Nflare-30min}
    \begin{tabular}{lcc}
      \hline
 & $N_{\mathrm{star}}$ & $N_{\mathrm{flare}}$  \\
      \hline 
(1) Original data \citep{Shibayama2013} & 279 & 1547 \\
(2) Stars having $R_{\mathrm{Gaia}}$ values \citep{Berger2018} among (1) & 245 & 1402 \\
(3) Stars identified as main sequence among (2) (cf. Figure \ref{fig:HR}(a)) & 136 (55.5\%) & 496 \\
(4) Stars identified as subgiants among (2) (cf. Figure \ref{fig:HR}(a)) & 108 (44.1\%) & 905 \\
(5) Stars identified as red giants among (2) (cf. Figure \ref{fig:HR}(a)) & 1 (0.4\%) & 1 \\
(6) Stars with $T_{\mathrm{eff, DR25}}=$ 5100 -- 6000 K among (3) & 106 & 419 \\
(7) Stars that was originally early K-dwarfs in \citet{Candelaresi2014} 
\\ but are newly identified as solar-type stars with $T_{\mathrm{eff, DR25}}$ and $R_{\mathrm{Gaia}}$. & 36 & 178 \\
(8) (6)$+$(7) & 142 & 597 \\
(9) Stars having $\Delta F/F$ values in \citet{McQuillan2014} among (8) & 113 & 527 \\
    \hline     
    \end{tabular}
      \end{center}
\end{table} 

 \begin{table}[t!]
\begin{center} 
  \caption{Number of solar-type superflare stars ($N_{\mathrm{star}}$) 
  and superflares on these stars ($N_{\mathrm{flare}}$) found from {\it Kepler} 1-min time cadence data 
  (cf. Figures \ref{fig:HR}(b) and \ref{fig:spotEene-sunstar1})
  }\label{tab:Nstar-Nflare-1min}
    \begin{tabular}{lcc}
      \hline
 & $N_{\mathrm{star}}$ & $N_{\mathrm{flare}}$  \\
      \hline 
(1) Original data \citep{Maehara2015} & 23 & 187 \\
(2) Stars having $R_{\mathrm{Gaia}}$ values \citep{Berger2018} among (1) & 18 & 68 \\
(3) Stars identified as main sequence among (2) (cf. Figure \ref{fig:HR}(b)) & 13 (2, 2)  \tablenotemark{a} & 55 (15, 9)  \tablenotemark{a} \\
(4) Stars identified as subgiants among (2) (cf. Figure \ref{fig:HR}(b)) & 5 (0, 2)  \tablenotemark{a} & 13 (0, 9)  \tablenotemark{a} \\
(5) Stars identified as red giants among (2) (cf. Figure \ref{fig:HR}(b)) & 0 & 0 \\
(6) Stars with $T_{\mathrm{eff, DR25}}=$ 5100 -- 6000 K among (3) & 10 & 51 \\
(7) Stars having $\Delta F/F$ values in \citet{McQuillan2014} among (6) & 8 & 48 \\
    \hline     
    \end{tabular}
      \end{center}
\tablenotetext{a}{
Numbers in parentheses show the stars identified as binary and
the stars that we have not conducted spectroscopic observations because they are too faint, respectively, 
on the basis of our APO3.5m spectroscopic observations in this study
(see also Figure \ref{fig:HR}(d) and Table \ref{tab:DR25-gaia}).
}
\end{table}

In Figure \ref{fig:HR}(d), we plot the target stars of the APO3.5m spectroscopic observation of this study,
which are equal to all the 23 superflare stars found from {\it Kepler} 1-min cadence data \citep{Maehara2015}.
The radius and temperature values in the vertical and horizontal axes are plotted with the basically same way as Figure \ref{fig:HR}(c),
but $R_{\mathrm{spec}}$ values are used for ``RV" binary stars if the stars have no $R_{\mathrm{Gaia}}$ values,
and $T_{\mathrm{eff}}$ estimated from spectroscopic data are used for all ``RV" binary stars.
As a result, 9 stars are classified as main-sequence, while 4 stars \footnote{
KIC4554830, KIC4742436, KIC6777146, and KIC8656342. 
} are as subgiants, among all the 13 ``single target stars.
Among the 5 ``binary" target stars, three stars are classified as main-sequence,
while two stars \footnote{KIC11551430A and KIC7093428.} are as subgiants.
Moreover, among the remaining 5 stars that we did not get spectroscopic data with enough S/N in this study (cf. Section \ref{sec:target-obs}) \footnote{
As also done for ``SB2" and ``VB" binary stars, $R_{\mathrm{gaia}}$ and $T_{\mathrm{eff, DR25}}$ values are used if $R_{\mathrm{gaia}}$ exists.
If not, $T_{\mathrm{eff, DR25}}$ and $T_{\mathrm{eff, DR25}}$ values are used.
}, three stars are classified as main-sequence, while two stars \footnote{KIC10528093 and KIC10745663.} are as subgiants.
In the following sections, we update statistical studies by using the data of the stars that we can newly classify as 
solar-type (G-type main-sequence) stars.

\subsection{Starspot size and energy of superflares}\label{sec:spot-fene}

\begin{figure}[htbp]
\gridline{\fig{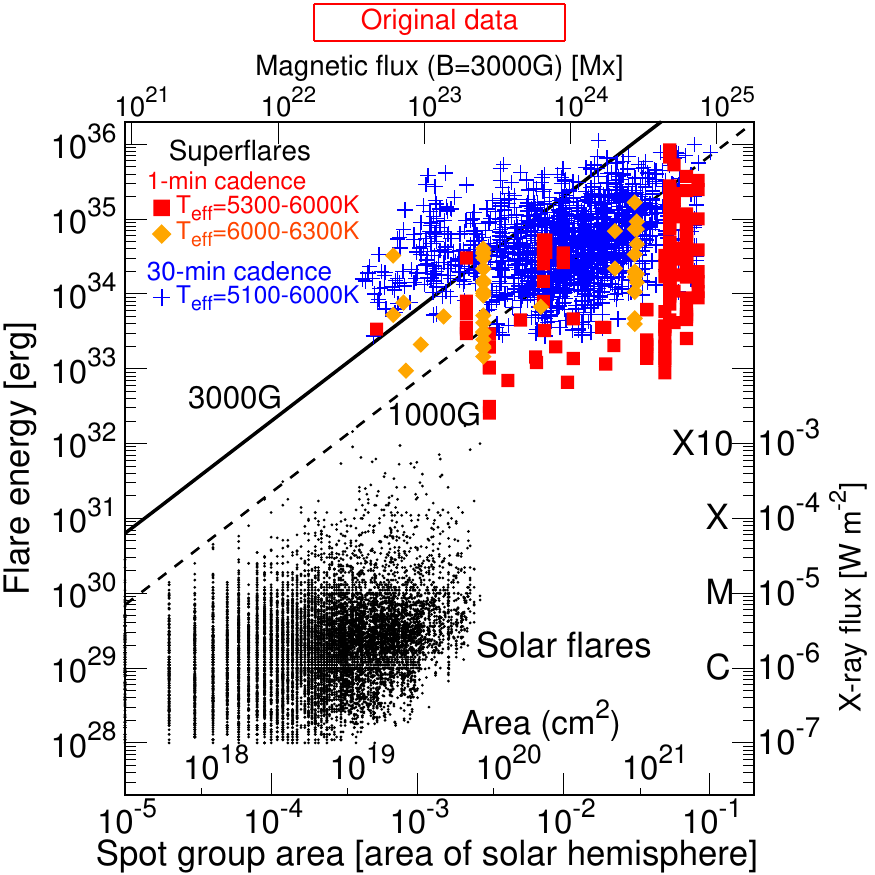}{0.5\textwidth}{(a)}
\fig{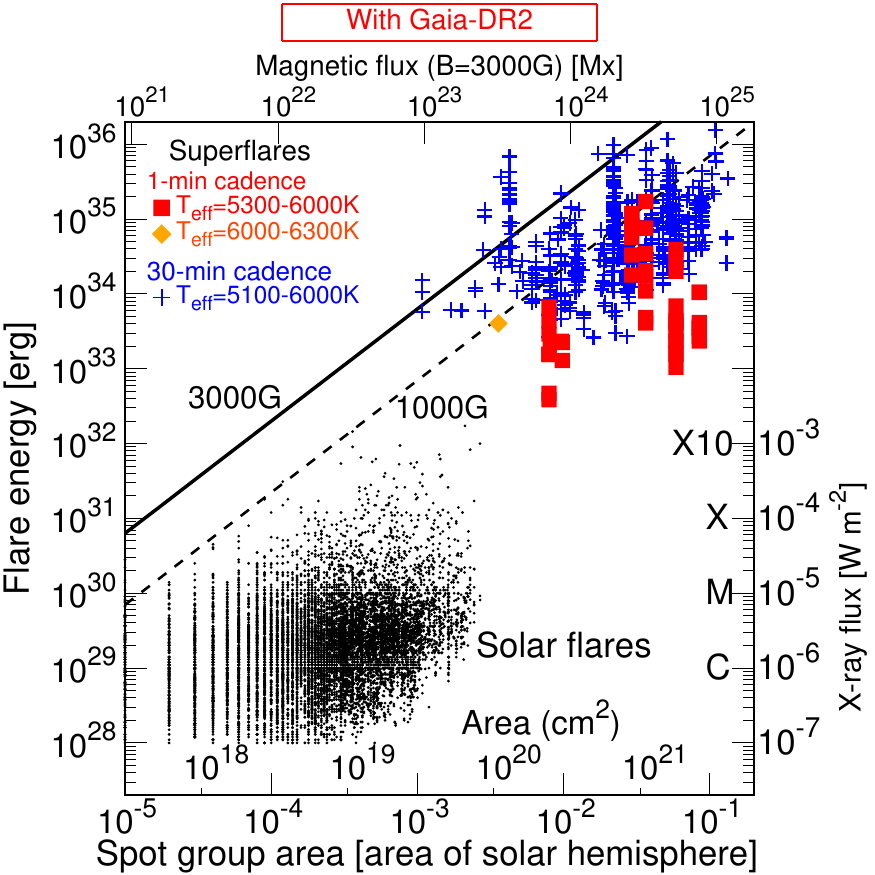}{0.5\textwidth}{(b)}}
\caption{
Scatter plot of flare energy ($E_{\mathrm{flare}}$) as a function of spot area ($A_{\mathrm{spot}}$) for solar flares and superflares.
The lower and upper horizontal axes indicate the area of the spot group 
in the unit of the area of the solar hemisphere ($A_{1/2\sun}\sim 3\times 10^{22}$ cm$^{2}$) and the magnetic flux for $B$=3000 G.
The vertical axis represents the bolometric energy released by each flare. 
The data of solar flares are completely the same as those in our previous studies 
(e.g., Figure 5 of \citealt{Maehara2015}).
We assumed that bolometric energies of B, C, M, X, and X10 class solar flares are 
$10^{28}$, $10^{29}$, $10^{30}$, $10^{31}$, and $10^{32}$ erg 
from observational estimates of typical solar flare energies (See the main text for the details).
The black solid and dashed lines correspond to the analytic relationship 
between the $E_{\mathrm{flare}}$ and $A_{\mathrm{spot}}$ from Equation (\ref{eq:Eflare-Emag}) for $B$=3000 G and 1000 G, respectively.
\\
\bf (a) \rm
The original data of superflares on solar-type (G-type main-sequence) stars presented in our previous studies 
(\citealt{Shibayama2013} \&~\citealt{Maehara2015}) are plotted.
Blue crosses indicate 1549 superflares (on 279 stars) detected from {\it Kepler} 30-min cadence data \citep{Shibayama2013},
while red squares and orange diamonds are those (187 flares on 23 stars) detected from 1-min cadence data \citep{Maehara2015}.
Among the 187 flares from 1-min cadence data, stars with $T_{\mathrm{eff}}$ = 6000 -- 6300 K, 
which are not included in the range of solar-type stars in this study ($T_{\mathrm{eff}}$ = 5100 -- 6000 K), 
are distinguished with orange diamonds.
\\
\bf (b) \rm
The data of superflares on solar-type stars updated in this study using $T_{\mathrm{eff, DR25}}$ and $R_{\mathrm{Gaia}}$ values.
Symbols are used with the same way as (a). 
In addition to the data of solar-type stars (blue crosses and red squares), 
the data of one main-sequence star with $T_{\mathrm{eff}}$ = 6000 -- 6300 K (KIC8508009) from {\it Kepler} 1-min cadence data is also calculated and plotted for reference.
\label{fig:spotEene-sunstar1}}
\end{figure}

Most of superflare stars show large amplitude brighteness variations, 
and they suggest that the surface of superflare stars are covered by large starspots (cf. Section \ref{sec:Ca-amp}). 
Figure \ref{fig:spotEene-sunstar1} shows the scatter plot of flare energy ($E_{\mathrm{flare}}$)
as a function of the spot group area ($A_{\mathrm{spot}}$).
The values of solar flares in Figure \ref{fig:spotEene-sunstar1} are 
the same as those in our previous studies (\citealt{Shibata2013}; \citealt{Notsu2013b}; \citealt{Maehara2015}). 
We estimate bolometric energies (white light flare (WLF) energies) of solar flares from {\it GOES} soft X-ray (SXR) flux values.
We use the relation that WLF energy ($E_{\mathrm{WL}}$) 
is in proportional to SXR flux ($F_{\mathrm{SXR}}$): $E_{\mathrm{WL}}\propto F_{\mathrm{SXR}}$,
on the basis of the results and detailed descriptions in Section 4.1 of \citet{Namekata2017}.
This relation is supported from the observational comparisons between WL and SXR data during flares (e.g., Figure 4 of \citealt{Namekata2017}),
and are related with the well-known relation between Hard X-ray (HXR) flux and SXR flux during a flare (``Neupert effect" : \citealt{Neupert1968}). 
As a result, we here assume that bolometric energies of B, C, M, X, and X10 class solar flares are 
$10^{28}$, $10^{29}$, $10^{30}$, $10^{31}$, and $10^{32}$ erg. 

In Figure \ref{fig:spotEene-sunstar1}, $A_{\mathrm{spot}}$ of superflare stars were estimated from the normalized amplitude of light variations ($\Delta F/F$) 
by using the following equation (\citealt{Shibata2013}; \citealt{Notsu2013b}):
\begin{eqnarray}\label{eq:aspot}
A_{\mathrm{spot}}=\frac{\Delta F}{F} A_{\mathrm{star}}\left[1-\left(\frac{T_{\mathrm{spot}}}{T_{\mathrm{star}}}\right)^{4}\right]^{-1} \ ,
\end{eqnarray}
where $A_{\mathrm{star}}$ is the apparent area of the star, and $T_{\mathrm{spot}}$ and $T_{\mathrm{star}}$
are the temperature values of the starspot and photosphere of the star.
In Figure \ref{fig:spotEene-sunstar1}(a), 
original values of superflares on solar-type stars reported in \citet{Shibayama2013} 
(from {\it Kepler} 30-min cadence data) 
and \citet{Maehara2015} (from {\it Kepler} 1-min cadence data) are plotted \footnote{
Figure \ref{fig:spotEene-sunstar1}(a) is basically the same as Figure 5 of \citet{Maehara2015},
but we plot it again because of the following two reasons.
It is helpful for readers to compare the original figure with the new one updated in this study.
Second, the solid lines corresponding to Equation (\ref{eq:Eflare-Emag}) were mistakenly plotted (the vertical axis values are factor 7 smaller than 
the correct values) in Figure 5 of \citet{Maehara2015}, and then it is better to show the revised figure here. 
} for comparisons with the results of this study. 
In these previous papers, we defined the amplitude ($\Delta F/F$) as the brightness range normalized by the average stellar brightness, 
in which the lower 99\%~of the distribution of the brightness difference from average, except for the flares, are included. 
$A_{\mathrm{star}}$ and $T_{\mathrm{star}}$ values were based on the stellar radius and temperature values used in these previous papers \footnote{
In \citet{Shibayama2013}, we used the values taken from the first {\it Kepler} Input Catalog (KIC: \citealt{Brown2011}). 
In \citet{Maehara2015}, we used those from the latest one (the revised KIC) at that time \citep{Huber2014}.
},
and we assumed $T_{\mathrm{spot}}$ = 4,000 K.

In this study, we newly updated both of the $E_{\mathrm{flare}}$ and $A_{\mathrm{spot}}$ values
by using the latest $T_{\mathrm{eff, DR25}}$ and $R_{\mathrm{Gaia}}$ values described in Section \ref{sec:dis-atmosHR}.
First, from the superflare stars that we previously reported from {\it Kepler} 30-min \& 1-min time cadence data 
(\citealt{Shibayama2013}; \citealt{Candelaresi2014}; \citealt{Maehara2015}) 
we selected the stars classified again as solar-type stars (main-sequence stars with $T_{\mathrm{eff}}$ = 5100 -- 6000 K) 
on the basis of  $T_{\mathrm{eff, DR25}}$ and $R_{\mathrm{Gaia}}$ values (cf. Figure \ref{fig:HR}) 
(Line (6) of Tables \ref{tab:Nstar-Nflare-30min} and \ref{tab:Nstar-Nflare-1min}).
Not only ``subgiants" and ``red giants", 
but also stars identified as ``cool main-sequence binary" stars in Figure \ref{fig:HR} are not included in the ``solar-type" stars classified in this study.

As for the selection from {\it Kepler} 30-min cadence data, 
we used the data not only from \citet{Shibayama2013} (described in Figure \ref{fig:HR} (a)) but also from \citet{Candelaresi2014}.
We investigated superflares on solar-type stars in \citet{Shibayama2013} 
and those on G,K,M-type stars in \citet{Candelaresi2014}.
Exactly the same {\it Kepler} dataset (the first 500 days : Quarters 0 -- 6) are used for the flare surveys in these both studies,
and the superflare data of \citet{Shibayama2013} (solar-type stars) exactly correspond to the subset of those of \citet{Candelaresi2014} (G,K,M-type stars).
Then in this paper, we newly selected solar-type superflare stars not only from the original solar-type superflare stars in \citet{Shibayama2013} (cf. Figure \ref{fig:HR}(a)), 
but also from the stars originally identified as K,M-type stars in \citet{Candelaresi2014}.
36 superflare stars originally identified as early K-type stars 
(with $T_{\mathrm{eff}}$ = 4900 -- 5100K) in \citet{Candelaresi2014} are newly categorized as solar-type stars 
with the revised $T_{\mathrm{eff, DR25}}$ and $R_{\mathrm{Gaia}}$ values (Line (7) of Tables \ref{tab:Nstar-Nflare-30min}).
We note that these superflare stars were not included in Figure \ref{fig:HR}(a), 
but are included in the following discussions of solar-type superflare stars in this paper.
As a result, 142 solar-type stars (with 597 superflares in total) from {\it Kepler} 30-min cadence data (\citealt{Shibayama2013}; \citealt{Candelaresi2014}) 
(Line (8) of Table \ref{tab:Nstar-Nflare-30min})
and 10 solar-type stars (with 51 superflares in total) from 1-min cadence data \citep{Maehara2015} are selected 
(Line (6) of Table \ref{tab:Nstar-Nflare-1min}).

We then recalculated $E_{\mathrm{flare}}$ and $A_{\mathrm{spot}}$ for these selected data.
The recalculation of $E_{\mathrm{flare}}$ was done by applying these updated $T_{\mathrm{eff, DR25}}$ and $R_{\mathrm{Gaia}}$ values
to the equations presented in Section 2.3 of \citet{Shibayama2013}.
As for $A_{\mathrm{spot}}$, we used the new methods presented in \citet{Maehara2017}.
$\Delta F/F$ values taken from \citet{McQuillan2014} were used, 
and we applied the following relation on the temperature difference between photosphere and spot ($T_{\mathrm{star}} - T_{\mathrm{spot}}$) 
deduced from \citet{Berdyugina2005}:
\begin{eqnarray}\label{eq:tspot-tstar}
\Delta T(T_{\mathrm{star}}) 
&=& T_{\mathrm{star}}-T_{\mathrm{spot}} \nonumber \\
&=& 3.58 \times 10^{-5}T_{\mathrm{star}}^{2} +0.249T_{\mathrm{star}} -808 \ .
\end{eqnarray}
As a result, we got the resultant values of the updated $E_{\mathrm{flare}}$ and $A_{\mathrm{spot}}$ values 
for 113 stars (with 527 flares in total) from {\it Kepler} 30-min cadence data (\citealt{Shibayama2013}; \citealt{Candelaresi2014}) 
(Line (9) of Table \ref{tab:Nstar-Nflare-30min})
and 8 stars (with 48 flares in total) from 1-min cadence data \citep{Maehara2015} 
(Line (7) of Table \ref{tab:Nstar-Nflare-1min}).
We note that only the stars having $\Delta F/F$ values in \citet{McQuillan2014} are selected here.
These resultant superflare values are plotted in Figure \ref{fig:spotEene-sunstar1}(b).

\begin{figure}[htbp]
\gridline{  \fig{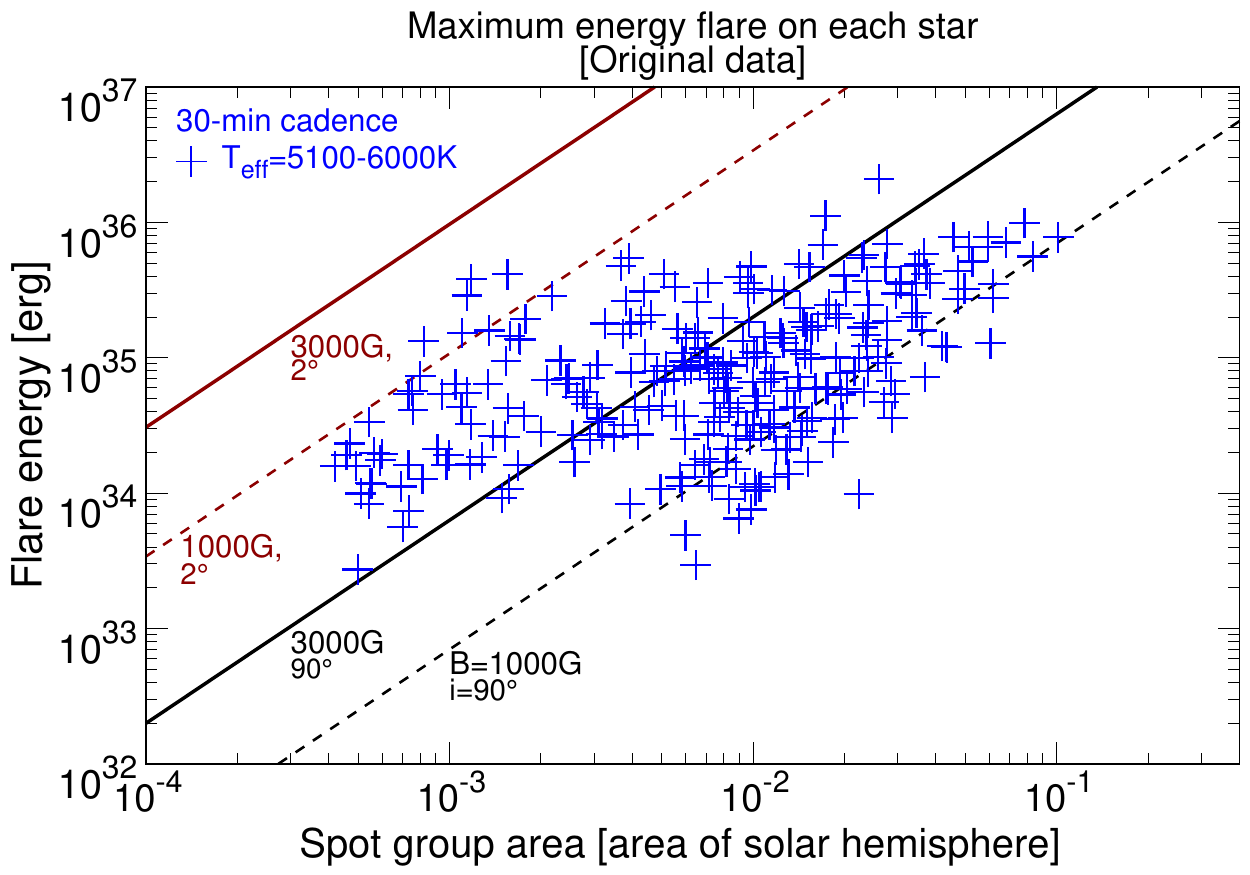}{0.5\textwidth}{(a)}
\fig{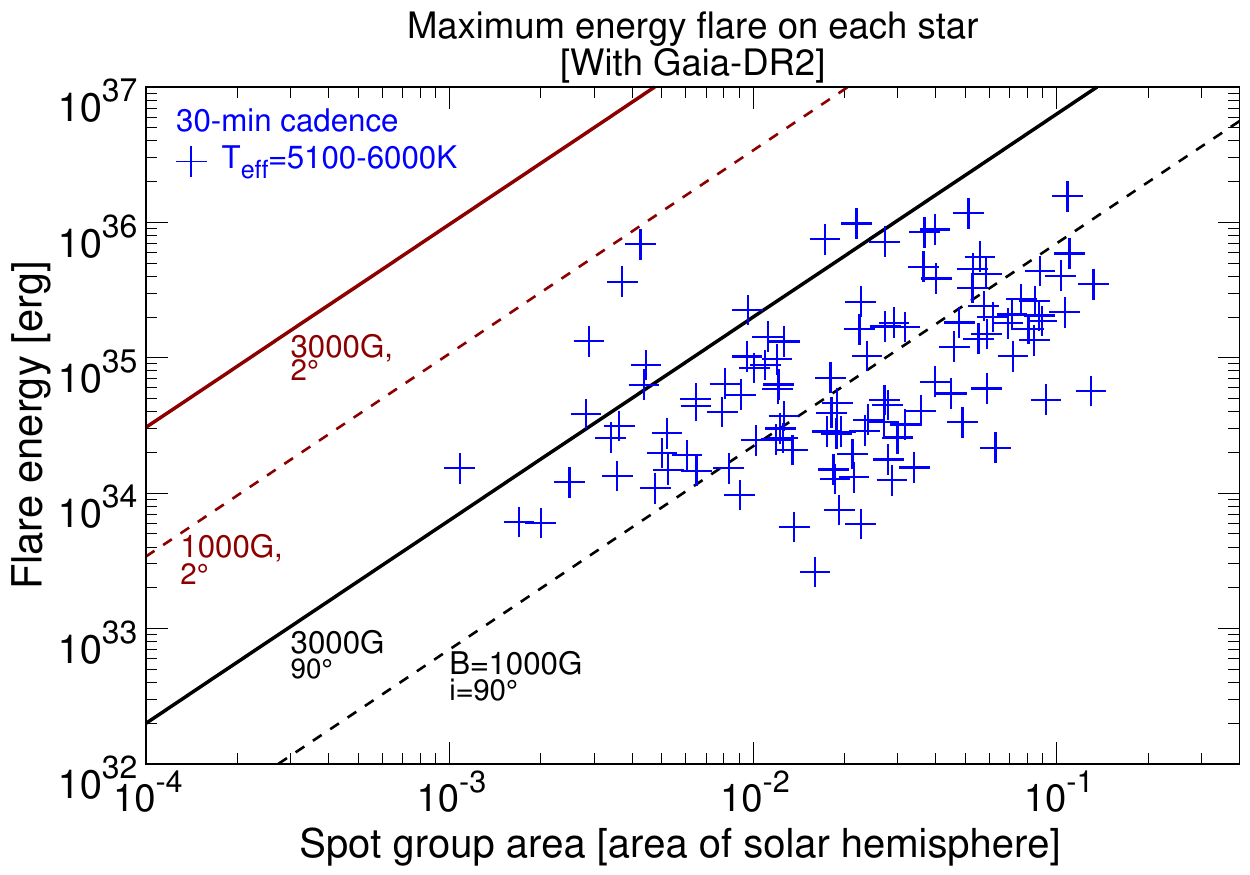}{0.5\textwidth}{(b)}}
  \caption{
\bf (a) \rm
Scatter plot of flare energy ($E_{\mathrm{flare}}$) as a function of spot area ($A_{\mathrm{spot}}$) for superflares (as Figure \ref{fig:spotEene-sunstar1}), 
but the maximum energy flare of each superflare star from {\it Kepler} 30-min cadence data
using the original data in \citet{Shibayama2013} are only plotted. 
The black solid and dashed lines correspond to the analytic relationship 
between the $E_{\mathrm{flare}}$ and $A_{\mathrm{spot}}$ from Equation (\ref{eq:Eflare-Emag}) with the inclination angle $i=90^{\circ}$ 
for $B$=3000 G and 1000 G, respectively.
The dark-red solid and dashed lines correspond to the same relationship with $i=2^{\circ}$ for $B$=3000 G and 1000 G, respectively.
\\
\bf (b) \rm
Same as (a),
but the maximum energy flare of each superflare star from {\it Kepler} 30-min cadence data
using the data updated with $T_{\mathrm{eff, DR25}}$ and $R_{\mathrm{Gaia}}$ in this study are only plotted.
   \label{fig:spot-fene-maxene}}
  \end{figure}

\begin{figure}[htbp]
\gridline{  \fig{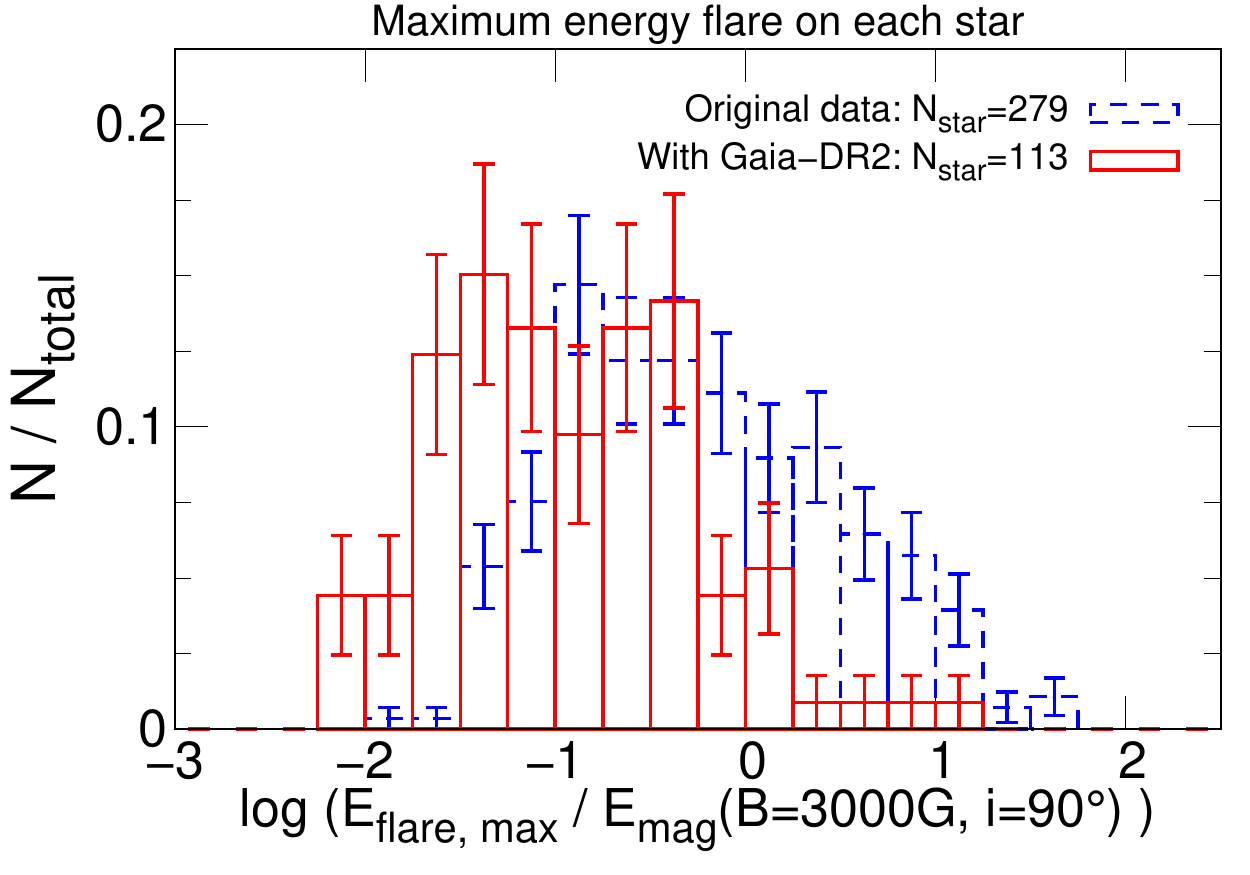}{0.55\textwidth}{}}
  \caption{
Histograms showing the distribution of the maximum energy flare of each superflare star ($E_{\mathrm{flare, max}}$) found from {\it Kepler} 30-min time cadence data.
The horizontal axis value is the fraction of $E_{\mathrm{flare, max}}$ to the upper limit magnetic energy ($E_{\mathrm{mag}}$).
$E_{\mathrm{mag}}$ values correspond to
the black solid lines in (a)\&(b) (B=3000 G and $i$=90$^{\circ}$) at $A_{\mathrm{spot}}$ value of each star.
The blue dashed line corresponds to the original superflare data shown in Figure \ref{fig:spot-fene-maxene}(a), 
while the red solid line corresponds to the updated data in Figure \ref{fig:spot-fene-maxene}(b). 
The error bars represent the 1-$\sigma$ uncertainty estimated from the square root of the number of stars in each bin. 
   \label{fig:spot-fene-hist}}
  \end{figure}

In Figure \ref{fig:spotEene-sunstar1},
the majority of superflares occur on the stars with large starspots, though there is a large scatter.
Flares are sudden releases of magnetic energy stored around starspots (cf. \citealt{Shibata2011}).
The total energy released by the flare ($E_{\mathrm{flare}}$) must be smaller than (or equal to) 
the magnetic energy stored around starspots ($E_{\mathrm{mag}}$). 
Our previous paper (e.g., \citealt{Shibata2013}) suggested that the upper limit of $E_{\mathrm{flare}}$ can be
determined by the simple scaling law:
\begin{eqnarray}\label{eq:Eflare-Emag}
E_{\mathrm{flare}} 
&\approx& fE_{\mathrm{mag}}\approx\frac{B^{2}L^{3}}{8\pi}\approx\frac{B^{2}}{8\pi}A_{\mathrm{spot}}^{3/2} \nonumber \\
&\approx& 7\times10^{32}(\mathrm{erg})\left(\frac{f}{0.1}\right)\left(\frac{B}{10^{3}\mathrm{G}}\right)^{2}
\left(\frac{A_{\mathrm{spot}}}{3\times 10^{19}\mathrm{cm}^{2}}\right)^{3/2} \nonumber \\
&\approx& 7\times10^{32}(\mathrm{erg})\left(\frac{f}{0.1}\right)\left(\frac{B}{10^{3}\mathrm{G}}\right)^{2}
\left(\frac{A_{\mathrm{spot}}/(2\pi R_{\sun}^{2})}{0.001}\right)^{3/2} \ ,
\end{eqnarray}
where $f$ is the fraction of magnetic energy that can be released as flare energy, $B$ and $L$ are the magnetic field strength and size of the spot,
and $R_{\sun}$ is the solar radius.
The black solid and dashed lines in Figure \ref{fig:spotEene-sunstar1} represent Equation (\ref{eq:Eflare-Emag}), 
and almost all the solar flare data located below these lines.
As for the original superflare data from our previous papers (Figure \ref{fig:spotEene-sunstar1}(a)),
many of the superflares locate below the solid line, but some of them locate above this line.
Our previous papers (e.g., \citealt{Notsu2013b} \& \citeyear{Notsu2015b}) considered that these stars above this line are expected to
have low inclination angle or have starposts around the pole region.
In contrast, as for the new data of superflares updated in this study (Figure \ref{fig:spotEene-sunstar1}(b)),
the number of flares above this line (Equation (\ref{eq:Eflare-Emag}) with $B$=3000 G) looks to be decreased and much higher fraction of superflares locate below this line.
In order to see this point more clearly,
we plot only the maximum energy flare of each superflare star in Figures \ref{fig:spot-fene-maxene} (a) \& (b), 
and compare them using histograms in Figure \ref{fig:spot-fene-hist}.
The number of stars above the black solid line (the upper limit line from Equation (\ref{eq:Eflare-Emag}) 
with B=3000 G and the inclination angle $i$=90$^{\circ}$)
is significantly decreased.
With these updates, it is more strongly supported that the upper limit of the energy released by the flare 
is not inconsistent with the magnetic energy stored around the starspots. 

\begin{figure}[htbp]
\gridline{\fig{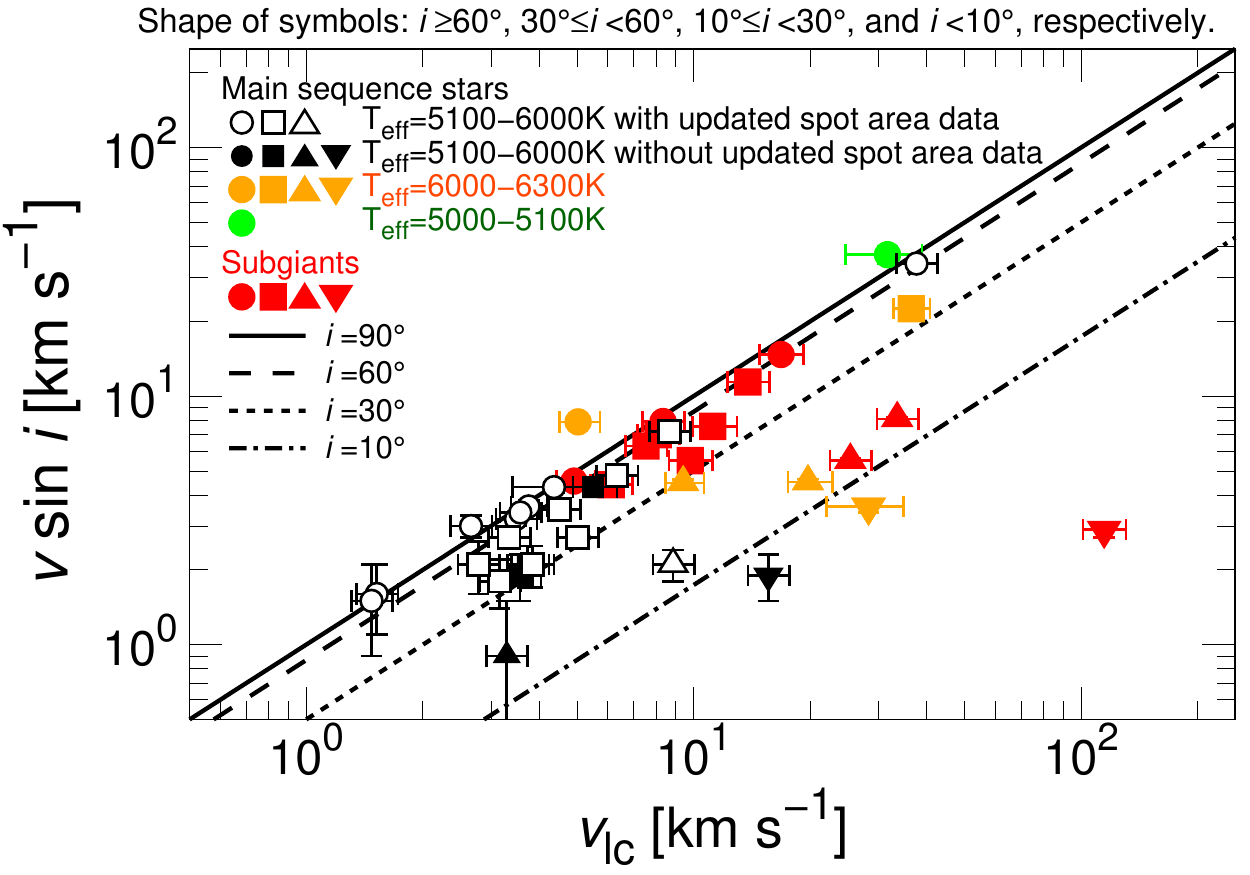}{0.6\textwidth}{}}
  \caption{
Projected rotational velocity ($v\sin i$) vs. the stellar rotational velocity ($v_{\mathrm{lc}}$)
estimated from the period of the brightness variation ($P_{\mathrm{rot}}$) and stellar radius ($R$). 
The datapoints are the basically the same as those of Figure \ref{fig:vlc-vsini} in Section \ref{sec:velocity},
but are limited to single superflare stars. 
The three single stars (KIC11652870, KIC4554830, and KIC11253827, which are shown with blue triangles in Figure \ref{fig:vlc-vsini}) that have only $v\sin i$ upper limit values in APO data 
(and no values from Subaru data) are removed from this figure since we cannot estimate their inclination angle values correctly.
They are classified with colors and shapes of the symbols on the basis of 
the stellar type classifications (see the main text for the details) and inclination angles (circles: $i\geq 60^{\circ}$, 
squares: $30^{\circ}<i\leq 60^{\circ}$, upward triangles: $10^{\circ}<i\leq 30^{\circ}$, and downward triangles: $i<10^{\circ}$), resepctively.
The solid line represents the case that our line of sight is vertical to the stellar rotation axis (i = 90$^{\circ}$; $v\sin i= v_{\mathrm{lc}}$). 
We also plot three different lines, which correspond to smaller inclination angles (i = 60$^{\circ}$, 30$^{\circ}$, 10$^{\circ}$). 
   \label{fig:vlc-vsini-HR}}
  \end{figure}

\begin{figure}[ht!]
\gridline{  \fig{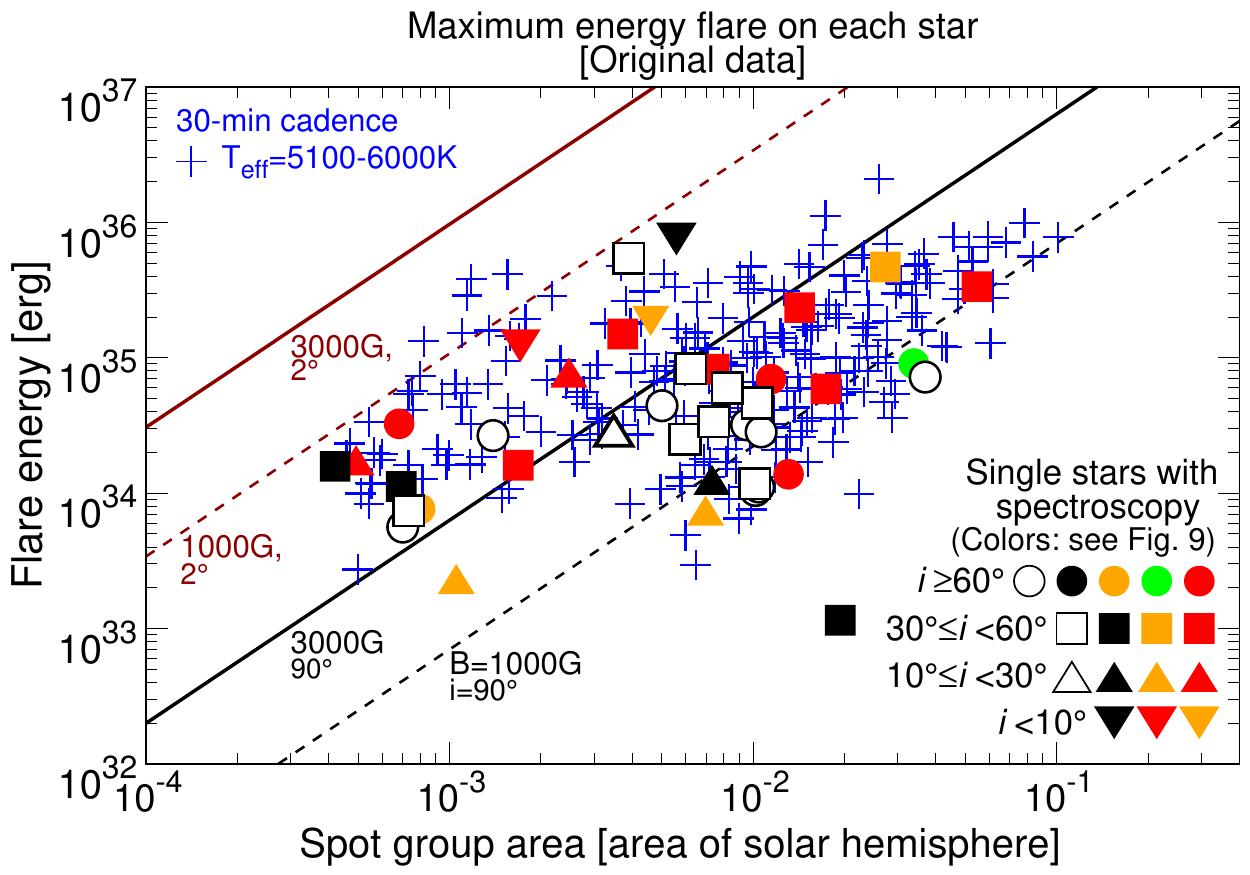}{0.5\textwidth}{(a)}
 \fig{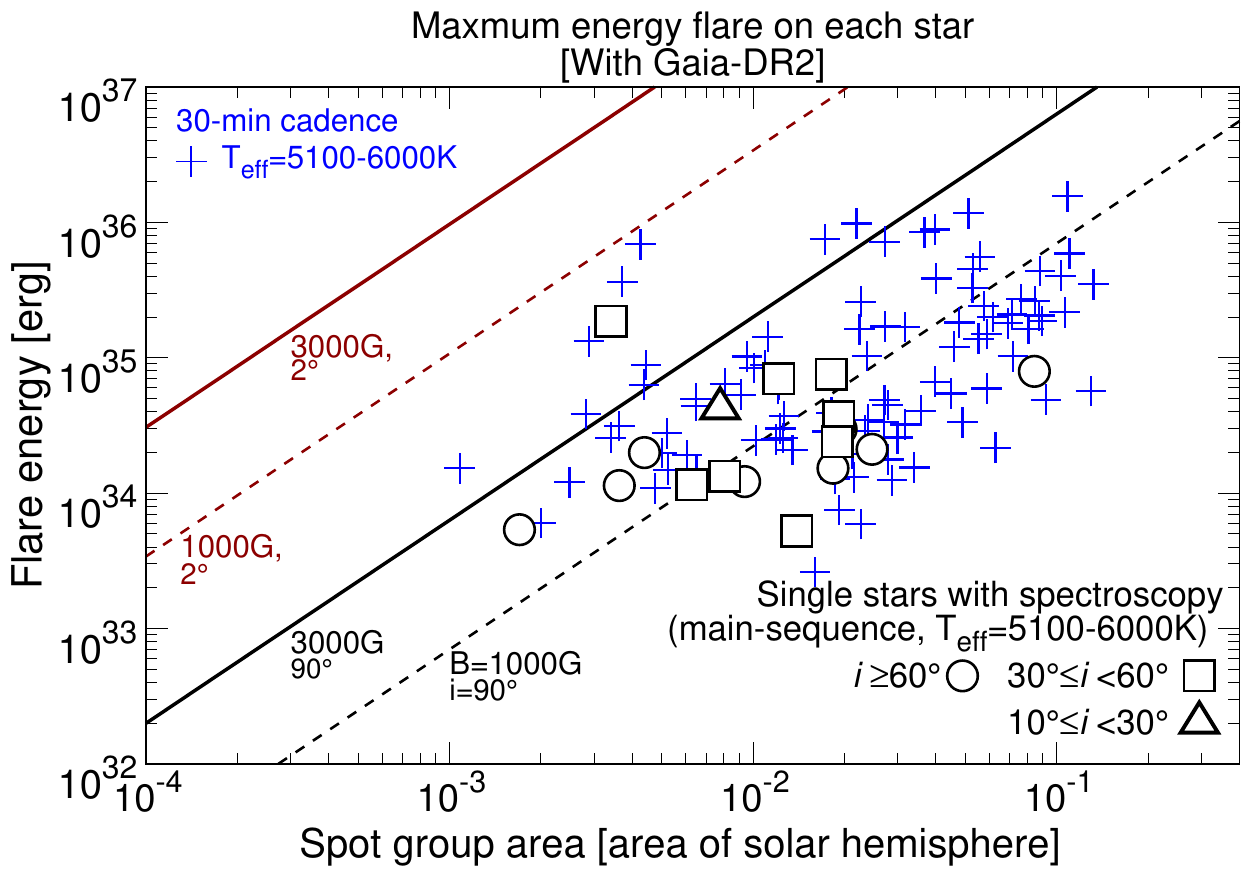}{0.5\textwidth}{(b)}}
  \caption{
\bf (a) \rm
Same as Figure \ref{fig:spot-fene-maxene}(a), but the data of maximum energy flares on single superflare stars that 
we have spectroscopically investigated (Figure \ref{fig:vlc-vsini-HR}) are overplotted with bigger points. 
The colors and shapes of the overplotted symbols are used with the basically the same way as Figure \ref{fig:vlc-vsini-HR}.
\\
\bf (b) \rm
Same as Figure \ref{fig:spot-fene-maxene}(b), but the maximum energy flares on single superflare stars that 
we have spectroscopically investigated (Figure \ref{fig:vlc-vsini-HR}) are overplotted using the updated values (see the main text for the details).
Different from Figure \ref{fig:spot-fene-incspec}(a), only the stars categorized as solar-type stars in this study are plotted.
The shapes of the overplotted symbols (open points) are used with the basically the same way as Figure \ref{fig:vlc-vsini-HR}.
   \label{fig:spot-fene-incspec}}
  \end{figure}

Next, we see this point (the difference of the number of superflare stars above Equation (\ref{eq:Eflare-Emag}) 
between the previous studies and this study) a bit more in detail by incorporating the results of our spectroscopic studies in Section \ref{sec:discussion-spec}.
In Figure \ref{fig:vlc-vsini} in Section \ref{sec:velocity}, 
we compared the projected rotational velocity ($v\sin i$) with the stellar rotational velocity ($v_{\mathrm{lc}}$) 
on the basis of our spectroscopic observations (\citealt{Notsu2015b} and this study), 
and stellar inclination angle ($i$) can be calculated using Equation (\ref{eq:inc}).
We plot again the data of single superflare stars in Figure \ref{fig:vlc-vsini-HR}.
In this figure, we newly classified solar-type stars by using spectroscopically-measured temperature ($T_{\mathrm{eff,spec}}$)
and stellar radius values ($R_{\mathrm{gaia}}$ or $R_{\mathrm{spec}}$) in Figures \ref{fig:HR}(c)\&(d).
White open symbols and black filled ones are solar-type stars (main-sequence stars with $T_{\mathrm{eff}}$ = 5100 -- 6000 K).
The former points are the stars having $A_{\mathrm{spot}}$ values deduced from $\Delta F/F$ in \citet{McQuillan2014} 
so that we can plot the data points in Figure \ref{fig:spot-fene-incspec}(b), 
but the latter ones are the stars without $\Delta F/F$ values in \citet{McQuillan2014} so that we cannot plot the data in Figure \ref{fig:spot-fene-incspec}(b).
Orange and green symbols correspond to the main-sequence stars with a bit hotter ($T_{\mathrm{eff}}$ = 6000 -- 6300 K) and a bit cooler
($T_{\mathrm{eff}}$=5000 -- 5100 K) temperature values, resepectively.
Red symbols are the stars classified as subgiants (cf. Section \ref{sec:dis-atmosHR}).
As a result, many of the low-inclination angle stars are not classified as solar-type stars in this study.
 
Then, in Figures \ref{fig:spot-fene-incspec}(a) \& (b), 
we overplotted the maximum energy flare data of these spectroscopically observed superflare stars 
on the data of Figures \ref{fig:spot-fene-maxene} (a) \& (b), respectively.
In Figure \ref{fig:spot-fene-incspec} (a), the starspot size and flare energy data of the overplotted stars are taken from those used in Figure \ref{fig:spot-fene-maxene} (a) 
\footnote{
We note that three spectroscopically investigated stars (KIC7420545, KIC6934317, and KIC8429280) that are not in Figure \ref{fig:spot-fene-maxene}(a)
are also included in this figure. These stars were not included in the data of \citet{Shibayama2013}, but 
have been spectroscopically investigated in our previous papers (see a footnote of Table 1 of \citealt{Notsu2015b}).
}. As for Figure \ref{fig:spot-fene-incspec}(b), we update starspot size and flare energy values of the overplotted stars
by recalculating them with $T_{\mathrm{eff,spec}}$ and stellar radius values ($R_{\mathrm{gaia}}$ or $R_{\mathrm{spec}}$) used in Figures \ref{fig:HR}(c) \&~(d).
The equations used for these recalculations are the same as those used for Figures \ref{fig:spotEene-sunstar1}(b) \&~\ref{fig:spot-fene-maxene} in the above.
\citet{Notsu2015b} suggested that many of the data points located in upper-left side of $A_{\mathrm{spot}}$ vs. $E_{\mathrm{flare}}$ diagram 
tend to have low inclination angle values.
This tendency can also be seen in Figure \ref{fig:spot-fene-incspec}(a), 
but most of these low-inclination angle stars are now classified as subgiant stars (i.e. red points in this figure).
In the updated Figure \ref{fig:spot-fene-incspec}(b), 
the data points locate in more right-bottom side of this diagram.
There is only one star above the black solid line (Equation (\ref{eq:Eflare-Emag})).
However, the number of stars is decreased in Figure \ref{fig:spot-fene-incspec}(b) compared with Figure \ref{fig:spot-fene-incspec}(a), 
and we need to increase the number of target stars with more spectroscopic observations in the future.

\begin{figure}[ht!]
    \gridline{\fig{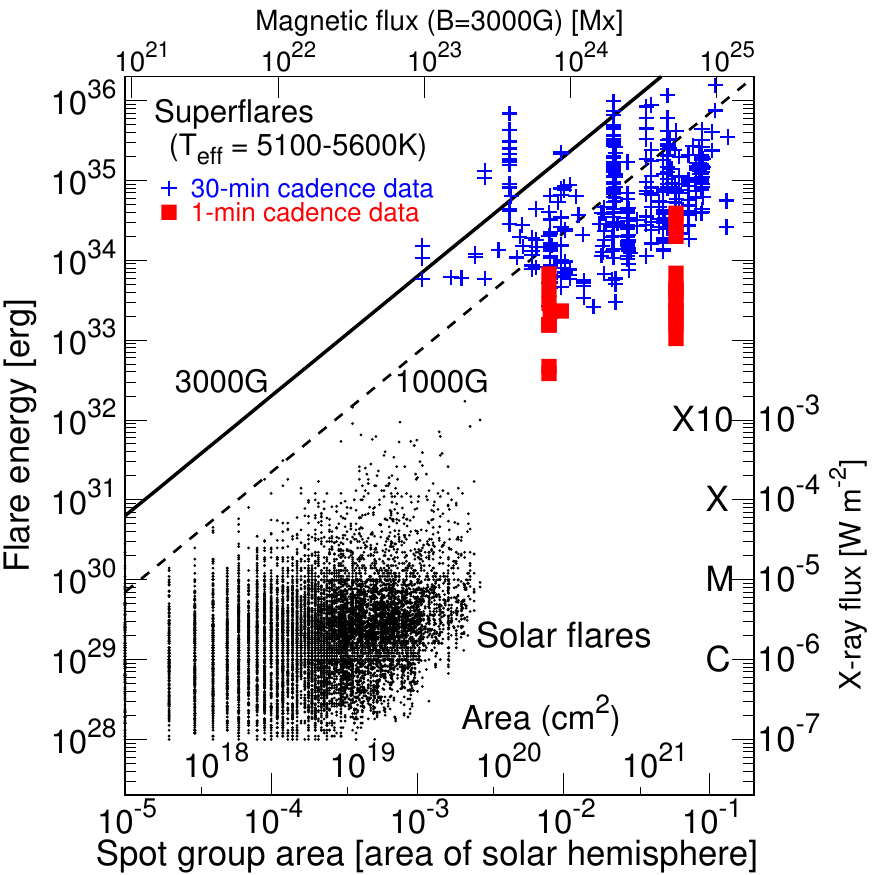}{0.5\textwidth}{(a)}
     \fig{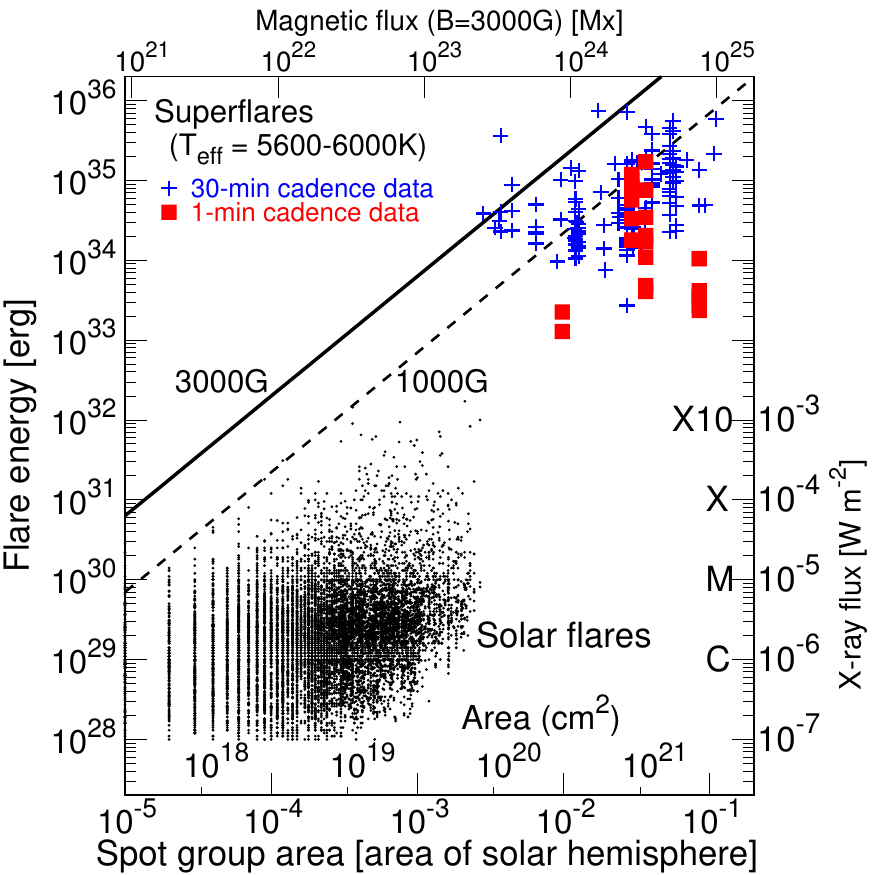}{0.5\textwidth}{(b)}}
\caption{
The scatter plot of flare energy ($E_{\mathrm{flare}}$) as a function of spot area ($A_{\mathrm{spot}}$) of solar flares and superflares
(as Figure \ref{fig:spotEene-sunstar1}(b)).
Blue plus marks and red filled squares are the data of superflares found from {\it Kepler} 30-min and 1-min time cadence data,
which are originally in Figure \ref{fig:spotEene-sunstar1}(b).
They are separated into (a) and (b) on the basis of the stellar temperature values: 
\bf (a) \rm $T_{\mathrm{eff}}$ = 5100 -- 5600 K and \bf (b) \rm $T_{\mathrm{eff}}$ = 5600 -- 6000 K.
\label{fig:spotEene-temp}}
\end{figure}

Finally in Figures \ref{fig:spotEene-temp}(a) \& (b), 
we plot again $A_{\mathrm{spot}}$ and $E_{\mathrm{flare}}$ values (taken from Figure \ref{fig:spotEene-sunstar1}(b)), 
separating the superflare data on the basis of the stellar temperature values.
The data of the stars with a temperature range ($T_{\mathrm{eff}}$ = 5600 -- 6000 K) 
close to the solar temperature are plotted in Figure \ref{fig:spotEene-temp}(b),
while those of late G-type main-sequence stars ($T_{\mathrm{eff}}$ = 5100 -- 5600 K) are in Figure \ref{fig:spotEene-temp}(a).
There are no big differences between Figures \ref{fig:spotEene-temp}(a) \& (b).
As a result, almost all the data points of superflares locate below the line of Equation (\ref{eq:Eflare-Emag}),
and we confirmed again that the upper limit of the energy released by flares 
is not inconsistent with the magnetic energy stored around the starspots. 
  
 \clearpage     

\subsection{Dependence of superflare energy and frequency on rotation period}\label{sec:Prot-fene} 

\begin{figure}[ht!]
  \gridline{ \fig{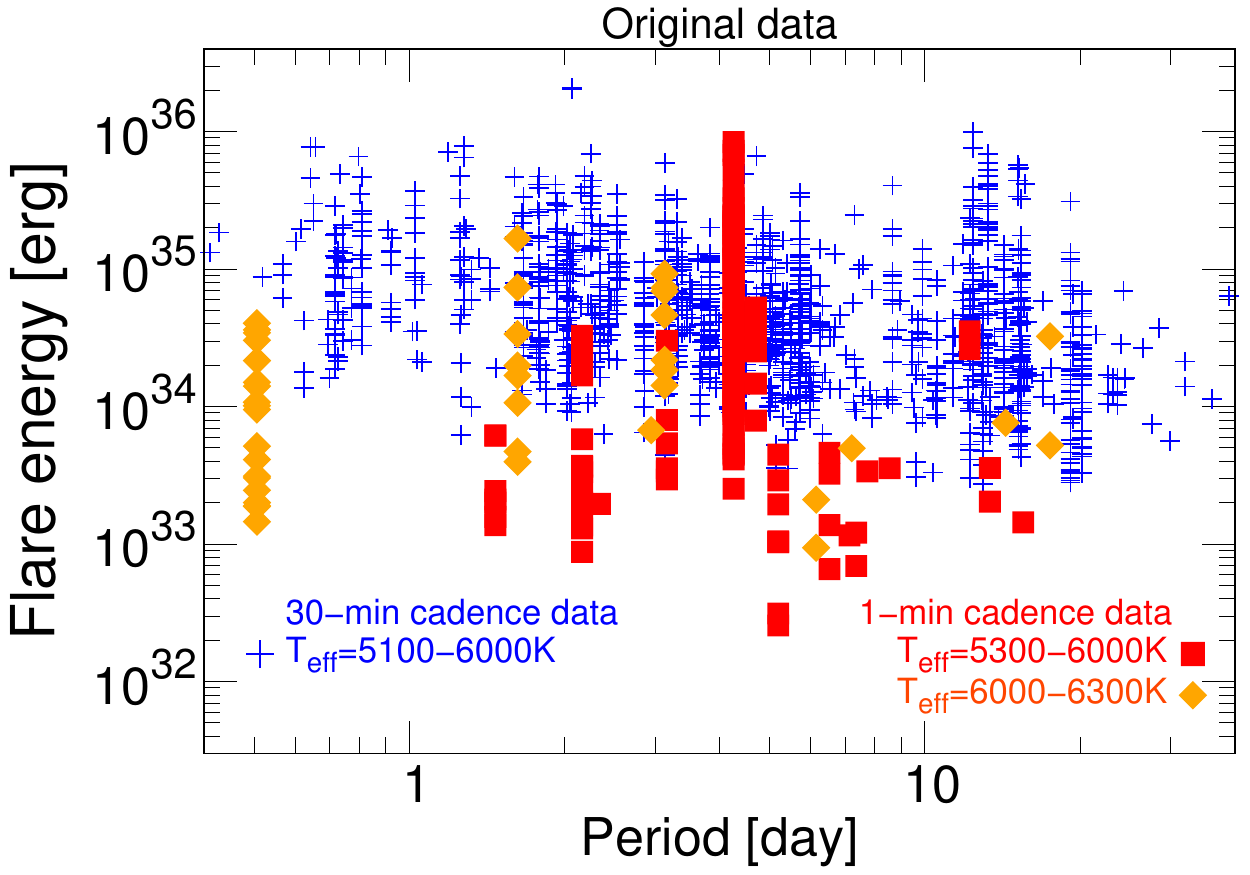}{0.5\textwidth}{(a)}
\fig{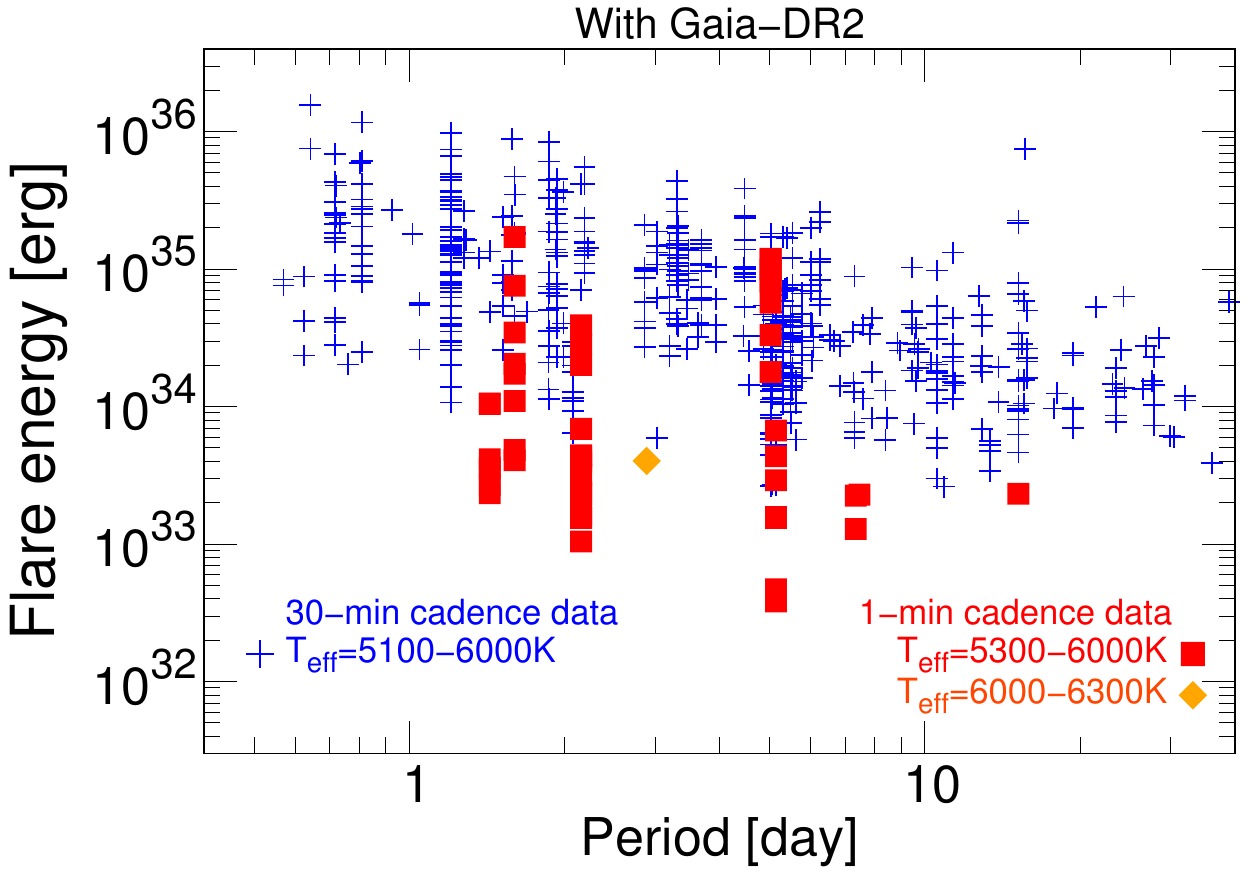}{0.5\textwidth}{(b)}}
\caption{
Scatter plot of the superflare energy ($E_{\mathrm{flare}}$) vs. the rotation period ($P_{\mathrm{rot}}$) of each star (estimated from the brightness variation period). 
Apparent negative correlations between  $P_{\mathrm{rot}}$ and the lower limit of $E_{\mathrm{flare}}$, 
which are mainly seen for blue cross points, result from the detection limit of our flare search method (cf. \citealt{Shibayama2013}).
\\ \bf (a) \rm
The original data of superflares on solar-type (G-type main-sequence) stars presented in our previous studies 
(\citealt{Shibayama2013} \&~\citealt{Maehara2015}) are plotted.
Blue crosses indicate superflares detected from {\it Kepler} 30-min cadence data \citep{Shibayama2013},
while red squares and orange diamonds are those detected from 1-min cadence data \citep{Maehara2015}.
Among the flares from 1-min cadence data, stars with $T_{\mathrm{eff}}$ = 6000 -- 6300 K, 
which are not included in the range of solar-type stars in this study ($T_{\mathrm{eff}}$ = 5100 -- 6000 K), 
are distinguished with orange diamonds.
\\ \bf (b) \rm
The data of superflares updated in this study using $T_{\mathrm{eff, DR25}}$ and $R_{\mathrm{Gaia}}$ values.
Symbols are used with the same way as (a). 
In addition to the data of solar-type stars (Blue crosses and red squares), 
the data of one main-sequence star with $T_{\mathrm{eff}}$ = 6000 -- 6300 K (KIC8508009) from {\it Kepler} 1-min cadence data is also calculated and plotted for reference.
\label{fig:ProtEene1}}
\end{figure}

Previous observations of solar-type stars (e.g., X-ray, UV, Ca II H\&K) have shown that 
the stellar magnetic activity level decreases as rotation period increases (e.g., \citealt{Noyes1984}; \citealt{Guedel2007}; \citealt{Wright2011}).
Since stellar age has a good correlation with rotation, young rapidly-rotating stars show the higher activity levels, 
and it was expected that they show more energetic flares more frequently compared with slowly-rotating stars like the Sun.
In our previous paper \citep{Notsu2013b}, we investigated the relationship between the superflare energy ($E_{\mathrm{flare}}$), 
and the rotation period ($P_{\mathrm{rot}}$), 
and suggested that the maximum superflare energy in a given rotation period bin 
does not have a clear correlation with the rotation period (Figure \ref{fig:ProtEene1}(a) of this paper),
while the average flare frequency in a given period bin has a correlation with the rotation 
(Figure 7(b) of \citealt{Notsu2013b} and Figure 2 of \citealt{Maehara2015}).
This suggestion is important since, against the above mentioned previous expectations, 
energetic superflares with $\sim$10$^{35}$ erg can occur on solar-type stars rotating 
as slow as the Sun ($P_{\mathrm{rot}}\sim$25 days), 
even though the frequency is low (once in a few thousand years), 
compared with rapidly-rotating stars. 

Next, we investigate again this relation by using superflare values updated with {\it Gaia}-DR2 stellar radius ($R_{\mathrm{Gaia}}$) data.
In Section \ref{sec:spot-fene}, we newly classified solar-type stars on the basis of $T_{\mathrm{eff, DR25}}$ and $R_{\mathrm{Gaia}}$,
and recalculated flare energy $E_{\mathrm{flare}}$.  
Figure \ref{fig:ProtEene1}(b) shows the relationship between this updated $E_{\mathrm{flare}}$ value with 
the rotation period $P_{\mathrm{rot}}$ of each solar-type superflare star. 
$P_{\mathrm{rot}}$ values plotted here are taken from \citet{McQuillan2014}.
Unlike the results of our previous studies (Figure \ref{fig:ProtEene1} (a)), 
Figure \ref{fig:ProtEene1}(b) suggests that the upper limit of $E_{\mathrm{flare}}$ in each period bin has a continuous decreasing trend with rotation period.
For example, stars rotating as slow as the Sun ($P_{\mathrm{rot}}\sim$25 days)
show superflares up to 10$^{35}$ erg, while rapidly-rotating stars with $P_{\mathrm{rot}}$=1 -- 3 days have more energetic superflares up to 10$^{36}$ erg. 

\begin{figure}[ht!]
  \gridline{  
   \fig{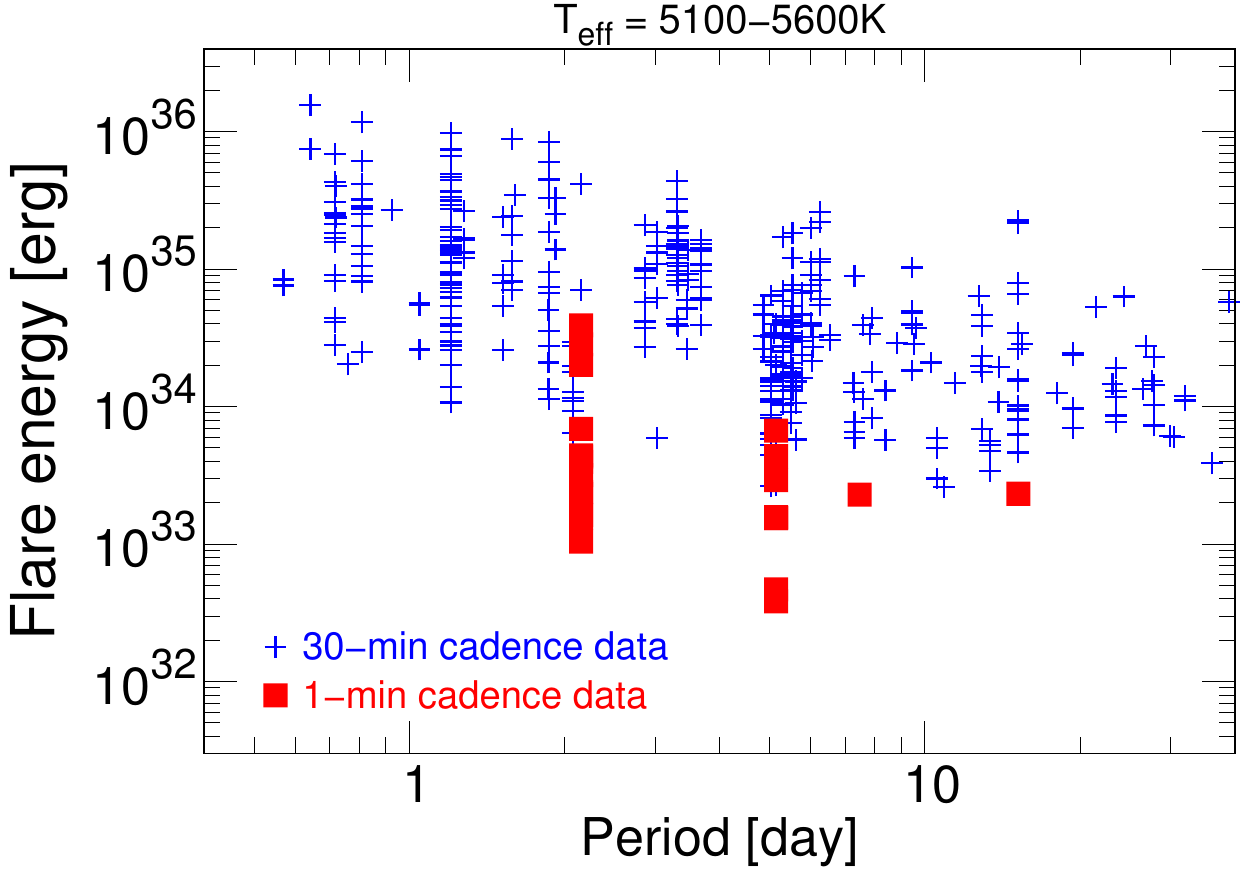}{0.5\textwidth}{(a)}
  \fig{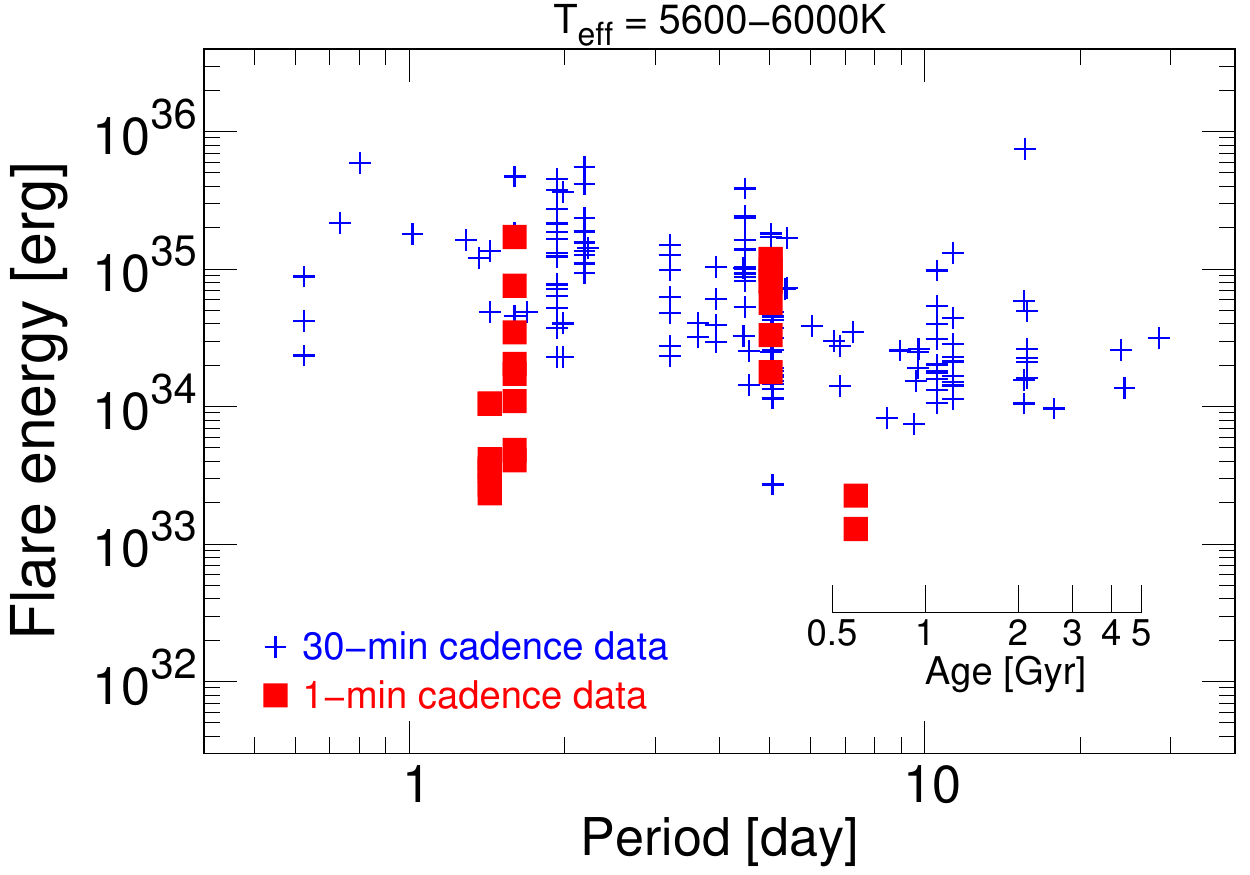}{0.5\textwidth}{(b)}
     }
\caption{
Scatter plot of the superflare energy ($E_{\mathrm{flare}}$) vs. the rotation period ($P_{\mathrm{rot}}$) of each star (as Figure \ref{fig:ProtEene1}(b)).
Blue plus marks and red filled squares are the data of superflares found from {\it Kepler} 30-min and 1-min time cadence data,
which are originally in Figure \ref{fig:ProtEene1}(b).
They are separated into (a) and (b) on the basis of the stellar temperature values: 
\bf (a) \rm $T_{\mathrm{eff}}$ = 5100 -- 5600 K and \bf (b) \rm $T_{\mathrm{eff}}$ = 5600 -- 6000 K.
Only in (b), we added the scale of stellar age ($t$) on the basis of the gyrochronology relation of solar-type star ($t\propto P_{\mathrm{rot}}^{0.6}$: \citealt{Ayres1997})
(See main text for the details).
\label{fig:ProtEene2}}
\end{figure}

Flare activity also depends on stellar temperature (\citealt{Candelaresi2014}; \citealt{Davenport2016}; \citealt{Doorsselaere2017}), 
and even among solar-type stars, cooler stars can have higher flare activities (cf. Figure 5 of \citealt{Shibayama2013}). 
Then the data of superflares plotted in Figure \ref{fig:ProtEene2} are separated into (a) and (b) on the basis of the stellar temperature values.
The data of the stars with a temperature range ($T_{\mathrm{eff}}$ = 5600 -- 6000 K) 
close to the solar temperature are plotted in Figure \ref{fig:ProtEene2}(b),
while those of late G-type main-sequence stars ($T_{\mathrm{eff}}$ = 5100 -- 5600 K) are in Figure \ref{fig:ProtEene2}(a).
In Figure \ref{fig:ProtEene2}(b), we also added the scale of stellar age ($t$) on the basis of the gyrochronology relation of solar-type star ($t\propto P_{\mathrm{rot}}^{0.6}$: \citealt{Ayres1997}) in order to compare the age of superflare stars with that of the Sun ($t$=4.6 Gyr). 
The scale of $t$ are only plotted in the limited age range $t$ = 0.5 -- 5 Gyr because of the following two reasons.
(1) As for young solar-type stars with $t\lesssim$ 0.5 -- 0.6 Gyr, a large scatter in the age-rotation relation has been reported from young cluster observations
(e.g., \citealt{Soderblom1993}; \citealt{Ayres1997}; \citealt{Tu2015}).
(2) As for old solar-type stars beyond solar-age ($t$=4.6 Gyr), a breakdown of gyrochronology relations has been recently reported (\citealt{vanSaders2016}; \citealt{Metcalfe2018}).
With this scale, 
for example, we can see that Sun-like stars with $P_{\mathrm{rot}}\sim$25 days are more than four times older ($t\sim$4.6 Gyr) than the stars 
with $P_{\mathrm{rot}}\sim$10 days ($t\sim$1 Gyr).

As a result of Figures \ref{fig:ProtEene2} (a) \& (b), we confirmed again the suggestions from Figures \ref{fig:ProtEene1} (b).
The upper limit of superflare energy in each period bin depends on the rotation period in both Figures \ref{fig:ProtEene2} (a) \&~(b),
and there is one order of magnitude difference between the maximum flare energy on young stars ($t<$0.5 Gyr) and Sun-like old stars ($t\sim$4.6 Gyr). 
We cannot judge whether there are any clear differences between these two figures 
because of the low-number statics.
It is necessary to increase the number of superflare data with the future observations (See also Section \ref{subsec:future}).

\begin{figure}[ht!]
  \gridline{ \fig{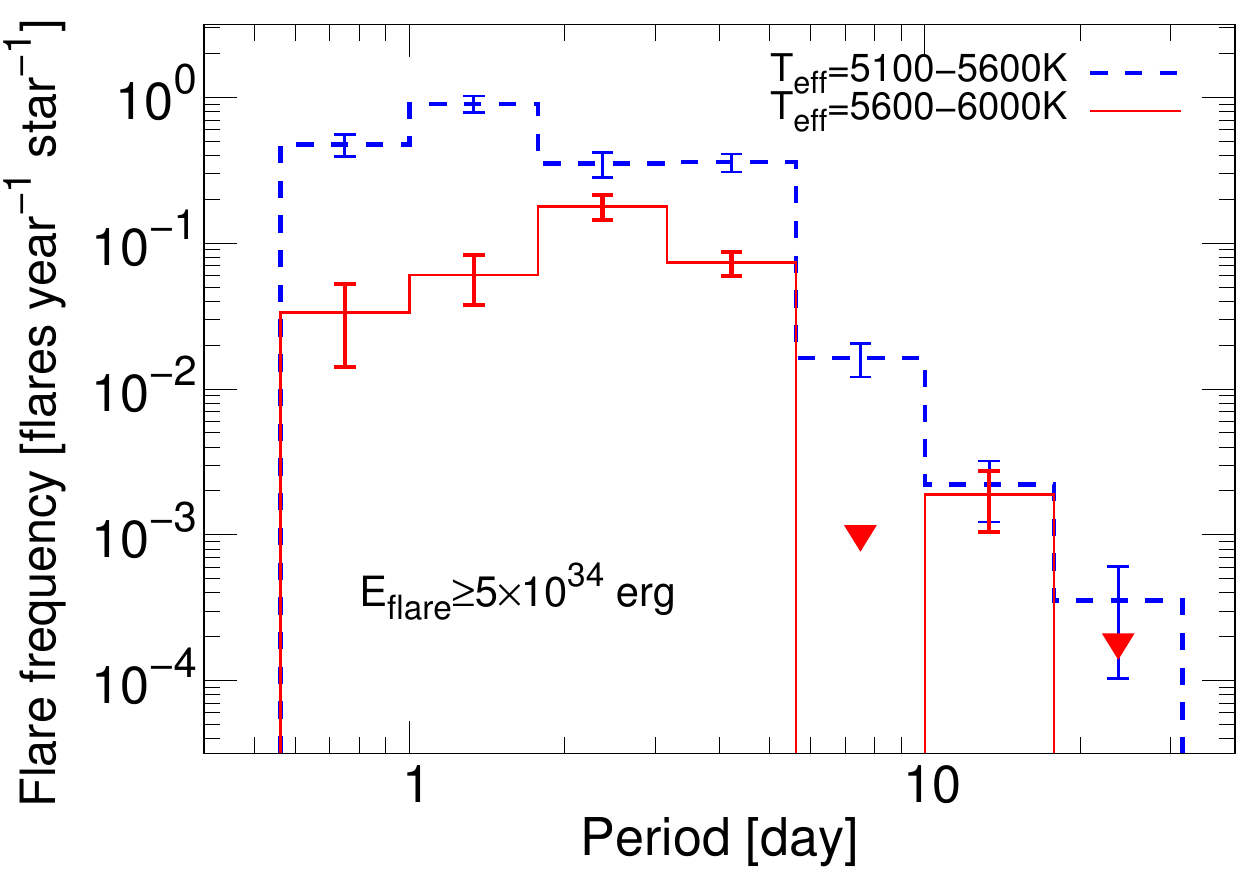}{0.55\textwidth}{}}
\caption{
Occurrence frequency distribution of superflares as a function of the rotation period ($P_{\mathrm{rot}}$),
using the data of superflares on solar-type stars 
that were originally found from {\it Kepler} 30-min cadence data (\citealt{Shibayama2013}; \citealt{Candelaresi2014}) 
but are updated with $T_{\mathrm{eff, DR25}}$ and $R_{\mathrm{Gaia}}$ in this study 
($E_{\mathrm{flare}}$ and $P_{\mathrm{rot}}$ data in Figure \ref{fig:ProtEene1}).
The vertical axis indicates the number of superflares with energy $\geq 5\times 10^{34}$ erg per star and per year.
The error bars represent the 1-$\sigma$ uncertainty estimated from the square 
root of the number of flares in each bin.
Unlike Figure 7 of \citet{Notsu2013b}, we do not take into account the error values of $E_{\mathrm{flare}}$ 
since this is now more reliable with {\it Gaia}-DR2 stellar radius values.
This vertical axis value is calculated by using 
the number of the solar-type stars in each $P_{\mathrm{rot}}$ bin 
($N_{P}$($P_{\mathrm{rot}}$) in Table \ref{tab:Nstar-amp-detected}) 
detected in \citet{McQuillan2014}. 
The potential errors from this $N_{P}$($P_{\mathrm{rot}}$) value are discussed in Appendix \ref{sec:Kepler-Prot-Nstar}.
Red solid lines correspond to the frequency values calculated from the solar-type stars 
with their temperature values limited to a range ($T_{\mathrm{eff}}$ = 5600 -- 6000 K) close to the solar temperature,
while blue dashed lines are those from the late G-type main-sequence stars with $T_{\mathrm{eff}}$ = 5100 -- 5600 K.
$T_{\mathrm{eff, DR25}}$ values are used for the temperature classifications here.
As for the red solid lines, in the case of no events in a period bin, 
the upper limit values are shown with the red downward triangle points assuming less than one event occur in each bin.
\label{fig:Prot-freq-evo0}}
\end{figure}

Then we see the relation between flare frequency and rotation period by using the updated superflare data. 
Figure \ref{fig:Prot-freq-evo0} shows that the average flare frequency in a given period bin tends to decrease 
as the period increases in the range of $P_{\mathrm{rot}}$ longer than a few days. The frequency is averaged from all of the superflare stars in the same period bin. 
This result is basically the same as that presented in \citet{Notsu2013b} with the original superflare data.
The frequency of superflares on young rapidly-rotating stars ($P_{\mathrm{rot}}$=1 -- 3 days) is $\sim$100 times higher 
compared with old slowly-rotating stars ($P_{\mathrm{rot}}\sim$25 days), 
and this indicates that as a star evolves (and its rotational period increases), the frequency of superflares decreases.
We can now interpret that this correlation between the rotation period (roughly corresponding to age) and flare frequency 
is consistent with the correlation between the rotation period (age) and previous measurement of the stellar activity level such as the average X-ray luminosity
(e.g., \citealt{Noyes1984}; \citealt{Guedel2007}; \citealt{Wright2011}).
Summarizing the results in this section, 
superflares with their energy $\lesssim 5\times10^{34}$ erg can occur 
on slowly-rotating old Sun-like stars similar to the Sun ($T_{\mathrm{eff}}$ = 5600 -- 6000 K, $P_{\mathrm{rot}}\sim$25 days and $t\sim$4.6 Gyr), 
even though the frequency and maximum flare energy are lower compared with young rapidly-rotating stars and cooler stars.

\subsection{Starspot size vs. rotation period of solar-type stars, and implications for superflare energy.}\label{sec:spot-Prot}

\begin{figure}[ht!]
  \gridline{ \fig{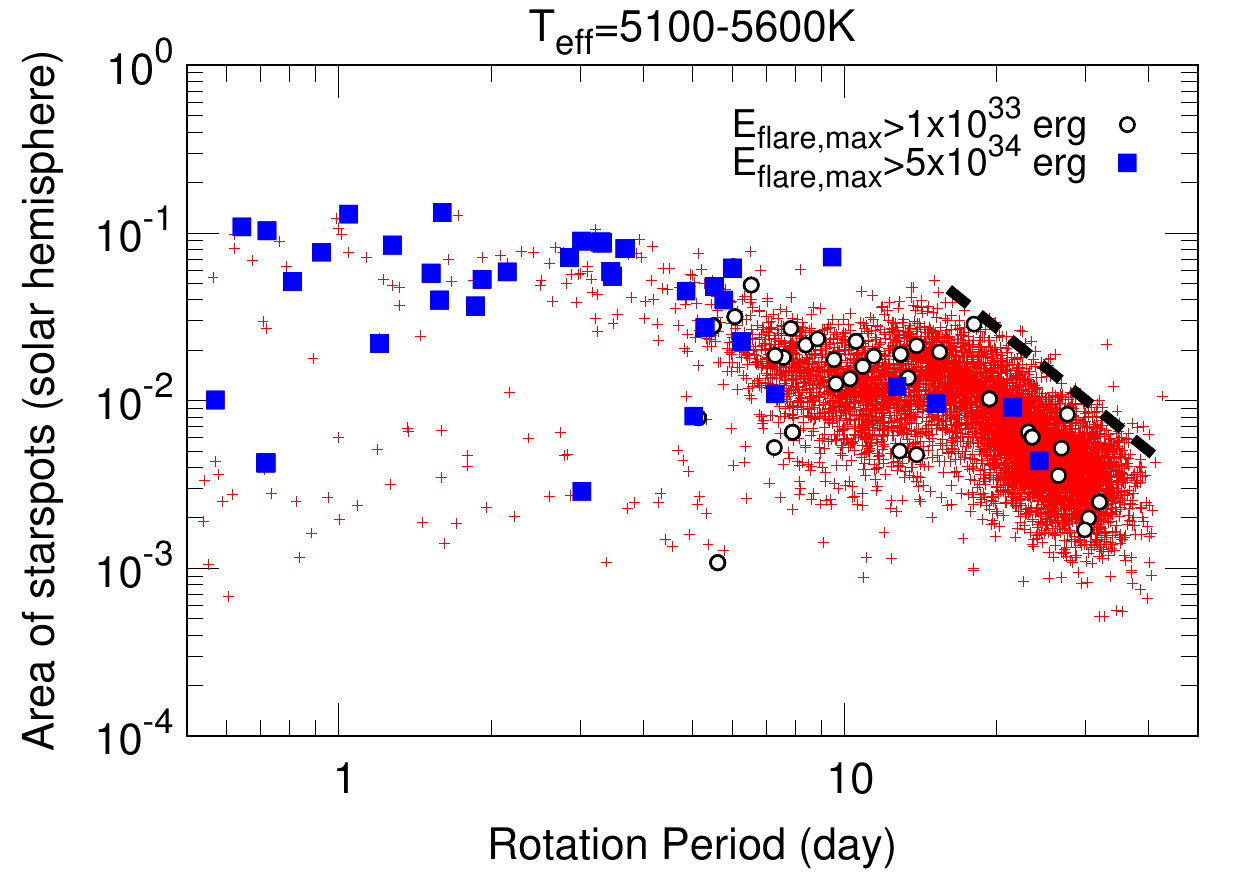}{0.5\textwidth}{(a)}
  \fig{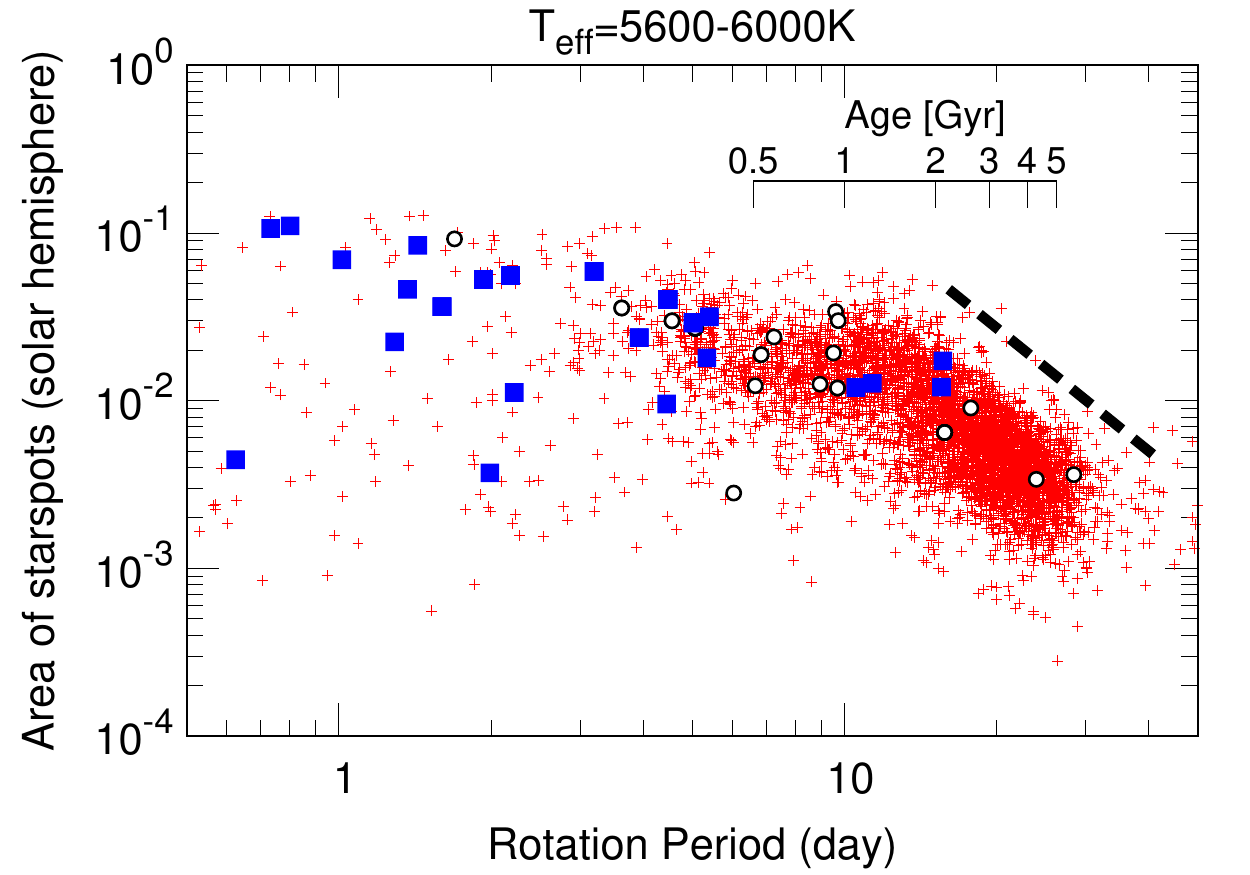}{0.5\textwidth}{(b)}}
\caption{
Scatter plot of the spot group area of solar-type stars ($A_{\mathrm{spot}}$) 
as a function of the rotation period ($P_{\mathrm{rot}}$), using the data updated with $T_{\mathrm{eff, DR25}}$ and $R_{\mathrm{Gaia}}$ values in this study.
The vertical axis represents $A_{\mathrm{spot}}$ in units of the area of solar hemisphere ($A_{1/2\sun}\sim 3\times 10^{22}$ cm$^{2}$).
Open circles and blue filled squares indicate solar-type stars that have superflares 
with the energy values of their most energetic flares $E_{\mathrm{flare,max}}>1\times 10^{33}$ erg and  $E_{\mathrm{flare,max}}>5\times 10^{34}$ erg, respectively.
Red small cross marks indicate solar-type stars without superflares.
The plotted data are separated into (a) and (b) on the basis of the temperature values: 
\bf (a) \rm $T_{\mathrm{eff}}$ = 5100 -- 5600 K and \bf (b) \rm $T_{\mathrm{eff}}$ = 5600 -- 6000 K.
The black dashed line in (a) is plotted by eye to roughly show the upper limit of the datapoints in the range of $P_{\mathrm{rot}}\gtrsim 14$ days.
The black dashed line in (b) is plotted at the same place as (a) in order for comparison with the results of (a).
Only in (b), we also added the scale of stellar age ($t$) on the basis of the gyrochronology relation of solar-type star ($t\propto P_{\mathrm{rot}}^{0.6}$: \citealt{Ayres1997})
(See the main text of Section \ref{sec:Prot-fene} for the details).
 \label{fig:aspot-prot}}
\end{figure}

Our previous paper \citet{Maehara2017} investigated the statistical properties of starspots on solar-type stars
by using the starspot size $A_{\mathrm{spot}}$ and rotation period $P_{\mathrm{rot}}$ estimated from the brightness variations of {\it Kepler} data.
Here we update these values by using {\it Gaia}-DR2 stellar radius, as also done for superflare stars in the above.
As done in \citet{Maehara2017}, we used $P_{\mathrm{rot}}$ and brightness variation amplitude $\Delta F/F$ values reported in \citet{McQuillan2014}.
We newly classified 49212 solar-type stars ($N_{\mathrm{data}}$ in Table \ref{tab:Nstar-amp-detected} 
in Appendix \ref{sec:Kepler-Prot-Nstar})
among the stars in \citet{McQuillan2014} by using $T_{\mathrm{eff, DR25}}$ and $R_{\mathrm{Gaia}}$ values,
as done for superflare stars in Section \ref{sec:spot-fene}.
$P_{\mathrm{rot}}$ and $\Delta F/F$ values of 11594 stars 
(cf. $N_{P}$(all) values in Table \ref{tab:Nstar-amp-detected}) are detected among these 49212 stars.
Then, we recalculated $A_{\mathrm{spot}}$ values from $\Delta F/F$ values, using these $T_{\mathrm{eff, DR25}}$ and $R_{\mathrm{Gaia}}$ values.
The method of $A_{\mathrm{spot}}$ calculation is exactly the same as those used for superflare stars 
in Section \ref{sec:spot-fene}.
The resultant values of  are plotted in Figure \ref{fig:aspot-prot}.
The data in Figure \ref{fig:aspot-prot} are separated into (a) and (b) on the basis of the temperature values.
The data of the stars with a temperature range ($T_{\mathrm{eff}}$ = 5600 -- 6000 K) 
close to the solar temperature are plotted in Figure \ref{fig:aspot-prot}(b),
while those of late G-type main-sequence stars ($T_{\mathrm{eff}}$ = 5100 -- 5600 K) are in Figure \ref{fig:aspot-prot}(a).
The stars showing superflares tend to have shorter rotation periods (younger ages) and larger starspot areas.

Figure \ref{fig:aspot-prot}(b) shows that
the largest area of starspots on Sun-like stars with $T_{\mathrm{eff}}$ = 5600 -- 6000 K in a given period bin 
has roughly a constant or very gentle decreasing trend around $A_{\mathrm{spot}}=5\times 10^{-2}$ -- $1\times 10^{-1}A_{1/2\sun}$ 
in the period range of $P_{\mathrm{rot}}\lesssim 12$ days (age: $t\lesssim$1.4 Gyr). 
However, in the period range of $P_{\mathrm{rot}}\gtrsim 12$ days ($t\gtrsim$1.4 Gyr), 
the largest starspot area on them clearly decreases as the rotation period increases.
As for the late G-type main-sequence stars with $T_{\mathrm{eff}}$ = 5100 -- 5600 K in Figure \ref{fig:aspot-prot}(a), 
the similar steep trends can be seen for maximum size of starspots, but the exact values are a bit different.
This steep decreasing trend starts around $P_{\mathrm{rot}}\sim 14$ days (Figure \ref{fig:aspot-prot}(a)). 
In the longer period range, the maximum area of spots on the stars with $T_{\mathrm{eff}}$ = 5600 -- 6000 K at a given period bin 
is roughly half compared with those of the stars with $T_{\mathrm{eff}}$ = 5100 -- 5600 K 
(see the black dashed line in Figure \ref{fig:aspot-prot}(b)).
These differences support that cooler stars can generate larger magnetic flux more effectively thanks to the development of 
the convection zone.
We discuss such temperature differences of the $P_{rot}$ vs. $A_{\mathrm{spot}}$ relations including more cooler (K and M-type) dwarf stars in more detail
in our next study (Maehara et al. 2019 in preparation).

The trends similar to those presented here was already reported in Figure 1(a) of \citet{Maehara2017}, but we did not discuss the temperature differences.
Moreover, the steep decreasing trend in the longer period range is now much more clear compared with this previous paper.
It might be because the error of $A_{\mathrm{spot}}$ values decrease thanks to the updates using {\it Gaia}-DR2 stellar radius ($R_{\mathrm{Gaia}}$) values. 
and the potential contamination of subgiants in the previous data can also be eliminated.
Then the steep decreasing trend of maximum spot area in the longer period range is more strongly supported.

In Section \ref{sec:Prot-fene}, we reported that 
maximum energy of superflares in a given period bin decreases as $P_{\mathrm{rot}}$ increases (Figures \ref{fig:ProtEene2}(a) \& (b)).
This decreasing trend of maximum flare energy can be related with 
the decreasing trend of maximum area of starspots in longer period regime in Figures \ref{fig:aspot-prot} (a) \& (b) described in this section.
This is because the maximum area of starspots determines well the upper limit of flare energy (cf. Section \ref{sec:spot-fene}).
For example, Figure \ref{fig:aspot-prot}(b) shows that maximum size of starspots on old ($t\sim$4.6 Gyr) Sun-like stars with
$T_{\mathrm{eff}}$ = 5600 -- 6000 K and $P_{\mathrm{rot}}\sim 25$ days is $\sim$1 \%~of the solar hemisphere.
This corresponds to flare energy of $10^{34}$ -- $10^{35}$ erg on the basis of Equation (\ref{eq:Eflare-Emag}), 
and the upper limit of superflare energy of such Sun-like stars in Figure \ref{fig:ProtEene2} (b) are roughly in the same range.

However, there is a difference in a bit strict sense between the decreasing trend of the maximum superflare energy 
in Figures \ref{fig:ProtEene2} (a) \& (b)
and that of the maximum area of starspots in Figures \ref{fig:aspot-prot} (a) \& (b).
The maximum superflare energy roughly continuously decrease as the rotation period increases (the star becomes older) in Figures \ref{fig:ProtEene2} (a) \& (b),
but the maximum area of starspots does not show such continuous decreasing trend in Figures \ref{fig:aspot-prot} (a) \& (b).
The maximum area of starspots is constant in the short period range ($P_{\mathrm{rot}}\lesssim 12-14$ days),
but it steeply decreases as the period increases in the longer range ($P_{\mathrm{rot}}\gtrsim 12-14$ days). 
This difference can suggest a possibility that the flare energy is determined not only by the starspot area but also by the other important factors,
though the starspot area is still a necessary condition determining flare energy (cf. Section \ref{sec:spot-fene}).
By analogy with the correlation between the flare activity and the magnetic structure of sunspot groups (cf. \citealt{Sammis2000}; \citealt{Maehara2017}), 
one of the possible factors might be the effect of the magnetic structure of starspots.
More complex spots can generate more frequent and more energetic flares according to solar observations.
If the magnetic structure (complexity) of spots also has a correlation with the rotation period, 
the upper limit of flare energy can depend on rotation period even if the starspot size is roughly constant.
We need to conduct more detailed studies on starspot properties to clarify such possibilities (See also the final paragraph of \citealt{Maehara2017}).
We must note here that we also need more superflare observations 
to more quantitatively discuss the above difference between Figures \ref{fig:ProtEene2} and \ref{fig:aspot-prot},
since the number of superflare events especially in Figure \ref{fig:ProtEene2}(b) is small.

In addition, the existence of constant and decreasing trends of maximum starspot coverage can be compared 
with the relation between soft X-ray flux and rotation period (e.g., \citealt{Wright2011}).
The soft X-ray fluxes of solar-type stars are also known to show the constant regime (or so-called  ``saturation" regime) 
in the period range of $P_{\mathrm{rot}}\lesssim$2--3 days,
but they decrease constantly as the $P_{\mathrm{rot}}$ values increase in the range of $P_{\mathrm{rot}}>$2--3 days.
The changing point of this soft X-ray trend ($P_{\mathrm{rot}}\sim$2--3 days) 
is different from that of maximum spotsize values ($P_{\mathrm{rot}}\sim 12-14$ days).
Although a detailed study on this point is beyond the scope of this paper, 
these similarities and differences can be interesting and helpful 
when considering the relation between stellar activity (including starspots, flares, X-ray steady emissions) and rotation 
in more detail.

\subsection{Superflare frequency on Sun-like stars and implications for the Sun}\label{fig:freq-sunlike}

\begin{figure}[t!]
\gridline{ \fig{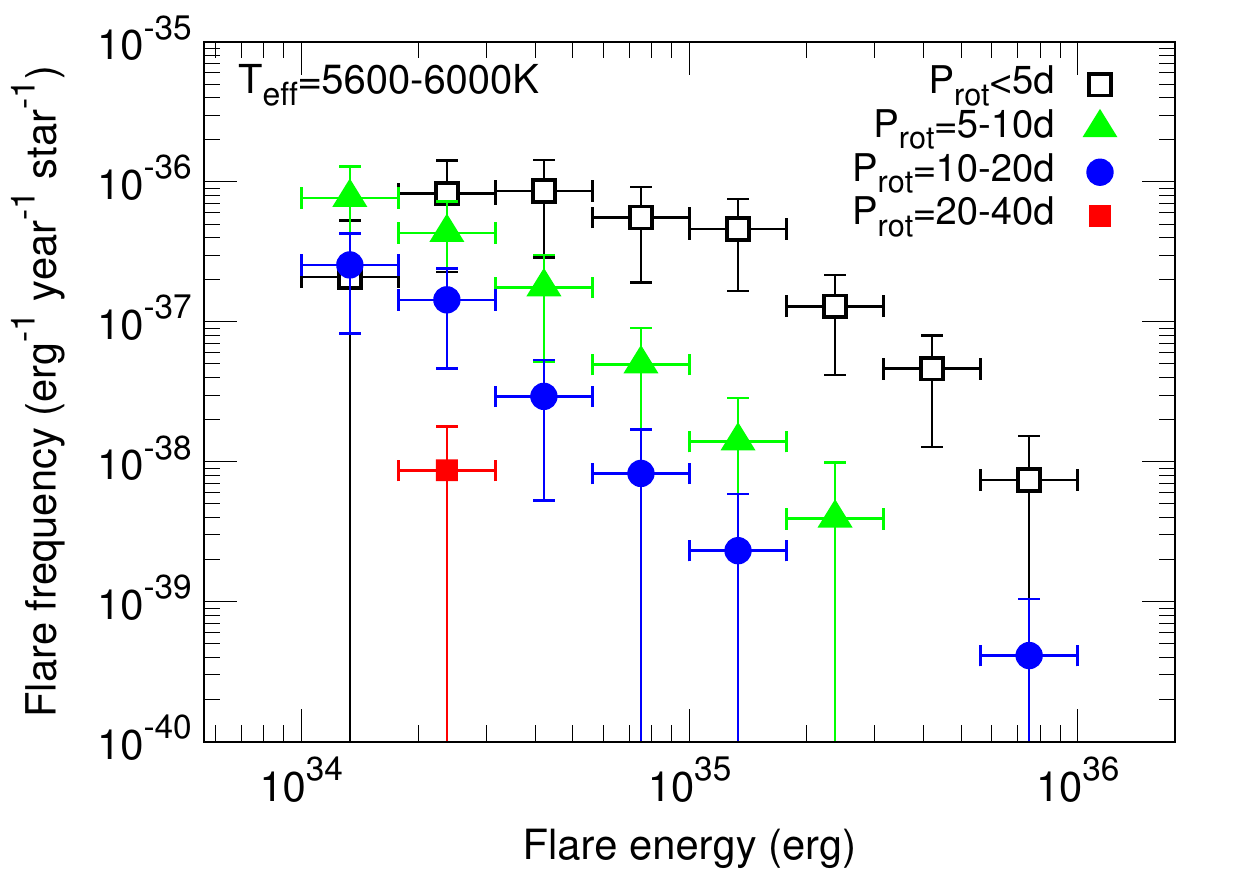}{0.6\textwidth}{}}
\caption{
Occurrence frequency distribution of superflares on solar-type stars with $T_{\mathrm{eff}}$ = 5600 -- 6000 K, 
using the superflare data that were originally found from {\it Kepler} 30-min cadence data (\citealt{Shibayama2013}; \citealt{Candelaresi2014}) 
but are updated using $T_{\mathrm{eff, DR25}}$ and $R_{\mathrm{Gaia}}$ in this study.
The flare energy values in the horizontal axis are limited to $10^{34}$ -- $10^{36}$ erg since 
flare frequency in the range of $\lesssim 10^{34}$ erg can be heavily affected from the detection limit.
The error bar in the horizontal axis direction corresponds to each energy bin.
The vertical axis indicates the number of superflares per star, per year, and per unit energy in each energy bin. 
Error bars in the vertical axis represent 1-$\sigma$ uncertainty of the frequency 
estimated from the $\sqrt{N+1}$ ($N$ : the number of detected flares in each energy bin) considering Poisson statistics.
Unlike the figures in our previous papers (e.g., \citealt{Shibayama2013}; \citealt{Maehara2015}), 
we do not take into account the error values of $E_{\mathrm{flare}}$ since this is now more reliable because of the {\it Gaia}-DR2 stellar radius values.
The symbols are classified with rotation period ($P_{\mathrm{rot}}$) values: 
open squares: $P_{\mathrm{rot}}<$ 5 days (age: $t<0.5$ Gyr), green triangles: $P_{\mathrm{rot}}$ = 5 -- 10 days ($t$= 0.5 -- 1 Gyr),
blue circles: $P_{\mathrm{rot}}$ = 10 -- 20 days ($t$ = 1 -- 3.2 Gyr), and the red filled square: $P_{\mathrm{rot}}$ = 20 -- 40 days ($t>$ 3.2 Gyr).
Age values here are taken from Figures \ref{fig:ProtEene2}(b) \& \ref{fig:aspot-prot}(b).
\label{fig:flare-freq-prot}}
\end{figure}

Figure \ref{fig:flare-freq-prot} represents the occurrence frequency distributions of superflares on 
solar-type stars with $T_{\mathrm{eff}}$ = 5600 -- 6000 K (``the temperature range close to the solar value"
in the above sections) and various rotation period ($P_{\mathrm{rot}}$) values, 
derived from the superflare data updated in the above of this study.
As also presented in our previous studies (e.g., \citealt{Shibayama2013}) with the original data of superflares,
we can see that there are the power-law distributions ($dN/dE\propto E^{-\alpha}$ with the index $\alpha\lesssim$2), 
and rapidly-rotating stars tend to have larger frequency values.
The upper limit values of flare energy roughly depend on rotation period, as already seen in Figure \ref{fig:ProtEene2}.
However, error values are relatively large especially for slowly-rotating stars,
because of the small number of analyzed data. 
For example, one blue circle point in the energy range $E_{\mathrm{flare}}=10^{35.75 \mathrm{-} 36.0}$ erg in Figure \ref{fig:flare-freq-prot}, 
which are a bit far from the other blue circle points, correspond to only one superflare event.
We must note here that these values are treated with cautions, and we need to increase the number 
of slowly-rotating superflare stars by including more new data in the future.

We also note here that as the same as in Figure \ref{fig:Prot-freq-evo0},
the flare frequency value in vertical axis of Figure \ref{fig:flare-freq-prot} is calculated by using 
the number of the solar-type stars in each $P_{\mathrm{rot}}$ bin ($N_{P}$($P_{\mathrm{rot}}$) in 
Table \ref{tab:Nstar-amp-detected} in Appendix \ref{sec:Kepler-Prot-Nstar}) 
detected in \citet{McQuillan2014}. 
The potential errors from this $N_{P}$($P_{\mathrm{rot}}$) value, which are discussed in Appendix \ref{sec:Kepler-Prot-Nstar}, 
should be kept in mind when we discuss the dependence of flare frequency on the rotation period on the basis of this Figure \ref{fig:flare-freq-prot}.
For example, flare frequency of Sun-like stars with $P_{\mathrm{rot}}$ = 20 -- 40 days could become factor two smaller in the largest error case, 
and more investigations with new data in the future are strongly needed. 

\begin{figure}[ht!]
  \gridline{\fig{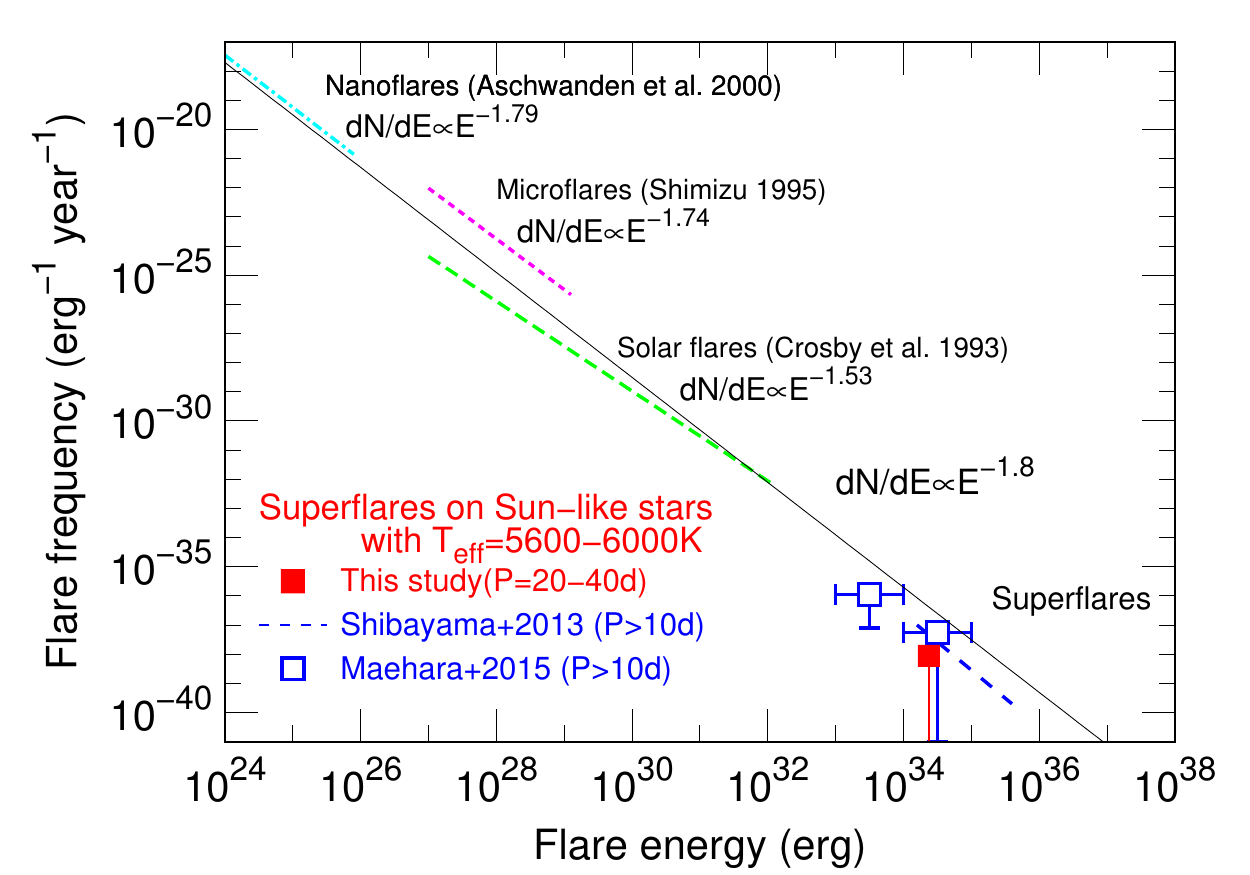}{0.7\textwidth}{}}
\caption{
The comparison between the frequency distribution of superflares and solar flares. 
The red square, blue dashed line, and blue open squares indicate the occurrence frequency distributions of superflares on Sun-like stars 
(slowly-rotating solar-type stars with $T_{\mathrm{eff}}$ = 5600 -- 6000 K). 
The red square corresponds to the updated frequency value of superflares on the stars with $P_{\mathrm{rot}}=20$ -- $40$ days,
which are calculated in this study and presented in Figure \ref{fig:flare-freq-prot}. 
Horizontal and vertical error bars are the same as those in Figure \ref{fig:flare-freq-prot}.
For reference, the blue dashed line and blue open squares are the values of superflares on the stars with $P_{\mathrm{rot}}>$10 days, 
which we presented in Figure 4 of \citet{Maehara2015}
on the basis of original superflare data using {\it Kepler} 30-min cadence data \citep{Shibayama2013} 
and 1-min cadence data \citep{Maehara2015}, respectively.
Definitions of error bars of the blue open squares are the same as those in Figure 4 of \citet{Maehara2015}.
Three dashed lines in upper-left side of this figure indicate the power-law frequency distribution of solar flares 
observed in hard X-ray \citep{Crosby1993}, soft X-ray \citep{Shimizu1995}, and EUV \citep{Aschwanden2000}.
Occurrence frequency distributions of superflares on Sun-like stars and solar flares are roughly on the same power-law line 
with an index of $-1.8$ (black solid line) for the wide energy range between $10^{24}$ and $10^{35}$ erg.
\label{fig:flare-freq-sunstar}}
\end{figure}

Our previous papers (\citealt{Shibayama2013}; \citealt{Maehara2015}) pointed out that the frequency distribution of superflares on 
Sun-like stars and those of solar flares are roughly on the same power-law line.
However, the definition of Sun-like stars in these previous studies ($T_{\mathrm{eff}}$ = 5600 -- 6000 K and $P_{\mathrm{rot}}>10$ days) 
include many stars much younger than the Sun. For example, stars with $P_{\mathrm{rot}}\sim10$ days have the age of $t\sim 1$ Gyr 
(cf. Figures \ref{fig:ProtEene2}(b) \& \ref{fig:aspot-prot}(b)).
It might be better to use only the data of stars rotating as slowly as Sun ($P_{\mathrm{rot}}\sim$25 days and $t\sim4.6$ Gyr).
Then in Figure \ref{fig:flare-freq-sunstar}, 
we newly plot the frequency value of superflares on Sun-like stars with $P_{\mathrm{rot}}=20$ -- $40$ days ($t>$3.2 Gyr) taken from Figure \ref{fig:flare-freq-prot}, 
in addition to the data of solar flares and superflares shown in Figure 4 of \citet{Maehara2015}.
As a result, the newly added data point of superflares on Sun-like stars are roughly on the same power law-line,
though the exact value of superflare frequency of stars with $P_{\mathrm{rot}}=20$ -- $40$ days ($t>$3.2 Gyr) is a bit smaller than 
those of stars with $P_{\mathrm{rot}}>10$ days ($t>$1 Gyr). 
From this figure, we can roughly remark that superflares with energy $>10^{34}$ erg 
would be approximately once every 2000 -- 3000 years on old Sun-like stars with $P_{\mathrm{rot}}\sim 25$ days and $t\sim4.6$ Gyr, 
though the error value is relatively large because of the small number of data of slowly-rotating stars.

We note here that several potential candidates of extreme solar flare events, which can be bigger than the largest solar flare in the past 200 years
($E_{\mathrm{flare}}\sim$10$^{32}$ erg), have been reported from the data over the recent 1000 -- 2000 years 
(e.g., \citealt{Usoskin2017} for review).
For example, significant radioisotope $^{14}$C enhancements have been reported in tree rings for the year 775AD and 994AD,
and they suggest extremely strong and rapid cosmic-ray increase events possibly caused by extreme solar flares (e.g., \citealt{Miyake2012}\&\citeyear{Miyake2013}).
Various potential low-latitude auroral drawings have been also reported from the historical documents around the world,
and they suggest the possibility that extreme solar flare events have occurred several times in the recent $\sim$1000 years
 (e.g., \citealt{Hayakawa2017a}\&\citeyear{Hayakawa2017b}).
In the future studies, the superflare frequency information as in Figure \ref{fig:flare-freq-sunstar} 
should be compared with these radioisotope and historical data in detail.

\subsection{Size frequency distribution of starspots and comparison with sunspots}\label{sec:freq-spots}

In addition to the relation between the rotation period and starspot area in Section \ref{sec:spot-Prot}, 
in the following we also investigate the size frequency distribution of large starspot groups on Sun-like stars and that of sunspots.
Our previous paper \citet{Maehara2017} already conducted this analysis,
but here we aim to investigate again whether the both sunspots and larger starspots can be related to the same physical processes,
by including the updates using {\it Gaia}-DR2 stellar radius values.

\begin{figure}[htbp]
  \gridline{ \fig{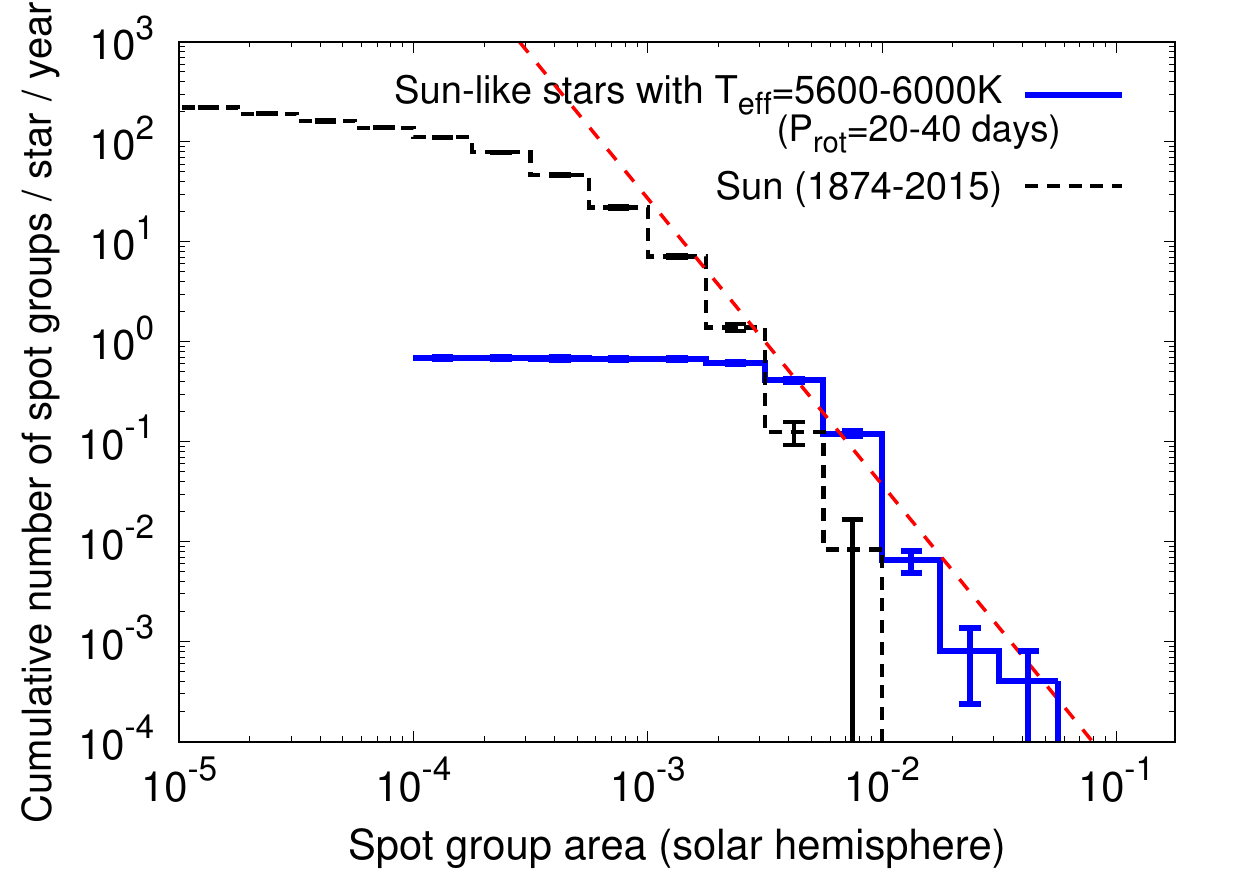}{0.6\textwidth}{}}
\caption{
Comparison between the appearance frequency vs. spot area distribution of starspots on Sun-like stars 
and that of sunspot groups. 
Blue solid lines indicate the cumulative appearance frequency of starspots on Sun-like stars 
($T_{\mathrm{eff}}$ = 5600 -- 6000 K and $P_{\mathrm{rot}}=20$ -- $40$ days).
The value is almost constant in the spot area range below $2\times 10^{-3} A_{1/2\sun}$ ($A_{1/2\sun}\sim 3\times 10^{22}$ cm$^{2}$), 
and this is because the brightness variations of the stars with small starspots could not be detected, as mentioned in \citet{Maehara2017}.
Black dashed lines indicate the cumulative appearance frequency of sunspot groups 
as a function of the maximum area of each sunspot group, 
which are exactly the same data presented in Figure 5 of \citet{Maehara2017} 
(Then as for the details of the data description, see \citealt{Maehara2017}). 
The vertical error bars of these two lines indicate the square root of the number of stars in each bin.
The red dashed line represents the power-law fit to the frequency distribution of starspots 
in the spot area range of $10^{-2.5}$ -- $10^{-1.0} A_{1/2\sun}$. 
The power-law index of the line is $-2.9\pm 0.5$.
 \label{fig:spot-freq-sunstar}}
\end{figure}

Figure \ref{fig:spot-freq-sunstar} shows the comparison between the cumulative appearance frequency of starspots on 
Sun-like stars ($T_{\mathrm{eff}}$ = 5600 -- 6000 K and $P_{\mathrm{rot}}$=20-40 days) and that of sunspot groups. 
The estimation method of the cumulative appearance frequency values is basically the same as that 
used for Figure 5 of \citet{Maehara2017}, and this was already described in detail in that paper.
Only one difference is that we use the updated $A_{\mathrm{spot}}$ values described in the above Section \ref{sec:spot-Prot}.
The occurrence frequency decreases as the area of sunspots or starspots increases.
The appearance rate of both the sunspot groups and that of starspots with the area $>10^{-2.5} A_{1/2\sun}$ 
is approximately once in a few years. 
The cumulative appearance frequency of starspots on Sun-like stars can be fitted 
by a power-law function with the power-law index of $-2.9\pm 0.5$ (red dashed line) 
for spot areas between $10^{-2.5}$ -- $10^{-1.0} A_{1/2\sun}$.
This power-law index is a bit steeper compared with those originally presented in \citet{Maehara2017}.

According to \citet{Bogdan1988}, the size distribution of individual sunspot umbral areas shows lognormal distribution. 
Although the overall size-frequency distribution of sunspot groups in Figure \ref{fig:spot-freq-sunstar} also shows the similar log-normal distribution, 
the distribution of sunspot groups for large sunspots is roughly on this power-law line for sunspot areas 
between $10^{-3.25}$ -- $10^{-2.5} A_{1/2\sun}$.
The appearance frequency of sunspots with spot areas of $\sim 10^{-2}A_{\mathrm{1/2\sun}}$ is about 10 times 
lower than that of starspots on Sun-like stars. 
This difference between the Sun and Sun-like stars might be caused 
by the lack of a ``super-active" phase on our Sun during the last 140 years \citep{Schrijver2012}.
These results shown here are basically the same as those originally presented in \citet{Maehara2017},
and we confirmed the similarity between the size distribution of sunspots and that of starspots.
This supports that the both sunspots and larger starspots could be related to the same physical processes.
The upper limit of starspot size values of Sun-like stars in Figure \ref{fig:spot-freq-sunstar} 
is $\sim$ a few $\times 10^{-2} A_{1/2\sun}$ and the appearance frequency of these spots is approximately once every 2000 -- 3000 years.

\clearpage

\section{Summary and Future prospects}\label{sec:summary}

This paper reports the latest view of superflares on solar-type stars found from our series of studies using {\it Kepler} data,
by including the recent updates using Apache Point Observatory (APO) 3.5m telescope spectroscopic observations 
and {\it Gaia}-DR2 data.

\subsection{Summary of  APO 3.5m telescope spectroscopic observations}\label{subsec:summary-apo3.5m}

In Sections \ref{sec:target-obs} -- \ref{sec:discussion-spec}, 
we described the results of our spectroscopic observations of {\it Kepler} solar-type superflare stars. 
We newly conducted APO 3.5m telescope spectroscopic observations of the 18 solar-type superflare stars 
found from {\it Kepler} 1-min (short) time cadence data \citep{Maehara2015}.

\begin{enumerate} 
\renewcommand{\labelenumi}{(\roman{enumi})}
\item 
More than half (43 stars) are confirmed to be ``single" stars, 
among the 64 solar-type superflare stars in total that have been spectroscopically investigated so far 
in this APO3.5m observation and our previous Subaru/HDS \citep{Notsu2015a} observations (Table \ref{tab:Nstar-Nsg-Nbin}).
\item  ``$v\sin i$" (projected rotational velocity) values are consistent 
with the rotational velocity values $v_{\mathrm{lc}}$ estimated from the brightness variation period of {\it Kepler} data (Figure \ref{fig:vlc-vsini}). 
\item There is a positive correlation between the amplitude of the brightness variation and 
Ca II (Ca II 8542\AA~and Ca II H\&K lines) index values (Figure \ref{fig:CaAmp}), 
and this suggests that there is a rough positive correlation between the starspot coverage from {\it Kepler} photometric data 
and the stellar average magnetic field.
\item The results of (ii)\&(iii) support the idea that 
the quasi-periodic brightness variation of {\it Kepler} solar-type superflare stars is caused by the rotation with large starspots.
\item Lithium abundaces of superflare stars suggest that {\it Kepler} solar-type superflare stars include many young stars 
but also include old stars like our Sun (Figure \ref{fig:Teff-ALi}).
\end{enumerate}

\subsection{Summary of the statistical properties of {\it Kepler} solar-type superflare stars incorporating {\it Gaia}-DR2 data}\label{subsec:summary-GaiaDR2}

Then in Section \ref{sec:dis-Kepler}, 
we investigated the statistical properties of {\it Kepler} solar-type superflare stars originally described in our previous studies 
(\citealt{Maehara2012}, \citeyear{Maehara2015}, \& \citeyear{Maehara2017}; \citealt{Shibayama2013}; \citealt{Notsu2013b}), 
by incorporating {\it Gaia}-DR2 stellar radius estimates (reported in \citealt{Berger2018}) and updating the parameters (e.g., flare energy, spot size).

\begin{enumerate} 
\renewcommand{\labelenumi}{(\roman{enumi})}
\item More than 40\%~of the original solar-type superflare stars in our previous studies are now classified as subgiant stars (Figure \ref{fig:HR}).

\item The bolometric energy released by flares ($E_{\mathrm{flare}}$) is not inconsistent with the magnetic energy ($E_{\mathrm{mag}}$) stored around the large starspots 
(Figures \ref{fig:spotEene-sunstar1}(b) \& \ref{fig:spotEene-temp}).

\item Our previous studies suggested that the maximum superflare energy in a given rotation period bin does 
not have a clear correlation with the rotation period, and superflares up to 10$^{35}$ erg could occur on slowly-rotating solar-type stars (Figure \ref{fig:ProtEene1}(a)).
This study suggests that the maximum superflare energy continuously decreases as the rotation period increases (as the star becomes older) 
(Figures \ref{fig:ProtEene1}(b) \& \ref{fig:ProtEene2}). 
Superflares with energies $\lesssim 5\times10^{34}$ erg occur 
on old, slowly-rotating Sun-like stars ($T_{\mathrm{eff}}$ = 5600 -- 6000 K, $P_{\mathrm{rot}}\sim$25 days, and age $t\sim$4.6 Gyr) 
approximately once every 2000 -- 3000 years (Figures \ref{fig:ProtEene2}(b) \& \ref{fig:flare-freq-sunstar}).
In contrast, superflares up to $\sim$10$^{36}$ erg can occur
on young rapidly-rotating stars ($P_{\mathrm{rot}}\sim$ a few days and $t\sim$ a few hundreds Myr) (Figure \ref{fig:ProtEene2}),
and the flare frequency of such young rapidly-rotating stars is $\sim$100 times higher compared with
the above old slowly-rotating Sun-like stars (Figures \ref{fig:Prot-freq-evo0} \& \ref{fig:flare-freq-prot}).

\item
The maximum area of starspots does not depend on the rotation period and are roughly constant or very gentle decreasing trend
around $A_{\mathrm{spot}} = 5\times 10^{-2}$ -- $1\times 10^{-1}A_{1/2\sun}$ ($A_{1/2\sun}\sim 3\times 10^{22}$ cm$^{2}$: solar hemisphere) 
when the star is young and rapidly-rotating.
However, as the star becomes older and its rotation slows down, it starts to have a steep decreasing trend at a certain point :
$P_{\mathrm{rot}}\sim$12 days ($t\sim$1.4 Gyr) for the stars with $T_{\mathrm{eff}}$ = 5600 -- 6000 K  (Figure \ref{fig:aspot-prot}(b)), 
and $P_{\mathrm{rot}}\sim$14 days for the stars with $T_{\mathrm{eff}}$ = 5100 -- 5600 K (Figure \ref{fig:aspot-prot}(a)). 
Maximum size of starspots on slowly-rotating Sun-like stars is $\sim$1 \%~of the solar hemisphere, 
and this is enough for generating superflares with the energy $\lesssim 5\times10^{34}$ erg described in (iii).

\item
These decreasing trends of the maximum flare energy (in (iii)) and the maximum starspot area (in (iv))
can be related with each other since the superflare energy can be explained by the starspot magnetic energy as in (ii).
However, there is also a difference between the two: the maximum area of starspots 
starts to steeply decrease at a certain $P_{\mathrm{rot}}$ value (as in (iv)), 
while the maximum flare energy continuously decrease as the rotation slows down (as in (iii)).
This can suggest a possibility that the flare energy is determined not only by the starspot area
but also by other important factors (e.g., spot magnetic structure).

\item The size distribution of starspots on Sun-like stars ($T_{\mathrm{eff}}$ = 5600 -- 6000 K and $P_{\mathrm{rot}}\sim$25 days) 
 between $A_{\mathrm{spot}}=10^{-2.5}$ -- $10^{-1.0} A_{1/2\sun}$ 
roughly locate on the extension line of the distribution of sunspot groups
between  $A_{\mathrm{spot}}=10^{-3.25}$ -- $10^{-2.5} A_{\mathrm{1/2\sun}}$ 
(Figure \ref{fig:spot-freq-sunstar}).
The upper limit of starspot size values on slowly-rotating Sun-like stars 
would be $\sim$a few $\times 10^{-2} A_{1/2\sun}$ and the appearance frequency of these spots is approximately once every 2000--3000 years.

\end{enumerate}
 
\subsection{Future prospects}\label{subsec:future}

In this paper, we have reported the current updates of our series of studies on superflares on solar-type stars.
However, we need more studies to clarify the properties of superflare stars on Sun-like stars 
and answer the important question ``Can our Sun have superflares ?".
For example, our spectroscopic observations so far have observed 64 solar-type superflare stars, 
but the number of Sun-like superflare stars ($T_{\mathrm{eff}}$ = 5600 -- 6000 K, $P_{\mathrm{rot}}\sim$25 days, and $t\sim4.6$ Gyr) 
that have been investigated spectroscopically and found to be ``single" stars
are now only a few ($\sim$1).
In the figures in Section \ref{sec:dis-Kepler}, the number of old slowly-rotating Sun-like superflare stars are now very small, 
and the current statistical discussions are not enough.
In this study, as for superflares found from {\it Kepler} 30-min cadence data, 
we only used the data originally found as solar-type superflare stars 
from {\it Kepler} data of the first $\sim$500 days (Quarter 0 -- 6) in our previous study (\citealt{Shibayama2013}; \citealt{Candelaresi2014}).
In our next study, we plan to increase the number of data by using the whole {\it Kepler} dataset of $\sim$1500 days (Quarter 0 -- 17) 
\footnote{
We note that \citet{Davenport2016} conducted the flare survey of F--M type stars using the whole {\it Kepler} 30-min cadence dataset,
but their data are not necessarily enough for investigating flare stars with smaller frequency (e.g., slowly-rotating Sun-like stars). 
}
Moreover, the data from the next missions such as {\it TESS} \citep{Ricker2015} and {\it PLATO} \citep{Rauer2014} are expected to bring us 
more superflare data on solar-type stars in near future. 
These two missions are helpful also for spectroscopic follow-up observations of superflare stars 
since these two missions observe more nearby stars (with their wider spatial coverage), compared with {\it Kepler}.

In addition to the statistical properties of superflares that we discussed in this paper,
there are so many topics that should be investigated in near future to fully understand the physics of superflares 
and their effects in the related research fields.
For example, mechanisms of white-light continuum emissions of superflares 
(e.g.,  \citealt{Kowalski2018}; \citealt{Heinzel2018}; \citealt{Namekata2017}), 
chromospheric line profiles during superflares (e.g., \citealt{Houdebine1993}; \citealt{Hawley2007}; \citealt{Honda2018}),
stellar mass ejections (e.g., CMEs) during superflares 
(e.g., \citealt{Aarnio2012}; \citealt{Leitzinger2014}; \citealt{Hudson2015}; \citealt{Osten2015}; \citealt{Takahashi2016}; 
\citealt{Harra2016}; \citealt{Vida2016}; \citealt{Moschou2017}; \citealt{Crosley2018}),
the impacts of superflares on planets  
(e.g., \citealt{Segura2010}; \citealt{Airapetian2016}; \citealt{Atri2017}; \citealt{Lingam2017}; \citealt{Riley2018}),
detailed comparisons with a history of solar activity over $\sim$1000 years 
(e.g., \citealt{Miyake2012}\&\citeyear{Miyake2013}; \citealt{Hayakawa2017a}\&\citeyear{Hayakawa2017b}; \citealt{Usoskin2017}), 
quasi-periodic pulsations (QPPs) of the brightness during superflares (e.g., \citealt{Pugh2016}; \citealt{McLaughlin2018}), 
complexities of superflare-generating starspots (e.g.,. \citealt{Maehara2017}; \citealt{Toriumi2017}),
how starspots are distributed on the surface of superflare stars (e.g., \citealt{Doyle2018}; \citealt{Roettenbacher2018}),
how starspots and plages are co-located in the case of active stars (e.g., \citealt{Morris2018}),
and lifetimes and formation/decay processes of large starspots in solar-type stars (e.g., \citealt{Shibata2013}; \citealt{Giles2017}; \citealt{Namekata2019}).
The superflare stars newly found from the above new missions ({\it TESS} and {\it PLATO} ) 
would be helpful for the detailed observations to investigate these topics
(e.g., spectroscopic or multi-wavelength observations of superflares themselves, and long-term changes of stellar activities).

\acknowledgments
This study is based on observations obtained with the Apache Point Observatory (APO) 3.5 m telescope, 
which is owned and operated by the Astrophysical Research Consortium.
We used observation time allocated to University of Washington.
We are grateful to APO 3.5m Observing Specialists (Candace Gray, Jack Dembicky, Russet McMillan, and Theodore Rudyk)
and other staffs of Apache Point Observatory and University of Washington for their large contributions in carrying our observations.
George Wallerstein and Charli Sakari kindly shared their observation time to take our data on 2017 October 15 (UT) as a time exchange.
We acknowledge with great thanks that Brett Morris helped us 
when the corresponding author Y.N. learned how to make observations at APO and how to analyze APO data. 
Thanks are also to Adam Kowalski and Petr Heinzel for general discussions.
We are indebted to Yoichi Takeda for providing us the TGVIT and SPTOOL programs developed by him.

This paper includes data collected by the {\it Kepler} mission. 
Funding for the {\it Kepler} mission is provided by the NASA Science Mission Directorate.
The {\it Kepler} data presented in this paper were obtained from the Multimission Archive at STScI. 
This paper also has made use of data from the European Space Agency (ESA)
mission {\it Gaia} (\url{https://www.cosmos.esa.int/gaia}), processed by
the {\it Gaia} Data Processing and Analysis Consortium (DPAC,
\url{https://www.cosmos.esa.int/web/gaia/dpac/consortium}). Funding
for the DPAC has been provided by national institutions, in particular
the institutions participating in the {\it Gaia} Multilateral Agreement.

This work originally started from the discussions 
during ``Superflare Workshop 2016 at Kyoto University: Superflares on Solar-type Stars and Solar Flares, and Their Impacts on Exoplanets and the Earth" 
supported by the International Research Unit of Advanced Future Studies of Kyoto University.
This work was also supported by JSPS KAKENHI Grant Numbers JP16J00320, JP16J06887, JP16H03955, JP17H02865, JP17K05400, and JP18J20048.

\vspace{5mm}
\facilities{APO/ARC3.5m(ARCES), Kepler, Gaia.}

\clearpage
\appendix

\section{Details of analyses and results of the APO3.5m spectroscopic observations}\label{sec:ana-results}
\subsection{Binarity}\label{subsec:ana-binarity}
\begin{figure}[htbp]
\gridline{\fig{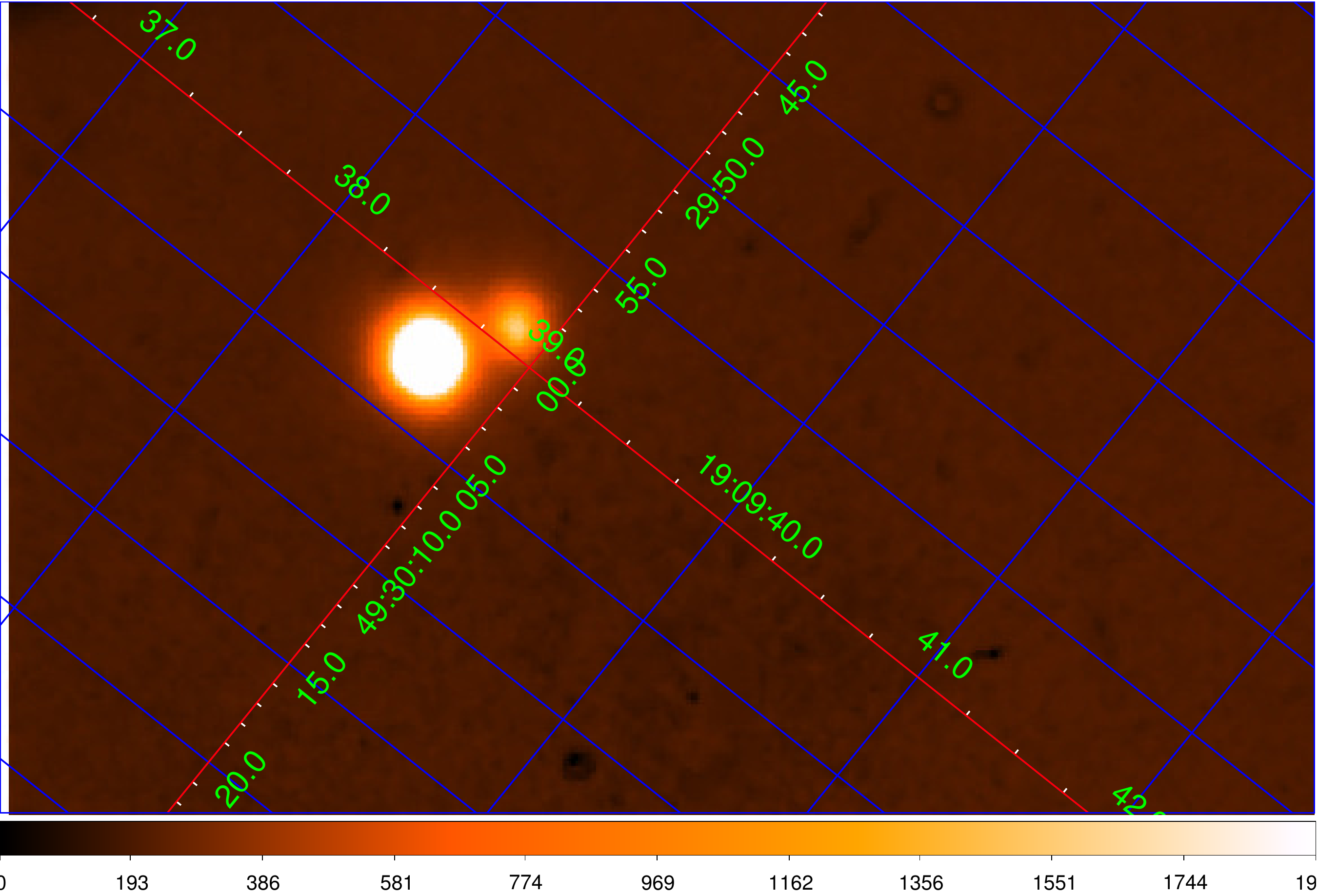}{0.45\textwidth}{(a)}}
\gridline{\fig{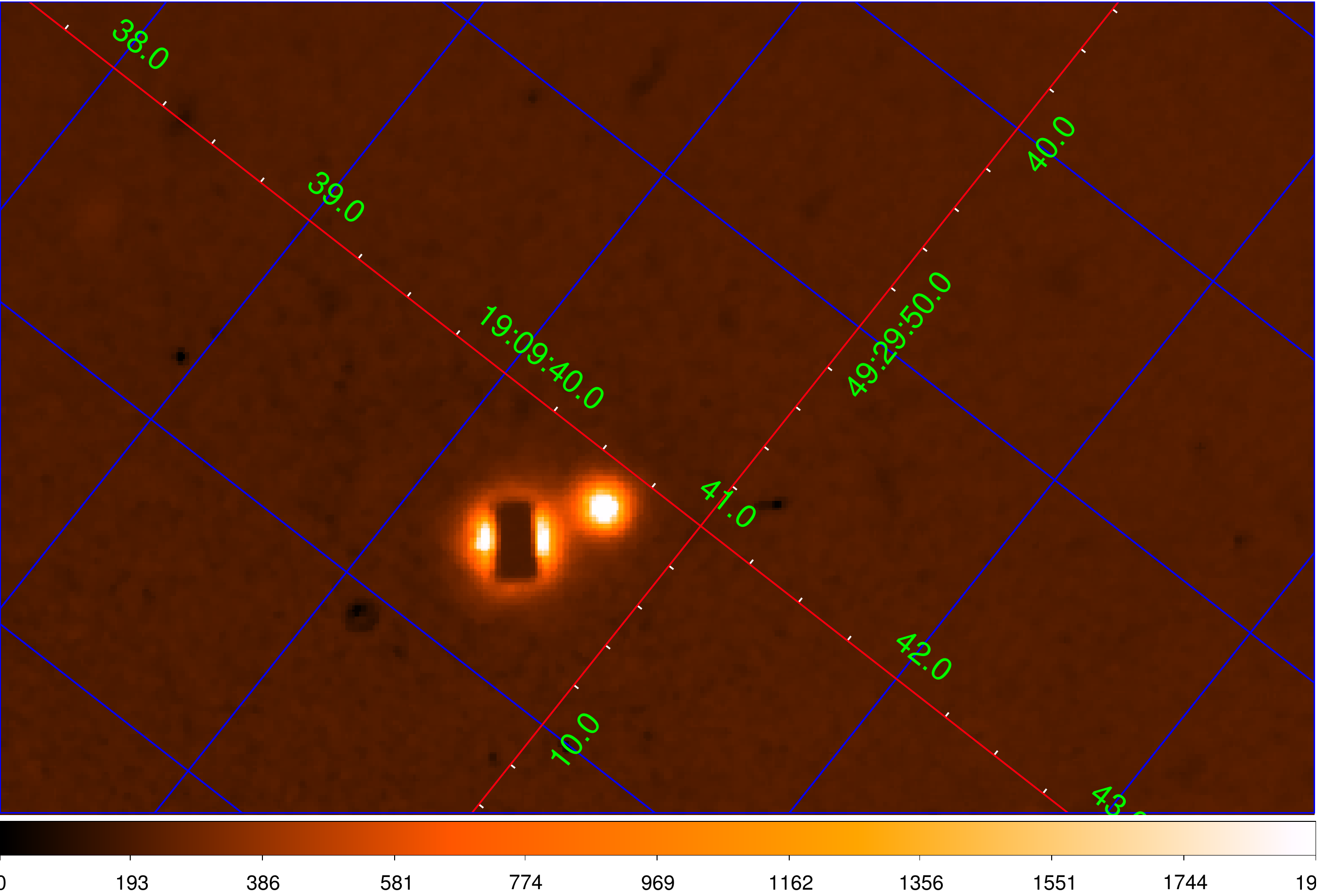}{0.45\textwidth}{(b)}
         \fig{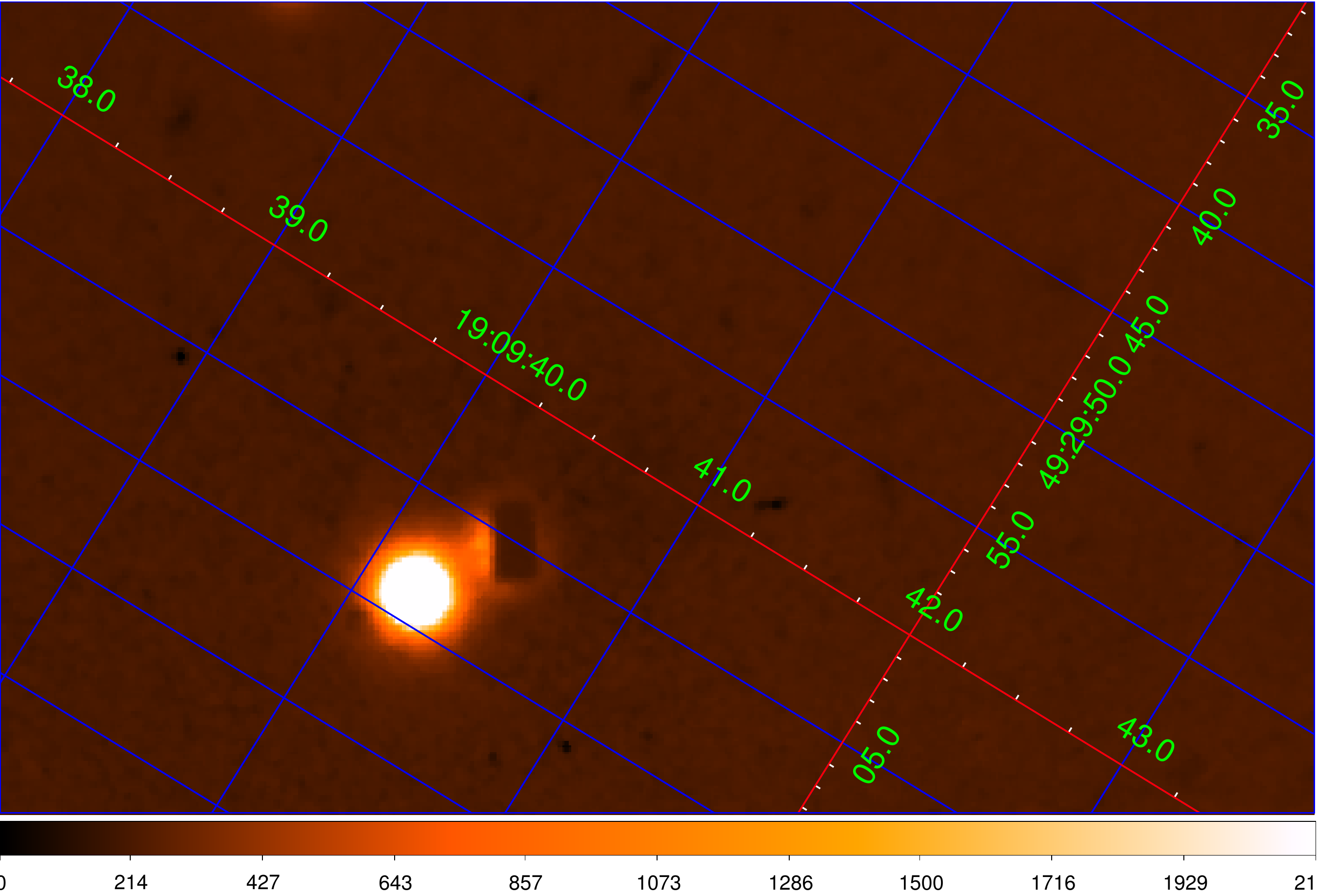}{0.45\textwidth}{(c)}}
\gridline{\fig{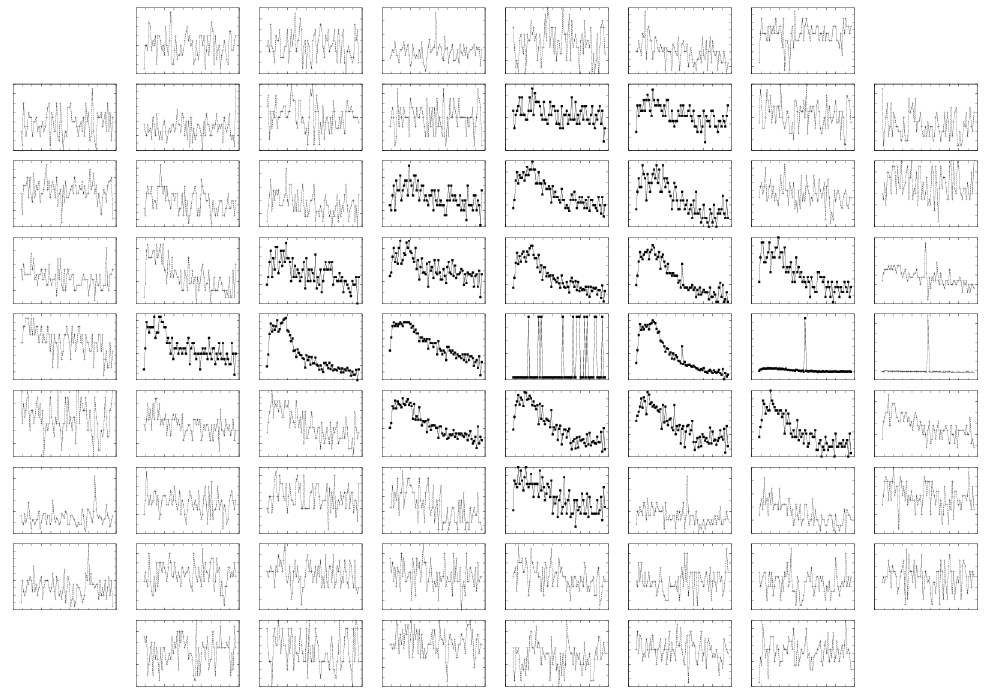}{0.45\textwidth}{(d)}
          \fig{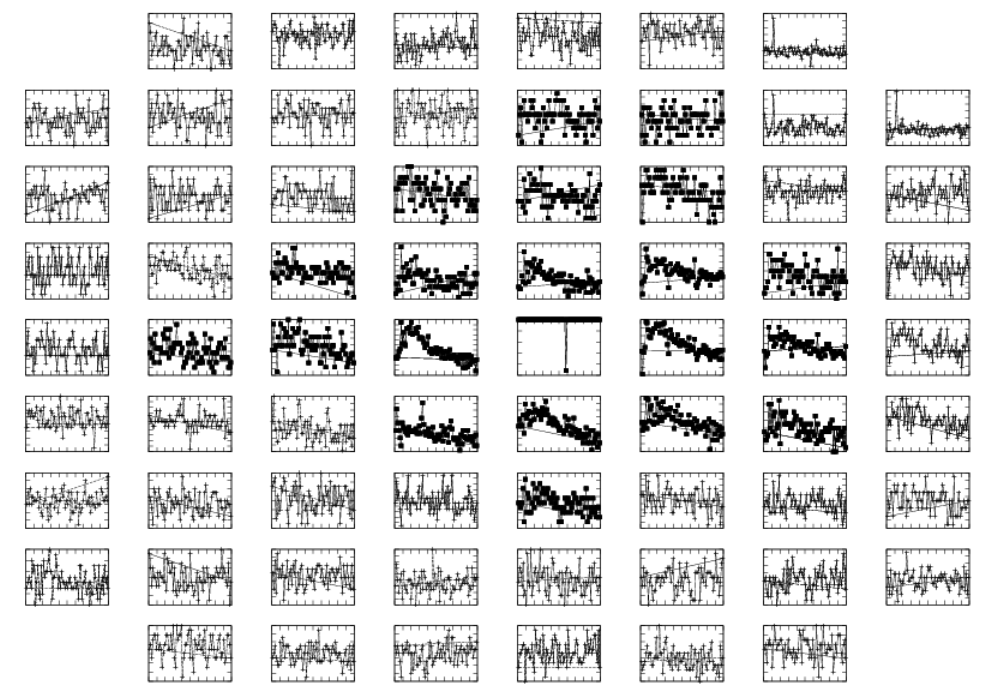}{0.45\textwidth}{(e)}}        
\caption{(a)(b)(c) Slit viewer images of the visual binary target KIC11551430. 
These images show that this star (the brighter primary star KIC11551430A, which we took spectroscopic data as in (b)) 
has a visual fainter companion star (KIC11551430B, which we took spectroscopic data as in (c)).
\\ (d)(e) Pixel count data around two typical superflares on KIC11551430 from the data of \citet{Maehara2015}. 
The peak time (BJD (Barycentric Julian Date) - 2,400,000) of flares in (d) \& (e) are 55019.673718 and 56196.445193, 
and the estimated bolometric energy of them are $2.2\times 10^{35}$ erg and $2.8\times 10^{34}$ erg, respectively.
The pixels in the center of the flares are saturated. \label{fig:SV-KIC11551430}}
\end{figure}

\begin{figure}[htbp]
\gridline{\fig{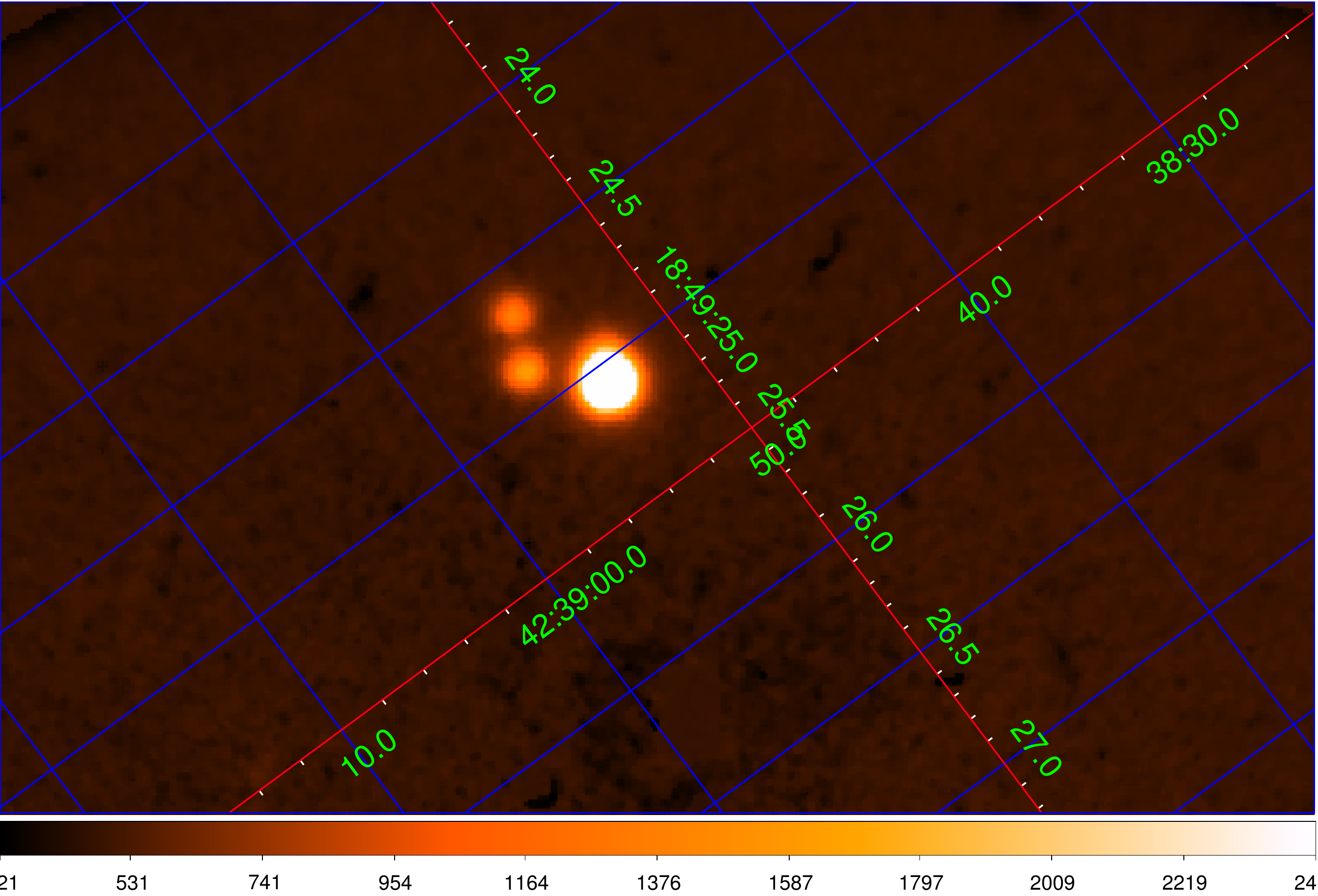}{0.45\textwidth}{(a)}
         \fig{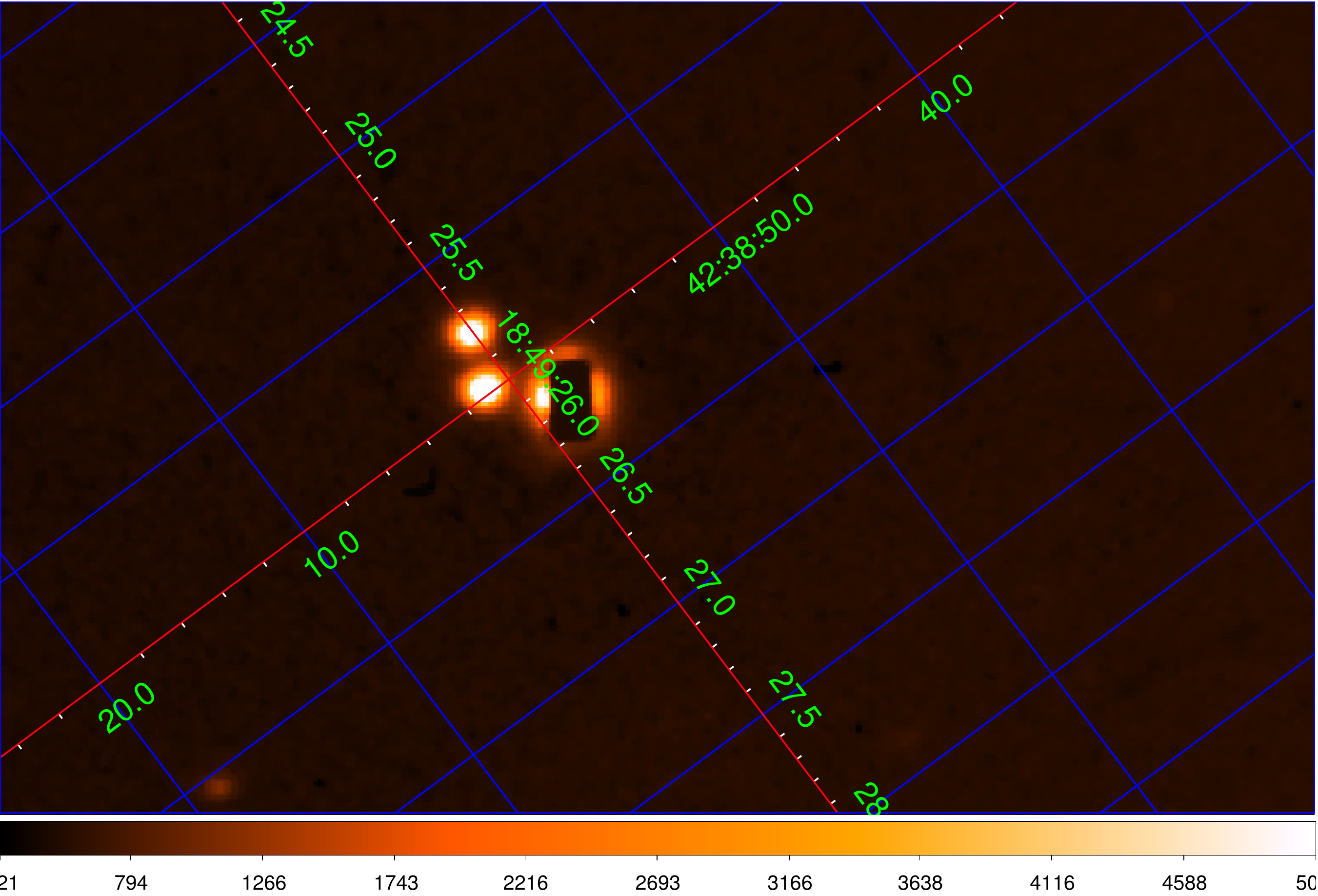}{0.45\textwidth}{(b)}}
\gridline{\fig{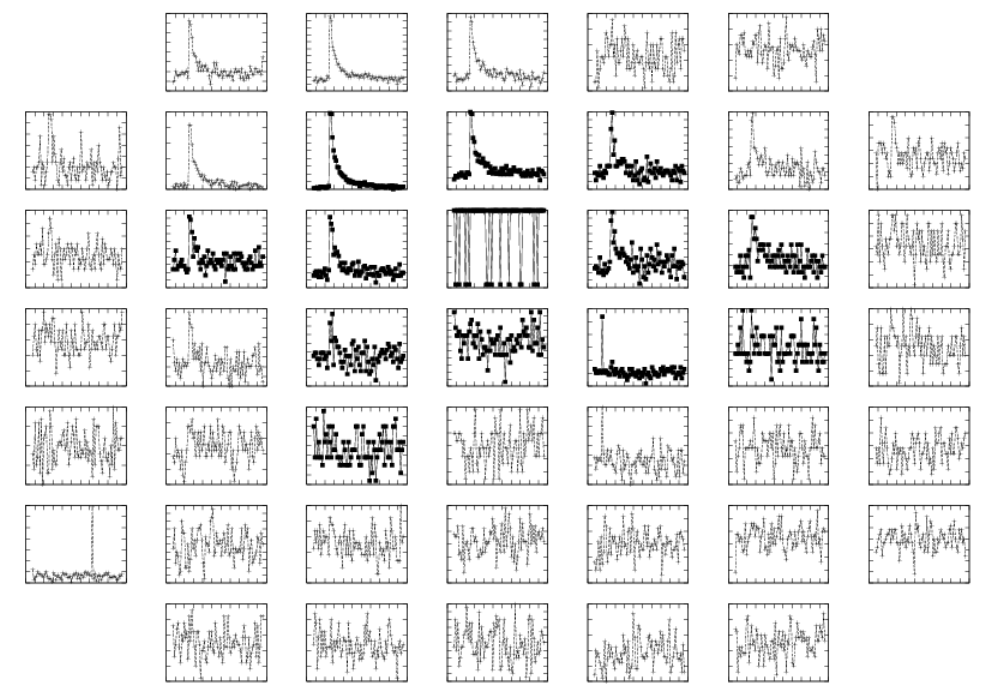}{0.45\textwidth}{(c)}
          \fig{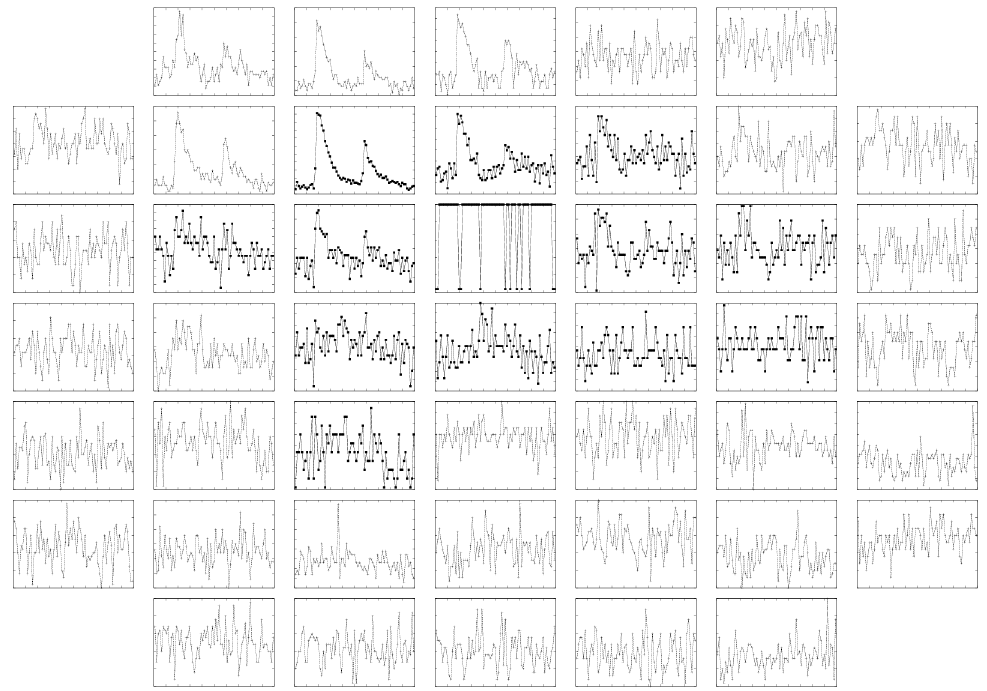}{0.45\textwidth}{(d)}}  
\caption{
(a)(b) Slit viewer images of the visual binary target KIC7093428. 
These images show that this star (the brighter primary star KIC7093428A, which we took spectroscopic data as in (b)) 
has two fainter visual companion stars (KIC7093428B and KIC7093428C).
\\ (c)(d) Pixel count data around two typical superflares on KIC7093428 from the data of \citet{Maehara2015}.
The peak time (BJD - 2,400,000) of flares in (c) \& (d) are 55102.453431 and 55104.090097, 
and the estimated bolometric energy of them are $4.0\times 10^{34}$ erg and $1.1\times 10^{34}$ erg, respectively.
The pixels in the center of the flares are saturated. 
\label{fig:SV-KIC7093428}}
\end{figure}

For the first step of our analyses, we checked the binarity of each superflare star, as we did in \citet{Notsu2015a}.
First, we checked slit viewer images of APO3.5m telescope.
Two stars (KIC11551430 and KIC7093428) have visual companion stars, 
as shown in Figures \ref{fig:SV-KIC11551430} (a) and \ref{fig:SV-KIC7093428} (a).

As for KIC11551430, we took spectra of both components (the brighter primary star: KIC11551430A and the fainter companion star KIC11551430B) 
of the visual binary system separately, as shown in Figures \ref{fig:SV-KIC11551430} (b) and (c). 
We checked again pixel count data of superflare events on KIC11551430 (Figures \ref{fig:SV-KIC11551430} (d) and (e)), 
which are originally in \citet{Maehara2015}, and confirmed that the center positions of the brightness during superflare events 
are roughly the same as those during quiescent state.
This is consistent with the possibility that superflares occur on the primary star KIC11551430A (cf. Figures S1 \& S2 of \citealt{Maehara2012}).

As for KIC7093428, we can see one primary brighter star (KIC7093428A) 
and two fainter companion stars (KIC7093428B and KIC7093428C) in Figure \ref{fig:SV-KIC7093428} (a).
We only took spectra of the main brighter star (KIC7093428A) as shown in Figure \ref{fig:SV-KIC7093428} (b), since companions stars (B \& C) are too faint.
We also checked again pixel count data of superflare events on KIC7093428 (Figures \ref{fig:SV-KIC7093428} (c) and (d)), 
which are originally in \citet{Maehara2015}.
We can see that the center positions of the brightness during superflare events are shifted compared with those during quiescent state.
This suggests the high possibility that superflares occur on the fainter companion stars KIC7093428B or KIC7093428C 
(cf. Figures S1 \& S2 of \citealt{Maehara2012}), not on the primary G-type star KIC7093428A.
Then KIC7093428 has ``VB" in the second column of Table \ref{tab:DR25-gaia}.

\begin{figure}[htbp]
\gridline{\fig{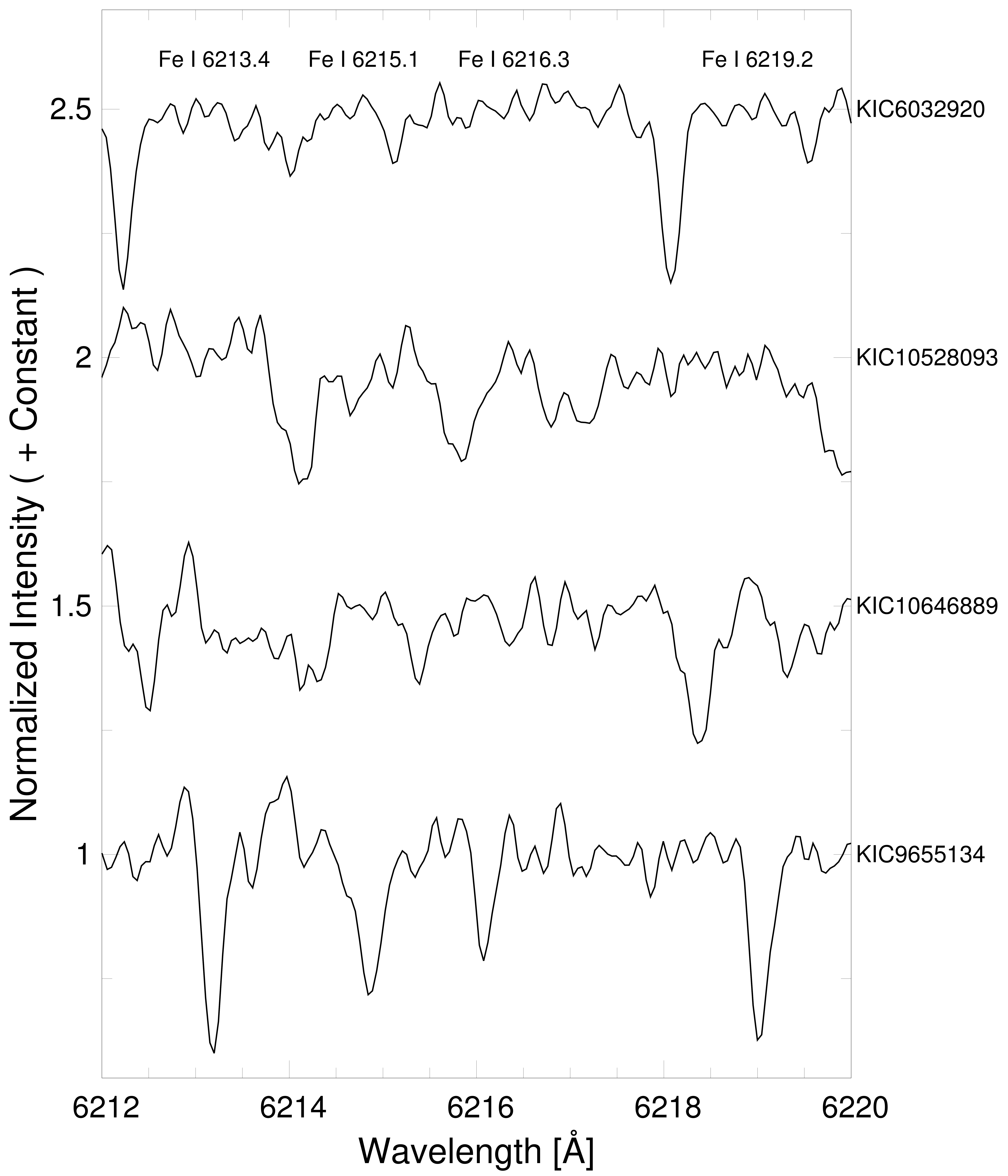}{0.5\textwidth}{}}
\caption{Example of photospheric absorption lines, including Fe I 6213, 6215, 6216, and 6219\AA~lines, 
of the fainter four superflare stars (KIC6032920, KIC10528093, KIC10646889, and KIC9655134). 
The wavelength scale is adjusted to the heliocentric frame.\label{fig:faintsp-Fe6212}}
\end{figure}

As described in Section \ref{sec:target-obs}, 
we selected 23 target stars and took spectra of 22 stars except for the too faint target star KIC10745663 ($K_{p}$=14.3 mag).
Among the 22 observed stars, the data quality of our spectroscopic data 
of the fainter four stars (KIC6032920, KIC10528093, KIC10646889, and KIC9655134) 
are not enough for the detailed following discussions. We plot examples of photospheric lines in Figure \ref{fig:faintsp-Fe6212} for reference.
We only remark here that they does not show any clear signs of binarity, on the basis of Figure \ref{fig:faintsp-Fe6212} 
and the slit viewer images of APO3.5m telescope.
In the following discussions of this paper, we only treat the 18 target stars that we got spectra of with enough S/N.

\begin{figure}[htbp]
\gridline{\fig{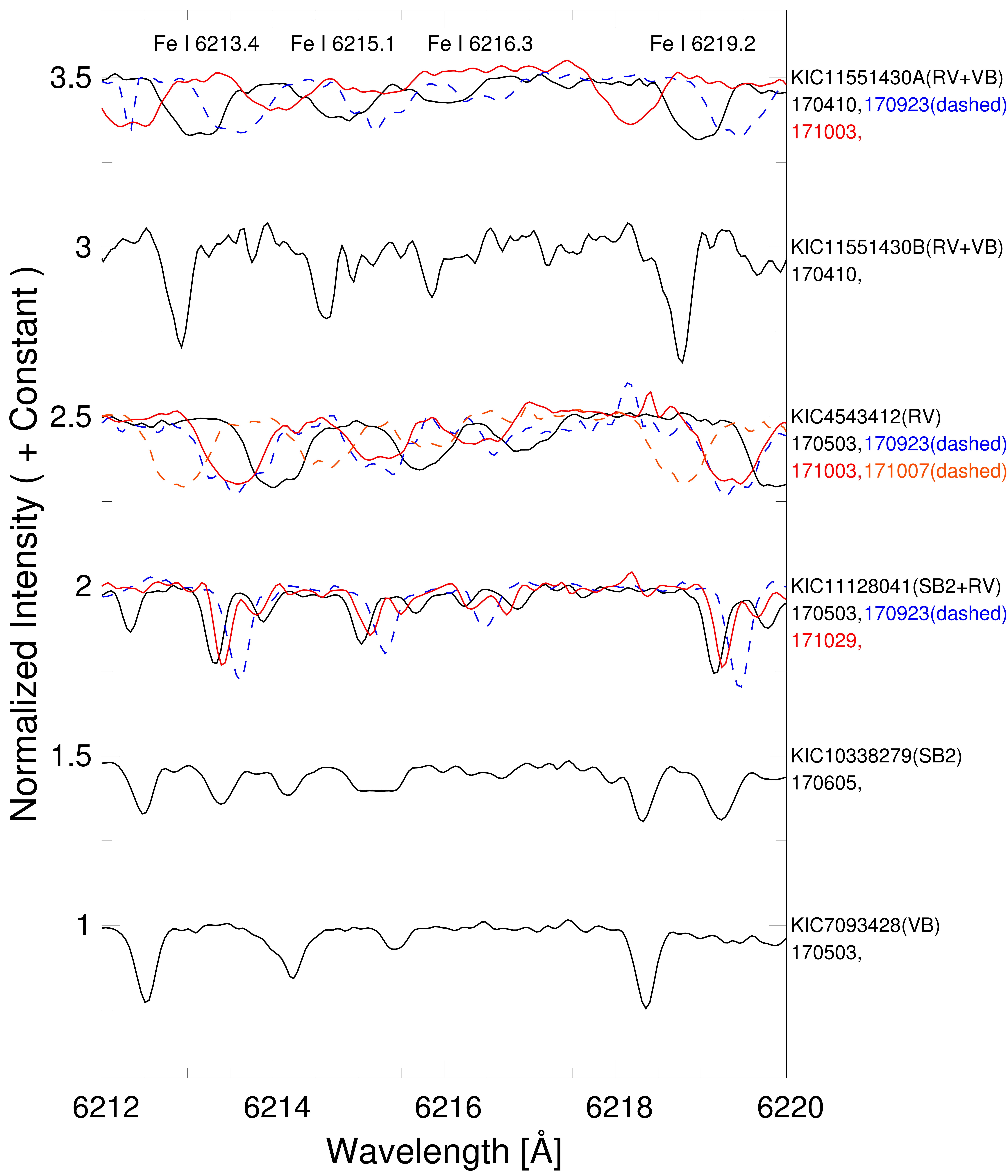}{0.6\textwidth}{}}
\caption{Example of photospheric absorption lines, including Fe I 6213, 6215, 6216, and 6219\AA~lines, of superflare stars 
that we consider as spectroscopic binary stars. The wavelength scale is adjusted to the heliocentric frame. 
Numbers below each star’s name show observation dates (cf. Table \ref{tab:obs-RV}).\label{fig:binsp-Fe6212}}
\end{figure}

Next we investigated the line profiles, and found two stars (KIC11128041 and KIC10338279) show double-lined profiles. 
In this process, we checked by eye the profiles of the many spectral lines, 
and the double-lined spectra of these two stars are shown in Figure \ref{fig:binsp-Fe6212}. 
Since these double-lined profiles are caused by the overlap of the radiation of multiple stars, 
we regard these two stars as double-lined spectroscopic binary stars. 
These two stars have ``SB2" in the second column of Table \ref{tab:DR25-gaia}.

Third, we investigated time variations of the line profiles between the multiple observations 
that are expected to be caused by the orbital motion in the binary system. 
This investigation was for the target stars that we observed multiple times (16 stars). 
In Section \ref{sec:target-obs}, we already measured the radial velocity (RV) of all the target stars 
that were not classified as visual binary stars or double-lined spectroscopic binary stars, 
and these values are listed in Table \ref{tab:obs-RV}.
We have also conducted spectroscopic observations of these four stars (KIC4742436, KIC4831454, KIC9652680, and KIC11610797) 
using Subaru telescope (\citealt{Notsu2015a}\&\citeyear{Notsu2015b}; \citealt{Honda2015}), 
and we also use these data in the RV investigation here (They are also listed in Table \ref{tab:obs-RV}).
As a result, KIC11551430A, KIC4543412, and KIC11128041 show RV changes as shown in Figure \ref{fig:binsp-Fe6212},
and these RV changes are larger enough compared with RV errors of APO data ($\lesssim$1 km s$^{-1}$).
We confirmed this typical RV error value ($\lesssim$1 km s$^{-1}$) 
by comparing our spectroscopic data of the single comparison stars that we observed multiple times in this study
and are listed in Table \ref{tab:Smwo-Sapo}.
These three stars have ``RV" in the second column of Table \ref{tab:DR25-gaia}.

\begin{figure}[ht!]
\gridline{\fig{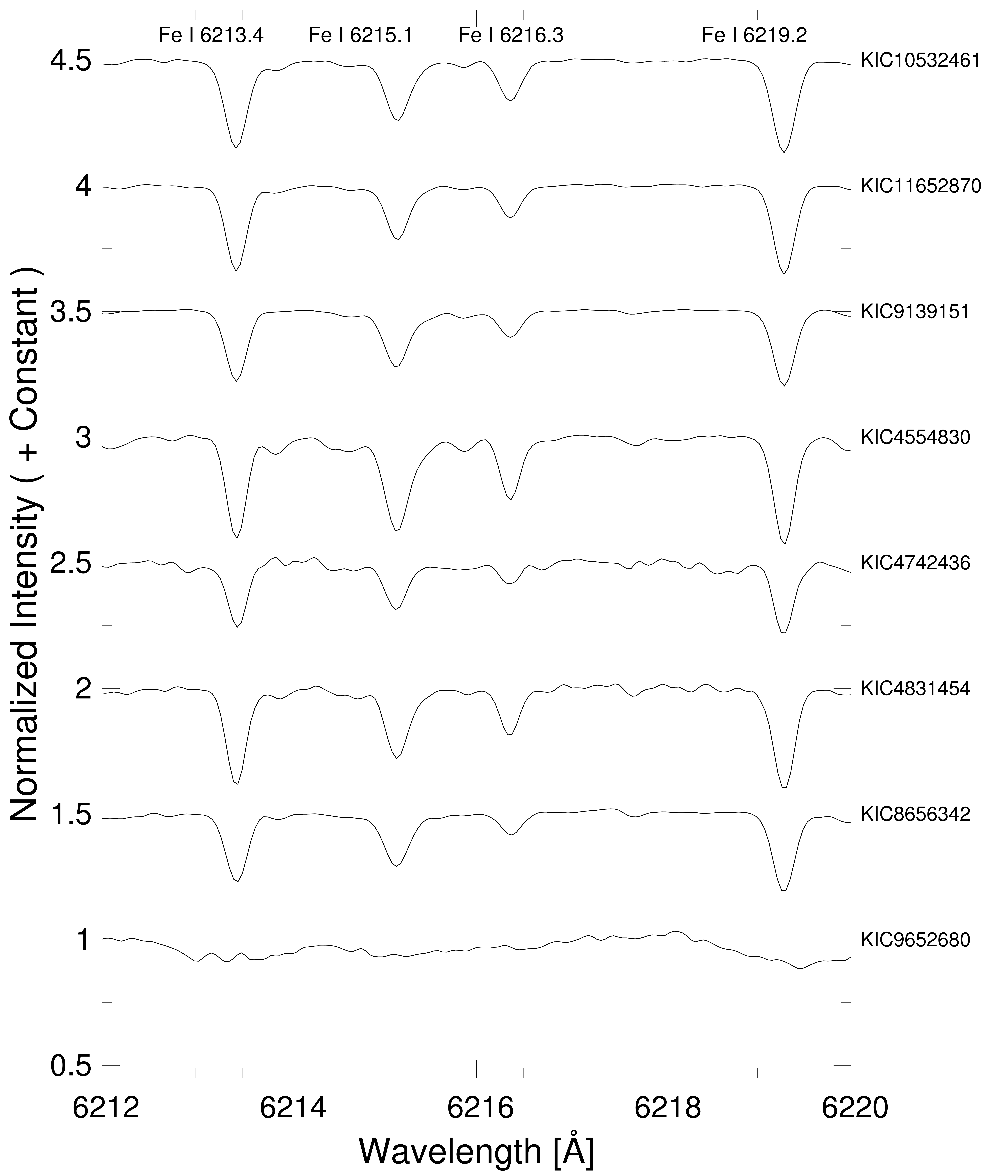}{0.5\textwidth}{(a)}
         \fig{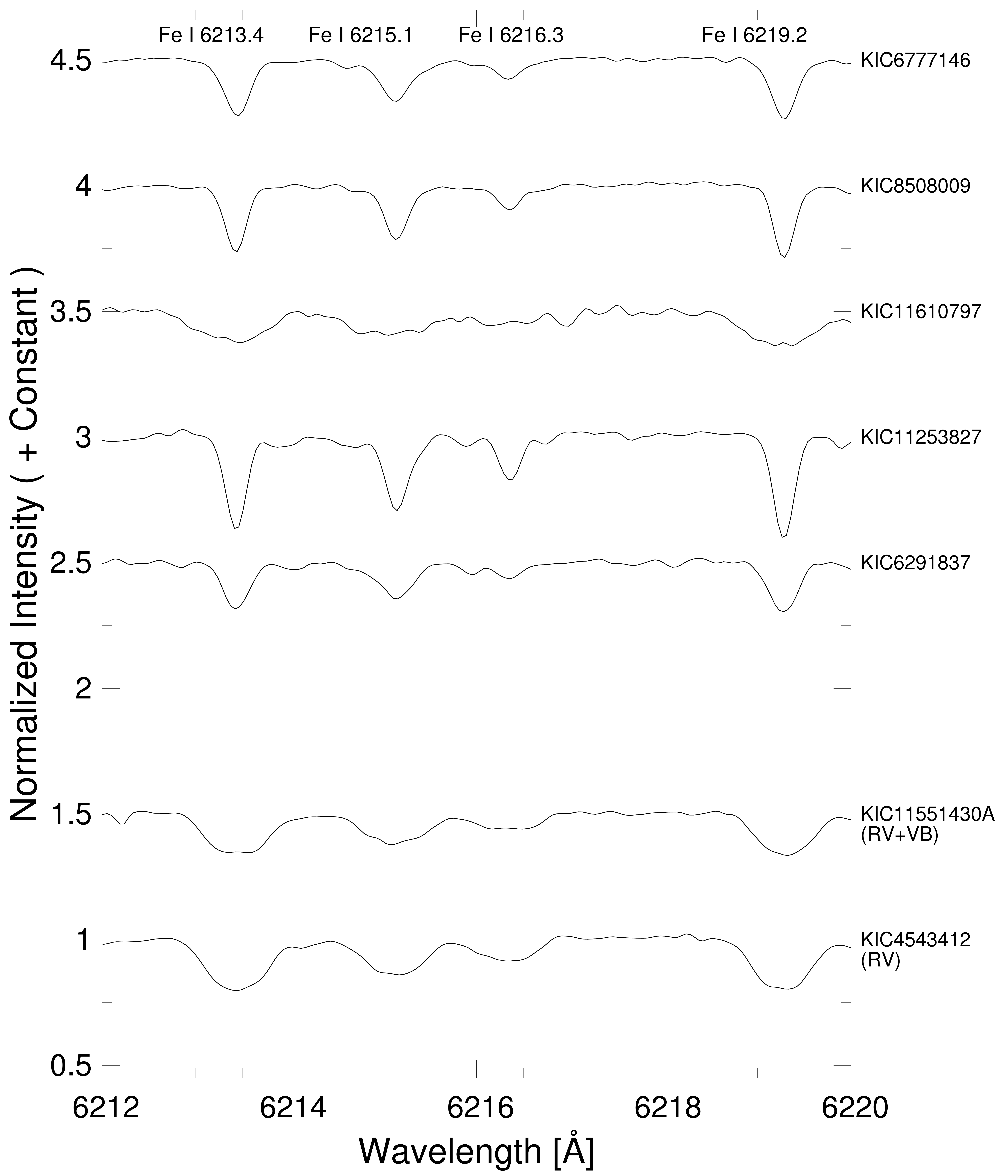}{0.5\textwidth}{(b)}}
\caption{
Example of photospheric absorption lines, including Fe I 6213, 6215, 6216, and 6219\AA~lines.
The wavelength scale of each spectrum is adjusted to the laboratory frame. 
In (a) and the upper part of (b), 13 ``single" superflare stars that show no evidence of binarity are plotted.
Co-added spectra are used here in case that the star was observed multiple times (See Table \ref{tab:obs-RV}).
In the bottom of (b), the co-added spectra of the two binary stars KIC11551430A and KIC4543412, which only show radial velocity shifts 
but do not show any double-lined profiles, are also plotted for reference. \label{fig:specFe}}
\end{figure}

In total, we regard 5 superflare stars as binary stars among the 18 target stars of which we got spectra with enough S/N.
The remaining 13 superflare stars does not show any evidence of binarity 
within the limits of our analyses, so we treat them as ``single stars" in this paper. They have ``no" in the second column of Tables \ref{tab:DR25-gaia}.
Spectra of photospheric lines, including Fe I 6212, 6215, 6216, and 6219\AA~lines, of the 13 ``single" superflare stars are 
shown in Figure \ref{fig:specFe}.
We observed 9 stars among the 13 ``single" stars multiple times, and we made co-added spectra of these 9 stars by conducting the following two steps. 
First, we shifted the wavelength value of each spectrum to the laboratory frame 
on the basis of the radial velocity value of each observation listed in Table \ref{tab:obs-RV}. 
Next, we added up these shifted spectra to one co-added spectrum. 
The co-added spectra are mentioned as ``comb" in Table \ref{tab:obs-RV}, 
and only co-added spectra of these 9 stars are used in Figure \ref{fig:specFe}.
Only the co-added spectra are used for the detailed analyses  in the following sections of this paper, 
when we analyze the spectral data of the 9 stars that we observed multiple times.

In addition to these 9 ``single" stars, we also made co-added spectra with the same methods 
for the two binary target stars (KIC11551430A and KIC4543412) that show radial velocity shifts but do not show double-lined profiles.
They are also shown in Figure \ref{fig:specFe}.
For reference, we also use these data when we estimate stellar parameters in the following.

\newpage
\subsection{Temperature, surface gravity, and metallicity}\label{subsec:ana-atmos}

We estimated the effective temperature $T_{\mathrm{eff}}$, surface gravity $\log g$, 
microturbulence $v_{\mathrm{t}}$, and metallicity [Fe/H] of the target superflare stars,
by using the method that is basically the same as the one we have used in \citeauthor{Notsu2015a} (\citeyear{Notsu2015a} \& \citeyear{Notsu2017}).
We measured the equivalent widths of $\sim200$ Fe I and Fe II lines, 
and used TGVIT program developed by Y. Takeda (\citealt{Takeda2002} \& \citeyear{Takeda2005b}).
For reference, \citet{Rich2017} also applied this method to their ARCES spectroscopic data, 
which were taken with the same wavelength resolution value ($R\sim32,000$) as our data,
and they confirmed the resultant values are consistent with the other previous studies.
The resultant atmospheric parameters ($T_{\mathrm{eff}}$, $\log g$, $v_{\mathrm{t}}$, and [Fe/H]) 
are listed in Table \ref{tab:spec-para}.

\begin{figure}[ht!]
\gridline{\fig{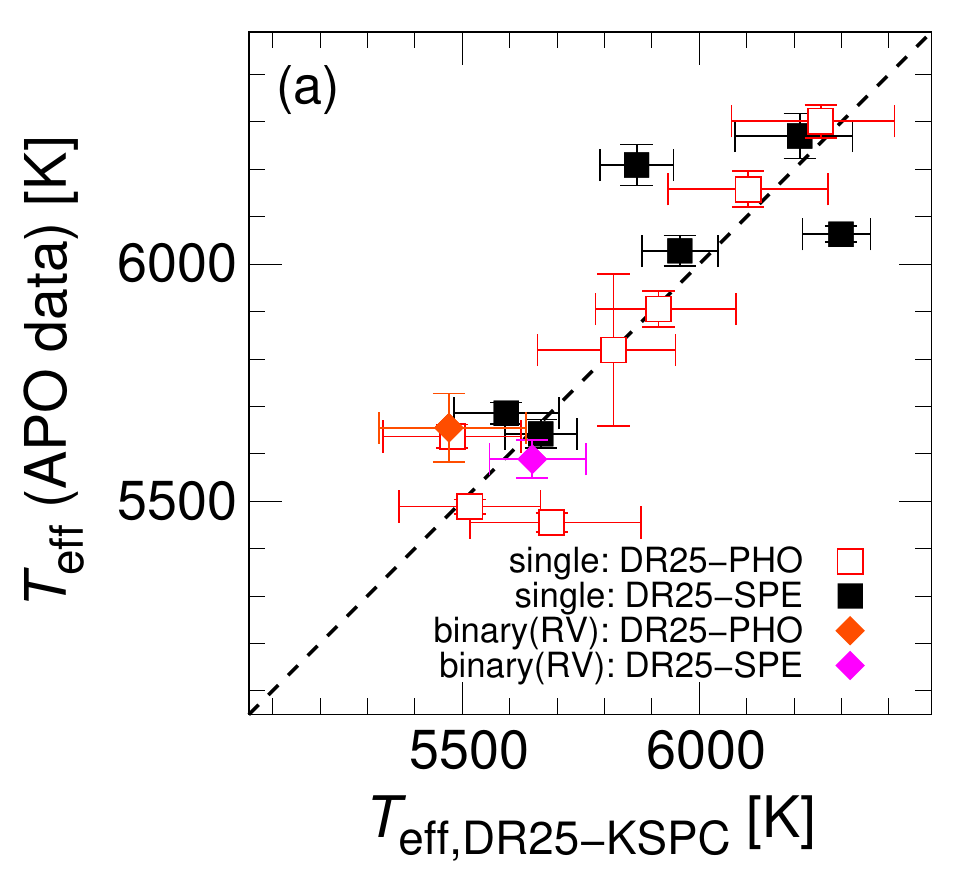}{0.333\textwidth}{}
         \fig{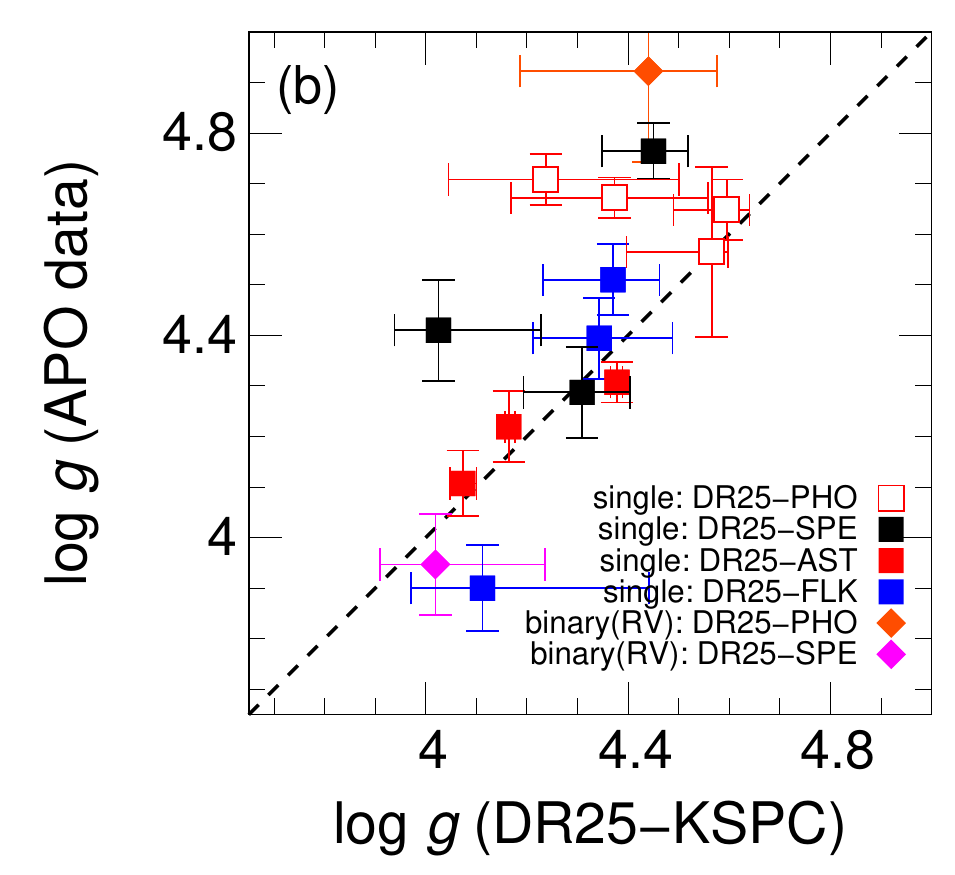}{0.333\textwidth}{}
         \fig{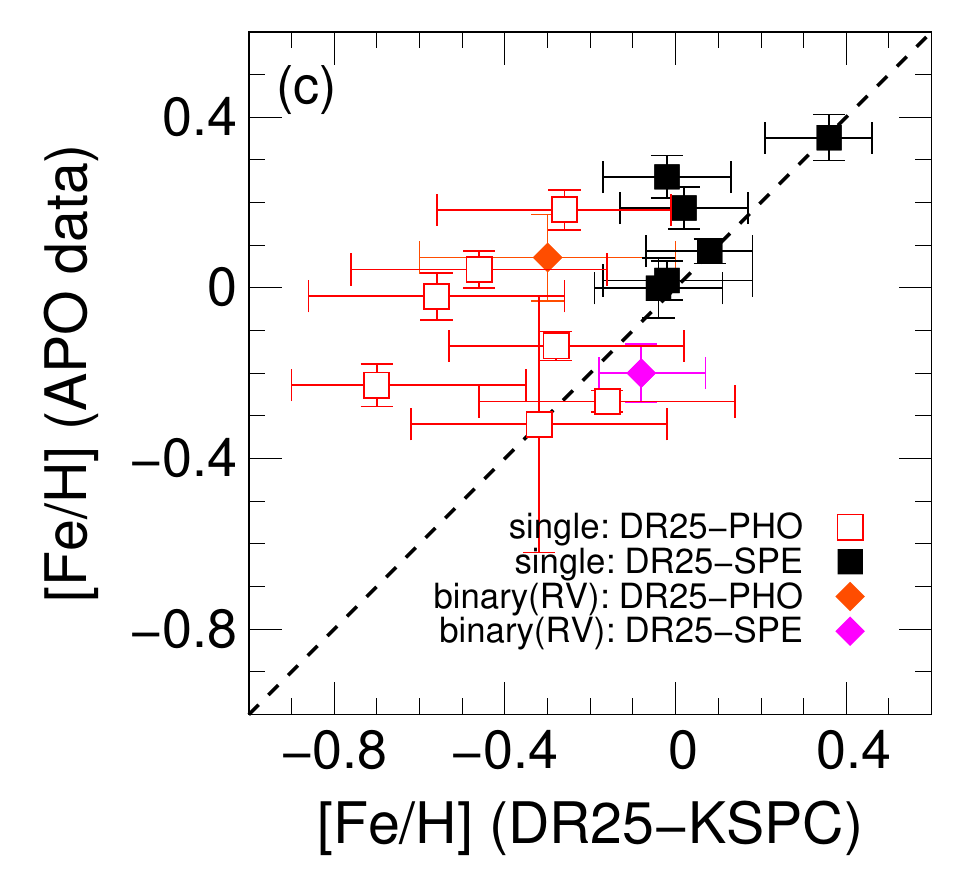}{0.333\textwidth}{}}
\caption{
Comparison between the atmospheric parameters ($T_{\mathrm{eff}}$, $\log g$, and [Fe/H]) that we estimated in this study, 
and those reported in the Data Release 25 {\it Kepler} Stellar Properties Catalog (DR25-KSPC: \citealt{Mathur2017}), respectively. 
The square points are the target superflare stars classified as single stars in Appendix \ref{subsec:ana-binarity}, 
while the diamond points correspond to the spectra of binary superflare stars 
that do not show any double-lined profiles (KIC11551430A and KIC4543412). 
Colors show the methods used to derive the atmospheric parameters in DR25-KSPC 
(PHO: photometry, SPE: Spectroscopy, FLK: flicker method (cf. \citealt{Bastien2016}), AST: Asteroseismolgy). 
See \citet{Mathur2017} for the details of this classification.\label{fig:DR25-spec}}
\end{figure}

We then compare these resultant atmospheric parameters with the values reported 
in the Data Release 25 {\it Kepler} Stellar Properties Catalog (DR25-KSPC: \citealt{Mathur2017}).
We show the results in Figure \ref{fig:DR25-spec}, where the data points are classified with colors 
on the basis of the methods used to derive the parameters in DR25-KSPC.
Our values seem to be in good agreement with the DR25-KSPC values, especially for $T_{\mathrm{eff}}$ (Figure \ref{fig:DR25-spec}(a)).
$\log g$ has some dispersion (Figure \ref{fig:DR25-spec}(b)), 
but this small dispersion does not cause essential problems when we consider whether the target stars are main-sequence stars or not.
As for [Fe/H], if we only consider the spectroscopic values in DR25-KSPC (``DR25-SPE" in Figure \ref{fig:DR25-spec}(c)),
the both values have better agreement and the difference does not affect on the overall discussions 
(e.g., whether the target stars are ``metal-rich/poor" or not).
Through this, we can remark that our spectroscopically derived values are good sources 
with which to discuss the actual properties of stars, as also mentioned for our previous Subaru spectroscopic data in Section 4.3 of \citet{Notsu2015a}.

\subsection{Stellar radius}\label{subsec:ana-rad}
As already listed in Table \ref{tab:DR25-gaia}, 
we have stellar radius values $R_{\mathrm{Gaia}}$ of {\it Kepler} stars deduced from {\it Gaia}-DR2 parallax values \citep{Berger2018}.
These values are listed again in Table \ref{tab:spec-para}, but five stars among our observed 18 target stars 
have no $R_{\mathrm{Gaia}}$ values reported in \citet{Berger2018}. 

\begin{figure}[ht!]
\gridline{\fig{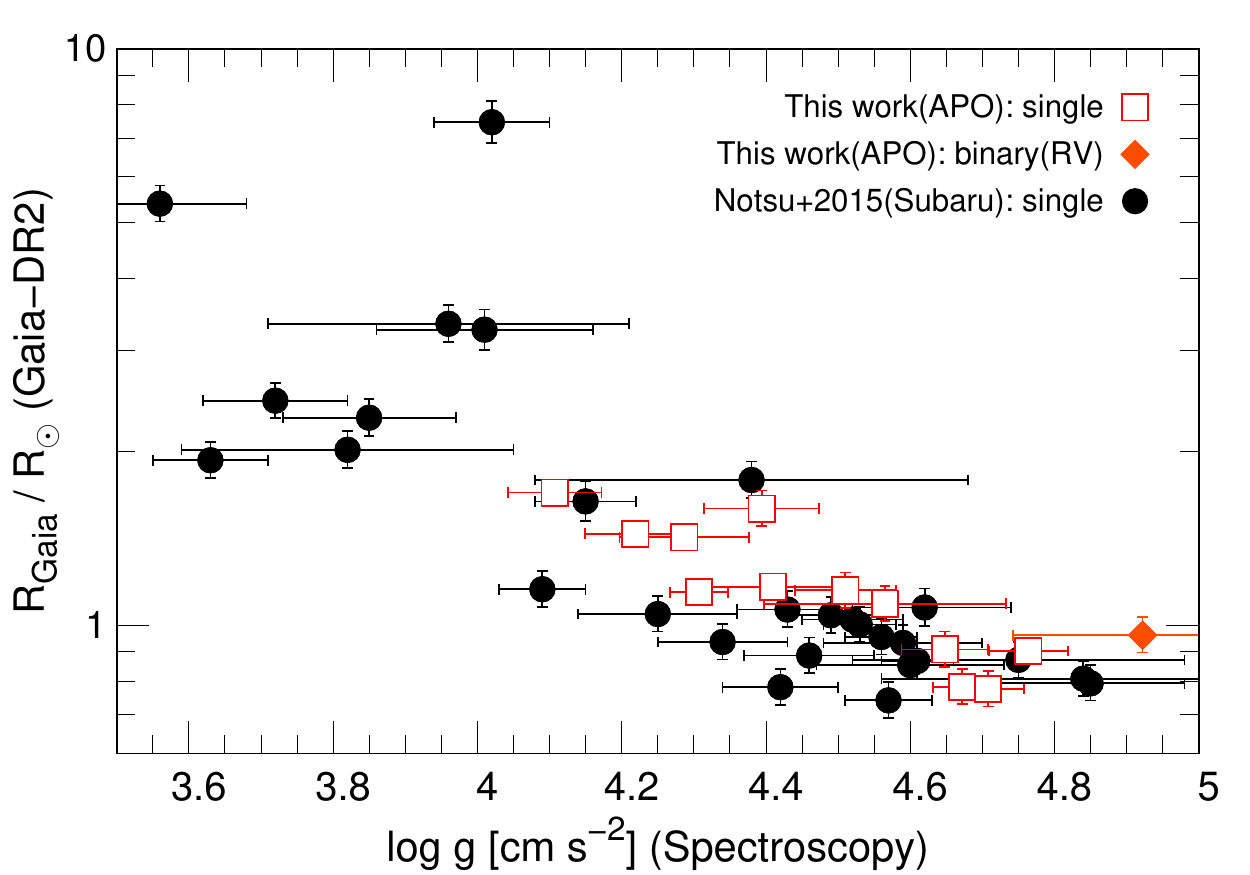}{0.5\textwidth}{}}
\caption{
Stellar radius values $R_{\mathrm{Gaia}}$ from {\it Gaia}-DR2 parallaxes \citep{Berger2018} as a function of
$\log g$ estimated from our spectroscopic observations. 
Only the stars that have $R_{\mathrm{Gaia}}$ are included in this figure.
The red open square points are the target superflare stars classified as single stars in Appendix \ref{subsec:ana-binarity}, 
and the orange diamonds correspond to the spectra of binary superflare stars 
that do not show any double-lined profiles (KIC4543412). 
The single superflare stars that we investigated using Subaru telescope (\citealt{Notsu2015a} \& \citeyear{Notsu2015b}),
excluding the four stars also investigated in this study (See footnote {\it f} of Table \ref{tab:DR25-gaia}),  
are plotted with black circles. 
\label{fig:logg-Rs}}
\end{figure}

In Figure \ref{fig:logg-Rs} we compare these $R_{\mathrm{Gaia}}$ values
with $\log g$ values estimated from our spectroscopic studies (\citealt{Notsu2015a} and this study). 
These two values are roughly correlated, and $R_{\mathrm{Gaia}}$ values look more sensitive to the boundary region 
between main-sequence stars and subgiants (around $\log g=4.0$ -- $4.5$ and $R_{\mathrm{Gaia}}=1.5$ -- $2.0R_{\sun}$).
Then in Section \ref{sec:dis-Kepler}, we mainly use these stellar radius $R_{\mathrm{Gaia}}$ values to strictly define main-sequence stars.

We also estimated stellar radius ($R_{\mathrm{spec}}$) values from the stellar atmospheric parameters ($T_{\mathrm{eff}}$ , $\log g$, and [Fe/H]),
by applying the latest PARSEC isochrones (\citealt{Bressan2012}; \citealt{Marigo2017}) \footnote{\url{http://stev.oapd.inaf.it/cgi-bin/cmd}}.
In this process, we used basically the same method as the one used in our previous study \citep{Notsu2015a}.
We selected all the data points that had possible sets of $T_{\mathrm{eff}}$, $\log g$, and [Fe/H] from the PARSEC isochrones, 
taking into account the error values of $T_{\mathrm{eff}}$ and $\log g$ ($\Delta T_{\mathrm{eff}}$ and $\Delta\log g$). 
For the six stars that have no suitable isochrones within their original error range of $T_{\mathrm{eff}}$ and $\log g$ values,
we then took into account larger error values as mentioned in footnote $c$ of Table \ref{tab:spec-para}. 
We note that the resultant values of these six stars can have relatively low accuracy.
We then selected the maximum and minimum $R_{\mathrm{spec}}$ value of each star, 
and determined the resultant $R_{\mathrm{spec}}$ value as a median of the maximum and minimum values.
The error values of  $R_{\mathrm{spec}}$ listed in Table \ref{tab:spec-para} are $\lesssim$ 20\% for most of the stars, as also mentioned in \citet{Notsu2015a}.

In the following, we use $R_{\mathrm{Gaia}}$ values as a first priority, and we use $R_{\mathrm{spec}}$ values 
only for the stars without $R_{\mathrm{Gaia}}$ values.

\subsection{Projected rotation velocity ($v\sin i$)}\label{subsec:ana-vsini}
We measured $v \sin i$ (stellar projected rotational velocity) of the target stars by using the method that is basically the same as 
that in our previous studies (\citealt{Notsu2015a} \& \citeyear{Notsu2017}).
This is originally based on the one described in \citet{Takeda2008}.
We took into account the effects of macroturbulence and instrumental broadening 
by considering a simple relationship among the line-broadening parameters (cf. \citealt{Takeda2008}), which can be expressed as
\begin{eqnarray}\label{eq:vm}
v_{\mathrm{M}}^{2}=v_{\mathrm{ip}}^{2}+v_{\mathrm{rt}}^{2}+v_{\mathrm{mt}}^{2} \ .
\end{eqnarray}
Here, $v_{\mathrm{M}}$ is the e-folding width of the Gaussian macrobroadening function 
$f(v)\propto\exp[−(v/v_{\mathrm{M}})^{2}]$, including instrumental broadening ($v_{\mathrm{ip}}$), rotation ($v_{\mathrm{rt}}$), and macroturbulence ($v_{\mathrm{mt}}$). 
We derived $v_{\mathrm{M}}$ by applying an automatic spectrum-fitting technique \citep{Takeda1995}, 
assuming the model atmosphere corresponding to the atmospheric parameters estimated in Appendix \ref{subsec:ana-atmos}. 
In this process, we used the MPFIT program contained in the SPTOOL software package developed by Y.Takeda. 
We applied this fitting technique to the 6212--6220\AA~region (shown in Figures \ref{fig:binsp-Fe6212} \& \ref{fig:specFe}) to derive $v\sin i$ values. 
This region has also been used in our previous studies (\citealt{Notsu2015a} \& \citeyear{Notsu2017}).

The instrumental broadening velocity $v_{\mathrm{ip}}$ was calculated using the following equation \citep{Takeda2008}:
\begin{eqnarray}\label{eq:ip}
v_{\mathrm{ip}}=\frac{3\times10^{5}}{2R\sqrt{\ln{2}}} \ ,
\end{eqnarray}
where $R(=\lambda/\Delta\lambda)$ is the wavelength resolution of the observation.
For estimating the $R$ value adopted here, we conducted Gaussian-fitting to emission lines in the 6180--6240\AA~region 
of the Th-Ar spectrum data. This region is around the 6212--6220\AA~region, where we conduct the above fitting process.
We finally got $R$=32,500 and we applied this to Equation (\ref{eq:ip}) in the following. 
The macroturbulence velocity $v_{\mathrm{rt}}$ was estimated by using the relation $v_{\mathrm{mt}}\sim0.42\xi_{\mathrm{RT}}$ \citep{Takeda2008}.
The term $\xi_{\mathrm{RT}}$ is the radial-tangential macroturbulence, and $\xi_{\mathrm{RT}}$ was estimated
using the relation reported in \citet{Valenti2005}: 
\begin{equation}\label{eq:xiRT}
\xi_{\mathrm{RT}} = \left(3.98 - \frac{T_{\mathrm{eff}}-5770\mathrm{K}}{650\mathrm{K}} \right) \ .
\end{equation}
As we described in \citet{Notsu2015a}, 
the choice of macroturbulence equation is often important, 
but we only use this Equation (\ref{eq:xiRT}) to estimate the resultant values of this study, as we have also done in our previous studies
(\citealt{Notsu2015a} \& \citeyear{Notsu2017}).
$v_{\mathrm{rt}}$ was then derived with the above equations, and we finally got $v\sin i$ with the relation 
$v_{\mathrm{rt}}\sim0.94 v\sin i$ \citep{Gray2005}.
The resultant $v\sin i$ values of the target superflare stars are listed in Table \ref{tab:spec-para}.
As error values of  $v\sin i$, we consider the systematic uncertainty with changing $\xi_{\mathrm{RT}}$ up to $\pm$25\%,
as described in \citet{Hirano2012}.  

The above estimation method of $v\sin i$ has been developed to be suitable for the Subaru/HDS spectroscopic data 
with the high spectral resolutions of $R\sim$ 55,000 -- 100,000 (\citealt{Hirano2012}; \citealt{Notsu2013a}; \citealt{Notsu2015a}).
It is not so appropriate that we apply this method without any modifications 
into our APO3.5m/ARCES spectroscopic data (the spectral resolution of only $\sim$32,500). 
It is difficult to estimate $v\sin i$ values as low as 2--3 km s$^{-1}$ only with APO spectra. 
On the other hand, in this study, we do not need to estimate $v\sin i$ values with the high precision of 2--3km s$^{-1}$ 
for the overall discussions in this study with Figures \ref{fig:vlc-vsini} and \ref{fig:vlc-vsini-HR} 
(e.g., whether $v\sin i \lesssim v_{\rm{lc}}$ is roughly achieved, and whether the target stars can have low inclination angle values or not).
There are twice (or more) differences of the spectral resolution values between this APO observation and the previous Subaru/HDS observations (\citealt{Hirano2012}; \citealt{Notsu2013a}; \citealt{Notsu2015a}),
and the instrumental velocity $v_{\mathrm{ip}}$ has twice (or more) larger values between them.
Then in this study (e.g., Table \ref{tab:spec-para}), we only report a rough upper limit value ``$v \sin i <4$ km s$^{-1}$" for the slowly-rotating stars with $v \sin i <4$ km s$^{-1}$. 
We also need to use the $v\sin i$ values of mildly slowly-rotating stars ($v\sin i\sim$ 5km s$^{-1}$) with cautions.

\subsection{Measurements of stellar activity indicators Ca II 8542\AA~and H$\alpha$ 6563\AA} \label{sec:measure-IRTHa}

\begin{figure}[ht!]
%%%%
\gridline{\fig{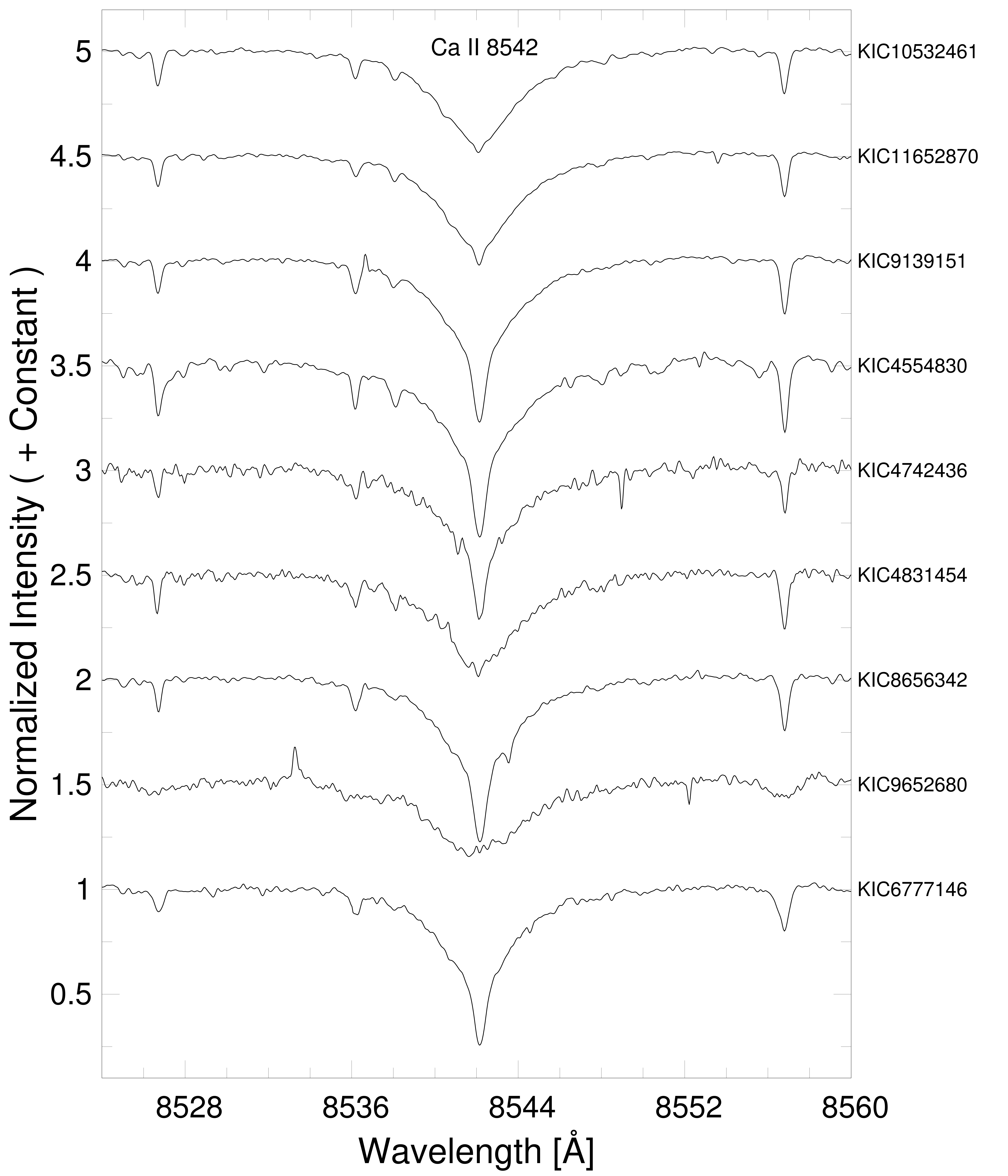}{0.5\textwidth}{(a)}
         \fig{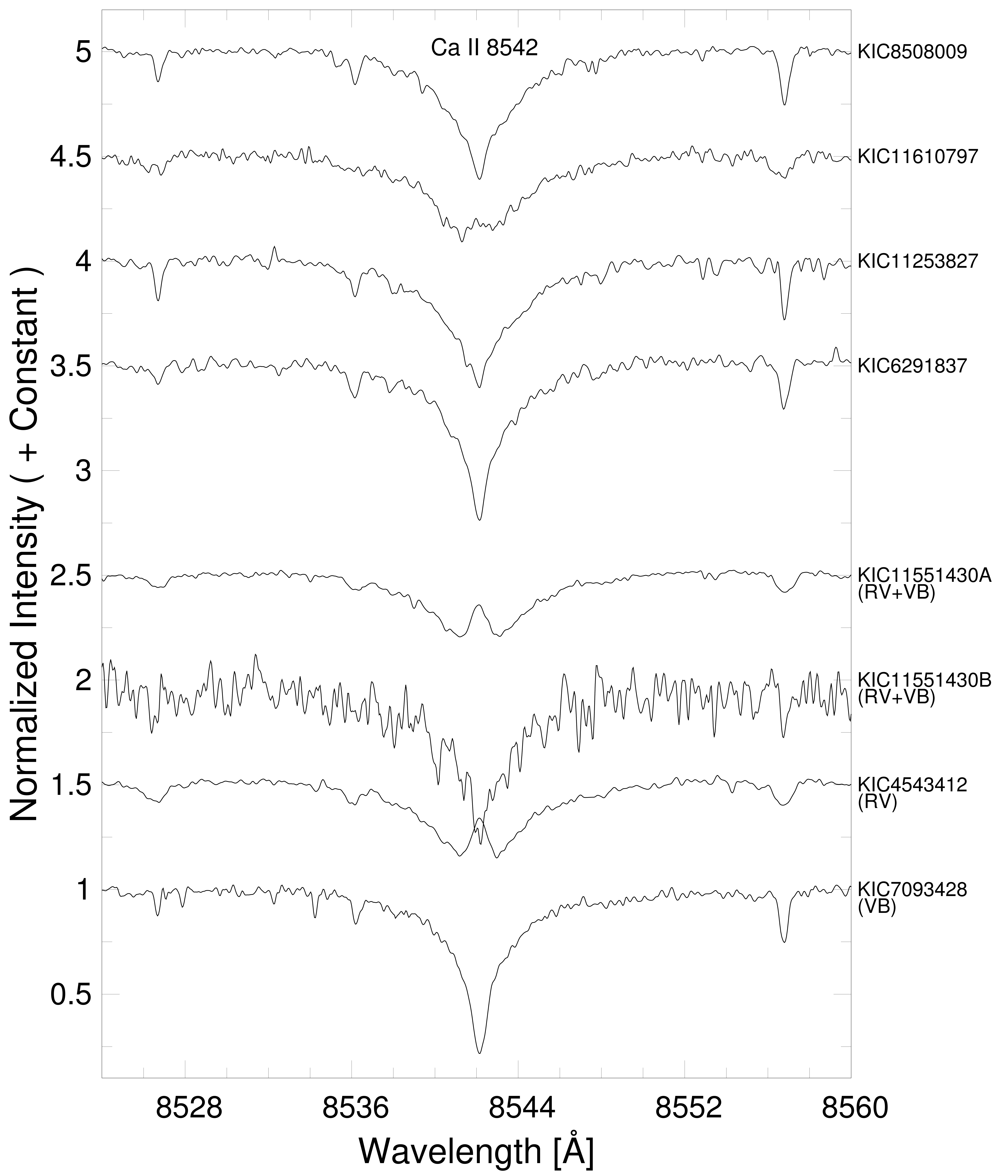}{0.5\textwidth}{(b)}}
\caption{Spectra around Ca II 8542\AA~line. The wavelength scale of each spectrum is adjusted to the laboratory frame. 
In (a) and the upper part of (b), 13 ``single" superflare stars that show no evidence of binarity are plotted.
Co-added spectra are used here in case that the star was observed multiple times (See Table \ref{tab:obs-RV}).
In the bottom of (b), the spectra of binary stars that do not show any double-lined profiles 
are plotted for reference.
As for KIC11551430A and KIC4543412 among them, which only show radial velocity shifts, the co-added spectra are used.
\label{fig:specCa8542}}
\end{figure}

\begin{figure}[ht!]
%%%%
\gridline{\fig{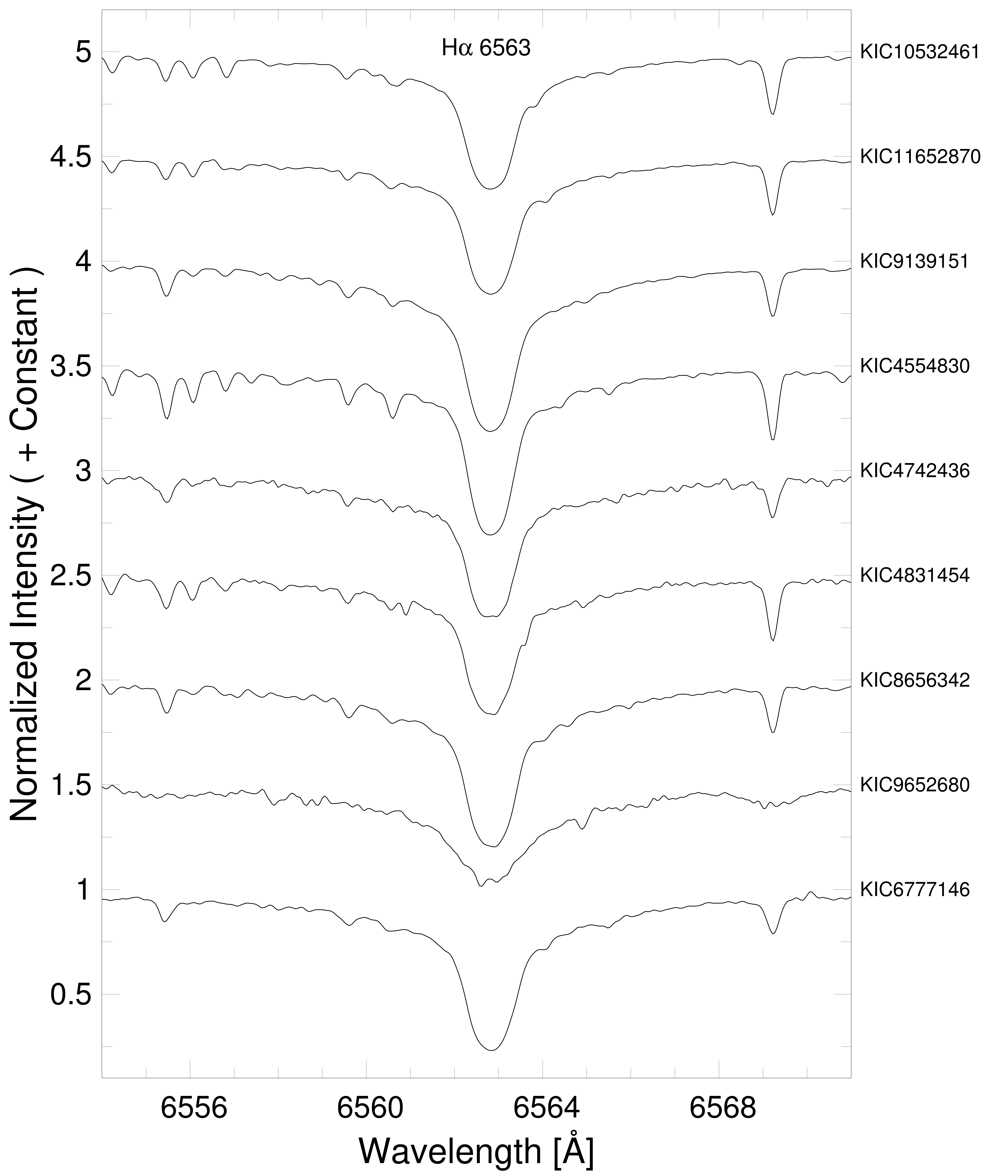}{0.5\textwidth}{(a)}
         \fig{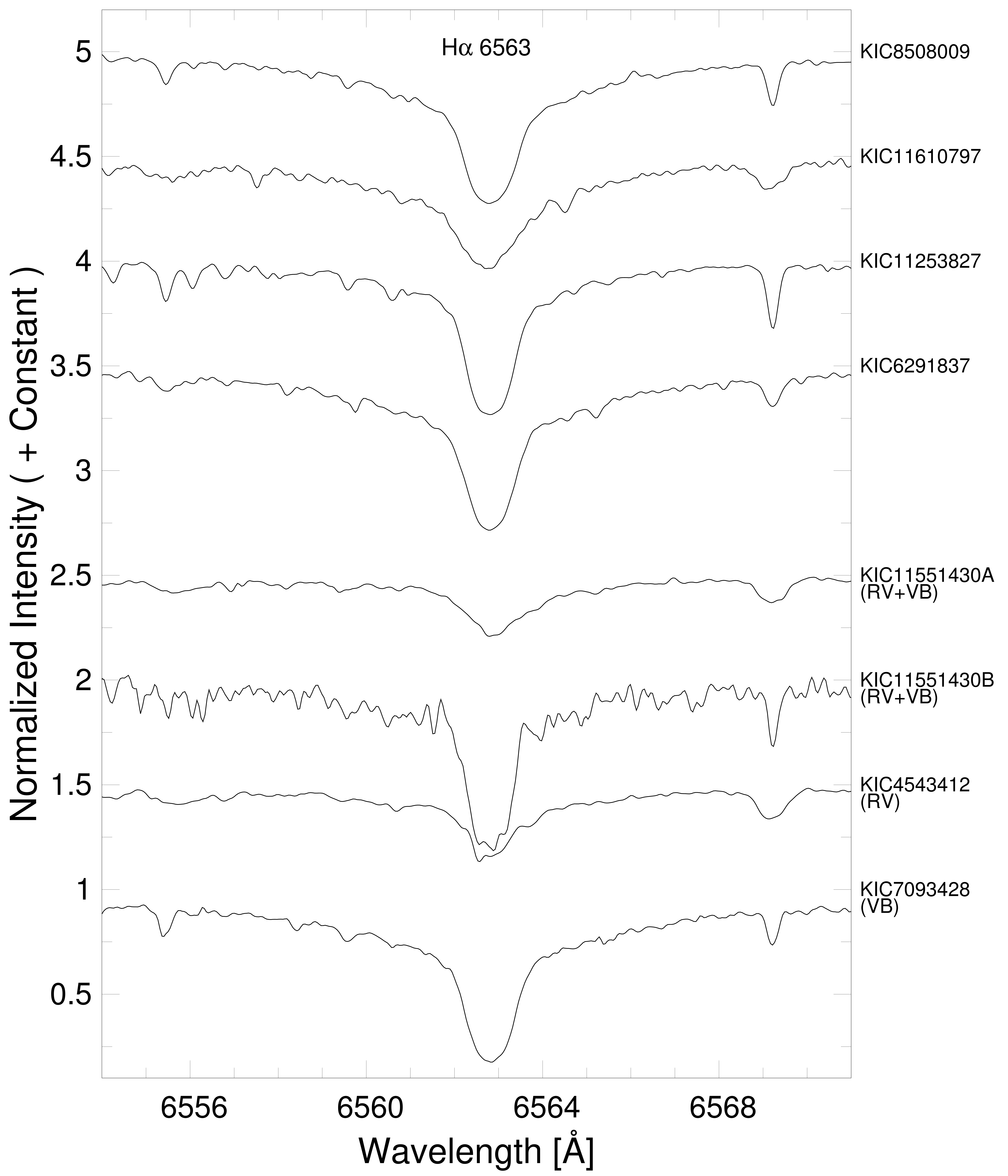}{0.5\textwidth}{(b)}}
\caption{Spectra around H$\alpha$ 6563\AA~line.
The wavelength scale of each spectrum is adjusted to the laboratory frame. 
In (a) and the upper part of (b), 13 ``single" superflare stars that show no evidence of binarity are plotted.
Co-added spectra are used here in case that the star was observed multiple times (See Table \ref{tab:obs-RV}).
In the bottom of (b), the spectra of binary stars that do not show any double-lined profiles 
are plotted for reference.
As for KIC11551430A and KIC4543412 among them, which only show radial velocity shifts, the co-added spectra are used.
\label{fig:specHa}}
\end{figure}

The observed spectra of the target superflare stars around Ca II 8542\AA~and H$\alpha$ 6563\AA~are shown 
in Figures \ref{fig:specCa8542} and \ref{fig:specHa}, respectively.
We measured the $r_{0}$(8542) and $r_{0}$(H$\alpha$) indexes, 
which are the residual core fluxes normalized by the continuum level at the line cores of Ca II 8542\AA~and H$\alpha$ 6563\AA, respectively. 
As we have already introduced in \citet{Notsu2013a}, these indexes are known to be indicators of stellar chromospheric activity 
(e.g., \citealt{Linsky1979b}; \citealt{Takeda2010}). 
As the chromospheric activity is enhanced, 
the intensity of these indicators becomes large since a greater amount of emission from the chromosphere fills in the core of the lines. 
The values of the $r_{0}$(8542) and $r_{0}$(H$\alpha$) indexes of the target superflare stars are listed in Table \ref{tab:act}. 
For reference, $r_{0}$(8542) values of the 28 comparison stars are also listed in Table \ref{tab:Smwo-Sapo}.

\subsection{Measurements of Ca II H\&K $S$-index} \label{sec:measure-Sindex}

\begin{figure}[ht!]
\gridline{\fig{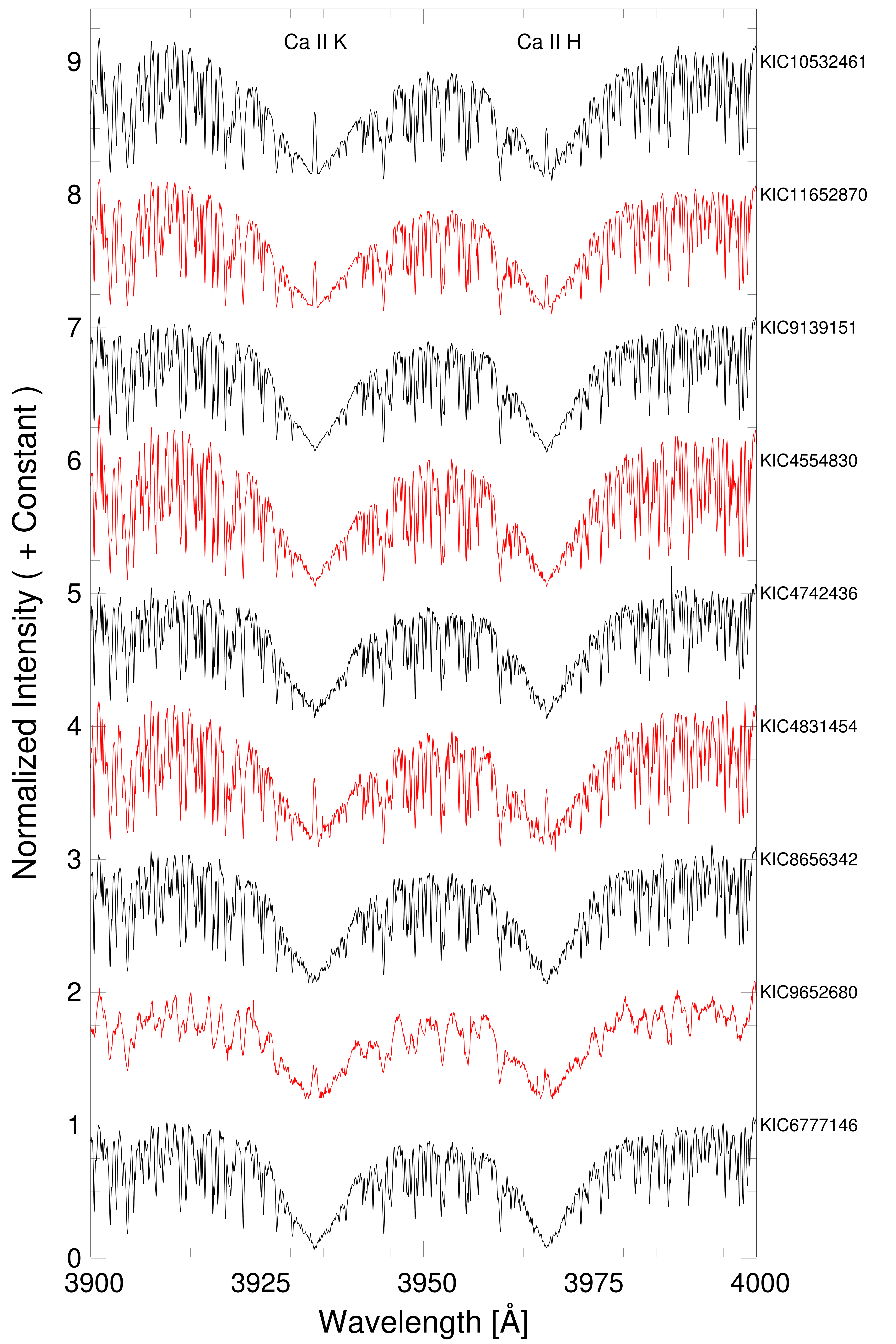}{0.5\textwidth}{(a)}
         \fig{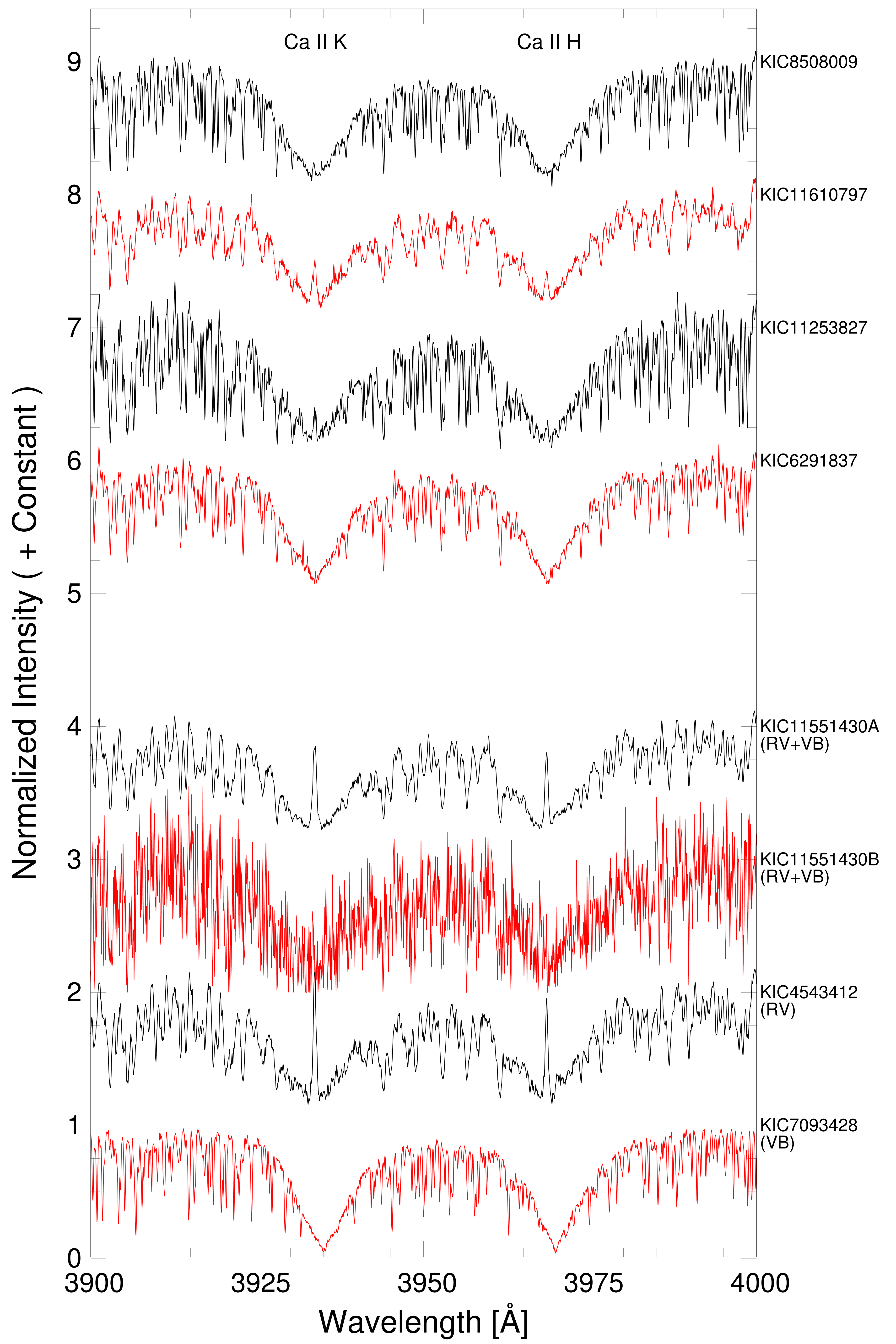}{0.5\textwidth}{(b)}}
\caption{
Spectra around Ca II H 3968\AA~and Ca II K 3934\AA~lines. 
The wavelength scale of each spectrum is adjusted to the laboratory frame. 
In (a) and the upper part of (b), 13 ``single" superflare stars that show no evidence of binarity are plotted.
Co-added spectra are used here in case that the star was observed multiple times (See Table \ref{tab:obs-RV}).
In the bottom of (b), the spectra of binary stars that do not show any double-lined profiles 
are plotted for reference.
As for KIC11551430A and KIC4543412 among them, which only show radial velocity shifts, the co-added spectra are used.
As for the spectra of KIC6291837 and KIC11551430B, datapoints heavily contaminated by cosmic-rays are removed to avoid confusions.
\label{fig:specHK}}
\end{figure}

The observed spectra of the target superflare stars around Ca II H 3968\AA~and Ca II K 3934\AA~lines 
are shown in Figure \ref{fig:specHK}. 
As described in Section \ref{sec:target-obs}, these spectra are normalized by using spectra of early-type standard stars, as also done in \citet{Morris2017}.
The emission in the cores of the Ca II H \& K lines is a widely-known indicator of the stellar chromospheric activity (e.g., \citealt{Hall2008} for review),
and this is more sensitive to activity level changes compared with Ca II 8542\AA~line (cf. \citealt{Takeda2010} \& \citeyear{Takeda2012}).
Ca II H \& K emission is often measured as ``$S$-index", 
which is the flux in the emission features normalized by two pseudocontinuum regions on either side of the absorption features 
(e.g., \citealt{Vaughan1978}; \citealt{Duncan1991}; \citealt{Isaacson2010}; \citealt{Mittag2013}; \citealt{Karoff2016}; \citealt{Morris2017}). 
We measured the $S$-index values of the target stars in the following of this section.

The $S$-index value can vary among other instruments for the same intrinsic flux, as described in Section 2.1 of \citet{Isaacson2010}.    
Following the method used in \citet{Isaacson2010} and \citet{Morris2017}, 
we calibrated $S$-index values of our APO observation data ($S_{\mathrm{APO}}$) 
against the $S$-index values already calibrated to the Mount Wilson Observatory (MWO) sample ($S_{\mathrm{MWO}}$) 
\footnote{\citet{Morris2017} already conducted the calibration of $S$-index measurements for APO data, 
but most of their observed stars are K-type stars. 
Since $S$-index is a value also depending on stellar colors \citep{Noyes1984}, 
we newly conduct the calibration using solar-type (G-type main-sequence) stars in this study.}.
In this calibration process, 
we use our observation data of the 28 bright solar-type comparison stars described in Section \ref{sec:target-obs} and listed in Table \ref{tab:Smwo-Sapo}.
All of these 28 stars have $S$-index values calibrated to the Mount Wilson Observatory (MWO) sample 
($S_{\mathrm{MWO}}$ values in Table \ref{tab:Smwo-Sapo}) that are reported in \citet{Isaacson2010} on the basis of 
California Planet Search (CPS) program spectroscopic observations.

We first measured the $S$-index from our APO data ($S_{\mathrm{APO}}$) of these comparison stars 
using the following equation (cf. \citealt{Morris2017}):
\begin{eqnarray}\label{eq:Sapo}
S_{\mathrm{APO}}=\frac{aH+bK}{cR+dV} \ ,
\end{eqnarray}
where $H$and $K$ are the recorded counts in a 1.09\AA~full-width at half-maximum triangular bandpasses
centered on the Ca II H and K lines at 3968.469\AA~and 3933.663\AA, respectively.
$V$ and $R$ are two 20\AA~wide reference bandpasses centered on 3901.07\AA~and 4901.07\AA.
The values of $a$, $b$, $c$, and $d$ should be selected so that $S_{\mathrm{APO}}$ has 
roughly equal flux contribution from the $H$ and $K$ emission lines and roughly 
equal flux contribution from the $R$ and $V$ pseudocontinuum regions in the APO spectra.
Here we finally had $a=b=c=d=1$ after trial and error in this study.

\begin{figure}[ht!]
\gridline{\fig{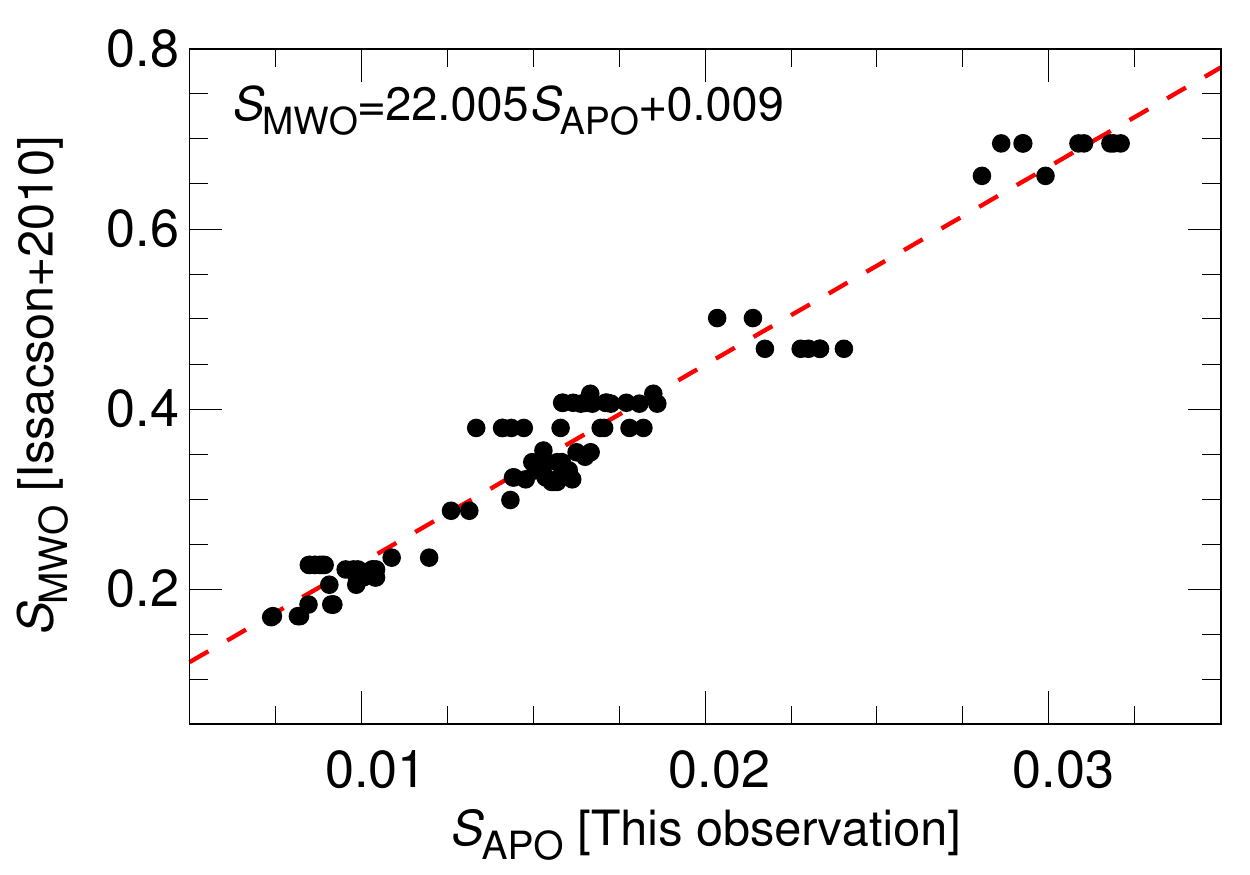}{0.5\textwidth}{}}
\caption{
Calibration of $S$-index values measured from the APO data in this observation ($S_{\mathrm{APO}}$) 
against the $S$-index values calibrated to the Mount Wilson Observatory (MWO) sample ($S_{\mathrm{MWO}}$).
$S_{\mathrm{APO}}$ values in this figure are from the spectroscopic data of multiple observations 
of the 28 bright solar-type comparison stars (Table \ref{tab:Smwo-Sapo}).
$S_{\mathrm{MWO}}$ values are the values reported in \citet{Isaacson2010}.
\label{fig:SapoSmwo}}
\end{figure}

Then in Figure \ref{fig:SapoSmwo}, we compared these APO S-index values ($S_{\mathrm{APO}}$) listed in Table \ref{tab:Smwo-Sapo}
with the $S_{\mathrm{MWO}}$ values reported in \citet{Isaacson2010},
and investigated the two constants of the following equation (cf. \citealt{Morris2017}): 
\begin{eqnarray}\label{eq:SmwoSapo}
S_{\mathrm{MWO}}=C_{1}S_{\mathrm{APO}}+C_{2} \ .
\end{eqnarray}
Since $S$-index varies over time for each star in the sample, the linear correlation between the $S_{\mathrm{APO}}$ and $S_{\mathrm{MWO}}$ 
can have some intrinsic spread as in Figure \ref{fig:SapoSmwo}.
Applying the least-square method to the data in Figure \ref{fig:SapoSmwo}, finally we got $C_{1}=22.005$ and $C_{2}=0.009$.
Such larger (e.g., $C_{1}\sim22$) conversion factors of $S$-indexes are commonly appeared in the researches 
using relatively high dispersion spectrograph. 
For example, \citet{Morris2017} using APO/ARCES data (the same instruments and same method as our study: $C_{1}\sim$21.1), 
\citet{Karoff2013} using NOT/FIES data ($R\sim$25,000 and $C_{1}\sim$16.6), and 
\citet{Isaacson2010} using Keck/HIRES data ($R\sim$52,000 and $C_{1}\sim$31.5).  
As described in the final paragraph of Section 3.6 of \citet{Karoff2013}, 
the relatively large $C_{1}$ value in our study can be mainly due to the higher spectral resolutions of APO data, 
compared with the original Mount Wilson observations \citep{Duncan1991}.
Using Equations (\ref{eq:Sapo}) \& (\ref{eq:SmwoSapo}) and these $C_{1}$ \& $C_{2}$ values,
we finally estimated the calibrated $S$-index value of each spectrum ($S_{\mathrm{HK}}\equiv S_{\mathrm{MWO}}$) from each $S_{APO}$ value.
The $S_{\mathrm{HK}}$ values from the observations of comparison stars are listed in Table \ref{tab:Smwo-Sapo},
and the resultant $S_{\mathrm{HK}}$ values of the target superflare stars from their spectral data (Figure \ref{fig:specHK})
are listed in Table \ref{tab:act}.

\subsection{Measurements of Ca II H\&K flux} \label{sec:measure-HKflux}
$S$-index is known to be stellar temperature dependent (e.g., \citealt{Noyes1984}), and it is a purely empirical quantity.
Then it can be advantageous to calculate chromospheric fluxes. 
Then in the following, we calculated Ca II H\&K flux values from the $S$-index values with stellar colors, 
using the formulation described in \citet{Mittag2013}. We summarize the method in the following.

The stellar surface fluxes emitted in the Ca II H\&K lines ($\mathcal{F}_{\mathrm{HK}}$) are expressed with $S$-index ($S_{\mathrm{HK}}$) as:
\begin{eqnarray}\label{eq:SHK-FHK}
\mathcal{F}_{\mathrm{HK}}=\frac{\mathcal{F}_{\mathrm{RV}}}{\alpha}S_{\mathrm{HK}} \ ,
\end{eqnarray}
where $\mathcal{F}_{\mathrm{RV}}$ is the surface flux in both continua ($V$ and $R$ region mentioned in Appendix \ref{sec:measure-IRTHa}),
and the factor $\alpha$ is a historical dimensionless conversion factor.
Following \citet{Mittag2013}, we use the value $\alpha=19.2$. 
$\mathcal{F}_{\mathrm{RV}}$ is estimated from stellar color $B-V$ 
using the following empirical equation applicable to main-sequence and subgiant stars:
\begin{eqnarray}\label{eq:FRV-BV}
\log\left(\frac{\mathcal{F}_{\mathrm{RV}}}{19.2}\right)=8.25-1.67(B-V) \ . 
\end{eqnarray}
As for the target {\it Kepler} superflare stars in Table \ref{tab:spec-para} 
and the 15 comparison stars (among all the 28 comparison stars) with remarks (1) or (2) in Table \ref{tab:Smwo-Sapo},
which were observed in \citet{Notsu2017},
$B-V$ values were calculated from $T_{\mathrm{eff}}$ and [Fe/H] using Equation (2) of \citet{Alonso1996}.
As for the remaining 13 stars, $B-V$ values reported in Table 1 of \citet{Isaacson2010} are used, and $T_{\mathrm{eff}}$ values used in the following 
are derived from $B-V$ values by using Equation (2) of \citet{Valenti2005}.

We then estimated chromospheric excess flux of Ca II H\&K lines ($\mathcal{F}^{'}_{\mathrm{HK}}$) by subtracting the photospheric flux contribution 
($\mathcal{F}_{\mathrm{HK,phot}}$) in the line center: 
\begin{eqnarray}\label{eq:FdHK}
\mathcal{F}^{'}_{\mathrm{HK}}= \mathcal{F}_{\mathrm{HK}}-\mathcal{F}_{\mathrm{HK,phot}} \ . 
\end{eqnarray}
$\mathcal{F}_{\mathrm{HK,phot}}$ were derived from $B-V$ by using the following equations in \citet{Mittag2013}:
\begin{eqnarray}
\log \mathcal{R}_{\mathrm{HK,phot}}&=&-4.898+1.918(B-V)^{2}-2.893(B-V)^{3} \label{eq:Rphot-BV} \\
\mathcal{R}_{\mathrm{HK,phot}}&=&\frac{\mathcal{F}_{\mathrm{HK,phot}}}{\sigma T^{4}_{\mathrm{eff}}} \label{eq:Rphot-Fphot} \ . 
\end{eqnarray}
$\mathcal{F}^{'}_{\mathrm{HK}}$ is frequently converted 
to the flux-related stellar activity index $\log R^{'}_{\mathrm{HK}}$ \citep{Linsky1979a}:
\begin{eqnarray}\label{eq:RdHK}
\mathcal{R}^{'}_{\mathrm{HK}} = \frac{\mathcal{F}_{\mathrm{HK}}-\mathcal{F}_{\mathrm{HK,phot}}}{\sigma T^{4}_{\mathrm{eff}}} 
= \frac{\mathcal{F}^{'}_{\mathrm{HK}}}{\sigma T^{4}_{\mathrm{eff}}} \ . 
\end{eqnarray}
With this normalization, we can compare the activity level of stars with different effective temperatures and colors.
The resultant values of $\mathcal{F}^{'}_{\mathrm{HK}}$ and $\mathcal{R}^{'}_{\mathrm{HK}}$ of the target superflare stars 
are listed in Table \ref{tab:act}.

\citet{Schrijver1987} introduced the concept of ``basal flux" ($\mathcal{F}_{\mathrm{HK,basal}}$), which is residual flux 
remaining in the core of the Ca II H\&K lines of inactive stars 
after their photospheric line contribution ($\mathcal{F}_{\mathrm{HK,phot}}$) is removed.
\citet{Mittag2013} investigated $\mathcal{F}^{'}_{\mathrm{HK}}$ values of very large number of main-sequence, subgiant, and giant stars, 
and measured the ``lower" boundary of $\mathcal{F}^{'}_{\mathrm{HK}}$ distribution as a function of $B-V$.
\citet{Mittag2013} then reported the empirical scaling between $\mathcal{F}_{\mathrm{HK,basal}}$  and $B-V$:
\begin{eqnarray}\label{eq:FHKbasal}
\log \mathcal{F}_{\mathrm{HK,basal}} = a+b(B-V)+c(B-V)^{2} \ , 
\end{eqnarray}
where coefficients $a$, $b$, and $c$ vary depending on $B-V$ and luminosity classes (V, IV, or III), and they are listed in Table 3 of \citet{Mittag2013}.
By subtracting this basal flux component $\mathcal{F}_{\mathrm{HK,basal}}$ from $\mathcal{F}^{'}_{\mathrm{HK}}$, 
we can get new ``pure" and universal activity indicators $\mathcal{F}^{+}_{\mathrm{HK}}$ and $\log \mathcal{R}^{+}_{\mathrm{HK}}$ defined by
\begin{eqnarray}\label{eq:RpHK}
\mathcal{R}^{+}_{\mathrm{HK}}
=\frac{\mathcal{F}^{'}_{\mathrm{HK}}-\mathcal{F}_{\mathrm{HK,basal}}}{\sigma T^{4}_{\mathrm{eff}}} 
=\frac{\mathcal{F}_{\mathrm{HK}}-\mathcal{F}_{\mathrm{HK,phot}}-\mathcal{F}_{\mathrm{HK,basal}}}{\sigma T^{4}_{\mathrm{eff}}} 
= \frac{\mathcal{F}^{+}_{\mathrm{HK}}}{\sigma T^{4}_{\mathrm{eff}}} \ . 
\end{eqnarray}
This $\log \mathcal{R}^{+}_{\mathrm{HK}}$ 
index allows comparisons of the activity levels of stars with different luminosity classes and different temperatures on the same scale \citep{Mittag2013}.
The resultant values of $\mathcal{F}^{+}_{\mathrm{HK}}$ and $\mathcal{R}^{+}_{\mathrm{HK}}$ of the target superflare stars and the comparison stars 
are listed in Tables \ref{tab:act} and \ref{tab:Smwo-Sapo}, respectively.

\begin{figure}[ht!]
\gridline{\fig{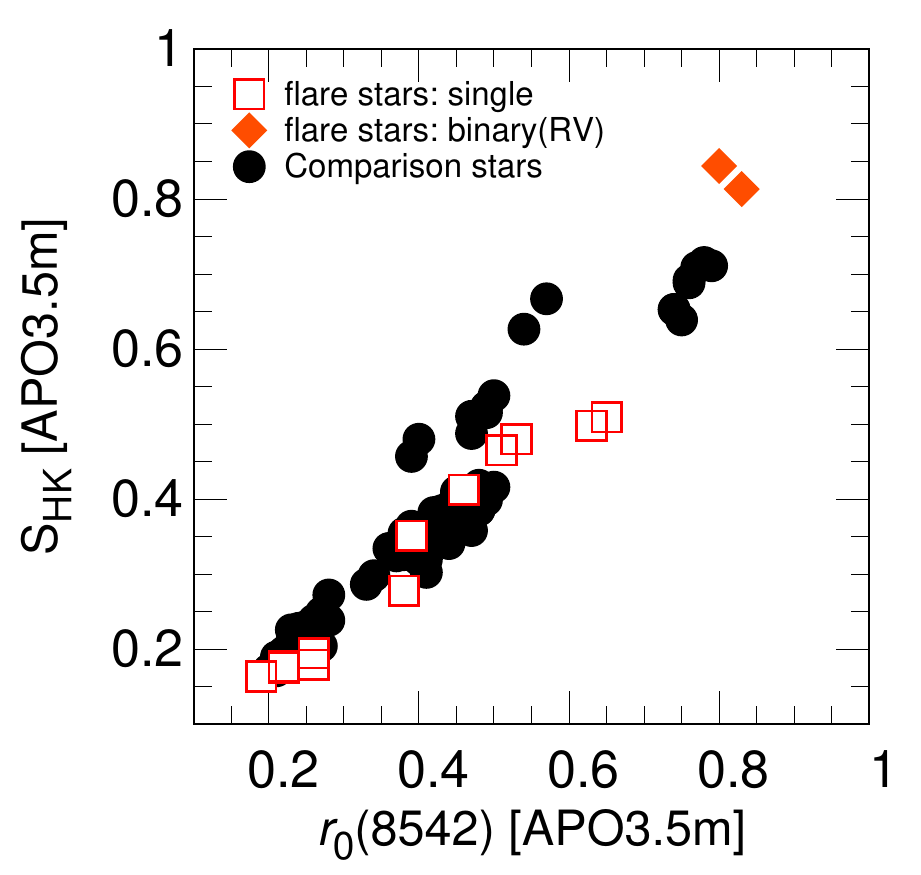}{0.333\textwidth}{(a)}
       \fig{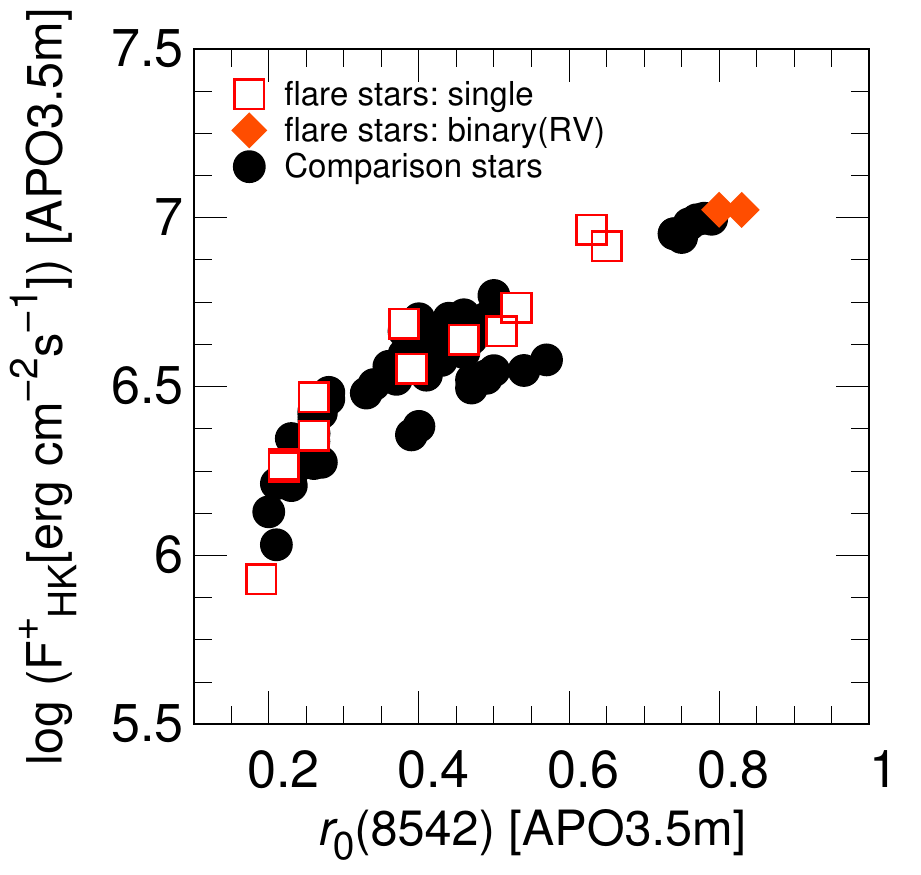}{0.333\textwidth}{(b)}
\fig{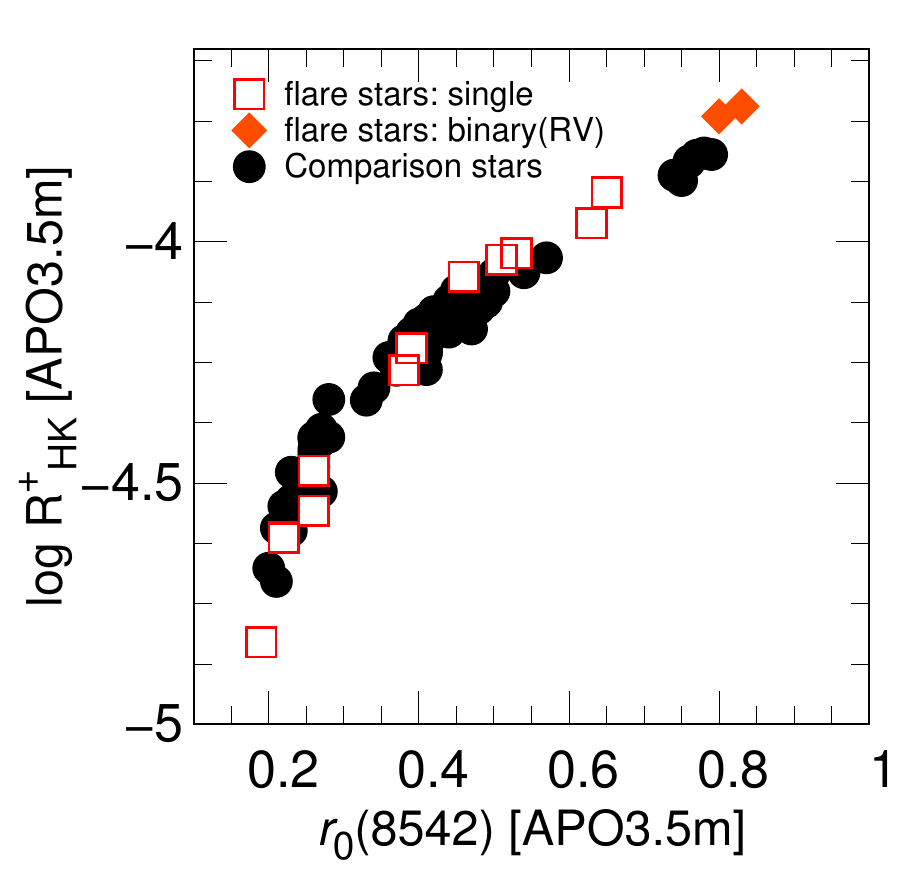}{0.333\textwidth}{(c)}}
\caption{
\bf (a) \rm $r_{0}(8542)$ vs. $S_{\mathrm{HK}}$
\bf (b) \rm $r_{0}(8542)$ vs. $\log \mathcal{F}^{+}_{\mathrm{HK}}$ 
\bf (c) \rm $r_{0}(8542)$ vs. $\log \mathcal{R}^{+}_{\mathrm{HK}}$
\\
The white square points are the target superflare stars classified as single stars in Appendix \ref{subsec:ana-binarity}, 
and the orange diamonds correspond to the spectra of binary superflare stars
that do not show any double-lined profiles (KIC11551430A and KIC4543412). 
Black circles show the values of comparison stars.
\label{fig:8542oaoHKapo}}
\end{figure}

In Figures \ref{fig:8542oaoHKapo} (a)--(c), 
we compared the resultant values of the Ca II H\&K indexes 
($S_{\mathrm{HK}}$ index, $\log\mathcal{F}^{+}_{\mathrm{HK}}$ index, and  $\log R^{+}_{\mathrm{HK}}$ index)
with those of $r_{0}$(8542) index (normalized intensity at the center of Ca II 8542\AA~line).
As seen in these figures, Ca II 8542 index and Ca II H\&K indexes have good correlations also for {\it Kepler} solar-type superflare stars, 
as expected from the previous studies (e.g.,\citealt{Takeda2012}; \citealt{Karoff2016}).
In particular, $\log R^{+}_{\mathrm{HK}}$ index (Figure \ref{fig:8542oaoHKapo} (c)) 
has the best correlation with Ca II 8542 index among these three Ca II H\&K indexes. 
As for this $\log R^{+}_{\mathrm{HK}}$ index, stellar temperature (color) dependence and the contribution from 
photospheric/basal fluxes are removed (Equation (\ref{eq:RpHK})).
In addition, Figure \ref{fig:8542oaoHKapo} (c) might suggest that $\log R^{+}_{\mathrm{HK}}$ index
is much more sensitive to the difference in the lower activity level region ($r_{0}$(8542)$\lesssim$0.3), compared with $r_{0}$(8542) index, 
as also suggested in \citet{Takeda2012}.
Then we mainly use this $\log R^{+}_{\mathrm{HK}}$ index when we discuss the measurement results of Ca II H\&K lines in in Section \ref{sec:Ca-amp}.

\subsection{Li abudances}\label{subsec:ana-Li}

\begin{figure}[ht!]
\gridline{\fig{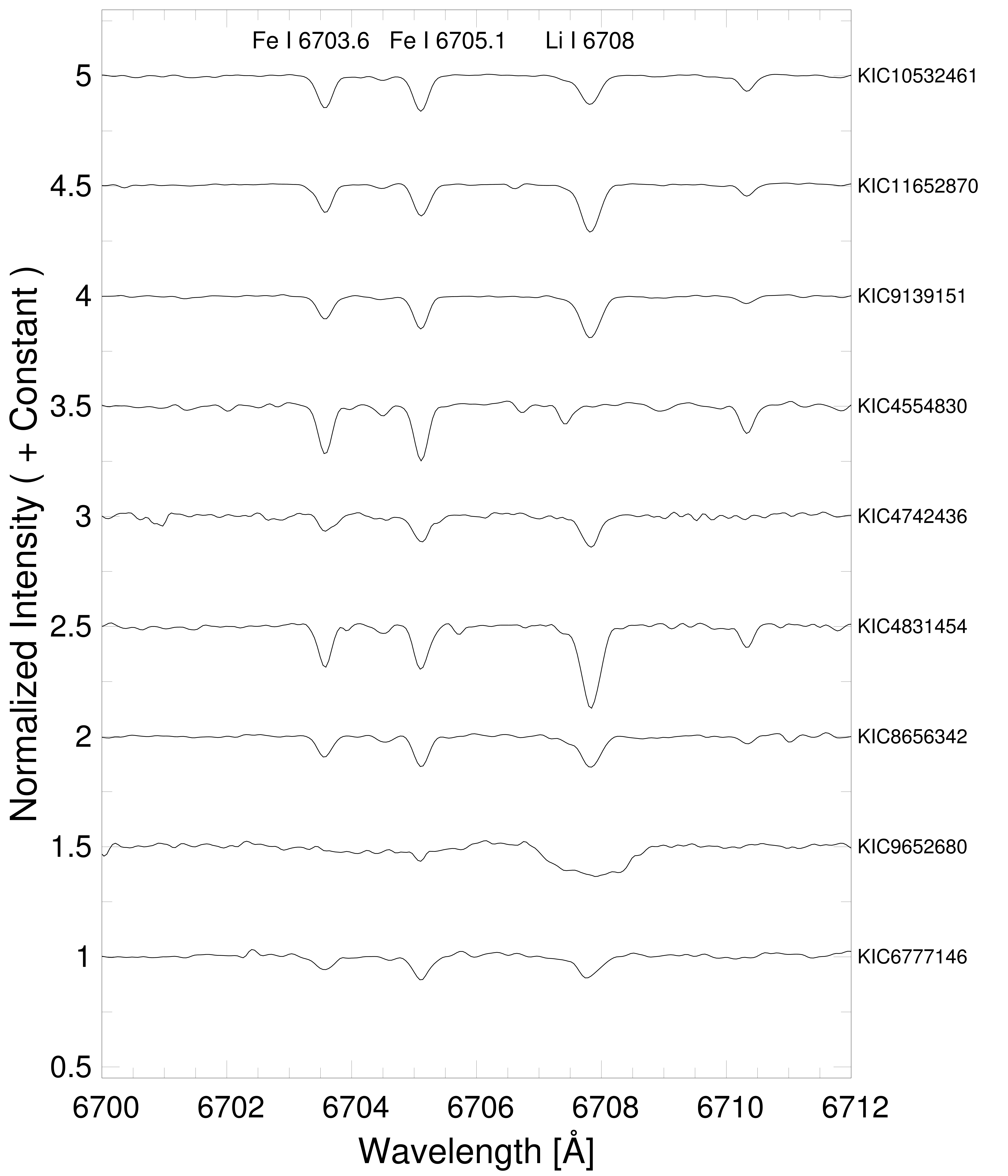}{0.5\textwidth}{(a)}
         \fig{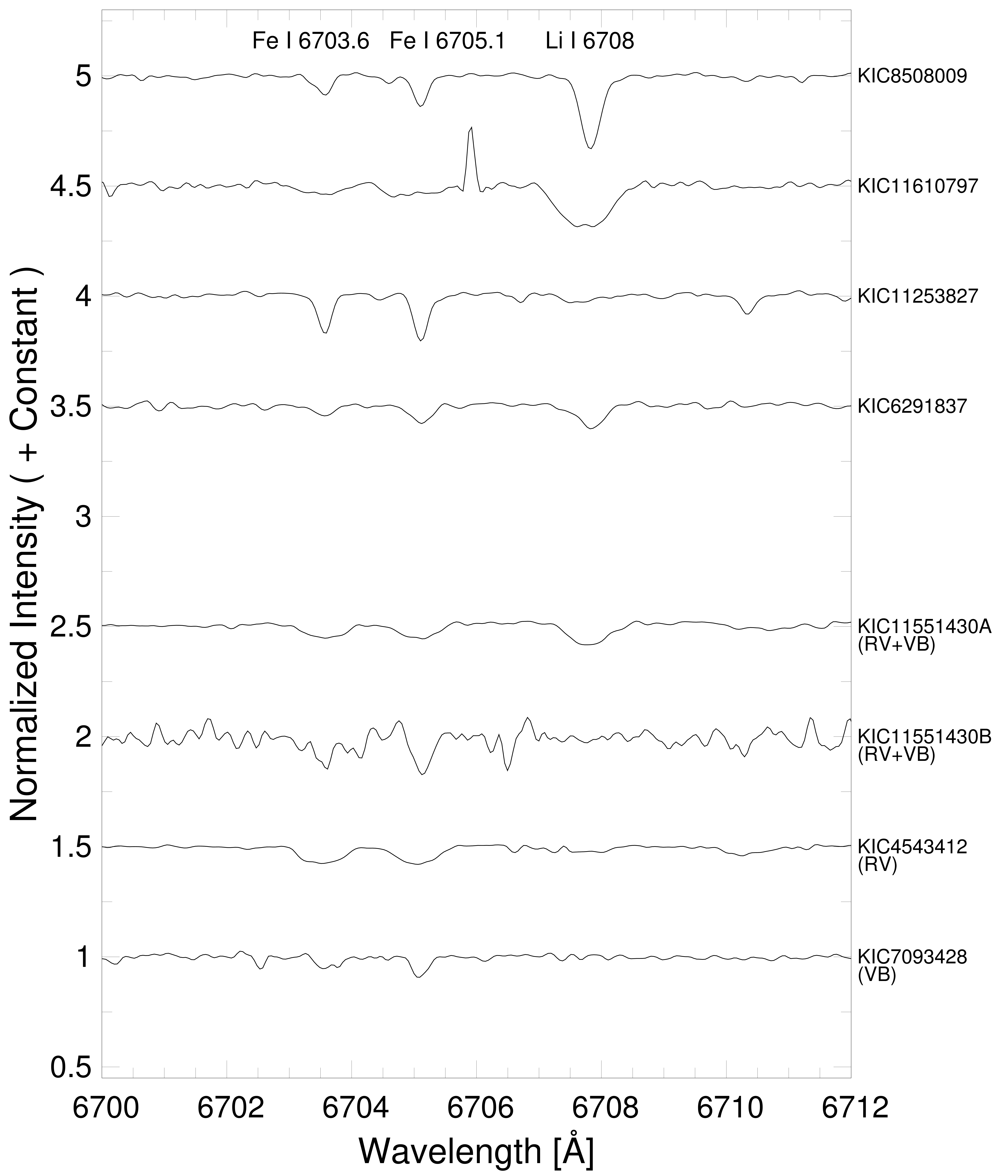}{0.5\textwidth}{(b)}}
\caption{Spectra around Li I 6708\AA~line.
The wavelength scale of each spectrum is adjusted to the laboratory frame. 
In (a) and the upper part of (b), 13 ``single" superflare stars that show no evidence of binarity are plotted.
Co-added spectra are used here in case that the star was observed multiple times (See Table \ref{tab:obs-RV}).
In the bottom of (b), the spectra of binary stars that do not show any double-lined profiles 
are plotted for reference.
As for KIC11551430A and KIC4543412 among them, which only show radial velocity shifts, the co-added spectra are used.
\label{fig:specLi}}
\end{figure}

The observed spectra of the target superflare stars around Li I 6708\AA~are shown in Figure \ref{fig:specLi}. 
We measured the Li abundances [$A$(Li)] of the stars 
using these spectra and the atmospheric parameters estimated in Appendix \ref{subsec:ana-atmos}. 
We used the automatic profile-fitting method that is basically the same as in our previous studies (\citealt{Honda2015}; \citealt{Notsu2017}). 
This is originally based on the method described in \citet{Takeda2005a}. 
We summarize the method in the following.

In the process of calculating $A$(Li), we used the MPFIT program contained in the SPTOOL software package also used above. 
We assumed local thermodynamic equilibrium (LTE) and 
derived Li abundances using the synthesis spectrum with interpolated model atmospheres 
taken from \citet{Kurucz1993}. We also assumed $^{6}$Li/$^{7}$Li=0 throughout this study. 
The line data around the Li I 6708\AA~region adapted here are the same as those used in Section 2.2 of \citet{Takeda2005a}. 
The estimated $A$(Li) values of the target stars are listed in Table \ref{tab:spec-para}.
For the stars where Li features are absent (e.g., below the detectable limit), 
we estimated the upper limit values of $A$(Li) by applying the method that we described in \citet{Honda2015} and \citet{Notsu2017}.

Our previous paper, \citet{Honda2015} and \citet{Notsu2017} discussed the typical errors of Li abundances 
by considering errors arising from multiple causes (errors linked to atmospheric parameters, 
uncertainties arising from profile fitting errors, and non-LTE effects). 
Since we conduct basically the same analyses as the above two papers, we here roughly assume 
that errors of $A$(Li) is $\sim$0.15 dex, as in \citet{Notsu2017} .

\section{Potential differences between the results form the {\it Kepler} stars 
and those from the ``real" sample of field stars}
\label{sec:Kepler-Prot-Nstar}

  \begin{table}[ht!]
\begin{center} 
  \caption{Number of stars that have $P_{\mathrm{rot}}$ and $\Delta F/F$ values ($N_{P}$) reported in \citet{McQuillan2014}.}\label{tab:Nstar-amp-detected}
    \begin{tabular}{lcccccc}
      \hline
$T_{\mathrm{eff, DR25}}$ & $N_{\mathrm{data}}$ \tablenotemark{a} & $N_{P}$(all) \tablenotemark{b} & 
 $N_{P}$($P_{\mathrm{rot}}<$5d) \tablenotemark{c} &  $N_{P}$($P_{\mathrm{rot}}=$5--10d) \tablenotemark{c} & 
 $N_{P}$($P_{\mathrm{rot}}=$10--20d) \tablenotemark{c} & 
 $N_{P}$($P_{\mathrm{rot}}\geq$20d) \tablenotemark{c} \\
      \hline 
5600--6000K & 28329 & 5065 & 314 (6.2\%) & 786 (15.5\%) & 2325 (45.9\%) & 1640 (32.4\%) \\
5100--5600K & 20883 & 6529 & 175 (2.7\%) & 745 (11.4\%) & 2234 (34.2\%) & 3375 (51.7\%) \\
5100--6000K & 49212 & 11594 & 489 (4.2\%) & 1531 (13.2\%) & 4559 (39.3\%) & 5015 (43.3\%) \\
    \hline     
    \end{tabular}
      \end{center}
\tablenotetext{a}{
Number of all newly classified solar-type stars among the stars in \citet{McQuillan2014},
by using $T_{\mathrm{eff, DR25}}$ and $R_{\mathrm{Gaia}}$ values      
}
\tablenotetext{b}{
Number of stars that have $P_{\mathrm{rot}}$ and $\Delta F/F$ values reported in \citet{McQuillan2014}.
}
\tablenotetext{c}{
Subgroups of $N_{P}$ categorized with $P_{\mathrm{rot}}$ values. 
The numbers in the parentheses are the fractions of $N_{P}$($P_{\mathrm{rot}}<$5d), $N_{P}$($P_{\mathrm{rot}}=$5--10d),
$N_{P}$($P_{\mathrm{rot}}=$10--20d), and $N_{P}$($P_{\mathrm{rot}}>$20d) to $N_{P}$(all), respectively. 
}
\end{table} 

  \begin{table}[ht!]
\begin{center} 
  \caption{$P_{\mathrm{rot}}$ distribution of field ordinary solar-type stars estimated from the gyrochronological relation 
  (cf. Equation (\ref{eq:Np-tgyro})).}\label{tab:Nstar-Age-Prot}
    \begin{tabular}{lcccc}
      \hline
$T_{\mathrm{eff, DR25}}$ & 
 $N_{\mathrm{star}}$($P_{\mathrm{rot}}<$5d)$/N_{\mathrm{all}}$ &
 $N_{\mathrm{star}}$($P_{\mathrm{rot}}=$5--10d)$/N_{\mathrm{all}}$ & 
 $N_{\mathrm{star}}$($P_{\mathrm{rot}}=$10--20d)$/N_{\mathrm{all}}$ & 
 $N_{\mathrm{star}}$($P_{\mathrm{rot}}\geq$20d)$/N_{\mathrm{all}}$  \\
      \hline 
5800K & 3\% ($<$0.26 Gyr) & 6\% (0.26--0.89 Gyr) 
& 21\% (0.89--3.0 Gyr) & 70\% ($\geq$3.0 Gyr) \\
5350K & 2\% ($<$0.17 Gyr) & 4\% (0.17--0.59 Gyr) 
& 14\% (0.59--2.0 Gyr) & 80\% ($\geq$2.0 Gyr) \\
    \hline     
    \end{tabular}
      \end{center}
\end{table} 

As for Figure \ref{fig:aspot-prot} in Section \ref{sec:spot-Prot}, 
we should note that there are many ``inactive" stars with $A_{\mathrm{spot}}<10^{-3}A_{1/2\sun}$ ($A_{1/2\sun}\sim 3\times
10^{22}$ cm$^{2}$: solar hemisphere). 
As shown in Table \ref{tab:Nstar-amp-detected},
approximately 76\%~((49612-11594)/49612 = 37618/49212) of the solar-type stars 
are not plotted in Figure \ref{fig:aspot-prot} since their brightness variation amplitude value is smaller than the detection limit.
This also means that these $\sim$76\% ``inactive" solar-type stars have no $P_{\mathrm{rot}}$ values in our sample, 
and this can cause biases when we discuss the relations of the superflare properties with the rotation period.
For example, in Figure \ref{fig:Prot-freq-evo0} (in Section \ref{sec:Prot-fene}), 
flare frequency distribution in each $P_{\mathrm{rot}}$ bin is calculated by using 
the number of the solar-type stars in each $P_{\mathrm{rot}}$ bin detected in \citet{McQuillan2014}
(e.g., $N_{P}$($P_{\mathrm{rot}}<$5d), $N_{P}$($P_{\mathrm{rot}}=$5--10d), $N_{P}$($P_{\mathrm{rot}}=$10--20d),
and $N_{P}$($P_{\mathrm{rot}}\geq$20d) in Table \ref{tab:Nstar-amp-detected}).
However, it is possible that these $N_{P}$ values does not show the actual $P_{\mathrm{rot}}$ distribution
of field ordinary solar-type stars because of the following two reasons.
(1) The above ``inactive" stars with no $P_{\mathrm{rot}}$ values 
are expected to be dominated by old slowly-rotating stars,
and the real number fraction of slowly-rotating stars to rapidly-rotating stars 
(e.g.,  the fraction of $N_{P}$($P_{\mathrm{rot}}\geq$20d) to the other 
$N_{P}$($P_{\mathrm{rot}}<$5d), $N_{P}$($P_{\mathrm{rot}}=$5--10d), and $N_{P}$($P_{\mathrm{rot}}=$10--20d) values) 
can become larger than the values used for the calculations in this study (e.g., Figure \ref{fig:Prot-freq-evo0}).
(2) In a strict sense, it might be possible that 
$P_{\mathrm{rot}}$ and activity level distribution of {\it Kepler} stars can be (slightly) different from
those of the field ordinary stars, because of the effects from the target selection 
(e.g., some active variables could be preferred in the Guest Observer programs).

In order to roughly evaluate the potential differences caused by the above points, 
we then roughly estimate the number fraction of solar-type stars in specific $P_{\mathrm{rot}}$ bins,
by using the empirical gyrochronology relation (e.g., \citealt{Mamajek2008}), 
which we have used in \citet{Maehara2017}.
The number of stars with $P_{\mathrm{rot}}\geq P_{0}$ [$N_{\mathrm{star}}(P_{\mathrm{rot}}\geq P_{0})$]
can be estimated from the duration of the main-sequence phase ($\tau_{\mathrm{MS}}$), 
the gyrochronological age of the star [$t_{\mathrm{gyro}}(P_{0})$], 
and the total number of samples ($N_{\mathrm{all}}$):
\begin{eqnarray}\label{eq:Np-tgyro}
N_{\mathrm{star}}(P_{\mathrm{rot}}\geq P_{0})=\left[1-\frac{t_{\mathrm{gyro}}(P_{0})}{\tau_{\mathrm{MS}}}\right]N_{\mathrm{all}} \ ,
\end{eqnarray}
assuming that the star formation rate around the {\it Kepler} field has been roughly constant over $\tau_{\mathrm{MS}}$.
Using Equations (12)--(14) of \citet{Mamajek2008}, 
we roughly estimated the age values ($t_{\mathrm{gyro}}$) of solar-type stars 
with $T_{\mathrm{eff}}\sim$ 5800 \& 5350 K ($B-V\sim$ 0.63 \& 0.78  from Equation (2) of \citealt{Valenti2005}) 
and $P_{\mathrm{rot}}\sim$5, 10, \& 20 days,  
as listed in the parentheses in Table \ref{tab:Nstar-Age-Prot}. 
Using these $t_{\mathrm{gyro}}$ values and Equation (\ref{eq:Np-tgyro}),
we estimated the number fraction of stars with $P_{\mathrm{rot}}$ 
 $<$5d, 5--10d, 10--20d, and $\geq$20d as listed in Table \ref{tab:Nstar-Age-Prot}.
As a result, there are differences between the number fractions of 
slowly/rapidly-rotating stars in the {\it Kepler} sample from \citet{McQuillan2014} (Table \ref{tab:Nstar-amp-detected})
and those estimated from the gyrochronology relation (Table \ref{tab:Nstar-Age-Prot}).
For example, the number fraction of $P_{\mathrm{rot}}\geq$20d 
among all the sample stars with $T_{\mathrm{eff}}=$5600--6000 K
has a roughly factor two difference:  
$N_{P}$($P_{\mathrm{rot}}\geq$20d)$/N_{P}$(all)$\sim$32\% (Table \ref{tab:Nstar-amp-detected}) and 
$N_{\mathrm{star}}$($P_{\mathrm{rot}}\geq$20d)$/N_{\mathrm{all}}\sim$70\% (Table \ref{tab:Nstar-Age-Prot}).
This means that the flare frequency value of $T_{\mathrm{eff}}=$5600--6000 K and $P_{\mathrm{rot}}\geq$20 days
in Figure \ref{fig:Prot-freq-evo0} can become factor two smaller than the real value.
We must note these potential errors of our study, 
and the exact values of the flare frequency estimated from our analyses should be treated carefully with cautions.
However, the overall dependences (e.g., the results that flare frequency clearly depends on the rotation period)
does not change and the differences are much smaller than one order of magnitude.
Moreover, the estimates from the gyrochronology can also include several errors. 
One example is that 
the age--rotation relation (gyrochronology relation) 
of young (e.g., $t\lesssim$ 0.5 -- 0.6 Gyr) and old (e.g., $t>$5.0 Gyr) solar-type stars 
can have a large scatter (e.g., \citealt{Soderblom1993}; \citealt{Ayres1997}; \citealt{Tu2015}) 
and breakdown (\citealt{vanSaders2016}; \citealt{Metcalfe2018}), respectively, 
as we have also mentioned in Section \ref{sec:Prot-fene}.
We also assumed that the star formation rate around the {\it Kepler} field has been roughly constant over $\tau_{\mathrm{MS}}$
in the above estimation, but this assumption is not necessarily correct.
Because of these points, 
more investigations on the above potential differences (as large as factor two or three) between the results from the {\it Kepler} stars 
and those from the ``real" sample of field stars are beyond the scope of this paper,
and we expect future studies using new data (e.g., {\it TESS} data mentioned in Section \ref{subsec:future}).   
We note again here that we should keep in mind these potential errors when discussing 
the relation of superflare frequency with the rotation period (e.g., Figure \ref{fig:Prot-freq-evo0} in Section \ref{sec:Prot-fene} 
and Figures \ref{fig:flare-freq-prot}\&\ref{fig:flare-freq-sunstar}  in Section \ref{fig:freq-sunlike}).

\end{document}